\documentclass[12pt,reqno]{article}
\usepackage{amsmath,graphicx}

\allowdisplaybreaks
\numberwithin{equation}{section}

\newcommand{\eqn}[1]{(\ref{#1})}
\newcommand{\ft}[2]{{\textstyle\frac{#1}{#2}}}
\newcommand{\tr}{\mbox{tr}}

\newcommand{\ggg}{{\sl g}}

\newcommand{\cK}{{\cal K}}

\newcommand{\cO}{{\cal O}}

\newcommand{\cN}{{\cal N}}

\newcommand{\cL}{{\cal L}}

\newcommand{\mG}{{\mit\Gamma}}
\newcommand{\mP}{{\mit\Psi}}
\newcommand{\mS}{{\mit\Sigma}}
\newcommand{\phan}[1]{{\textstyle\phantom{#1}}}
\font\cmss=cmss12 
\def\1{\hbox{{1}\kern-.25em\hbox{l}}}
\def\bfZ{\relax{\hbox{\cmss Z\kern-.4em Z}}}

\def\a{\alpha}
\def\b{\beta}
\def\d{\delta}
\def\e{\epsilon}           
\def\g{\gamma}

\def\l{\lambda}

\def\o{\omega}
\def\s{\sigma}                    

\def\G{\Gamma}

\def\O{\Omega}

\def\del{\partial}

\def\ch{{\cal H}}

\def\cl{{\cal L}}

\def\cn{{\cal N}}
\def\co{{\cal O}}

\def\eq{\begin{equation}}
\def\be{\begin{equation}}
\def\beq{\begin{equation}}
\def\eqe{\end{equation}}
\def\ee{\end{equation}}
\def\eeq{\end{equation}}
\def\eqa{\begin{eqnarray}}
\def\bea{\begin{eqnarray}}
\def\eqae{\end{eqnarray}}
\def\ena{\end{eqnarray}}
\def\eea{\end{eqnarray}}

\def\nn{\nonumber\\}

\def\under#1#2{\mathop{\null#2}\limits_{#1}}

\begin{document}

\begin{flushright} \small
hep-th/yymmnnn \\ ITP--UU--08/05 \\ SPIN--08/05
\end{flushright}
\bigskip

\begin{center}
 {\large\bfseries Lectures on instantons}
\\[5mm]
Stefan Vandoren$^1$ and Peter van Nieuwenhuizen$^2$ \\[3mm]
  {\small\slshape
$^1$ Institute for Theoretical Physics and Spinoza Institute \\
 Utrecht University, 3508 TD Utrecht, The Netherlands \\
 {\upshape\ttfamily s.vandoren@phys.uu.nl}\\[3mm]
$^2$ C.N. Yang Institute for Theoretical Physics \\
State University of New York at Stony Brook,
NY 11790, USA \\
 {\upshape\ttfamily vannieu@insti.physics.sunysb.edu}}

\end{center}
\vspace{5mm}

\hrule\bigskip

\centerline{\bfseries Abstract} \medskip

This is a selfcontained set of lecture notes on instantons in (super)
Yang-Mills theory in four dimensions and in quantum mechanics. First the
basics are derived from scratch: the regular and singular one-instanton
solutions for Yang-Mills theories with gauge groups $SU(2)$ and $SU(N)$,
their bosonic and fermionic zero modes, the path integral instanton
measure, and supersymmetric Yang-Mills theories in Euclidean space.
Then we discuss applications: the $\theta$-angle of QCD, the solution of
the $U(1)$ problem, the way Higgs fields solve the large-instanton
problem, and tunneling and phase transitions in quantum mechanics and
in nonabelian gauge theories. These lecture notes are an extension of a
review on Yang-Mills and D-instantons written in 2000 by both authors and
A.Belitsky \cite{BelVanNie00a}.

\bigskip

\hrule\bigskip

\tableofcontents

\section{Introduction}

In the last decades enormous progress has
been made in understanding nonperturbative effects, both in
supersymmetric field theories and in superstring theories.  By
non-perturbative effects we mean effects due to solitons and
instantons, whose masses and actions, respectively, are inversely
proportional to the square of the coupling constant \cite{Rev}.
Typical examples of solitons are the kink, the vortex, and the
magnetic monopole in field theory, and some D-branes in
supergravity or superstring theories.  In supersymmetric field
theories these solutions preserve half of the supersymmetry
and saturate BPS bounds.  As for instantons, we have the
Yang-Mills (YM) instantons in four dimensions \cite{Belavin,Hooft1,Hooft2}, or
tunnelling phenomena in quantum mechanics with a double-well
potential as described by the kink, see e.g. \cite{EGildener},
and there are various kinds of instantons in string theory, for
example the D-instantons \cite{Dinst}. Also instantons
preserve half the number of supersymmetries in supersymmetric
field theories. Instantons can also be defined in field theories
in dimensions higher than four \cite{IH4}, but we discuss in this
chapter mainly the case of four dimensions.

Instantons in ordinary (i.e., nongravitational) quantum field
theories are by definition solutions of the classical field
equations in Euclidean space with finite action.\footnote{In
gravity there are various definitions of instantons:  Einstein
spaces with selfdual Weyl tensors, selfdual Riemann tensors,
solutions of the Einstein equations with/without finite action
etc.  Since in gravity spacetime is part of the solution, one
usually considers spacetime topologies which are different from
that of $\bf{R}^4$.
A selfdual Riemann tensor leads to an Einstein space $(R_{\mu\nu}
= \Lambda g_{\mu\nu})$ whose Einstein-Hilbert action is either
infinite (if the cosmological constant $\Lambda$ is nonvanishing),
or it only gets contributions from the Gibbons-Hawking boundary
term \cite{GHawkingGW}.   In general, the semiclassical
approximation of the Einstein-Hilbert action is not well defined
due to the unboundedness of the action inside the path integral.
To cure this, one probably has to discuss gravitational instantons
inside a full theory for quantum gravity.  For instanton solutions in flat space but
using curvilinear coordinates (for example $S^4$, or cylindrical
coordinates) see \cite{AAAbrikosov}.} Only for a finite classical
action $S_{\rm cl}$ is the factor $\exp[-\frac{1}{\hbar}S_{{\rm cl}}]$ in the path integral
nonvanishing. We shall consider
instantons in nonabelian gauge theories in flat spacetime (there
are no instantons in abelian gauge theories in flat space), both regular
instantons (which actually have a singularity at $|x|^2 =
\infty$) and singular instantons (which have a singularity at a
point $x = x_0$ but not at $|x|^2 = \infty$).  A singular gauge
transformation maps the first into the second, and
vice-versa\footnote{For the "regular solution", $A^a_\mu$ is
finite on $\bf{R}^4 \cup \infty = S^4$ everywhere, but this does
not mean that it is regular. It is finite only because one can use two
different patches to cover $S^4$, and $A^a_\mu$ is regular in each
patch. If one maps infinity to the origin by a space-inversion
transformation $(x^\mu = y^\mu /y^2)$, then one finds a
singularity at the origin.  In this sense the "regular solution"
is singular. We further clarify this issue in the next section.}.
Around a given instanton solution, there are the quantum
fluctuations. The action contains terms with 2, 3, 4 \dots
quantum fields, and one can perform perturbation theory around the
instanton.  The terms quadratic in quantum fields yield the
propagators, which are complicated background-dependent
expressions, and the terms cubic and higher in quantum fields
yield the vertices.  However, there is a subtlety with an
instanton background:  there are zero modes. A zero mode is by
definition a solution of the linearized field equations for the
fluctuations which is {\bf normalizable}.  (It is an eigenfunction
of the quantum field operator with eigenvalue zero).  In a trivial
vacuum there are no zero modes:  there are, of course, solutions
of the linearized field equations, but they are not normalizable.
We must treat the zero modes in instanton physics separately from
the nonzero modes; for example, they have their own measure in the
path integral.  The nonzero modes live in the space orthogonal to
the zero modes and in this space one can invert the linearized
field equations for the fluctuations and construct propagators,
and do perturbation theory.

Instantons describe tunnelling processes in Minkowski space-time
from one vacuum at time $t_1$ to another vacuum at time $t_2$. The
simplest model which exhibits this phenomenon
is a quantum mechanical point particle
with a double-well potential having two vacua, or a periodic
potential with infinitely many vacua.  Classically there is no trajectory for a particle to
travel from one vacuum to the other, but quantum mechanically
tunnelling occurs. The tunnelling amplitude can be computed in the
WKB approximation, and is typically exponentially suppressed.  In
the Euclidean picture, after performing a Wick rotation, the
potential is turned upside down, and it is possible for a particle
to propagate between the two vacua, as described by the classical
solution to the Euclidean equations of motion.  The claim is then
that the contributions from instantons in Euclidean space yield a
good approximation of the path integral in Minkowski space.  We shall prove this for the case of quantum mechanics.

Also in YM theories, instantons are known to describe tunnelling
processes between different vacua, labeled by an integer winding
number, and lead to the introduction of the CP-violating
$\theta$-term in the action \cite{CDG,JR}. It was hoped that instantons could
shed some light on the mechanism of quark confinement.  Although
this was successfully shown in three-dimensional gauge theories
(based on the Georgi-Glashow model) \cite{Pol}, the role of
instantons in relation to confinement in four dimensions is less
clear. Together with the non-perturbative chiral $U(1)$ anomaly in
an instanton background, which leads to baryon number violation
and a solution of the $U(1)$ problem \cite{Hooft1,Hooft2},
instantons are used in phenomenological applications to QCD and the
Standard Model.  To avoid confusion, note that the triangle chiral
anomalies in perturbative field theories in Minkowski space-time
are canceled by choosing suitable multiplets of fermions. There
remain, however, chiral anomalies at the non-perturbative level.
It is hard to compute
the non-perturbative terms in the effective action which lead to a
breakdown of the chiral symmetry by using methods in Minkowski
space-time. However, by using instantons in Euclidean space, one
can relatively easily determine these terms.  The nonperturbative
chiral anomalies are due to fermionic zero modes which appear in
the path integral measure (in addition to bosonic zero modes). One
must saturate the Grassmann integrals over these zero modes, and
this leads to correlation functions of composite operators with
fermion fields which do violate the chiral $U (1)$ symmetry. The
new non-perturbative terms are first computed in Euclidean space,
but then continued to Minkowski space where they give rise to new
physical effects \cite{Hooft2}. They have the following generic
form in the effective action (we suppress here possible flavor or
adjoint indices that the fermions can carry)
\begin{eqnarray}
\label{EffAct} S_{\rm eff} \propto {\rm e}^{\left\{- \frac{8
\pi^2}{\ggg^2} \left( 1 + \cO (\ggg^2) \right) + i \theta
\right\}}
 ( \bar\lambda \bar\lambda )^n  \ ,
\end{eqnarray}
where $2n$ is the number of fermionic zero modes ($n$ depends on the
representation of the fermions and the gauge group). The prefactor
is due to the classical instanton action and is clearly
non-perturbative. The terms indicated by $\cO (\ggg^2)$ are due to
standard radiative corrections computed by using Feynman graphs in
an instanton background. The term $\left( \bar\lambda \bar\lambda
\right)^n$ involving antichiral spinors $\bar\lambda$ is produced if
one saturates the integration in the path integral over the
fermionic collective coordinates and violates in general the chiral
symmetry. On top of \eqn{EffAct} we have to add the contributions
from anti-instantons, generating $\left( \lambda \lambda \right)^n$
terms in the effective action, where $\lambda$ denotes chiral
spinors. As we shall discuss, for Majorana spinors in Euclidean
space the chiral and anti-chiral spinors are independent, but in
Minkowski space-time they are related by complex conjugation, and
one needs the sum of instanton and anti-instanton contributions to
obtain a hermitean effective action.

We shall also apply the results of the general formalism to
supersymmetric gauge theories, especially to the $\cN = 4$ $SU(N)$
supersymmetric Yang-Mills (SYM) theory.  Here $\cN$ stands for the number of
supersymmetries. Instantons in $\cN = 1,2$ models have been
extensively studied in the past, see e.g.\cite{Dadda-DV} for an early
reference, and still are a topic of current
research. For the $\cN = 1$ models, one is mainly interested in
the calculation of the superpotential and the gluino condensate
\cite{vnovikov,n=1}. In some specific models, instantons also
provide a mechanism for supersymmetry breaking \cite{n=1}, see
\cite{amatikonishi} for a review on these issues. In the case of
$\cN = 2$, there are exact results for the prepotential \cite{SW}
based only on general symmetry principles and electric-magnetic
duality; the prepotential acquires contributions from all
multi-instanton sectors. These predictions were successfully
tested against direct field theoretical calculations in the
one-instanton sector in \cite{test-SW}, and for a two-instanton
background in \cite{2inst-SW}.  More recently, new techniques were
developed to perform multi-instantons calculations in
\cite{NNekrasov}. Finally, the nonperturbative structure of
$\cN =4$ SYM has been studied thoroughly in the context of the AdS/CFT
correspondence \cite{ads/cft}. SYM instantons in the limit of large number of
colors were succesfully shown to reproduce the D-instanton contributions to
certain correlation functions, both for single instantons
\cite{BGKR,DorKhoMatVan98} and for multi-instantons \cite{DorHolKhoMatVan99}.
Other correlation functions were studied in \cite{GK,K}. For a recent 
review of instantons in supersymmetric gauge theories, see 
\cite{Bianchi:2007ft}.

The material is organized as follows.  In section 2, we discuss the
winding number of gauge fields, and we present the standard
one-instanton solution in $SU(2)$ and in $SU(N)$.  This already
raises the question how to embed $SU(2)$ into $SU(N)$, and we
discuss the various embeddings.  In section 3 we discuss instanton
solutions in general: we solve the duality condition and find
multi-instanton solutions which depend on their position and their
scale.  We concentrate on the one-instanton solutions, and first
determine the singular solutions, but then we make a (singular)
gauge transformation and obtain the regular solutions. In section 4
we start the study of  ``collective coordinates", the parameters on
which the most general instanton solutions depend.  We show that the
number of collective coordinates is given by an index theorem for
the Dirac operator in an instanton background.  We then give a
derivation of this index theorem, and conclude that a $k$-instanton
solution in $SU(N)$ has $4Nk$ bosonic collective coordinates, $2Nk$
fermionic collective coordinates for fermions in the adjoint
representation, and $k$ fermionic collective coordinates for
fermions in the defining (fundamental, vector) representation. In
section 5 we explicitly construct the zero modes for gauge group
$SU(N)$ in a one-instanton background.  First we construct the
bosonic zero modes; these are associated to the collective
coordinates for translations, dilatations and gauge orientations.
Next we derive the explicit formula for the general solution of the
fermionic zero modes of the Dirac equation in a one-instanton
background, first for $SU(2)$ and then for $SU(N)$.

In section 6 we construct the one-instanton measure for the bosonic
and fermionic collective coordinates.  We explain in detail the
normalization of
 the zero modes since it is crucial for the construction of the measure.  We
convert the integration over the coefficients of the bosonic zero
modes to an integration over the corresponding bosonic collective
coordinates by the Faddeev-Popov trick, but for fermionic zero modes
we do not need this procedure because in this case the coefficients
of fermionic zero modes are already the fermionic collective
coordinates.  In section 7 we discuss the one-loop determinants in
the background of an instanton, arising from integrating out the
quantum fluctuations. We then apply this to supersymmetric theories,
and we use an index theorem to prove that the determinants for all
supersymmetric YM theories cancel each other. Furthermore, we
compute the complete nonperturbative $\beta$ function for
supersymmetric Yang-Mills theories by assuming that the measure for
the zero modes does not depend on the renormalization scale $\mu$.
However, since it is not known to which regularization scheme this
procedure corresponds, this result cannot be checked by standard
perturbative calculations. In section 8 we discuss the ${\cal N}=4$
supersymmetric Yang-Mills theory in Euclidean space and its
instantons.

The remaining sections contain applications. Section 9 discusses the
problem of large instantons and its solution in terms of Higgs
fields and spontaneous symmetry breaking.  Section 10 gives a
detailed discussion how instantons can describe tunnelling.  In
section 11 we use a quantum mechanical model with a double-well
potential to discuss the phase transition from a false vacuum to the
true vacuum by bubble formation.  Section 12 contains the strong CP
problem, the mystery that the $\theta$ angle is so small.  Section
13 discusses that instantons solve the $U(1)$ problem and in section
14 we finally discuss how instantons lead to baryon decay.

In a few appendices we set up our conventions and give a detailed
derivation of some technical results in order to make the material
self-contained.  In appendix A we provide details of the calculation
of the winding number.  In appendix B we discuss the 't Hooft
tensors and the spinor formalism in Euclidean space.  In appendix C
we calculate the volume of the moduli space of gauge orientations.
Finally, in appendix D we show that conformal boosts and Lorentz
rotations do not lead to additional zero modes.

\section{Winding number and embeddings} 

We start with some elementary facts about instantons in $SU (N)$
Yang-Mills theories. The action, continued to Euclidean space, is
\begin{equation} \label{YMaction}
S = - \frac{1}{2\ggg^2} \int\, {\rm d}^4
x\, \tr\, F_{\mu\nu} F_{\mu\nu}\ ;\qquad F_{\mu\nu} = F^a_{\mu\nu}
T_a\ .
\end{equation}
The generators $T_a$ are traceless anti-hermitean
$N$ by $N$ matrices satisfying $\left[ T_a, T_b \right] =
f_{ab}{}^c T_c$ with real structure constants and ${\rm tr}
( T_a T_b ) = - \ft12 \delta_{ab}$. For instance, for
$SU(2)$ one has $T_a = - \ft{i}2 \tau_a$, where $\tau_a$ are the
Pauli matrices and $f^a{}_{bc}=\epsilon^a{}_{bc}$. Notice that with these conventions the action
is positive. Further
conventions are $D_\mu Y = \partial_\mu Y + [ A_\mu, Y ]$ for any
Lie algebra valued field $Y$, and $F_{\mu\nu} = \partial _\mu
A_\nu - \partial_\nu A_\mu + [ A_\mu, A_\nu ]$, so that
$F_{\mu\nu} = [ D_\mu, D_\nu]$. The Euclidean metric is
$\delta_{\mu\nu} = {\rm diag} (+,+,+,+)$. In \eqn{YMaction}, the
only appearance of the coupling constant is in front of the
action. The group metric $g_{ab}=\delta_{ab}$ is an invariant tensor\footnote{\rightskip=-45pt From $tr [T_c , T_a T_b ] = tr ([T_c, T_a ] T_b + tr T_a [T_c, T_b])$ it follows that $g_{ab}$ is an invariant tensor: transforming its indices by an adjoint transformation with parameter $\l^c$ yields zero: $\d g_{ab} = \l^c f_{ca}{}^d g_{db} + \l^c f_{cb}{}^d g_{ac} =0$.}, so
we may raise and lower indices with $\delta^{ab}$ and $\delta_{ab}$.  Thus we
may also write $[T_a , T_b ] = f_{abc} T_c$, and from now on we
shall write group and Lorentz indices either as covariant indices
or as contravariant indices, depending on which way is most
convenient.

By definition, a Yang-Mills instanton is a solution of the classical Euclidean
equations of motion with finite action. The classical equations of motion read
\begin{equation} \label{eom}
D_\mu F_{\mu\nu} = 0 \ .
\end{equation}
To find solutions with finite action, we require that the field strength tends
to zero at infinity faster than $|x|^{-2}\equiv r^{-2}$, hence the gauge fields
asymptotically approach  a pure gauge\footnote{
Another way of satisfying the finite action requirement is to first formulate
the theory on a compactified ${\bf R}^4$, by adding and identifying points at
infinity. Then the topology is that of the four-sphere, since
${\bf R}^4 \cup \infty \simeq S^4$. The stereographic map from
${\bf R}^4 \cup \infty$ to $S^4$ preserves the angles, and is therefore
conformal. Also the YM action is conformally invariant, implying that the
action and the field equations on ${\bf R}^4 \cup \infty$ are the same as on
$S^4$ (the metric on the sphere is $g_{\mu\nu}=\delta_{\mu\nu}(1+x^2)^{-2}$).
The finiteness requirement is satisfied when the gauge potentials can be smoothly
extended from ${\bf R}^4$ to $S^4$. The action is then finite because $S^4$ is
compact and $A_\mu$ is well-defined on the whole of the four-sphere.  \label{Comp}}
\begin{equation} \label{puregauge}
A_\mu \stackrel{|x|^2\rightarrow \infty}{=} U^{-1} \partial_\mu U \ ,
\end{equation}
for some $U \in SU(N)$.  To prove that gauge fields are pure gauge
if the curvature $F_{\mu\nu}$ vanishes, is easy.  Using $U \del_\mu
U^{-1} = - \del_\mu UU^{-1}$ we must solve for $U$ from $\del_\mu U
=- A_\mu U$, whose solution is the path-ordered integral $U= \exp [-
\int^x A_\mu (y) {\rm d}y^\mu]$.  This expression does not depend on
the path chosen because $F_{\mu\nu} =0$.  (Note, however, that if
two gauge field configurations, say $A^I_\mu$ and $A^{II}_\mu$,
yield the same curvature, $F_{\mu\nu} (A^I) =F_{\mu\nu} (A^{II})$,
they need not be gauge equivalent.  A simple example proves this.
Consider \eqa A^I_\mu = \left\{ - \ft12 By T_3 , \ft12
BxT_3, 0,0 \right\} ;\qquad  A^{II}_\mu = \left\{ A^{II}_1 =
\sqrt{B} T_1, A^{II}_2 = \sqrt{B} T_2 ,0,0 \right\}\ , \eqae where
$B$ is a constant and $T_a$ are the generators of $SU(2)$ with
structure constants $f_{ab}{}^c = \e_{abc}$.  Clearly $F_{12} (A^I )
= BT_3$ and also $F_{12} (A^{II})= BT_3$ while all other components
of $F_{\mu\nu}$ vanish.  To prove that $A^I_\mu$ cannot be written
as $U^{-1} (\del_\mu + A^{II}_\mu)U$ we note that if there was such
a group element $U$, it should satisfy $U F_{\mu\nu} U^{-1} =
F_{\mu\nu}$, hence $U$ should commute with $T_3$.  This implies that
$U$ would be given by $\exp (f(x) T_3)$ for some real function
$f(x)$.  Then $- \frac12 By T_3 = \del_x f\, T_3 + {\rm e}^{-fT_3}
{\sqrt B}\, T_1 {\rm e}^{-fT_3}$ which has no solution.  One can
also calculate a Wilson loop $W = \tr P \exp \oint Adl$.  This
expression is gauge invariant, and if one chooses as loop a square
in the $x-y$ plane with sides $L_1$ and $L_2$, one finds \eqa W^I =
B L_1 L_2 T_3 ; \qquad W^{II} = 2 \sqrt{B} (L_1 T_1 + L_2 T_2) \eqae
If $A^I_\mu$ and $A^{II}_\mu$ were gauge equivalent, $W^I$ should
have been equal to $W^{II}$).

There is actually a way of classifying fields which satisfy the boundary condition in (\ref{puregauge}).
It is known from homotopy theory  that all gauge fields with
vanishing field strength at infinity can be classified into
sectors characterized by an integer number called the Pontryagin
class, or the winding number, or instanton number, or the
topological charge
\begin{equation} \label{winding} k = - \frac{1}{16\pi^2} \int\,
{\rm d}^4 x\, \tr\, F_{\mu\nu}{^*\!F_{\mu\nu}} \ ,
\end{equation}
where ${^*\!F_{\mu\nu}} = \ft12\epsilon_{\mu\nu\rho\sigma}
F_{\rho\sigma}$ is the dual field strength, and $\epsilon_{1234} =
1$. Note that it is not necessary that these gauge fields satisfy
the field equations, only that their field strength vanishes
sufficiently fast at $r = \infty$. The derivation of this result
can be found in Appendix \ref{Winding}.  As part of the proof, one
shows that the integrand in \eqn{winding} is the divergence of a
current
\begin{equation} K_\mu = - \frac{1}{8\pi^2} \epsilon_{\mu\nu\rho\sigma}
\tr\, A_\nu \left( \partial_\rho A_\sigma + \ft23 A_\rho A_\sigma
\right) \ .
\end{equation}
The four-dimensional integral in \eqn{winding} then reduces to an
integral over a three-sphere at spatial infinity, and one can use
\eqn{puregauge} to show that the integer $k$ counts how many times
this spatial three-sphere covers the gauge group three-sphere $S^3 \approx SU(2)
\subset SU(N)$. In more mathematical terms, the integer $k$
corresponds to the third homotopy group $\pi_3(SU(2)) = {\bf Z}$.
So $k$ as defined in \eqn{winding} does not depend on the values
of the fields in the interior, but only on the fields at large
$|x|^2$. This can also directly be seen:  under a small variation
$A_\mu \rightarrow A_\mu + \delta A_\mu$ one has $F_{\mu\nu}
\rightarrow F_{\mu\nu} + D_\mu \delta A_\nu - D_\nu \delta A_\mu$,
and partial integration (allowed when $\delta A_\mu$ is only
nonzero in a region in the interior) yields $\delta A_\nu D_\mu
{^*\!F_{\mu\nu}}$ which vanishes due to the Bianchi identity
$D_{[\mu}  F_{\nu\rho ]} = 0$.   (To prove this Bianchi identity
one may use  $F_{\nu\rho} = [D_\nu , D_\rho ]$.  In $[D_\mu ,
[D_\nu , D_\rho ]] + [D_\nu , [D_\rho , D_\mu ]] + [D_\rho ,
[D_\mu , D_\nu ]]$ there are then 12 terms which cancel pairwise.)

Since we require instantons to have finite action, they satisfy
the above boundary conditions at infinity, and hence they are
classified by  $k$, which we call the instanton number.  Gauge
potentials leading to field strengths with different instanton
number can not be related by continuous gauge transformations.
This follows from the fact that the instanton number is a gauge
invariant quantity. In a given topological sector, the field
configuration which minimizes the action is a solution of the
field equations.  (It is a priori not obvious that there exist
field configurations that minimize the action, but we shall
construct such solutions, thereby explicitly proving that they
exist). We now show that, in a given topological sector, the
solution to the field equations that minimizes the action has
either a selfdual or anti-selfdual field strength
\begin{equation}
F_{\mu\nu} = \pm {^*\!F_{\mu\nu}} = \pm \ft12
\epsilon_{\mu\nu\rho\sigma} F_{\rho\sigma} \ . \end{equation}
This equation is understood in Euclidean space, where $(^*)^2 = 1$. In
Minkowski space there are no real solutions to the selfduality
equations since $(^*)^2 = - 1$.   As seen from \eqn{winding},
\textbf{instantons (with selfdual field strength) have $k>0$ whereas
anti-instantons (with anti-selfdual field strength) have $k < 0$}.
 (Recall that $tr T_a T_b$ is negative).  To see that minimum action solutions are indeed selfdual or
anti-selfdual, we perform a trick similar to the one used for
deriving the BPS bound for solitons: we write the action as the square
of a sum plus a total derivative term
\begin{eqnarray}
S &=& - \frac{1}{2\ggg^2} \int\, {\rm d}^4x \, \tr\, F^2 = -
\frac{1}{4\ggg^2} \int\, {\rm d}^4 x \, \tr\, (F \mp {^*\!F})^2
\mp \frac{1}{2\ggg^2} \int\, {\rm d}^4x \, \tr\, F  \; {^*\!F} \nonumber\\
&\geq& \mp \frac{1}{2\ggg^2} \int\, {\rm d}^4x \, \tr\, F  \;
{^*\!F} = \frac{8\pi^2}{\ggg^2} ( \pm k )\ .
\end{eqnarray}
We used that $\tr {^*\!F} {^*\!F} = \tr FF$ and omitted Lorentz indices to
simplify the notation.  The equality is satisfied if and only if
the field strength is (anti-) selfdual. The value of the action is
then $S_{\rm cl} = (8\pi^2/\ggg^2)|k|$, and has the same value for
the instanton as for the anti-instanton. However, we can also add a theta-angle term to the action, which
reads
\begin{equation}
S_\theta =- i \frac{\theta}{16\pi^2} \int\, {\rm d}^4x \, \tr \,
F_{\mu\nu}{^*\!F}^{\mu\nu} = i\theta k = \pm i\theta |k| \ .
\end{equation}
The plus or minus sign corresponds to the instanton and
anti-instanton respectively, so the theta-angle distinguishes
between them. In Minkowski spacetime this term is the same
because both ${\rm d}^4x$ and $F_{\mu\nu}{^*\!F}^{\mu\nu}$ produce a factor $i$
under a Wick rotation. We give a more detailed
treatment of the theta-angle term and its applications in
Section 12.

It is interesting to note that the energy-momentum tensor for a
selfdual (or anti-selfdual) field strength always
vanishes\footnote{  Note that
 ${T}_{12}$ is proportional to $\tr (F_{13} F_{23} + F_{14}
F_{24})$, which is equal to minus itself due to the selfduality
relations $F_{12} = F_{34} , F_{13} =- F_{24}$ and $F_{14} =
F_{23}$.  Similarly ${T}_{11}$ vanishes because it is proportional
to the trace of $(F^2_{12} + F^2_{13} + F^2_{14}) - (F^2_{23} + F^2_{24} +
F^2_{34})$.}
\begin{equation} {T}_{\mu\nu} =-
\frac{2}{\ggg^2} \tr \left\{ F_{\mu\rho}F_{\nu\rho} - \ft14
\delta_{\mu\nu} F_{\rho\sigma} F_{\rho\sigma} \right\} = 0 \ .
\end{equation}
(Because in Euclidean space $T_{44} = -\frac{1}{\ggg^2} \tr\,  ( \vec{E}^2 -
\vec{B}^2)$, the Euclidean ``energy" $T_{44}$ need not be positive
definite).
This agrees with the observation that the instanton action $\int
{\rm d}^4 x \, \tr \, F^2 = \int {\rm d}^4 x \, \tr \,  {^*\!F} F$
is metric independent in curved space.
The vanishing of the energy-momentum tensor
is consistent with the fact that instantons are topological in
nature. It implies that instantons do not curve Euclidean space,
as follows from the Einstein equations.

An explicit construction of finite action solutions of the Euclidean
classical equations of motion was given by Belavin et al.
\cite{Belavin}.   We shall derive this solution, and others, in
section 3, but to get oriented we present it here, and discuss some
of its properties.  The gauge configuration for one instanton ($k =
1$) in $SU(2)$ contains the matrices $\sigma_{\mu\nu}$ or
$\bar\sigma_{\mu\nu}$.  One often writes it in terms of the 't Hooft
$\eta$ tensors, related to $\bar\s_{\mu\nu}$ by $\bar\s_{\mu\nu} = i
\eta^a{}_{\mu\nu} \tau_a$ where $\tau_a$ are the generators of
$SU(2)$. We discuss these tensors in Appendix B. The regular
one-instanton solution reads then
\begin{eqnarray} \label{1inst}
&& A_\mu^a (x; x_0,\rho) = 2 \frac{\eta^a{}_{\mu\nu} (x - x_0)^\nu}
{( x - x_0 )^2 + \rho^2}\ , \nonumber\\
&& A_\mu \equiv A^a_\mu \left( {\tau_a \over 2i} \right) = -
{\bar\s_{\mu\nu} (x - x_0)^\nu \over (x - x_0)^2 + \rho^2}\ ,
\end{eqnarray}
where $x_0$ and $\rho$ are arbitrary parameters
called collective coordinates. They correspond to the position and
the size of the instanton. The above expression solves the
selfduality equations for any value of the collective coordinates.
Notice that it is regular at $x = x_0$, as long as $\rho \neq 0$.
The real antisymmetric eta-symbols are defined as follows
\begin{eqnarray}\label{eta-tensors}
\eta^a{}_{\mu\nu} = \epsilon^a{}_{\mu\nu} \qquad
&\mu,\nu = 1, 2, 3 \, , & \qquad \eta^a{}_{\mu 4} = -
\eta^a{}_{4\mu} = \delta^a_\mu\ ,\nonumber\\ {\bar
\eta}^a{}_{\mu\nu} = \epsilon^a{}_{\mu\nu} \qquad &\mu,\nu = 1, 2,
3 \, , & \qquad {\bar \eta}^a{}_{\mu 4} = - {\bar \eta}^a{}_{4\mu}
= -\delta^a_\mu\ .
\end{eqnarray}
The $\eta$ and $\bar\eta$-tensors are selfdual and anti-selfdual respectively,
for fixed index $a$. They form a basis for antisymmetric four
by four matrices, and we have listed their properties in Appendix
\ref{HooftSpinor}.  They are linear combinations of the Euclidean Lorentz generators
$L_{\mu\nu}$, namely $\eta^a{}_{\mu\nu}=(J^a+K^a)_{\mu\nu}$ and
${\bar \eta}^a{}_{\mu\nu}=(J^a-K^a)_{\mu\nu}$,
where $J^a=\epsilon^{abc} L_{bc}$ and $K^a=L_{a4}$, and $(L_{mn})_{\mu\nu} = \d_{m\mu} \d_{n\nu} - \d_{m\nu} \d_{n\mu}$ with $m,n = 1,4$.   In this subsection we use $\eta$ tensors, but
in later sections we shall use the matrices $\sigma_{\mu\nu}$ and
$\bar\sigma_{\mu\nu}$.

The field strength corresponding to this gauge potential is (use
\eqref{eta-eta1})
\begin{equation}
F^a_{\mu\nu} = - 4 \eta^a{}_{\mu\nu} \frac{\rho^2}
{[( x - x_0 )^2 + \rho^2]^2} \ ,
\end{equation}
and it is selfdual. Thus \eqref{1inst} is a solution of the classical field
equations.
Far away, $A^a_\mu$ becomes proportional to
the inverse radius ${1 \over r}$ so that it contributes a finite
amount to the integral for the winding number which is of the form
$\int A^3 (r^3 {\rm d} \O)$, while $F_{\mu\nu}$ becomes proportional
to ${1 \over r^4}$, yielding a finite action.  However, $A^a_\mu$ itself
vanishes at $r \rightarrow \infty$, hence we have a smooth
configuration on $S^4$. Notice that the special point $\rho = 0$,
corresponding to zero size instantons, leads to zero field
strength and corresponds to pure gauge. Strictly speaking, this point must
therefore be excluded from the instanton moduli space of collective
coordinates. Finally one
can compute the value of the action by integrating the density
\begin{equation}
\label{inst-dens} \tr\, F_{\mu\nu} F^{\mu\nu} = - 96\,
\frac{\rho^4}{[( x - x_0 )^2 + \rho^2]^4} \ .
\end{equation}
Using the integral given at the end of Appendix \ref{HooftSpinor}, one finds
that this solution corresponds to $k = 1$.

One may show by direct calculation that the regular one-anti-instanton solution
is also given by \eqref{1inst} but with ${\bar \eta}^a{}_{\mu\nu}$. (In the
proof one uses that the first formula in (B.5) also holds for
${\bar \eta}^a{}_{\mu\nu}$).

We shall also derive the one-instanton solution in the singular gauge.  In terms of $\eta$ symbols it
reads
\begin{equation} \label{inst-sing} A^a_\mu = 2\,
\frac{\rho^2{\bar \eta}^a_{\mu\nu} (x - x_0)_\nu} {( x - x_0 )^2
[( x - x_0 )^2 + \rho^2]} = - \bar\eta^a_{\mu\nu} \partial_\nu\, \ln
\left\{ 1 + \frac{\rho^2}{( x - x_0 )^2} \right\} \ .
\end{equation}
This gauge potential is singular for $x = x_0$, where it
approaches a pure gauge configuration as we shall show in the next
section, $A_\mu \stackrel{x\rightarrow x_0}{=} U \partial_\mu U^{-
1}$. The gauge transformation $U$ is singular and relates the
regular gauge instanton \eqn{1inst} to the singular one
\eqn{inst-sing} at all points. The field strength in singular
gauge is then (taking the instanton at the origin, $x_0 = 0$,
otherwise replace $x \rightarrow x - x_0$)
\begin{equation}
F^a_{\mu\nu} = - \frac{4\rho^2}{( x^2 + \rho^2)^2} \left\{
\bar\eta^a_{\mu\nu} - 2 \bar\eta^a_{\mu\rho} \frac{x_\rho
x_\nu}{x^2} + 2 \bar\eta^a_{\nu\rho} \frac{x_\rho x_\mu}{x^2}
\right\} \ .
\end{equation}
Notice that despite the presence of the anti-selfdual
eta-tensors $\bar\eta$, this field strength is still selfdual, as
can be seen by using the properties of the eta-tensors given in
\eqref{eta-eta1}.  The singular gauge is frequently used, because,
as we will see later, zero modes fall off more rapidly at large
$x$ in the singular gauge.   One can compute the winding number
again in singular gauge. Then one finds that there is no
contribution coming from infinity. Instead, all the winding is
coming from the singularity at the origin. The singular solution
is singular at $x_0$, so one would expect that the regular
solution is singular at infinity.  This may seem puzzling since we
saw that the regular solution was smooth on $S^4$.  However, to
decide whether a configuration is smooth at $r \rightarrow
\infty$, one should first transform the point at infinity to the
origin and then study how the transformed configuration behaves
near the origin.  Making the coordinate transformation $x^\mu =
y^\mu / y^2$ or $x^\mu =-y^\mu / y^2$ , not forgetting that a
vector field transforms as
$A'_\mu (y) = (\partial x^\nu / \partial y^\mu ) A_\nu (x)$, one
finds that the transformed regular $k=1$ solution is indeed
singular at the origin\footnote{  This coordinate transformation in ${\bf R}^4$
can be viewed as a product of two conformal projections, one from the plane to
the coordinate patch on the sphere $S^4$ containing the south pole, and the
other from the other coordinate patch on $S^4$ with the north pole back to the
plane. The transformed metric is $g'_{\mu\nu}(y)=\delta_{\mu\nu}/y^4$, so
conformally flat. Then the action for the $A'_\mu(y)$ in $y$-coordinates is
again the usual flat space action in \eqref{YMaction}, and the transformed
instanton solution is an anti-instanton solution.}.
In fact, it is equal to the singular
$k=-1$ solution with $\rho$ replaced by ${1 \over \rho}$.

At first sight it seems that  there are five collective
coordinates for the $k=1$ solution. There are however extra
collective coordinates corresponding to the gauge orientation. One
can act with an $SU(2)$ matrix on the solution \eqn{1inst} to
obtain another solution,
\begin{equation} \label{1inst-angles}
A_\mu ( x; x_0, \rho, \vec\theta ) = U^{-1}(\vec\theta )
\, A_\mu(x; x_0,\rho)\, U  ( \vec \theta ) \ , \qquad U \in  SU(2) \ .
\end{equation}
with constant $\vec{\theta}$.  One might think that these
configurations should not be considered as a new solution since
they are gauge equivalent to the expression given above.   This is
not true, however, the reason being that, after we fix the gauge,
we still have left a rigid $SU(2)$ symmetry which acts as in
\eqn{1inst-angles}. So in total there are eight collective
coordinates, also called moduli. In principle, one could also act
with the (space-time) rotation matrices $SO(4)$ on the instanton
solution, and construct new solutions. However,  these rotations
can be undone by suitably chosen gauge transformations
\cite{JackReb}. Actually, the Yang-Mills action is not only
invariant under the Poincar\'e algebra (and the gauge algebra),
but it is also invariant under the conformal algebra which
contains the Poincar\'e algebra and further the generators for
dilatations (D) and conformal boosts $(K_\mu)$.   As shown in
Appendix D, for the Euclidean conformal group $SO(5,1)$, the
subgroup $SO(5)$ consisting of $SO(4)$ rotations and a combination of
conformal boosts and translations ($R^\mu \equiv K^\mu + \rho^2 P^\mu$),
leaves the instanton invariant up to gauge transformations. This
leads to a 5 parameter instanton moduli space $SO(5,1)/SO(5)$,
which is the Euclidean version of the five-dimensional anti-de
Sitter space $AdS_5$. The coordinates on this manifold correspond
to the four positions and the size $\rho$ of the instanton. On top
of that, there are still three gauge orientation collective
coordinates, yielding a total of eight moduli for the $k=1$ instanton in
$SU(2)$.

Instantons in $SU(N)$ can be obtained by embedding $SU(2)$ instantons into
$SU(N)$. For instance, a particular embedding is given by the
following $N$ by $N$ matrix
\begin{equation} \label{1instSUN} A_\mu^{SU(N)} = \left(
\begin{array}{cc} 0 & 0             \\ 0 & A_\mu^{SU(2)}
\end{array} \right) \ . \end{equation}
where the instanton resides
in the $2 \times 2$ matrix on the lower right. Of course this is
not the most general solution, as one can choose different
embeddings, see below.

One can act with a general $SU(N)$ element on the solution
\eqn{1instSUN} and obtain a new one. Not all elements of $SU(N)$
generate a new solution. There is a stability group that leaves
\eqn{1instSUN} invariant, acting only on the zeros, or commuting
trivially with the $SU(2)$ embedding. Such group elements should
be divided out, so we consider, for $N > 2$,
\begin{equation}
A_\mu^{SU(N)} = U\, \left( \begin{array}{cc} 0 &
0    \\ 0 & A_\mu^{SU(2)} \end{array} \right) \,
U^\dagger, \qquad U \; \in \; \frac{SU(N)}{SU(N - 2) \times U(1)}
\ . \label{coset}
\end{equation}
One can now count the number of
collective coordinates. From counting the dimension of the coset
space in \eqn{coset}, one finds there are $4 N - 5$ parameters.
Together with the position and the scale of the $SU(2)$ solution,
we find in total $4N$ collective coordinates for a one-instanton
solution in $SU(N)$.  It is instructive to work out the example of
$SU(3)$. Here we use the eight Gell-Mann matrices
$\{\lambda_\alpha\}, \alpha = 1,\dots,8$. The first three
$\lambda_a, a = 1, 2, 3$, form an $SU(2)$ algebra and are used to
define the $k = 1$ instanton by contracting \eqn{1inst} or \eqn{inst-sing} with
$\lambda_a$. The generators $\lambda_4,\dots,\lambda_7$ form two
doublets under this $SU(2)$, so they act on the instanton and can be used to generate new
solutions. This yields four more collective coordinates.  Then
there is $\lambda_8$, corresponding to the $U(1)$ factor in
\eqn{coset}. It commutes with the $SU(2)$ subgroup spanned by
$\lambda_a$, and so it belongs to the stability group leaving the
instanton invariant.  So for $SU(3)$ and $k=1$, there are seven
gauge orientation zero modes, which agrees with $4N-5$ for $N=3$.

The embedding of instanton solutions as a $2 \times 2$ block
inside the $N \times N$ matrix representation of $SU(N)$ is not
the only embedding possible.  For example, one can also use the $3
\times 3$ matrix representation $T_a$ of $SU(2)$, and put the
instanton inside a $3 \times 3$ block of the ${\bf N}$ of $SU(N)$.
This ${\bf 3}$ of $SU(2)$ is sometimes called ``the other $SU(2)$
in $SU(3)$", but it is simply the adjoint representation of
$SU(2)$, which is also the defining representation of $SO(3)$, and
is given by $(T_a)_{ij}=\epsilon_{iaj}$,
\begin{eqnarray}
T_1 =  \left( \begin{array}{cccc} 0 & 0 & 0 \\ 0 & 0 & -1 \\ 0 & 1 & 0
\end{array} \right) ;\quad T_2 = \left( \begin{array}{cccc} 0 & 0 & 1 \\ 0 & 0
& 0 \\ -1 & 0 & 0 \end{array} \right) ;\quad T_3 = \left(
\begin{array}{cccc} 0 & -1 & 0 \\ 1 & 0 & 0 \\ 0 & 0 & 0
\end{array} \right) \label{possible}
\end{eqnarray}
This representation has the same structure constants $f_{abc} =
\epsilon_{abc}$ as the representation $T_a = \tau_a / (2i)$, but
now $\tr \{T_a T_b\} =-2\delta_{ab}$, four times larger.

In fact, going back to the construction of the instanton, we note
that ${\bf any}$ representation $T_a$ of $SU(2)$ yields an
instanton solution for  $SU(N)$ as long as it fits inside the $N \times N$ matrices of $SU(N)$ \cite{Wilczek}
\begin{eqnarray}
A_\mu = 2 \eta^a_{\mu\nu} T_a {x^\nu \over x^2 + \rho^2}\ .
\end{eqnarray}
The ${\bf 2}$ of $SU(2)$ with $T_a = {\tau_a \over 2i}$ yields \eqn{1instSUN},
but any other representation yields another embedding.

For $SU(3)$ there are only two possibilities.  We can embed the instanton
using the ${\bf 2}$ of $SU(2)$; this yields \eqn{1instSUN}.  But we can also
use the matrices $T_a$ given in \eqn{possible} as the first 3 generators of
$SU(3)$.  For $SU(N)$ we can use any spin $j$ representation of $SU(2)$
provided it fits inside the $N \times N$ matrices.  Since the action and
winding number are proportional to the trace $\tr \; T_a T_b$, which is
proportional to
the quadratic Casimir operator $j (j+1)$ times the dimension $2 j+1$ of the spin $j$
representation\footnote{  Use $\d^{ab} tr T_a T_b = - tr C_2 (R) = - (2 j+1) C_2 (R)$ where the quadratic Casimir operator for the representation $R$ with spin $j$ is given by $C_2 (R) =- \d^{ab} T(R)_a T(R)_b = j (j+1)$.}, we see that we get instanton solutions with winding number
$k = \pm \ft23 j (j+1) (2j+1)$.  For $j = 1/2$ this reduces to $k= \pm 1$.
For the first few $SU(N)$ the results are as follows
\begin{eqnarray}
&SU(3): & k = \pm 1; k = \pm 4 \; (j= 1/2 \; {\rm and} \; j=1) \nonumber\\
&SU(4): & k = \pm 1 ; k = \pm 4 ; k = \pm 10 \; (j= 1/2 , 1, 3/2) \nonumber\\
            && k = \pm 2 \; ({\rm two} \; j = 1/2 \;
{\rm in \; block \; form}) \nonumber\\
&SU(5): & k= \pm 1, \pm 4 , \pm 10, \pm 20 \quad (j= \ft12 , 1, \ft32 , 2)
\nonumber\\
          && k = \pm 2, \pm 5 \; (j = \ft12 \oplus \ft12 \; {\rm and}\; j=
\ft12 \oplus 1)\ .
\end{eqnarray}
All these instanton solutions with winding number $|k| > 1$ still
are (anti-) selfdual, so they still have minimal action, determined
by the winding number, so the same as $k$ instantons embedded as
$2 \times 2$ matrices but far apart. Two instantons far apart and
each of the form \eqn{1inst} repel each other (as opposed to
an instanton and anti-instanton) with an interaction energy
proportional to $1/r$. Bringing $k$ instantons together
such that they sit all at the same point, gives solutions of the
kind above. So far apart there is a small positive interaction,
but when they are brought together the interaction energy
vanishes. Hence, there must be domains of attraction in between.
This already shows that the interaction of instantons is a
complicated problem \cite{Wilczek}. In fact, one can deform these
single-instanton solutions such that a multi-instanton solution is
obtained in which the single-instantons do not attract or repel
each other. In other words, in such a multi-instanton solution the
positions, sizes and gauge orientations of the single instantons
are collective coordinates.

For the general multi-instanton solution, the dependence on all
collective coordinates is in implicit form given by the ADHM
construction \cite{ADHM}. For a recent review, see \cite{ndorey}.
In the next section we will obtain explicit formulas for the
dependence on $5k$ collective coordinates.  Explicit formulas for
the dependence on all collective coordinates only exist for the $k=2$
instanton solution \cite{ADHM,Osb,Christ:1978jy,ndorey} and the $k=3$
instanton solution \cite{KorepinS}.

We end this section with some remarks on embeddings into other gauge
groups \cite{Bern-et-al}. For $k=1$ and gauge group $SO(N)$, it is known that
there are $4N-8$ collective coordinates. This can be understood as
follows. The one-instanton solution is constructed by choosing an
embedding of $SO(4)=SU(2)\times SU(2)$ generated by $\eta^a_{\mu\nu}$
and ${\bar \eta}^a_{\mu\nu}$, and putting the instanton in one of the
$SU(2)$ groups. The stability group
of this instanton is $SO(N-4)\times SU(2)$, so we obtain (for $N>4$)
\begin{equation}
A_\mu^{SO(N)} = U\, \left( \begin{array}{cc} 0 &
0             \\ 0 & A_\mu^{SU(2)} \end{array} \right) \,
U^\dagger, \qquad U \; \in \; \frac{SO(N)}{SO(N - 4) \times SU(2)}
\ . \label{coset-orthogonal}
\end{equation}
The number of collective coordinates of such solutions follows from
the dimension of the coset (which is $4N-13$). Including the positions
and size of the $SU(2)$ instanton, we arrive at $4N-8$ for the total number
of collective coordinates. Notice that for $N=6$, we can use the isomorphism
between $SO(6)$ and $SU(4)$. For both countings, we arrive at 16 moduli.

Similarly, we can analyze the symplectic gauge groups $USp(2N)$. Here
we can simply choose the lower diagonal $SU(2)=USp(2)$ embedding
inside $USp(2N)$ for a $k=1$ instanton. The stability group of this embedding is now $USp(2N-2)$, so
for we have the following instanton solution:
\begin{equation}
A_\mu^{Sp(N)} = U\, \left( \begin{array}{cc} 0 &
0             \\ 0 & A_\mu^{SU(2)} \end{array} \right) \,
U^\dagger, \qquad U \; \in \; \frac{USp(2N)}{USp(2N - 2)}
\ . \label{coset-symplectic}
\end{equation}
The dimension of $USp(2N)$ is $N(2N+1)$,\footnote{  The dimension of $U(2 N)$ is $4N^2$ and the generators have the form $\left( \begin{array}{ll} a_1+ is_1 & b \\ -b^\dagger & a_2 +is_2 \end{array} \right)$ where $a_i$ is antisymmetric and $s_i$ is symmetric.  Complex symplectic matrices $M= \left( \begin{array}{ll} A & B \\ C & D \end{array} \right)$ satisfy $M^T \O + \O M=0$ where $\O = \left( \begin{array}{ll} 0 & I \\ -I & 0 \end{array} \right)$.  The restriction that the unitary generators be also symplectric leads to $N^2 + N (N-1)$ constraints $(D + A^T =0$ and $C - C^T = B - B^T =0)$.} and so the total number of collective coordinates that follows from this construction is $5+(4N-1) = 4(N+1)$,
which is the correct number \cite{Bern-et-al}. For $N=2$, we have the
isomorphism $USp(2)=SO(5)$, which in both countings leads to 12 collective
coordinates.

For higher instanton number, not all instantons can be constructed from a
properly chosen embedding. There the ADHM formalism must be used. We just mention here that
the total number of collective coordinates is $4kN, 4k(N-2)$ and $4k(N+1)$
for the gauge groups $SU(N), SO(N)$ and $USp(2N)$ respectively. The geometric
relation between instanton moduli spaces and quaternionic manifolds (whose
dimension is always a multiple of four) can e.g. be found in
\cite{Vand-quat}.

\subsection{Some remarks on nonselfdual instanton solutions}

Note that we have not shown that all solutions of \eqn{eom}
with finite action are given by selfdual (or anti-selfdual) field
strengths. In principle there could be configurations which are
extrema of the action, but are neither selfdual nor
anti-selfdual\footnote{  It is possible to construct
solutions for $SU(2)$ that are not selfdual, but not with finite action. An
example is $A_\mu=-\ft12 {\sigma}_{\mu\nu}\frac{x_\nu}{r^2}$. Its field strength is $F_{\mu\nu} = {1 \over 2} \s_{\mu\nu} / r^2 + {1 \over 2} (x^\mu \s_{\mu\rho} - x^\nu \s_{\mu\rho} ) x^\rho / r^4$.  One
can check that it satisfies the second order equation of motion
\eqn{eom} (both $\partial_\mu F_{\mu\nu}$ and $[A_\mu,F_{\mu\nu}]$ vanish),
but this configuration is not selfdual since
$F_{\mu\nu}-{}^*F_{\mu\nu}= \ft12 {\sigma}_{\mu\nu}\frac{1}{r^2}$.
Because this field strength does not tend to zero fast enough at
infinity, the action evaluated on this solution diverges
logarithmically.}.
For the gauge group $SU(2)$ this has been a long standing question.
The first result was established in \cite{BLS,BL,T} where it was shown that
for gauge groups $SU(2)$ and $SU(3)$, nonselfdual solutions cannot be
local minima, hence if they exist, they should correspond to saddle points.
The existence of nonselfdual solutions
with finite action and gauge group $SU(2)$ was first established in
\cite{SSU}, for $k=0$, and later for $k\neq 0$ in \cite{SS}. For gauge group
$SU(3)$ some results have been obtained in \cite{Burzlaff,Schiff}.
The situation seems to be quite complicated, and no elegant and simple
framework to address these issues has been found so far. For bigger gauge
groups, it is easier to construct non-selfdual (or anti-selfdual) solutions. This becomes clear in the example
of $SO(4) = SU(2)\times SU(2)$. If we associate a selfdual instanton to
the first factor, and an anti-selfdual instanton to the second factor, the
total field strength satisfies the equations of motion \eqn{eom} but
is neither selfdual nor anti-selfdual.
Even simpler is the example of $SU(4)$. By choosing two commuting $SU(2)$
subgroups, we can embed both an $SU(2)$ instanton and an anti-instanton
inside $SU(4)$,
\begin{equation}
A_\mu^{SU(4)} = \left( \begin{array}{cc} A_\mu^+ &
0             \\ 0 & A_\mu^- \end{array} \right)\ , \label{inst-antiinst}
\end{equation}
where $A_\mu^{\pm}$ denotes the (anti-) selfdual $SU(2)$ gauge potentials
with topological charges $k^{\pm}$.
Clearly the total field strength is neither selfdual nor
anti-selfdual, but satisfies the second order equations of motion.
The instanton action is finite and the total
topological charge is $k^+-k^-$.

From the embedding \eqref{inst-antiinst} one can generate more solutions
by acting on the gauge potential with a global gauge transformation
$U\in SU(4)$.  In this way, one
generates new exact and nonselfdual solutions which are not of the
form \eqref{inst-antiinst}.

For $SU(N)$ gauge groups, one has even more possibilities. One can embed $k_+$ instantons and $k_-$
anti-instantons on the (block)-diagonal of $SU(N)$, as long as
$2(k_++k_-)\leq N$.  If we take both $k_+ > 0$ and $k_- > 0$, the solution is
clearly not selfdual or anti-selfdual and the
instanton action, including the theta-angle, is given by
\begin{equation}
S=\frac{8\pi^2}{\ggg^2}(k_++k_-) + i\theta (k_+-k_-)\ .
\end{equation}
In a supersymmetric theory, these solutions will not preserve any
supersymmetry. This is interesting in the context of the AdS/CFT
correspondence that relates ${\cal N}=4$ SYM theory to type IIB superstrings.
In \cite{BCPVRV}, it is shown that these non-selfdual Yang-Mills instantons
are related to non-extremal (non BPS) D-instantons in IIB supergravity.

\section{Regular and singular instanton solutions}

To find explicit instanton solutions, we solve the selfduality (or
anti-selfdualty) equations $F_{\mu\nu} = \:  {^*\!F_{\mu\nu}}$
where ${^*\!F_{\mu\nu}} = \ft12 \epsilon_{\mu\nu\rho\sigma}
F_{\rho\sigma}$ with $\mu, \nu =1, 4$ and $\epsilon_{1234} =
\epsilon^{1234} =1$.  Since $D_\mu \: {^*\!F_{\mu\nu}}$ vanishes
identically due to the Bianchi identity, we then have a solution
of the field equations, $D_\mu F_{\mu\nu} = 0$.  The main idea is
to make a suitable ansatz, and then to check that it yields
solutions.  The ansatz is (we restrict ourselves for the moment to the gauge
group $SU(2)$)
\begin{eqnarray}
A_\mu (x) = \alpha \; \sigma_{\mu\nu} \partial_\nu \ln \phi (x^2)\ ,
\label{idea}
\end{eqnarray}
where $\alpha$ is a real constant to be fixed and $\sigma_{\mu\nu}$ is the $2 \times 2$ matrix representation of the Lorentz generators in Euclidean space.  Since we shall be using these matrices $\sigma_{\mu\nu}$ a lot, we first discuss their properties in some detail, and then we shall come back below ({\ref{indian}) to the construction of instanton solutions.

\subsection{Lorentz and spinor algebra}

In Euclidean space, a suitable $4 \times 4$ matrix representation of the Dirac matrices is given by
\begin{eqnarray}
\gamma^\mu =  \left( \begin{array}{ cc} 0  & - i
(\sigma^\mu)^{\alpha \beta'} \\  i ( \bar\sigma^\mu)_{\alpha'
\beta}  & 0 \end{array} \right)\ , \qquad \begin{array}{ll} \sigma^\mu
= (\vec\tau , i I) \\ \bar\sigma^\mu = ( \vec\tau , -i I)
\end{array}\ , \label{sppace}
\end{eqnarray}
where $\vec\tau$ are the Pauli matrices. We use slashes instead of
dots on the spinor indices to indicate that we are in Euclidean
space. All four Dirac matrices are hermitian, and satisfy $\{
\gamma^\mu, \gamma^\nu \} = 2 \delta^{\mu\nu}$.  The matrix
$\gamma^5$ is diagonal
\begin{eqnarray}
\gamma^5 \equiv \gamma^1 \gamma^2 \gamma^3 \gamma^4 = \left( \begin{array}{cc} I & 0 \\ 0 & - I \end{array} \right)\ ,
\end{eqnarray}
and chiral spinors correspond to projections with $\ft12(1 \pm \gamma_5)$ which yield the upper or lower two components of a nonchiral four-component spinor.

Since we are in Euclidean space, it does not matter whether we
write the index $\mu$ as a contravariant or covariant index.  In
Minkowski space this representation (with $\gamma^4$ replaced by
$\gamma^0$ where $\gamma^4 = i \gamma^0$, so that $(\gamma^k)^2
=+1$ but $(\gamma^0)^2 =-1)$ is used for two-component spinor
formalism.  Four-component spinors are then decomposed into
two-component spinors as $\psi = {\lambda^\alpha \choose
\bar\chi_{\dot\alpha}}$, and this explains the position of the
spinor indices on $\sigma^\mu$ and $\bar\sigma^\mu$ in
\eqn{sppace}.  The Euclidean Lorentz generators ($SO(4)$
generators) acting on 4-component spinors are $M_{\mu\nu} = \ft14
(\gamma_\mu \gamma_\nu - \gamma_\nu \gamma_\mu)$ and satisfy the
Euclidean Lorentz algebra
\begin{eqnarray}
[M_{\mu\nu}, M_{\rho\sigma} ] = \delta_{\nu\rho} M_{\mu\sigma} -
\delta_{\nu\sigma} M_{\mu\rho} - \delta_{\mu\rho} M_{\nu\sigma} +
\delta_{\mu\sigma} M_{\nu\rho}\ .
\end{eqnarray}
However, this representation is reducible: the upper and lower components
of $\psi$ form separate representations
\begin{eqnarray}
M_{\mu\nu} = \frac{1}{2}\left( \begin{array}{cc}
(\s^{\mu\nu})^\alpha{}_\beta & 0 \\ 0 &
(\bar\sigma^{\mu\nu})_{\alpha'}{}^{\beta'} \end{array} \right)\ .
\end{eqnarray}
In terms of $\sigma^\mu$ and $\bar\sigma^\mu$ we then find the following two inequivalent spinor representations of $SO(4) : M_{\mu\nu} = \ft12 \sigma_{\mu\nu}$ and $M_{\mu\nu} = \ft12 \bar\sigma_{\mu\nu}$, where
\begin{eqnarray}
\sigma^{\mu\nu} = \ft12 (\sigma^\mu \bar\sigma^\nu - \sigma^\nu
\bar\sigma^\mu )\ ; \qquad \bar\sigma^{\mu\nu} = \ft12 (\bar\sigma^\mu \sigma^\nu - \bar\sigma^\nu \sigma^\mu)\ .
\label{mine}
\end{eqnarray}
(It is customary not to include the factor $\ft12$ in $M_{\mu\nu} = \ft12 \sigma_{\mu\nu}$ into the definition of $\sigma^{\mu\nu}$).

The matrices $\sigma_{\mu\nu}$ and $\bar\sigma_{\mu\nu}$ satisfy some properties which we shall need repeatedly.  First of all, they are anti-selfdual and selfdual, respectively
\begin{eqnarray}
\sigma_{\mu\nu} = - \ft12 \epsilon_{\mu\nu\rho\sigma}
\sigma_{\rho\sigma}\ ; \qquad  \bar\sigma_{\mu\nu} = \ft12
\epsilon_{\mu\nu\rho\sigma} \bar\sigma_{\rho\sigma}\ .
\label{contrib}
\end{eqnarray}
This follows most easily by noting that the matrices $\gamma_\mu$
satisfy $\gamma_{[\mu} \gamma_{\nu]} = -
\ft12\epsilon_{\mu\nu\rho\sigma} \gamma_\rho \gamma_\sigma
\gamma_5$ where $\g_{\mu\nu}\equiv \g_{[\mu} \g_{\nu]} = \frac12
(\g_\mu \g_\nu - \g_\nu \g_\mu)$.   For example, $\gamma_1
\gamma_2 =- \gamma_3 \gamma_4 \gamma_5$ because $\epsilon_{1234} =
+ 1$. From this (anti)-selfduality one derives another useful
property
\begin{eqnarray}
\epsilon_{\mu\nu\rho\sigma} \sigma_{\sigma\tau} = \delta_{\mu\tau} \sigma_{\nu\rho} - \delta_{\nu\tau} \sigma_{\mu\rho} + \delta_{\rho\tau} \sigma_{\mu\nu}\ .
\label{yang1}
\end{eqnarray}
It is easiest to prove \eqn{yang1} by substituting \eqn{contrib} into the left-hand side, and decomposing the product of two $\epsilon$-tensors into a sum of products of Kronecker tensors. \noindent Another proof is based on the ``Schouten identity" which is the observation that a totally antisymmetric tensor with 5 indices vanishes in 4 dimensions (because there are always at least two indices equal).  Writing the left-hand side of (\ref{yang1}) as $\epsilon_{\mu\nu\rho\alpha} \delta_\beta{}^\tau \sigma_{\alpha \beta}$ and using the Schouten identity
\begin{equation}\label{schouten}
\epsilon_{\mu\nu\rho\alpha} \delta_\beta{}^\tau =
\epsilon_{\beta\nu\rho\alpha} \delta_\mu{}^\tau +
\epsilon_{\mu\beta\rho\alpha} \delta_\nu{}^\tau +
\epsilon_{\mu\nu\beta\alpha} \delta_\rho{}^\tau +
\epsilon_{\mu\nu\rho\beta} \delta_\alpha{}^\tau \ ,
\end{equation}
the identity \eqn{contrib} can be used to prove the property (\ref{yang1}) (the last term in \eqn{schouten} yields minus the contribution of the term on the left-hand side).  In a similar way one may prove
\begin{eqnarray}
\epsilon_{\mu\nu\rho\sigma} \bar\sigma_{\sigma\tau} =- \delta_{\mu\tau} \bar\sigma_{\nu\rho} + \delta_{\nu\tau} \bar\sigma_{\mu\rho} - \delta_{\rho\tau} \bar\sigma_{\mu\nu} .
\end{eqnarray}
The extra overall minus sign is due to the extra minus sign in the selfduality relation in \eqn{contrib}.

Further identities are the commutator of two Lorentz generators, and the
anticommutator which is proportional to the unit matrix in spinor space
\begin{eqnarray}
&& [ \sigma_{\mu\nu} , \sigma_{\rho\sigma} ] = 2 \delta_{\nu\rho}
\sigma_{\mu\sigma} + \mbox{three more terms}\ , \nonumber\\
&& \{ \sigma_{\mu\nu} , \sigma_{\rho \sigma} \} = 2 (\delta_{\mu\sigma} \delta_{\nu\rho} - \delta_{\mu\rho} \delta_{\nu\sigma} ) + 2
\epsilon_{\mu\nu\rho\sigma}\ .
\label{easyway}
\end{eqnarray}
One easy way to prove or check these identities is to use $4
\times 4$ Dirac matrices; for example $\{ \gamma_1 \gamma_2 ,
\gamma_3 \gamma_4 \} = 2 \gamma_5$ and $\{ \gamma_{12} ,
\gamma_{13} \} =0$ but $\{ \gamma_{12} , \gamma_{12} \} =-2$ and
$[\gamma_{12}, \gamma_{13} ] =-2 \gamma_{23}$.  Because $\gamma_5
= \left( \begin{array}{cc} I & 0 \\ 0 & -I \end{array} \right)$,
it is clear that the $\bar\sigma_{\mu\nu}$ satisfy the same
commutation and anticommutation relations, but with a different
sign for the $\epsilon$ symbol.  In particular,
\begin{eqnarray}
\{ \bar\sigma_{\mu\nu} , \bar\sigma_{\rho\sigma} \} = 2 ( \delta_{\mu\sigma} \delta_{\nu\rho} - \delta_{\mu\rho} \delta_{\nu\sigma} ) - 2 \epsilon_{\mu\nu\rho\sigma}\ .
\end{eqnarray}

All these identities can also be derived using two-component
spinor formalism for vectors.  For example, a vector $v^\mu$ is
written as $v^{\alpha \alpha'} \equiv v^\mu \sigma_\mu{}^{\alpha
\alpha'}$, and then one may use such identites as
\begin{equation}
\delta_\mu{}^\nu \sim \delta_{\alpha \alpha'}{}^{\beta \beta'}
\sim \delta_\alpha{}^\beta  \delta_{\alpha'}{}^{\beta'}; \qquad
\delta_{\mu\nu} \sim \epsilon_{\alpha\beta} \epsilon_{\alpha'
\beta'}\ .
\end{equation}
If one never introduces any vector indices at all but only uses
spinor indices, this spinor formalism turns about all identities
into trivialities, but we prefer to also keep vector indices
around.  The other extreme is to expand $\sigma_{\mu\nu}$ and
$\bar\sigma_{\mu\nu}$ into Pauli matrices $\tau_a$ as
$\sigma_{\mu\nu} = i \bar\eta^a_{\mu\nu} \tau_a$ and
$\bar\sigma_{\mu\nu} = i \eta^a_{\mu\nu} \tau_a$ where
$\eta^a_{\mu\nu}$ and $\bar\eta^a_{\mu\nu}$ are constructed from
$\epsilon_{aij}$ and $\delta_{ai}$ tensors, as in
\eqn{eta-tensors}. A whole calculus of these ``'t Hooft-tensors"
can be set-up, and is often used. We discuss it in appendix B. We
shall not limit ourselves to one of these extremes; proofs are
given either by using $2$-component spinors or $4 \times 4$ Dirac
matrices, depending on which approach is simplest for a given
problem.

The index structure of the ansatz for $A_\mu$ in \eqn{idea} merits a short
discussion.  A Lie-algebra valued gauge field $A_\mu$ has indices $i , j$ for a
representation $R$ of an $SU(N)$ group.  For $SU(2)$ the generators in the
defining representation are the Pauli matrices $\tau^a$ divided by $2i$, hence
$A_\mu = (A_\mu)^i{}_j = A^a_\mu \left( {\tau_a \over  2i} \right)^i{}_j$.
The ansatz for the instanton can then be written as
 \begin{equation}
 (A_\mu)^i{}_j = (\sigma_{\mu\nu})^i{}_j\, x^\nu f(x^2)\ .
 \end{equation}
The indices $\mu,\nu$ are Lorentz indices, but the indices $i , j$
are $SU(2)$ indices.  Hence the matrix $(\s_{\mu\nu})^i{}_j$
carries simultaneously spacetime indices and internal $SU(2)$
indices. The matrices $\sigma_{\mu\nu}$ are indeed proportional to
$\tau_a, \sigma_{\mu\nu} = i \bar\eta^a_{\mu\nu} \tau_a$,  as one
may check for specific values of $\mu$ and $\nu$, using
\begin{eqnarray}
(\sigma_{\mu\nu})^i{}_j &=& \ft12 \left\{ (\sigma_\mu)^{i \beta'}(
\bar\sigma_\nu)_{\beta' j}-(\sigma_\nu)^{i \beta'} ( \bar\sigma_\mu)_{\beta' j}
\right\}\nonumber\\
(\sigma_\mu)^{i \beta'} &=& \{ \vec\tau , i \} \quad ,
(\bar\sigma_\mu)_{\beta' j} = \{ \vec\tau , -i \}\ .
\end{eqnarray}
The matrices $\eta^a_{\mu\nu}$ and $\bar\eta^a_{\mu\nu}$ are
actually invariant tensors of a particular $SU(2)$ group. There
are three groups $SU(2)$: the gauge group $SU(2)_g$ and the rotation
group $SO(4)=SU(2)_L\times SU(2)_R$ generated by $\eta^a_{\mu\nu}$ and
${\bar \eta}^a_{\mu\nu}$.
The tensor $\eta^a_{\mu\nu}$ is invariant under the combined $SU(2)_g$ gauge
transformations acting on the index $a$ generated by $\epsilon_{abc}$, and
the $SU(2)_L$ Lorentz transformations generated by $\eta^b_{\rho\sigma}$.
Indeed, under infinitesimal variations with parameter
$\lambda^a$ we find, using (\ref{eta-eta1}),
\begin{equation}
\delta \eta^a=\epsilon^{abc}\eta^b\lambda^c_g+\ft12
\lambda^c_L[\eta^c,\eta^a]=0\qquad \mbox{if $\lambda^a_g=\lambda^a_L$}\ .
\end{equation}
Furthermore, $\eta^a_{\mu\nu}$ is separately invariant
under the $SU(2)_R$ subgroup of the Lorentz group generated by
${\bar \eta}^b_{\rho\sigma}$; this follows from
$[\eta^a,{\bar \eta}^b]=0$. In fact, $\eta^a=L^a{}_4+\ft12
\epsilon^{abc}L_{bc}$ and ${\bar \eta}^a=-L^a{}_4+\ft12
\epsilon^{abc}L_{bc}$ from which $[\eta^a,{\bar \eta}^b]=0$ easily follows.\footnote{  The $4 \times 4$ matrices $L^a{}_4$ and $L_{bc}$ have entries $(L^a{}_4)_{\mu\nu} = \d^a_\mu \d_{4 \nu}$ and $(L_{bc})_{\mu\nu} = \d_{b\mu} \d_{c\nu} - \d_{c\mu} \d_{b\nu}$.  They form the defining representation of the Euclidean Lorentz algebra.}

Spinor indices are raised and lowered by $\epsilon$-tensors
following the northwest-southeast convention: $v^{\alpha'} =
\epsilon^{\alpha' \beta'} v_{\beta'}$ and $v^\alpha =
\epsilon^{\alpha \beta} v_\beta$.  So $( \bar\sigma^\mu )^{\beta'
\alpha} = \epsilon^{\beta' \delta'} \epsilon^{\alpha \gamma} (
\bar\sigma^\mu )_{\delta' \gamma}$. There are various definitions
of these $\epsilon$ tensors in the literature; we define
  \begin{eqnarray}
\epsilon^{\alpha \beta} =- \epsilon^{\alpha' \beta'} .
  \end{eqnarray}
Note that numerically $\epsilon^{\alpha \beta} = \epsilon_{\alpha
\beta}$ but also $\epsilon^{\alpha' \beta'} = \epsilon_{\alpha'
\beta'}$ because one needs two $\epsilon$ tensors to raise or
lower both indices of an $\epsilon$ tensor. We fix the overall sign
by $\epsilon^{\alpha \beta}=\epsilon^{ij}$ where $\epsilon^{12}=1$. A crucial
relation in the spinor formalism which we shall frequently use is
\begin{equation}
{\bar \sigma}_{\mu,\alpha'\,i}=\sigma_{\mu,i\,\alpha'}\ ,
\end{equation}
where we recall that $\sigma_{\mu,i\,\alpha'}=\sigma_\mu^{j\,\beta'}
\epsilon_{ji}\epsilon_{\beta'\alpha'}$.

Using 2-component spinor indices for vectors,
  \begin{eqnarray}
 (\bar\sigma^\mu)_{\alpha' \alpha} A_\mu \equiv A_{\alpha' \alpha} \; {\rm and}
  \;  (\bar\sigma_\nu )_{\beta' j} x^\nu \equiv x_{\beta' j}\ ,
 \label{p16}
  \end{eqnarray}
the ansatz for the instanton solution in \eqn{idea} with spinor indices for $A_\mu$ becomes
 \begin{eqnarray}
&&  (\bar\sigma^\mu)_{\alpha' \alpha} (A_\mu)^i{}_j \equiv
A_{\alpha' \alpha}{}^i{}_j =
 ( \bar\sigma^\mu )_{\alpha' \alpha} (\sigma_{\mu\nu})^i{}_j x^\nu f ( x^2 )
 \nonumber\\
  && =  \left\{  \delta^{\beta'}_{\alpha'}  \delta^i_\alpha  x_{\beta' j}-
  \epsilon_{\alpha' \beta'}  \epsilon_{\alpha j} x^{i \beta'} \right\} f (x^2)
  =  \left\{ \delta^i_\alpha x_{\alpha' j} + \epsilon_{\alpha j}
  x^i{}_{\alpha'} \right\} f(x^2)\ .
  \label{pright}
 \end{eqnarray}
The trace over $(ij)$ clearly vanishes, and this fixes the relative sign. We
worked out the matrix $(\bar\sigma^\mu)_{\alpha' \alpha}
 (\sigma_{\mu\nu})^i{}_j$ using
  \begin{eqnarray}
  \bar\sigma^\mu_{\alpha' \alpha} \sigma^{i \beta'}_\mu = 2 \delta^{\beta'}
  _{\alpha'} \delta^i_\alpha\ ,
  \label{pleft}
   \end{eqnarray}
and
 \begin{equation}
 \bar\sigma^\mu_{\alpha' \alpha} ( \bar\sigma_\mu)_{\beta' j} =
 \bar\sigma^\mu_{\alpha' \alpha} (\sigma_\mu)_{j \beta'}=
 \bar\sigma^\mu_{\alpha' \alpha} \sigma^{k \gamma '}_\mu
\epsilon_{k j} \epsilon_{\gamma' \beta'} = 2 \epsilon_{\alpha'
\beta'} \epsilon_{\alpha j}\ .
 \label{indian}
  \end{equation}

\subsection{Solving the selfduality equations}

Let us now come back to the construction of instanton solutions.  Substituting the ansatz for $A_\mu$ in \eqn{idea} into the definition of $F_{\mu\nu}$ yields with \eqn{easyway}
\begin{eqnarray}
 F_{\mu\nu} &=& \alpha \sigma_{\nu\rho} \partial_\mu \partial_\rho \ln \phi - \alpha \sigma_{\mu\rho} \partial_\nu \partial_\rho \ln \phi + \alpha^2 [ \sigma_{\mu\rho} , \sigma_{\nu\sigma} ] (\partial_\rho \ln \phi) (\partial_\sigma \ln \phi ) \nonumber\\
&=& ( \alpha \sigma_{\nu\rho} \partial_\mu \partial_\rho \ln \phi - \mu \leftrightarrow \nu ) + 2 \alpha^2 (\sigma_{\mu\sigma} \partial_\nu \ln \phi \; \partial_\sigma \ln \phi - \mu \leftrightarrow \nu) \nonumber\\
&-& 2 \alpha^2 \sigma_{\mu\nu} (\partial \ln \phi )^2\ .
\end{eqnarray}
We want to solve the equation $F_{\mu\nu} = {^*\!F_{\mu\nu}}$.
The dual of $F_{\mu\nu}$ can be written as an expression without any $\epsilon$ tensor by using the identities for $\epsilon_{\mu\nu\rho\sigma} \sigma_{\sigma\tau}$ and $\epsilon_{\mu\nu\rho\sigma} \sigma_{\rho\sigma}$ in \eqn{contrib} and \eqn{yang1}.  One finds
\begin{eqnarray}
{^*\!F_{\mu\nu}} &=& \ft12 \epsilon_{\mu\nu\rho\sigma} F_{\rho\sigma} = \alpha \epsilon_{\mu\nu\rho\sigma} \sigma_{\sigma\alpha} \partial_\rho \partial_\alpha \ln \phi \nonumber\\
&+& 2 \alpha^2 \epsilon_{\mu\nu\rho\sigma} \sigma_{\rho\beta} \partial_\sigma \ln \phi \; \partial_\beta \ln \phi - \alpha^2 \epsilon_{\mu\nu\rho\sigma} \sigma_{\rho\sigma} (\partial \ln \phi )^2 \nonumber\\
&=& \sigma_{\nu\rho} ( \alpha \partial_\rho \partial_\mu \ln \phi - 2 \alpha^2 \partial_\rho \ln \phi \; \partial_\mu \ln \phi ) - \mu \leftrightarrow \nu \nonumber\\
&+& \sigma_{\mu\nu} ( \alpha \partial^2 \ln \phi)\ .
\end{eqnarray}
Equating $F_{\mu\nu}$ to ${^*\!F_{\mu\nu}}$ yields two equations for $\phi$, namely one for the terms with $\s_{\nu\rho}$ and the other for the terms with $\s_{\mu\nu}$
\begin{eqnarray}
&& \alpha \partial_\mu \partial_\rho \ln \phi - 2 \alpha^2 \partial_\mu \ln \phi \; \partial_\rho \ln \phi = \alpha \partial_\mu \partial_\rho \ln \phi - 2 \alpha^2 \partial_\mu \ln \phi \; \partial_\rho \ln \phi\ , \nonumber\\
&& \qquad   -2 \alpha^2 (\partial \ln \phi)^2 = \alpha \partial^2
\ln \phi \ .
\end{eqnarray}
The first equation is identically satisfied (for that reason we equated
$F_{\mu\nu}$ to $+ ^*\!F_{\mu\nu}$), while the second equation can be rewritten
as $\partial^2 \ln \phi  + 2 \alpha (\partial \ln \phi)^2 =0$.  For $\alpha = \ft12$ it simplifies to $\partial^2 \phi / \phi =0$.  (Setting $\alpha = \ft12$ is not a restriction because rescaling $\phi \rightarrow \phi^{1/2 \alpha}$ achieves the same result).

Setting $\phi = {1 \over x^2}$ yields for $x \not= 0$ a solution: $\partial^2 \phi / \phi = x^2 \partial_\mu (-2 x^\mu / x^4) =0$.  However, this is also a solution at $x=0$ because $\partial^2 x^{-2}$ is proportional to a delta function (note that the dimensions match) and $x^2 \delta^4 (x) =0$
\begin{eqnarray}
\partial^2 {1 \over x^2} =- 4 \pi^2 \delta^4 (x)\ .
\end{eqnarray}
To check the coefficient, we integrate over a small ball, which
includes the point $x=0$; we obtain then $\int \partial^2 {1 \over
x^2} {\rm d}^4 x = \int r^3  {\rm d}r{\rm d} \Omega_\mu
\partial_\mu {1 \over x^2}  = \int r^3 {\rm d} \Omega_\mu (-2 x^\mu
/ x^4)  =- 4 \pi^2$. (The surface of a sphere in 4 dimensions is
$2 \pi^2$).

We have thus found a selfdual solution
\begin{eqnarray}
A_\mu (x) = \ft12 \sigma_{\mu\nu} \partial_\nu \ln
\left[1+\frac{\rho^2}{(x-a)^2}\right]\ .
\end{eqnarray}
We have added unity to $\phi$ in order that $A_\mu(x)$ vanishes for large
$|x|$. A more general solution is given by $A_\mu (x) = \ft12 \sigma_{\mu\nu}
\partial_\nu \ln \phi$ with
\begin{eqnarray}
\phi =1 + \sum^k_{i=1} {\rho^2_i \over (x- a_i)^2}\ ,
\end{eqnarray}
which also solves $\partial^2 \phi / \phi =0$.  These are a class
of $k$-instanton solutions, parameterized by $5k$ collective
coordinates. In particular for $k=1$ we find the one-instanton solution
\begin{eqnarray}
A^{\rm sing}_\mu (x) &=& \ft12 \sigma_{\mu\nu} \partial_\nu \ln \left[ 1 + {\rho^2 \over (x-a)^2}  \right]\ , \nonumber\\
 &=& -  \sigma_{\mu\nu} { \rho^2 (x-a)_\nu \over (x-a)^2 ((x-a)^2 +
 \rho^2)}\qquad (k=1,{\rm singular})\ .
\label{clearlyy}
\end{eqnarray}
This solution is clearly singular at $x=a$, but one can remove the singularity
at $x=a$ by a singular gauge transformation (which maps the singularity to
$x^2 = \infty$).  To determine this gauge transformation we first study the
structure of the singularity.  Near $x=a$ the singular solution becomes
\begin{eqnarray}
A^{\rm sing}_\mu (x) \approx - \sigma_{\mu\nu} {(x-a)_\nu \over
(x-a)^2}\ , \label{hass}
\end{eqnarray}
which is a pure gauge field with $U(x-a)$ in (A.9)
\begin{eqnarray}
U^{-1} \partial_\mu U= - \sigma_{\mu\nu} {(x-a)_\nu \over
(x-a)^2}\ ;\qquad U(x) = {x_4 + i x_k \sigma_k \over \sqrt{x^2}} = i
\bar\sigma_\mu x_\mu / \sqrt{x^2} . \label{thatis}
\end{eqnarray}
Note that $U$ is unitary, and $U^{-1}$ equals $- i \s_\mu x_\mu /
\sqrt{x^2}$, which follows from the property $\s_\rho \bar\s_\mu +
\s_\mu \bar\s_\rho = 2 \delta_{\rho\mu}$.

From \eqn{clearlyy} and \eqn{hass} it follows that we can write $A^{\rm sing}_\mu$ as
\begin{eqnarray}
A^{\rm sing}_\mu (x) = {\rho^2 \over (x-a)^2 + \rho^2} U^{-1}
\partial_\mu U\ .
\end{eqnarray}
It is now clear that an opposite gauge tranformation removes the singularity at $x=0$
\begin{eqnarray}
A^{\rm reg}_\mu  (x) &=& U ( \partial_\mu + A^{\rm sing}_\mu ) U^{-1} = \partial_\mu UU^{-1} \left( -1 + {\rho^2 \over (x-a)^2 + \rho^2} \right) \nonumber\\
&=&  (U \partial_\mu U^{-1}) {(x-a)^2 \over (x-a)^2 + \rho^2}\ .
\end{eqnarray}
The expressions $U^{-1} \partial_\mu U$ and $U \partial_\mu U^{-1}$ are closely related; in fact, one finds by direct evaluation
\begin{eqnarray}
U \partial_\mu U^{-1} =- \bar\sigma_{\mu\nu} {(x-a)_\nu \over
(x-a)^2}\ .
\end{eqnarray}
Thus the regular one-instanton solution is given by
\begin{eqnarray}
A^{\rm reg}_\mu =- \bar\sigma_{\mu\nu} {(x-a)_\nu \over (x-a)^2 +
\rho^2}\qquad (k=1, {\rm regular})\ .
\end{eqnarray}

Of course, the singular and the regular solution are both selfdual, because
self-duality is a gauge-invariant property, but the field strengths differ by
 a gauge transformation.  Setting $a=0$ for simplicity, one finds for the
field strengths in the regular and singular gauge
\begin{eqnarray}
\begin{array}{ll}
F^{\rm reg}_{\mu\nu} = 2 \bar\sigma_{\mu\nu}
\displaystyle{\rho^2 \over [x^2 + \rho^2 ]^2} &  (k=1, {\rm regular})\ ,\\
F^{\rm sing}_{\mu\nu} = U^{-1} F^{\rm reg}_{\mu\nu}\, U =
-\displaystyle{\frac{i x^\rho \sigma_\rho}{\sqrt {x^2}}\, {2
\bar\sigma_{\mu\nu} \rho^2 \over (x^2 + \rho^2)^2}\,\frac{i
x^\sigma \bar\sigma_\sigma}{\sqrt {x^2}}} & (k=1, {\rm singular})\ .
\end{array}
\end{eqnarray}
It is clear that $F^{\rm reg}_{\mu\nu}$ is selfdual because
$\bar\sigma_{\mu\nu}$ is selfdual, but  also $F^{\rm
sing}_{\mu\nu}$ is selfdual\footnote{  Using some further
identities which follow from the results for $[\gamma_{\mu\nu} ,
\gamma_\rho ]$ and $\{ \gamma_{\mu\nu} , \gamma_\rho \}$
\begin{eqnarray*}
&& \bar\sigma_\mu \sigma_{\nu\rho} = \delta_{\mu\nu} \bar\sigma_\rho
- \delta_{\mu\rho} \bar\sigma_\nu - \epsilon_{\mu\nu\rho\sigma}
\bar\sigma_\sigma\ ;\qquad  \bar\sigma_{\mu\nu} \bar\sigma_\rho =
\delta_{\nu\rho} \bar\sigma_\mu - \delta_{\mu\rho} \bar\sigma_\nu -
\epsilon_{\mu\nu\rho\sigma} \bar\sigma_\sigma\ , \nonumber\\
&& \sigma_{\mu\nu} \sigma_\rho = \delta_{\nu\rho} \sigma_\mu -
\delta_{\mu\rho} \sigma_\nu + \epsilon_{\mu\nu\rho\sigma}
\sigma_\sigma\ ;\qquad \sigma_\rho \bar\sigma_{\mu\nu} =
\delta_{\rho\mu} \sigma_\nu - \delta_{\rho\nu} \sigma_\mu +
\epsilon_{\rho\mu\nu\sigma} \sigma_\sigma\ ,
\end{eqnarray*}
one finds for the $k=1$ singular solution
\begin{eqnarray*}
F^{\rm sing}_{\mu\nu} =  {2 \rho^2 \over (x^2 + \rho^2)^2} \left(
- 2 {x_\mu x_\rho \over x^2} \sigma_{\rho\nu} + 2 {x_\nu x_\rho
\over x^2} \sigma_{\rho\mu} + \sigma_{\mu\nu} \right)
\end{eqnarray*}
In this form the selfduality is no longer manifest. \label{s-ident}} as is clear from
acting with $\e^{\mu\nu\rho\sigma}$ on $\bar\sigma_{\mu\nu}$.

The action for the one-instanton solution is, of course, proportional to the winding number
\begin{equation}
S = - {1 \over 2\ggg^2} \int \tr\, F_{\mu\nu} F_{\mu\nu}\, {\rm
d}^4 x  = - {1 \over 2\ggg^2} \int \tr\, F_{\mu\nu}
{}^*\!F_{\mu\nu}\, {\rm d}^4 x = {8 \pi^2 \over \ggg^2}\ .
\end{equation}
The same result is obtained by direct evaluation of this integral.

The anti-instanton (the solution with $k=-1$) is closely related
to the instanton solution.  Recall that we derived the instanton
solution by making the ansatz $A_\mu = \alpha \sigma_{\mu\nu}
\partial_\nu \ln \phi$, evaluating $F_{\mu\nu}$ and
$^*\!F_{\mu\nu}$ in terms of $\sigma_{\mu\nu}$ matrices, and then
setting $F_{\mu\nu} = \:^*\!F_{\mu\nu}$.  For the anti-instanton
solution we make the ansatz $A_\mu = \beta \bar\sigma_{\mu\nu}
\partial_\nu \ln \phi$.  The expression for  $F_{\mu\nu}$ is
unchanged (except that $A_\mu$ contains $\bar\sigma_{\mu\nu}$
instead of $\sigma_{\mu\nu}$), but  the $\bar\sigma_{\mu\nu}$ are
selfdual instead of anti-selfdual, hence the expression for
$\epsilon_{\mu\nu\rho\sigma} \bar\sigma_{\sigma\tau}$ has opposite
signs from $\epsilon_{\mu\nu\rho\sigma} \sigma_{\sigma\tau}$.  The
equation with $\partial_\mu \partial_\rho \ln \phi$ again cancels
if $F_{\mu\nu} = - \:^*\!F_{\mu\nu}$, which leads to opposite
winding number $(k=-1)$.  The other equation is again $\partial^2
\ln \phi + 2 \beta (\partial \ln \phi)^2 =0$, hence $\beta =
\ft12$ and again $\phi = 1 + \sum^N_{i=1} {\rho^2_i \over
(x-a_i)^2}$.  This yields for the singular-gauge anti-instanton
solution
 \begin{eqnarray}
A^{\rm sing}_\mu =-  \bar\sigma_{\mu\nu} {\rho^2 (x-a)_\nu \over
(x-a)^2 [(x-a)^2 + \rho^2 ]}\ ,\qquad (k= -1, \; {\rm singular})
\label{expression}
\end{eqnarray}
Setting again temporarily $a=0$, we find near $x=0$
\begin{equation}
A^{\rm sing}_\mu \approx - \bar\sigma_{\mu\nu} x_\nu / x^2 = U
\partial_\mu U^{-1} \ ,
\end{equation}
with the same $U=i{\bar \sigma}^\mu x_\mu/{\sqrt {x^2}}$ as
before. Similarly as for the instanton, we have
\begin{eqnarray}
A^{\rm sing}_\mu &=& U \partial_\mu U^{-1} {\rho^2 \over x^2 + \rho^2}  \nonumber\\
A^{\rm reg}_\mu &=&  U^{-1} (\partial_\mu + A^{\rm sing}_\mu ) U \nonumber\\
&=& \partial_\mu U^{-1} U \left( {\rho^2 \over x^2 + \rho^2} -1 \right) \nonumber\\
&=& U^{-1} \partial_\mu U \left( {x^2 \over x^2 + \rho^2} \right)\
. \label{setting}
\end{eqnarray}
Using the expression for $U^{-1} \partial_\mu U$ in \eqn{thatis} one finds
\begin{eqnarray}
A^{\rm reg}_\mu = - \sigma_{\mu\nu} {(x-a)_\nu \over (x-a)^2 +
\rho^2} \qquad (k=-1, {\rm regular}) \label{overr}
\end{eqnarray}
The curvatures for the anti-instanton solution are obtained by interchanging $\sigma_{\mu\nu}$ and $\bar\sigma_{\mu\nu}$ in the instanton solution
\begin{eqnarray}
F^{\rm reg}_{\mu\nu} =2 \sigma_{\mu\nu} {\rho^2 \over [(x-a)^2 +
\rho^2]^2} \qquad (k=-1, {\rm regular})\ .
\end{eqnarray}
So, the only difference between the instanton and anti-instanton
solutions is the exchange between $\sigma_{\mu\nu}$ and
$\bar\sigma_{\mu\nu}$ in $F_{\mu\nu}$ and $A_\mu$. For the
instanton solution, $F^{{\rm reg}}_{\mu\nu}$ and $A_\mu^{{\rm reg}}$ depend on
${\bar \sigma}_{\mu\nu}$, but $A_\mu^{{\rm sing}}$ depends on
$\sigma_{\mu\nu}$, and $F_{\mu\nu}^{{\rm sing}}$ also depends on
$\sigma_{\mu\nu}$ (setting $a=0$ again for notational simplicity),
\begin{equation}
F^{\rm sing}_{\mu\nu} = U\, F^{\rm reg}_{\mu\nu}\, U^{-1} =
\displaystyle{\frac{i x^\rho {\bar \sigma}_\rho}{\sqrt {x^2}}\, {2
\sigma_{\mu\nu} \rho^2 \over (x^2 + \rho^2)^2}\,\frac{-i
x^\sigma \sigma_\sigma}{\sqrt {x^2}}} \quad (k=-1, {\rm singular})\ .
\end{equation}
If one evaluates the product of the $\sigma$ matrices as in footnote
\ref{s-ident}, one finds an expression for $F^{\rm sing}_{\mu\nu}$
in which the anti-selfduality is no longer manifest.

\section{Collective coordinates, the index theorem and fermionic zero modes}

We found in section 2 one-instanton solutions $(k=1)$ in $SU(N)$ with
$4N$ parameters.  The question arises whether these are all the solutions. To
 find this out, one can consider small deformations of the solution,  $A_\mu +
\delta A_\mu$, and study when they preserve selfduality.
Expanding to first order
in the deformation, and using that the variation of a curvature is the
covariant derivative of the variation of the gauge field,
this leads to the condition
\begin{equation} \label{SDdef}
 D_\mu \delta A_\nu - D_\nu \delta A_\mu = {^*\!(D_\mu} \delta A_\nu - D_\nu
\delta A_\mu ) \ ,
\end{equation}
where the covariant derivative depends only
 on the classical solution but not on $\delta A_\mu$.  In addition we require
 that the new solution is not related to the old one by a gauge transformation.
 This can be achieved by requiring that the small deformations are orthogonal
to any small gauge transformation $D_\mu \Lambda$, for any function
$\Lambda$, i.e.
\begin{equation}
\int\,\, {\rm d}^4x \, \tr\, \{(D_\mu \Lambda) \delta A_\mu\} = 0 \ .
\label{xxxx}
\end{equation}
This certainly rules out deformations of the form $\delta A_\mu=D_\mu\Lambda$.
After partial integration the orthogonality requirement leads to the usual
gauge condition in the background field formalism
\begin{equation}
\label{backgr-gauge} D_\mu \delta A^\mu = 0 \ .
\end{equation}

At this point the reader may start feeling uneasy because the conditions \eqn{SDdef} and \eqn{xxxx} may seem too strong.  First of all, the deformation should be a solution but need not be (anti-) selfdual.  Furthermore, the field equation for the fluctuations consists of the sum of a classical piece and a piece from the gauge fixing term, so that,  requiring each part to vanish separately may seem too restrictive.  However, one can prove the following general result
\cite{Brown}. Arbitrary solutions of the fluctuations around an (anti-)
instanton which are square-integrable so that they do not
change the winding number, are themselves also
(anti-) selfdual and transversal. To prove this property, note that the
field equations for the fluctuations read
$D_\mu F^{\mu\nu} (A+\delta A)+D_\nu (D^\mu \delta A_{\mu})=0$.
The second term comes from the gauge-fixing term. Taking the $D^\nu$
derivative, the first term vanishes while the second term yields
$D^2 (D^\mu \delta A_{\mu})=0$, hence $D^\mu \delta A_{\mu}$ on-shell.
The terms in the classical action which are quadractic in the fluctuations
can be written as $-\frac{1}{8}
(f_{\mu\nu}-{}^*f_{\mu\nu})^2$ where $f_{\mu\nu} = D_\mu \d A_\nu - D_\nu \d A_\mu$.  The minimum of the action yields a solution, hence $f_{\mu\nu}={}^*f_{\mu\nu}$ on-shell.  Thus imposing \eqn{SDdef} and \eqn{backgr-gauge} is not too restrictive.

The requirement that $\d A_\mu$ be square integrable is due to the fact that the inner product of zero modes $\d A_\mu$ will later give us the metric or moduli space, which in turn will give us the integration measure of the moduli space.  Also, for the index theorem which will be used to determine the number of zero modes, one needs the $L^2$ norm for fluctuations.  It is remarkable that the zero modes which satisfy the differential equations in (\ref{SDdef}) and (\ref{backgr-gauge}) are all square integrable.

In references \cite{Brown,Bern-et-al} the solutions of
\eqn{SDdef} subject to the condition \eqn{backgr-gauge} were studied using the
Atiyah-Singer index theorem. Index theory turns out to be a useful tool when
counting the number of solutions to a certain linear differential equation of
the form $\hat D T = 0$, where $\hat D$ is some differential operator and $T$
is a tensor. We will elaborate on this in the next subsection and also when
studying fermionic collective coordinates. The ultimate result of
\cite{Bern-et-al} is that there are $4Nk$ solutions, leading indeed to  $4N$
collective coordinates for $k=1$ \cite{Brown}.
An assumption required to apply index
theorems is that the space has to be compact. One must therefore compactify
Euclidean space to a four-sphere $S^4$, as was already discussed in footnote
\ref{Comp}.

\subsection{Bosonic collective coordinates and the Dirac operator}

In this section we will make more precise statements about  the
number of solutions to the selfduality equations by relating it to the index
of the Dirac operator.   The
problem is to determine the number of solutions to the (anti-)selfduality
equations with topological charge $k$.  For definiteness we consider
anti-instantons, so we look for deformations which satisfy an
anti-selfduality equation.

As explained in the last subsection, we study
deformations of a given classical solution $A_\mu^{\rm cl} + \delta A_{\mu}$.
Let us define $\phi_\mu \equiv \delta A_{\mu}$ and
$f_{\mu\nu} \equiv D_{\mu} \phi_\nu - D_\nu \phi_\mu$. The covariant
derivative here contains only $A_\mu^{\rm cl}$.  The constraints can then be
written as
\begin{eqnarray}
\bar\sigma_{\mu\nu} D_\mu \phi_\nu =0 \ ; \qquad D_\mu \phi_\mu =0\ ,
\end{eqnarray}
which are $3+1$ relations. Indeed, more explicitly,
$({\bar \sigma}_{\mu\nu})_\alpha{}^{\alpha'}D_\mu\phi_\nu$ are 3 Lie-algebra
valued expressions because $\alpha,\alpha'=1,2$ and
${\rm tr}{\bar \sigma}_{\mu\nu}=0$.
To prove the first relation, multiply by $\bar\sigma_{\rho\sigma}$ and
take the trace.  Since the trace of $[ \bar\sigma_{\rho\sigma} ,
\bar\sigma_{\mu\nu} ]$ vanishes, while $\{ \bar\sigma_{\rho\sigma} ,
\bar\sigma_{\mu\nu} \} = 2 ( \delta_{\mu\sigma} \delta_{\rho\nu} -
\delta_{\nu\sigma} \delta_{\rho\mu}) - 2 \epsilon_{\rho\sigma\mu\nu}$, one
finds $D_\sigma \phi_\rho - D_\rho \phi_\sigma - \epsilon_{\mu\nu\rho\sigma}
D_\mu \phi_\nu =0$, which is the anti-selfduality condition (3 relations).
Both equations can be written as one simple equation as follows:
\begin{eqnarray}
\bar\sigma_\mu \sigma_\nu D_\mu \phi_\nu =0\ ,
\end{eqnarray}
because $\bar\sigma_\mu \sigma_\nu = \delta_{\mu\nu} + \bar\sigma_{\mu\nu}$,
and the spinor structures of $\delta_{\mu\nu}$ and $\bar\sigma_{\mu\nu}$ are
independent.

Introducing two-component spinor notation with
\begin{eqnarray}\label{cov-spin-der}
 \bar{\rlap{\,/}D} = \bar\sigma_{\mu , \alpha' \beta} D_\mu =
\bar{D}_{\alpha' \beta}\ ;\qquad \sigma_\nu^{\alpha \beta'} \phi_\nu =
\Phi^{\alpha \beta'}\ ,
\end{eqnarray}
the deformations of an anti-instanton can be written as follows
\begin{eqnarray}
\bar{\rlap{\,/}D} \Phi = \bar{D}_{\alpha' \beta} \Phi^{\beta  \gamma'} = 0\ .
\label{aanti}
\end{eqnarray}
Note that $\Phi^{\beta  \gamma'}$ is in the adjoint representation, so
\eqref{aanti} stands for $\partial_{\alpha'\beta}\Phi^{\beta  \gamma'}
+[A_{\alpha'\beta},\Phi^{\beta  \gamma'}]$  $= 0$.
Using the explicit representation of the matrices $\sigma_\mu$ in
\eqref{Eucl-sigma}, we can represent the quaternion $\Phi$ by
\begin{equation}
\Phi = \left( \begin{array}{cc} a & b^* \\ b & - a^*
\end{array} \right) \ ,
\end{equation}
with $a$ and $b$ complex adjoint-valued functions. Then \eqn{aanti} reduces
to two spinor equations, one for
\begin{equation}
\lambda = \left( {a \atop b} \right)\ ; \qquad \bar{\rlap{\,/}D} {\lambda} =
0 \ , \label{Dirac-spinor}
\end{equation}
and one for $i \sigma^2 \lambda^* = \left( \begin{array}{cc}  b^* \\  -a^* \end{array} \right)$. Conversely, for each spinor solution
$\lambda$ to the Dirac equation, one may show that also $i\sigma^2\lambda^*$
is a solution.  (Use $(\bar\sigma^\mu)^* =- \sigma^2 \bar\sigma^\mu
\sigma^2)$.  Indeed, if $\lambda$ yields a deformation $(\delta A_1,\delta A_2,
\delta A_3, \delta A_4)$, then $i \sigma^2 \lambda^*$ corresponds to the
deformation $(\delta A_1',\delta A_2',
\delta A_3', \delta A_4')$ with $\delta A_1'=-\delta A_3, \delta A_3'=
\delta A_1, \delta A_2'=\delta A_4$ and $\delta A_4'=-\delta A_2$. They are not
related by a Lorentz transformation because the coordinates $x^\mu$ are not
transformed. Thus given $\lambda$, we obtain two linearly independent
deformations of the (anti-) instanton.
As we already stressed, the spinors $\lambda$ are in the adjoint
representation.  We shall discuss other representations later.

Given a solution $\lambda$ of the spinor equation, one can still construct
{\bf two} other solutions of the deformation of the anti-instanton, which
differ by a factor $i$
\begin{eqnarray}
\Phi^{(1)} = \left( \begin{array}{cc} a & b^* \\  b & -a^* \end{array} \right)
\; , \;  \Phi^{(2)} = \left( \begin{array}{cc} ia & -ib^* \\ ib & ia^*
\end{array} \right)\ .
\end{eqnarray}
The reason we do not count $i \lambda$ as a different solution for the
spinors but treat $\Phi^{(1)}$ and $\Phi^{(2)}$ as independent has to do
with reality properties: $\delta A^a_\mu$ should be real, and $\Phi^{(1)}$
and $\Phi^{(2)}$ yield different variations $\delta A_\mu$.
Namely, $a = \phi_3 + i \phi_4$ and $b = \phi_1 + i \phi_2$, so
\begin{eqnarray}
&& \Phi^{(1)} :\ \delta A_4 = \phi_4\ , \delta A_3 = \phi_3\ , \delta A_1 =
\phi_1\ , \delta A_2 = \phi_2\ , \nonumber\\
&& \Phi^{(2)} :\ \delta A_4 =  \phi_3\ , \delta A_3 = - \phi_4\ ,
\delta A_1 =- \phi_2\ , \delta A_2 = \phi_1\ .
\end{eqnarray}
It may seem miraculous that we find a second solution without any hard work,
but closer inspection reveals that no miracle is at work: under the
substitutions $\delta A_1\rightarrow \delta A_2,
\delta A_2\rightarrow - \delta A_1, \delta A_3\rightarrow \delta A_4,
\delta A_4\rightarrow -\delta A_3$, one of the anti-selfduality equations is
exchanged with the gauge condition, and the other two duality equations
get interchanged. Also for solitons this way of counting zero modes is
encountered:  for example for vortices one complex fermion zero mode
corresponds to two real bosonic zero modes \cite{RNWimmer}.

In fact, because $\Phi^{(2)} = \Phi^{(1)} i \sigma_3$, one
might wonder whether $\Phi^{(3)} = \Phi^{(1)} (-i \sigma_1)$ and $\Phi^{(4)} =  \Phi^{(1)} (-i \sigma_2)$ yield further solutions.  One obtains
\begin{eqnarray}
\Phi^{(3)} = \left( \begin{array}{cc}   -ib^* & -ia \\ ia^* & -ib \end{array} \right) \; ; \; \Phi^{(4)} = \left( \begin{array}{cc} b^* & -a \\ -a^* & -b \end{array} \right)
\end{eqnarray}
which are just the $\Phi^{(1)}$ constructed from $\sigma_2 \lambda^*$ and $i \sigma_2 \lambda^*$. So there are no further independent solutions \cite{Brown}.  Therefore, the number of solutions for $\Phi$ is twice the number of solutions for a single two-component adjoint spinor. So, the problem of counting the number of bosonic collective coordinates is now translated to the computation of the Dirac index, which we discuss next.

\subsection{Fermionic moduli and the index theorem} 

Both motivated by the counting of bosonic collective coordinates, as
discussed in the last subsection, and by the interest of coupling Yang-Mills
theory to fermions, we study the Dirac equation in the background of an
anti-instanton. We start with a massless four-component complex (Dirac)
fermion
$\psi$, in an arbitrary representation (adjoint, fundamental, etc) of an
arbitrary gauge group
\begin{equation} \gamma_\mu D_\mu  \psi \, = \, \rlap{\,/}D  \psi = 0 \ .
\end{equation}
We recall that a Dirac spinor can be decomposed into its chiral and anti-chiral components
\begin{equation} \psi = \left(  \begin{array}{cc} \lambda^\alpha \\ \bar\chi_{\alpha'} \end{array} \right)\ ;\qquad
\lambda \equiv \ft12 \left( 1 + \gamma^5 \right) \psi \ , \qquad {\bar\chi} \equiv \ft12 \left( 1 - \gamma^5 \right) \psi \ .
\end{equation}
We use the Euclidean representation for the Clifford algebra discussed before
\begin{equation} \label{gamma-sigma} \gamma^\mu = \left( \begin{array}{cc} 0 & -i \sigma^{\mu\, \alpha \beta'} \\ i
\bar\sigma^\mu_{\alpha'\beta} & 0 \end{array} \right) \ , \qquad \gamma^5 =
 \gamma^1\gamma^2\gamma^3\gamma^4 = \left( \begin{array}{cc} 1 & 0 \\ 0 & - 1
\end{array} \right) \ .
\end{equation}
In Euclidean space the Lorentz group decomposes according to $SO(4) = SU(2)
\times SU(2)$. The spinor indices $\alpha$ and $\alpha'$ correspond to the
doublet representations of these two $SU(2)$ factors. As opposed to the case
of Minkowski space, $\lambda^\alpha$ and ${\bar \chi}_\alpha'$ are not in
complex-conjugate
representations.  The Dirac equation then becomes
\begin{equation} \label{DiracEq} \not\!\!\bar{D}  \lambda = 0 \ ,
\qquad \not\!\!{D} \bar\chi = 0 \ ,
\end{equation}
where ${\not\!\!{D}}$ and ${\not\!\!\bar{D}}$ are two-by-two matrixes,
see \eqref{cov-spin-der}, and $\l$ and $\bar\chi$ are independent complex
two-component spinors.  We
now show that in the presence of an anti-instanton, \eqn{DiracEq} has zero
modes for $\lambda$, but not for $\bar\chi$. Conversely, in the background of
an instanton, ${\not\!\!{D}}$ has zero modes, but ${\not\!\!\bar{D}}$ has not.  A zero mode is by definition a solution of the linearized field equations for
the quantum fluctuations {\bf which is normalizable}.  The fermionic fields are
treated as quantum fields (there are no background fermionic fields), so
normalizable solutions of \eqn{DiracEq} are zero modes.

The argument goes as follows.  Given a zero mode $\bar\chi$ for
$\rlap{\,/}D$, it also satisfies ${\not\!\!\bar{D} \not\!\!{D}
\bar\chi = 0}$.  In other words, $\ker \rlap{\,/}D \subset
\ker \left\{ \not\!\!\bar{D}  \not\!\!{D} \right\}$ where ker
denotes the kernel.  Next we evaluate
\begin{equation}
\not\!\!\bar{D} \not\!\!{D} = \bar\sigma_\mu \sigma_\nu D_\mu
D_\nu = D^2 + {1 \over 2} \bar\sigma_{\mu\nu} F_{\mu\nu} \ ,
\label{dictionary}
\end{equation}
where we have used
$\bar\sigma_\mu \sigma_\nu + \bar\sigma_\nu \sigma_\mu = 2
\delta_{\mu\nu}$, and $\bar\sigma_{\mu\nu}$ was defined in
\eqn{mine}.  But notice that the anti-instanton field strength is
anti-selfdual whereas the tensor $\bar\sigma_{\mu\nu}$ is
selfdual, so the second term vanishes.  From this it follows that
$\bar\chi$ satisfies $D^2 \bar\chi = 0$.  Now we can multiply $D^2
\bar\chi$ with its conjugate $\bar\chi^*$ and integrate to get,
after partial integration and assuming that the fields go to zero
at infinity\footnote{ Normalizability of zero modes requires that
$\bar\chi$ tends to zero faster than $1/r^2$ (usually like $1/r^3$
or sometimes $1/r^4$).  Then the boundary term with $\bar\chi^\ast
D_\mu \bar\chi$ indeed vanishes.}, $\int\, {\rm d}^4x \left| D_\mu
\bar\chi \right|^2 = 0$. From this it follows that $\bar\chi$ is
covariantly constant, $D_\mu \bar\chi =0$, and so $F_{\mu\nu}\,
\bar\chi =0$.  Since $F_{\mu\nu}\, \bar\chi = F^a_{\mu\nu}\, T_a\,
\bar\chi$, with $T_a$ the generators of the gauge group $SU(2)$ in a
representation $R$, we conclude that $F^a_{\mu\nu} (x)\, T_a\,
\bar\chi (x)$ must vanish at all points $x$.  Since
$F_{\mu\nu}^a$ is proportional to $\eta_{a \mu\nu}$ (or
$\bar\eta_{a\mu\nu}$), and $\eta_{a\mu\nu} \eta_{b\mu\nu}$ is
proportional to $\delta_{ab}$, we find that $T_a \,\bar\chi (x)$
vanishes for all $a$ and all $x$. Then $D_\mu {\bar \chi}=0$ reduces
to $\partial_\mu {\bar \chi}=0$, and this implies that ${\bar \chi}=0$.
We conclude that $\rlap{\,/}D \bar\chi$ has no square-integrable solutions.
Stated differently, $- D^2$ is a positive definite operator and has no zero
modes. Note that this result is independent of the representation of the
fermion.

For the $\lambda$-equation, we have $\not\!\!{D} \not\!\!\bar{D}  \lambda = 0$,
 i.e. $\ker \not\!\!\bar{D} \subset \ker \left\{ \not\!\!{D} \not\!\!\bar{D}
\right\}$, and we obtain
\begin{equation} \label{DbarD} \not\!\!{D} \not\!\!\bar{D} =
D^2 + {1 \over 2} \sigma_{\mu\nu} F_{\mu\nu} \ .
\end{equation}
This time the second term does not vanish in the presence of an
anti-instanton, so zero modes cannot be ruled out. In fact, there do exist
fermionic zero modes, because we shall construct them.
Knowing that ${\not\!\!{D}}$ has
no zero modes, one easily concludes that $\ker {\not\!\!\bar D} =
\ker \left\{ \not\!\! D \not\!\!\bar D \right\}$ and $\ker {\not\!\! D}
= \ker \left\{ \not\!\!\bar D \not\!\! D \right\}=0$.

For massive spinors no zero modes are possible.  To prove this one may repeat
the same steps as for massless spinors, but now one finds that
$\bar{\rlap{\,/}D}  \l = i m \bar\chi$ and ${\rlap{\,/} D} \bar\chi = -i m \l$,
and iteration yields $\not\!\!\bar{D} \not\!\! D   \bar\chi =
m^2 \bar\chi$.  The crucial observation is that $m^2$ is positive, while $\not\!\!\bar D \not\!\! D$ is negative definite.
Hence, no zero modes exist for massive spinors.

Now we can count the number of solutions using index theorems.
The index of the Dirac operator is defined as
\begin{equation}
{\rm Ind} \not\!\!\bar D =   \dim \ker  \not\!\!{D}
\not\!\!\bar{D} - \dim \ker \not\!\!\bar{D}  \not\!\!{D}\ .
\end{equation}
This index will give us the number of zero modes,
since the second term is zero and since any renormalizable solution of
$\not\!\!{D}\not\!\!\bar{D}  \lambda = 0$ satisfies $\not\!\!\bar{D}
\lambda = 0$ as we have shown. There are several ways to compute
its value.  We begin by writing the index as follows
\begin{equation} \label{Ind-M}
{\rm Ind} \not\!\!\bar{D} =  \lim_{M^2\rightarrow 0}\, Tr
\left\{ \frac{M^2}{-\not\!\!{D} \not\!\!\bar{D} + M^2} -
\frac{M^2}{ -\not\!\!\bar{D} \not\!\!{D} + M^2} \right\} \ ,
\end{equation}
where $M$ is an arbitrary parameter. The trace
Tr stands for a sum over group indices and spinor indices, and includes
an integration over space-time. We shall discuss that this
expression (before taking the limit) is independent of $M$.
This implies that the operators
${\not\!\!{D} \not\!\!\bar{D}}$ and ${\not\!\!\bar{D}
\not\!\!{D}}$ not only have the same spectrum but also the same density
of states
for non-zero eigenvalues\footnote{  One can also (as is customary in the
literature) place the system in a large box to discretize
the spectrum, and let the boundary conditions for the
eigenfunctions of ${\not\!\! D \not\!\!\bar D}$ determine the
boundary conditions for the eigenfunctions of ${\not\!\!\bar D
\not\!\! D}$, and vice-versa, such that the non-zero eigenvalues are the
same. Such a treatment for the kink has been worked out in detail in
\cite{RNWimmer}. However, in the limit of infinite volume, the densities of
states can become different, as we shall discuss.}.
That they have the same non-zero eigenvalues is clear:
if $\psi$ is
an eigenfunction of ${\not\!\!\bar{D} \not\!\!{D}}$, then
${\not\!\!{D}} \psi$ is an eigenfunction of ${\not\!\!{D}
\not\!\!\bar{D}}$ with the same nonvanishing eigenvalue and
${\not\!\!{D}}\psi$ does not vanish. Conversely, if $\psi$ is an
eigenfunction of ${\not\!\!{D} \not\!\!\bar{D}}$ with nonzero eigenvalue,
then ${\not\!\!\bar{D}} \psi$ does not vanish and is an eigenfunction
of ${\not\!\!\bar{D} \not\!\!{D}}$ with the same nonvanishing
eigenvalue.

To show that \eqref{Ind-M} is independent of $M^2$, we rewrite
the index  in terms of four-dimensional Dirac
matrices,
\begin{equation}
I(M^2) \equiv {\rm Ind}
\not\!\!\bar{D} = Tr  \left\{ \frac{M^2}{- \not\!\!{D}^2 + M^2}\,\gamma_5 \right\} \ ,
\end{equation}
where now $\rlap{\,/}D = D_\mu \gamma^\mu$.
\eqa
\rlap{\,/}D_{4 \times 4} = \left( \begin{array}{cc} 0 & \rlap{\,/}D_{2 \times 2} \\ \bar{\rlap{\,/}D}_{2 \times 2} & 0 \end{array} \right)
\eqae
We rewrote the trace of the two terms in (\ref{Ind-M}) over a two-dimensional spinor space as the trace of one term over a four-dimensional spinor space.  It has been argued that
independence of $M^2$ follows by taking
the $M^2$-derivative (see \cite{Brown}),
\begin{equation}
\frac{\partial}{\partial M^2}I(M^2)=
- Tr \left\{ \frac{\not\!\!{D}^2}{(- \not\!\!{D}^2+M^2)^2}\gamma_5\right\} \ .
\end{equation}
Using that $\gamma_5$ anticommutes with $\not\!\! D$ and that
the trace is cyclic, we find
\begin{equation}
- Tr \,\frac{\not\!\! D\not\!\! D \gamma_5}{A}= Tr \,\frac{\not\!\! D \gamma_5\not\!\! D}{A}=
Tr \frac{\not\!\! D\not\!\! D\gamma_5}{A}\ ,
\end{equation}
where $A=(- \not\!\!{D}^2 + M^2)^2$. Hence $Tr (\not\!\!{D}^2 \gamma_5/A)$
would seem to vanish and this would prove that $I(M^2)$ is independent of
$M^2$. The problem with this
proof is that one can give a counter example:
one can repeat all the steps for the supersymmetric kink, and this would then
imply that the densities for chiral and anti-chiral fermion modes are
equal. However, one can directly calculate these densities for the
supersymmetric kink, and one then finds that they are different
\cite{Rebhan:2002yw}
\begin{equation}
\Delta \rho(k^2)=-\frac{2m}{k^2+m^2}\ ,
\end{equation}
where $m$ is the mass of the fluctuating fields far away from the kink.
Applied to the case of instantons, the situation was considered in
\cite{Brown}. In \cite{Weinberg:1979ma,Weinberg:1981eu,Rebhan:2006fg} it
was noted that the proof of
\cite{Brown} was incomplete. Cyclicity of the trace (on which the proof
in \cite{Brown} that $\Delta \rho(k^2)$ vanishes is based), breaks down due
to the presence of massless fluctuating fields\footnote{ One would expect
that at the regularized level the trace is cyclic but one may expect
that one should also regularize infrared aspects of the problem.   Consider for example quantum mechanics for a harmonic oscillator with mass term $\ft12 m^2q^2$.  Define $a= \sqrt{{m \over 2}} q+ i p/ \sqrt{2m}$ and $a^\dagger = \sqrt{{m \over 2}} q - ip / \sqrt{2m}$.  For $m$ tending to zero, the vacuum is annihilated by $p + \co (m)$  but the vacuum becomes non-normalizable when $m$ vanishes.  Still, at finite $m$, $tr [p,q]=0$.   }. One can
directly compute $I(M^2)$, using a more detailed index theorem
\cite{Weinberg:1981eu,Rebhan:2006fg}, and then finds that
the densities of chiral and antichiral fermionic modes in
an instanton background are equal,
\begin{equation}
\Delta \rho (k^2)=0 \ \qquad {\rm for\,\,\, instantons}\ .
\end{equation}

Given that  the density of states
of the operator ${\not\!\!{D} \not\!\!\bar{D}}$ in \eqref{Ind-M} is the same
as the density of states of the operator ${\not\!\!\bar{D}
\not\!\!{D}}$, there is a pairwise cancellation in
\eqn{Ind-M} coming from the sum over eigenstates with non-zero
eigenvalues, both for the discrete and continuous spectrum. So the only
contribution is coming from the zero
modes, for which the first term in \eqref{Ind-M} simply gives one for each zero
mode, and the second term vanishes because there are no zero
modes. The result is then clearly an integer, namely $\dim \left\{ \ker
\not\!\!\bar{D} \right\}$.  Since $I(M^2)$ is independent of $M^2$, one
can evaluate it in the large $M^2$ limit instead of the small
$M^2$ limit. The calculation is then identical to the calculation of the chiral
anomaly, which we now review.

The chiral anomaly is equal to the regulated trace of the matrix
$\gamma_5$. It can be written as
\begin{eqnarray}
{\rm Ind}  \not\!\!\bar{D} = \tr\, \int {\rm d}x < x \mid {M^2 \over -\rlap{\,/}D \bar{\rlap{\,/}D} + M^2} - {M^2 \over -\bar{\rlap{\,/}D} \rlap{\,/}D + M^2}
\mid x >\ ,
\label{replace}
\end{eqnarray}
where tr denotes the trace over group indices and spinor indices.
Because $\g_5={\mbox{diag}}(+1,+1,-1,-1)$ one finds in
\eqn{replace} a relative minus sign between the first and the second term.
We have chosen a quantum mechanical representation for the trace in a Hilbert
space spanned by the eigenfunctions $|x>$ of the position operator.
The operators $D_\mu$ depend on the operators $\hat{x}^\mu$ and the operators
$\hat{p}_\mu$.  When $\hat{x}$ reaches $| x>$, it becomes a $c$-number $x$.  Similarly $\hat{p}_\mu | x> = - {\hbar \over i} {\partial \over \partial x^\mu} | x>$.  The latter statement follows by contracting with a complete set of momentum eigenstates, using that $\:\;$ $< k | \hat{p}_\mu = \hbar k_\mu < k |$ because $\hat{p}_\mu$ is hermitian
\begin{eqnarray}
&& < k \mid \hat{p}_\mu \mid x> = \hbar k_\mu <k \mid x  > =
\hbar k_\mu {{\rm e}^{-ikx} \over (2 \pi)^2} \nonumber\\
&& =- {\hbar \over i} {\partial \over \partial x^\mu} {{\rm e}^{-ikx} \over (2 \pi)^2} =- {\hbar \over i} {\partial \over \partial x^\mu} < k \mid x> = <k \mid {-\hbar \over i} {\partial \over \partial x^\mu} \mid x>\ .
\end{eqnarray}
So, from now on we will replace the operators $D_\mu ( \hat{x} , \hat{p}_x )$ by $D_\mu (x, - {\hbar \over i} {\partial \over \partial x})$.  These ${\partial \over \partial x}$ act on the $x$ in $|x>$ and do not act on $| k>$.

Let us now insert a complete set of eigenstates of $\bar{\rlap{\,/}D} \rlap{\,/}D$ and $\rlap{\,/}D \bar{\rlap{\,/}D}$, respectively.  The index becomes then
\begin{eqnarray}
{\rm Ind} \; \bar{\rlap{\,/}D} = \tr\, \sum_{m,n} \int {\rm d}x <x \mid n_L >< n_L \mid \cO_L \mid m_L >< m_L \mid x> - \; \mbox{same with} \; L \leftrightarrow R \ ,\nonumber\\ &&
\end{eqnarray}
where $\cO_L$ is the first operator in (\ref{replace}) and $\cO_R$ the second.  As we already discussed the eigenfunctions $< x |n_L > = \varphi^{(L)}_n (x)$ and $< x | n_R >= \varphi^{(R)}_n (x)$ have the same nonvanishing eigenvalues $\lambda_n$ and the same densities.

So the eigenfunctions with nonzero eigenvalues do not contribute to the index.  (Note that it does not make sense to look for eigenfunctions of $\rlap{\,/}D$ or $\bar{\rlap{\,/}D}$ because these operators change the helicity of the spinors).  There are in general a finite number of zero modes in the $L$ sector but none in the $R$ sector.  Hence
\begin{eqnarray}
{\rm Ind} \; \bar{\rlap{\,/}D} &=& \int {\rm d}^4 x \left( \sum_n
\varphi^{(L)}_n (x) \varphi^{(L)}_m (x)^* - \sum_m \varphi^{(R)}_n (x)
\varphi^{(R)}_m (x)^* \right) {M^2 \delta_{mn} \over \lambda^2_n + M^2}
\nonumber\\
&+& \sum_\alpha \int {\rm d}^4x \varphi^{(L)}_\alpha (x) \varphi^{(L)}_\alpha (x)^* = n^{(L)}\ ,
\label{sector}
\end{eqnarray}
where $\varphi^{(L)}_\alpha (x)$ are the (square-integrable) zero modes, and $n^{(L)}$ is the number of these.   A sum over spinor indices is taken in \eqn{sector}.

To actually compute the index (namely, to compute the integer $n^{(L)}$), we use momentum eigenstates instead of eigenfunctions of $ \rlap{\,/}D \bar{\rlap{\,/}D}$ and $\bar{\rlap{\,/}D}  \rlap{\,/}D$
 \begin{eqnarray}
{\rm Ind} \; \bar{\rlap{\,/}D} = \int {\rm d}^4 x\, \int {\rm d}^4k
\int {\rm d}^4k^\prime \; \tr\,  <x \mid k^\prime >< k^\prime \mid {M^2 \over - \rlap{\,/}D^2 + M^2} \gamma_5 \mid k ><k \mid x>\ , \nonumber\\ &&
\end{eqnarray}
where we recall that $\gamma_5 = \left( \begin{array}{cc} 1& 0 \\ 0 & -1 \end{array} \right)$ and $\rlap{\,/}D = \left( \begin{array}{cc} 0 & -i  \rlap{\,/}D \\ i \bar{\rlap{\,/}D}  & 0 \end{array} \right)$.

As we have discussed, the operator $D_\mu = {\partial \over \partial x^\mu} + [A_\mu (x) , \cdot ]$
acts on the coordinates $x$ in $| x>$ but not on the $k$ in $| k><k |$, and
the trace $\tr$ sums over the group indices and the spinor indices of $\gamma^\mu$ in $\rlap{\,/}D =
\gamma^\mu D_\mu$ and $\gamma_5$.  Using $<k |x> = {\rm e}^{-ikx} / (2 \pi)^2$
and pulling these plane waves to the left, the derivatives $\partial_\mu$ act
on the $c$-numbers $x$ in ${\rm e}^{-ikx}$ and are replaced by
$\partial_\mu -ik_\mu$.  The matrix element $<k^\prime | M^2 (- \rlap{\,/}D^2
+ M^2)^{-1} \gamma_5 \mid k>$ is equal to $<k^\prime \mid k>$ times the
operator $[ M^2 / (- \rlap{\,/}D^2 + M^2) ] \gamma_5$ and
$<k^\prime |k>= \delta^4 (k-k^\prime)$.  When the plane wave $e^{-ikx}$ has been pulled all the way to the left, the plane waves
${\rm e}^{ik^\prime x}$ and ${\rm e}^{-ikx}$ in $<x|k'>={\rm e}^{ik'x}/
(2\pi)^2$ and $<k|x>={\rm e}^{-ikx}/(2\pi)^2$ cancel each other, and one is
left with
\begin{eqnarray}
{\rm Ind} \; \bar{\rlap{\,/}D} = \int {\rm d}^4 x \int {{\rm d}^4 k \over
(2 \pi)^4} \tr\, \left\{ {M^2 \over - (-i \rlap{/}k + \rlap{\,/}D)^2 + M^2}
\gamma_5\right\}\ .
\end{eqnarray}
The denominator can be written as
\begin{eqnarray}
{1 \over (k^2 +M^2) - (-2i k \cdot D + D_\mu D_\mu +
\ft12 \gamma_\mu \gamma_\nu F_{\mu\nu})}\ ,
\end{eqnarray}
and we can exhibit the $M^2$ dependence by rescaling $k_\mu = M \kappa_\mu$,
yielding
\begin{equation}
{\rm Ind} \; \bar{\rlap{\,/}D} = \int {\rm d}^4 x\,  M^4 \int
{d^4 \kappa \over (2 \pi)^4}\,\tr\,\left\{ {1 \over (\kappa^2 +1) - \left( - {2i \kappa_\mu D_\mu \over M} + {D_\mu D_\mu \over M^2} + \ft12 {\gamma_\mu \gamma_\nu F_{\mu\nu} \over M^2} \right) } \gamma_5\right\}\ .
 \end{equation}
Expanding the denominator, only terms due to expanding two, three or four
times can contribute in the limit $M \rightarrow \infty$, but only the terms
with at least four Dirac matrices can contribute to the trace due to the
matrix $\gamma_5$.  Thus we only need retain the square of $\ft12 \gamma_\mu
\gamma_\nu F_{\mu\nu}$, and the index becomes
\begin{eqnarray}
{\rm Ind} \; \bar{\rlap{\,/}D} &=& \int {\rm d}^4 x \int {{\rm d}^4 \kappa
\over (2 \pi )^4} {1 \over (\kappa^2 +1)^3} \tr\,\left(\ft12 F_{\mu\nu} \gamma_{\mu\nu} \ft12 F_{\rho\sigma} \gamma_{\rho\sigma} \gamma_5\right) \nonumber\\
&=& \int {\rm d}^4 x {2 \pi^2 \over (2 \pi)^4} \int^\infty_0 {r^3 {\rm d}r \over (r^2 + 1)^3} \ft14 (\tr\,  T_a T_b) (\tr \gamma_{\mu\nu} \gamma_{\rho\sigma} \gamma_5) F^a_{\mu\nu} F^b_{\rho\sigma} \nonumber\\
&=& \int {\rm d}^4 x {1 \over 32 \pi^2} (\tr\, T_aT_b) (\epsilon_{\mu\nu\rho\sigma} F^a_{\mu\nu} F^b_{\rho\sigma})\ ,
 \end{eqnarray}
where we used that $\int {\rm d} \Omega_\mu = 2 \pi^2$ and $\int^\infty_0
{r^3 {\rm d}r \over (r^2 + 1)^3} = \ft14$.  Note that both a trace over group indices and a trace over spinor indices has been taken.  The result for the index is twice the product of the winding number in \eqn{winding} and a group theory factor
\begin{equation}
{\rm Ind} \; \bar{\rlap{\,/}D} = 2 \left( {1 \over 32 \pi^2} \int {\rm d}^4 x
F^a_{\mu\nu}\:^* F^b_{\mu\nu} \right) \tr\, T_aT_b\ .
\end{equation}
For a representation $R$ of $SU(N)$ for the fermions, we
define $\tr\,{T_{a}^{R}T_{b}^{R}} =- \delta_{ab} T(R)$.  By definition one has
$T(R) = \ft12$ for the fundamental representation, and then $T(R) =N$
for the adjoint representation\footnote{ To compute $T(R)$ for
the adjoint representation, write the carrier space for the
adjoint representation of $SU(N)$ as $u^i \bar{v}_j-\frac{1}{N}\delta^i{}_j
(u^k{\bar v}_k)$. Then, for $i\neq j$, $T^{\rm
adj}_a u^i \bar{v}_j = (T^{(f)}_a)^i{}_{i'} u^{i'}  \bar{v}_j +
(T_a^{f^*})_j{}^{j\prime} u^i \bar{v}_{j\prime}$.  For a diagonal
generator $A$ of the fundamental representation of $SU(N)$ with entries
$(i \alpha_1 , \ldots , i
\alpha_N)$ with real $\alpha_j$ one has $A u^i = i \alpha^i u^i \;
{\rm and} \; A u^i \bar{v}_j = (i \alpha^i - i \alpha^j ) u^i
\bar{v}_j$, so $A(u^i{\bar v}_j-\frac{1}{N}\delta^i_ju^k{\bar v}_k)=(i\alpha^i-i\alpha^j)(u^i{\bar v}_j-\frac{1}{N}\delta^i_ju^k{\bar v}_k)$ and
$\sum \alpha_i =0$. Hence $\tr\,
A^2 =- \sum^N_{i=1} (\alpha^i)^2$ for the fundamental
representation, but $\tr A^2 =- \sum_{i,j}
(\alpha^i - \alpha^j)^2$ for the adjoint representation.  The
latter sum can also be written as
\begin{eqnarray*}
\sum^N_{i,j=1} (\alpha_i - \alpha_j)^2 = \left( \sum \alpha_i^2 \right) N- 2\left( \sum \alpha_i \right) \left( \sum \alpha_j \right) + \left( \sum \alpha^2_j
\right) N = 2\left( \sum \alpha^2_i \right) N\ .
\end{eqnarray*}
So $T (R^{\rm adj}) = 2 N T (R^f)$.}. Hence, finally,
\begin{eqnarray}
{\rm Ind} \; \bar{\rlap{\,/}D}  &=& |k| \qquad
\mbox{for the fundamental representation,} \nonumber\\
&=& 2N|k|  \qquad \mbox{for the adjoint representation}\ .
\label{adjoint}
\end{eqnarray}
(For an anti-instanton, $k$ is negative. The factor 2 corresponds to
our earlier observation that $i\sigma_2\lambda^*$ is also a zero mode if
$\lambda$ is a zero mode.)
Furthermore, as shown in the last subsection, an (anti-) instanton in $SU(N)$
has twice as many bosonic collective coordinates as there are fermionic zero
modes in the adjoint representation.  This proves that there are $4 N k$
bosonic collective coordinates for an instanton with winding number $k$ and
gauge group $SU(N)$.

\section{Construction of zero modes}

In two later sections we will show how to set up and do (one-loop)
perturbation theory around an (anti-) instanton. This will require the
reduction of the path integral
measure over instanton field configurations to an integral over the moduli
space of collective coordinates. In order to achieve this we need to know
the explicit form of the bosonic and fermionic zero modes. This is the content
of this section. We follow closely \cite{Bernard}.

\subsection{Bosonic zero modes and their normalization}

In order to construct the bosonic zero modes and discuss perturbation theory,
we first
decompose the fields into a background part and quantum fields
\begin{equation} \label{pert} A_\mu = A_\mu^{\rm cl} (\gamma) + A_\mu^{\rm qu}
\ .
\end{equation}
Here $\gamma_i$ denote a set of collective coordinates,
and, for gauge group $SU(N)$, $i = 1, \dots, 4Nk$. Before we make the
expansion of the action, we should first fix the gauge and introduce ghosts,
$c$,
and anti-ghosts, $b$. We choose the background gauge condition
\begin{equation}
D_\mu^{\rm cl} A_{\mu}^{\rm qu} = 0 \ .
\end{equation}
The gauge-fixing term is then $\cl_{\rm {fix}} = -{1 \over g^2}\tr
(D_\mu A^{{\rm qu}}_\mu)^2$ and the ghost action is $\cl_{\rm {ghost}} = -
b^a (D_\mu (A^{\rm cl}_\mu ) D_\mu (A^{\rm cl}_\mu + A^{\rm qu}_\mu ) c)^a$.
The action, expanded through quadratic order in the quantum fields, is of the form
\begin{equation} \label{gf-action}
S = \frac{8\pi^2}{\ggg^2}  \mid k \mid + \frac{1}{\ggg^2} \tr\, \int\, {\rm d}^4 x \left\{ A_\mu^{\rm qu}\, M_{\mu\nu} \, A_\nu^{\rm qu} + 2 b\, M^{\rm gh}\, c \right\}\ ,
\end{equation}
with $M^{\rm gh} = D^2$ and
\begin{eqnarray}
M_{\mu\nu} &=& \left( D^2 \delta_{\mu\nu} - D_\nu D_\mu + F_{\mu\nu} \right) + D_\mu D_\nu \equiv M_{\mu\nu}^{(1)} + M_{\mu\nu}^{(2)}\ , \nonumber\\
&=& D^2 \delta_{\mu\nu} + 2 F_{\mu\nu}\ ,
\end{eqnarray}
where we have dropped the subscript ${\rm cl}$.   Here, $M^{(1)}$ stands
for the quadratic operator coming from the classical action, and
$M^{(2)}$ is due to the gauge fixing term\footnote{ To arrive at
this expression for $M_{\mu\nu}^{(1)}$, use that $F_{\mu\nu} =
F^{{\rm cl}}_{\mu\nu} + (D^{\rm cl}_\mu A^{\rm qu}_\nu - D^{\rm cl}_\nu
A^{\rm qu}_\mu )
+ [A^{\rm qu}_\mu , A^{\rm qu}_\nu ]$ and note that $- {1 \over 2g^2} \tr
2 F^{\rm cl}_{\mu\nu} [A^{\rm qu}_\mu , A^{\rm qu}_\nu ] = {1 \over g^2} \tr
A^{\rm qu}_\mu [F^{\rm cl}_{\mu\nu} , A^{\rm qu}_\nu ]$.}.  (Recall that
$F_{\mu\nu}$ acts on $A^{\rm qu}_\nu$ as $[F_{\mu\nu} , A^{\rm
qu}_\nu ]$).  In an expansion as in \eqn{gf-action}, one
encounters  zero modes (i.e.\ normalizable eigenfunctions of the
operator $M_{\mu\nu}$ with zero eigenvalues).  They are of the
form
\begin{equation} \label{zeromode}
Z^{(i)}_\mu \equiv
\frac{\partial A_\mu^{\rm cl}}{\partial \gamma_i} + D_\mu^{\rm
cl} \Lambda^i \ ,
\end{equation}
where the gauge parameter $\Lambda^i$  is
chosen to keep $Z_\mu$ in the background gauge, so that
\begin{equation} \label{BGC}
D_\mu^{\rm cl} Z_\mu^{(i)} = 0 \ .
\end{equation}
The first term in \eqn{zeromode} is a solution of
$M^{(1)}$ (i.e. an eigenfunction with zero eigenvalue), as follows from
taking the
derivative with respect to
$\gamma_i$ of the field equation.  Namely, ${\delta S^{\rm cl}/
\delta A^{\rm cl}_\mu} =0$ for all $\gamma_i$, so
\eqa \label{zmode}
0 = {\partial \over \partial \gamma_i} {\delta S^{\rm cl} \over \delta A^{\rm cl}_\mu (x)} &=& \int {\delta^2 S^{\rm cl} \over \delta A^{\rm cl}_\nu (y) \delta A^{\rm cl}_\mu (x)} \partial_{\gamma_i} A^{\rm cl}_\nu (y) {\rm d}^4 y \ .
\eqae
The term $D_\mu \Lambda$ is also a solution
of $M^{(1)}$, since it is a pure gauge
transformation.\footnote{ This is also easy to prove by direct
calculation: $(D^2 \delta_{\mu\nu} - D_\nu D_\mu +
F_{\mu\nu}) D_\nu \Lambda$ is equal to $D_\nu [D_\nu ,
D_\mu ] \Lambda + F_{\mu\nu} D_\nu \Lambda$, and this vanishes
since $[ D_\nu , D_\mu ] = F_{\nu\mu}$ and $D_\nu F_{\nu\mu}
=0$.   More generally, ${\delta S^{\rm cl} / \delta  A^{\rm
cl}_\nu} \sim D^{\rm cl}_\mu F^{\rm cl}_{\mu\nu}$ is
gauge-covariant, hence $D^{\rm cl}_\mu F_{\mu\nu}^{\rm cl}
(A_\rho + D_\rho \Lambda) - D^{\rm cl}_\mu F_{\mu\nu}^{\rm cl}
(A_\rho) = [D^{\rm cl}_\mu F^{\rm cl}_{\mu\nu}
, \Lambda ]$ which vanishes on-shell (field equations transform
into field equations).  Hence ${\delta^2 S^{\rm cl} \over \delta
A^{\rm cl}_\mu \delta A^{\rm cl}_\rho} (D_\rho
\Lambda)  = M^{(1)}_{\mu\nu} D_\nu \Lambda$ vanishes.} The sum of
the two terms is also a solution of $M^{(2)}$, because $\Lambda$
is chosen such that $Z_\mu$ is in the background gauge.
As we shall show, the solutions in \eqref{zeromode} are normalizable, hence
they are zero modes.
Due to these zero modes, we cannot integrate over all quantum
fluctuations, since the corresponding determinants would vanish
and yield divergences in the path integral. They must therefore be
extracted from the quantum fluctuations, in a way we will describe
in a more general setting in the next subsection. It will turn out
to be important to compute the matrix of inner products
\begin{equation} \label{norms} U^{ij} \equiv \langle Z^{(i)} |
Z^{(j)} \rangle \equiv - \frac{2}{\ggg^2} \int \, {\rm d}^4x \, \tr\,
\left\{ Z_\mu^{(i)} Z^{\mu (j)} \right\}  = {1 \over g^2} \int
Z^{(i) a}_\mu Z^{(j) a}_\mu {\rm d}^4 x\ .
\end{equation}
We put a factor ${1 \over g^2}$ in front of the usual $L^2$ inner product because the metric $U^{ij}$ will be used to construct a measure $(\det U^{ij})^{1/2}$ for the zero modes, and this measure is also needed if one considers the quantum mechanics of zero modes $\g_i (t)$.  The action for these time-dependent $\g_i (t)$ is $U^{ij} \g_i \dot{\g}_j$ with the same prefactor ${1 \over g^2}$ as in the Yang-Mills gauge action.

We now evaluate this matrix for the anti-instanton.  For the four translational
zero modes, one can easily keep the zero mode in the background
gauge by choosing $\Lambda^i = A_\nu^{\rm cl}$. Indeed,
\begin{equation} Z_{\mu}^{(\nu)} = \frac{\partial A_\mu^{\rm
cl}}{\partial x_0^\nu} + D_\mu A_{\nu}^{\rm cl} = - \partial_\nu
A_\mu^{\rm cl} + D_\mu A_{\nu}^{\rm cl} = F_{\mu\nu}^{\rm cl} \
, \end{equation} which satisfies the background gauge condition.
The norms of these zero modes are
\begin{equation}
U^{\mu\nu} = \frac{8\pi^2|k|}{\ggg^2} \delta^{\mu\nu} = S_{\rm cl}\,
\delta^{\mu\nu}\ .
\end{equation}
As indicated, this result actually holds for
any $k$, and arbitrary gauge group.

Next we consider the
dilatational zero mode corresponding to $\rho$ and limit ourselves
to $k = -1$.  Taking the derivative with respect to $\rho$ leaves
the zero mode in the background gauge, so we can set $\Lambda^\rho
= 0$.  In the singular gauge of \eqn{expression} we have
\begin{equation} Z_\mu^{(\rho)} = - 2 \,\frac{\rho\,
\bar\sigma_{\mu\nu} \, x_\nu}{(x^2 + \rho^2)^2} \ . \end{equation}
(To show that \eqn{BGC} is satisfied, note that $(\partial/
\partial x^\mu) Z^{(\rho)}_\mu =0$ since $\bar\sigma_{\mu\nu}$ is
antisymmetric, while $[A^{\rm cl}_\mu , Z^{(\rho)}_\mu ] =0$ since
both involve $\bar\sigma_{\mu\nu} x^\nu$).   Using \eqn{AntiComm}
and \eqn{Integral},  one easily computes that \begin{equation}
U^{\rho \rho} = \frac{16\pi^2}{\ggg^2} = 2 S_{\rm cl} \ .
\end{equation} In regular gauge one finds $Z^{(\rho)}_\mu = {2
\rho \sigma_{\mu\nu} x^\nu \over (x^2 + \rho^2)^2}$ which has
clearly the same norm.  This result can also be derived from
${\partial \over \partial \rho} A^{\rm reg}_\mu (k=-1) = {\partial
\over \partial \rho} U^{-1} (\partial_\mu + A^{\rm sing}_\mu
(k=-1)) U$ and the identity $U^{-1} \bar\sigma_{\mu\nu} x^\nu U =-
\sigma_{\mu\nu} x^\nu$.

The gauge-orientation zero modes can be obtained from
\eqn{1inst-angles}. By expanding $U(\theta) = \exp( \theta^a T_a)$
infinitesimally in \eqn{1inst-angles} we get to lowest order in
$\theta$ (the case of general $\theta$ will be discussed shortly)
\begin{equation} \label{angle-zm} \frac{\partial A_\mu}{\partial
\theta^a} =  \left[ A_\mu, T_a \right] \ , \end{equation} which is
not in the background gauge (the matrices $T_a$ are in the fundamental
representation). To satisfy \eqn{BGC} we have to add
appropriate gauge transformations, which differ for different
generators of $SU(N)$. First, for the $SU(2)$ subgroup
corresponding to the instanton embedding, we add, for the
singular gauge,
\begin{equation}
\label{L-su2} \Lambda_a = - \frac{\rho^2}{x^2+\rho^2} \, T_a \ ,
\end{equation} and find that \begin{equation} \label{su2zm}
Z_{\mu\,(a)} = D_\mu \left[ \frac{x^2}{x^2 + \rho^2} T_a
\right] \ .
\end{equation}
(using $\partial_\mu {x^2 \over x^2 +
\rho^2} = - \partial_\mu {\rho^2 \over x^2 + \rho^2}$).     One
can now show, using \eqn{eta-eta3}, that the zero mode \eqn{su2zm}
is in the background gauge, and its norm reads\footnote{ A few
details may be helpful.  One finds for this zero mode in the singular gauge,
using \eqref{expression} and \eqref{obeyingg},
\begin{eqnarray*}
Z^{(a)}_\mu = 2  x_\mu \rho^2 (x^2 + \rho^2)^{-2} T_a + 2 \eta_{b\mu\nu} \epsilon_{bac} T_c x_\nu \rho^2 / (x^2 + \rho^2)^2 .
\end{eqnarray*}
It is covariantly transversal: $\partial_\mu$ acting on the first term plus
the commutator of $A^{\rm sing}_\mu$ with the second term vanishes upon
using (\ref{eta-eta3}).  (The commutator of the first term with $A^{\rm sing}_\mu$ is proportional to $(\bar\s_{\mu\nu} x^\nu)x^\mu$ and vanishes).  The norm is due to integrating the sum of the square
of the first term and the second term, using (\ref{Integral}) with $n=1$ and
$m=4$.  All terms which contribute to $Z^{(a)}_\mu$, namely ${\partial \over
\partial \theta^a} A^{\rm sing}_\mu$ and $\partial_\mu \Lambda^{a, {\rm sing}}$
 and $[A^{\rm sing}_\mu , \Lambda^{a, {\rm sing}} ]$ fall off as $1/r^3$ for
large $|x|$, and $Z^{(a)}_\mu$ itself is nonsingular at $x=0$.

In regular gauge one finds from \eqn{setting}
\begin{eqnarray*}
\partial_\gamma A^{\rm reg}_\mu = \partial_\gamma U^{-1} (\partial_\mu +A^{\rm sing}_\mu ) U = U^{-1}  \partial_\gamma A^{\rm sing}_\mu U\ ,
\end{eqnarray*}
and the transversality condition becomes
\begin{eqnarray*}
&& U^{-1} D_\mu (A^{\rm sing}) U [ U^{-1} \partial_\gamma A^{\rm sing}_\mu U + U^{-1} D_\mu (A^{\rm sing} ) U U^{-1} \Lambda^{a, {\rm sing}} U ] \nonumber\\
&& = D_\mu (A^{\rm reg}) [ \partial_\gamma A^{\rm reg}_\mu + D_\mu (A^{\rm reg}_\mu ) U^{-1}  \Lambda^{a, {\rm sing}} U] =0\ .
\end{eqnarray*}
Hence, $\Lambda^{a, {\rm reg}} = U^{-1} \Lambda^{a, {\rm sing}} U$, and now all contributions to $Z^{(a)}_\mu$ in the regular gauge fall only off as $1/r$.  Only their sum $Z^{(a), {\rm reg}}_\mu$ falls off as $1/r^3$, just as $Z^{(a), {\rm sing}}_\mu$.  It is clearly simpler to work in the singular gauge, because then all integrals separately converge.}
\begin{equation}
U_{ab} = \frac{4\pi^2}{g^2}\rho^2\delta_{ab}
=\ft12 \delta_{ab}  \rho^2 S_{\rm cl} \ .
\end{equation}

We need the gauge-orientation zero modes for arbitrary values of $\theta$
because this is needed for the group (Haar) measure.  They are obtained as
follows.
By differentiating $U(\theta)$ and using that $U^{-1} {\partial \over \partial
 \theta^\alpha} U$ is equal to $e_\alpha{}^a (\theta) T_a$, where the function
$e_\alpha{}^a (\theta)$ is called the group vielbein (with $\alpha$ a curved
and $a$ a flat index according to the usual terminology\footnote{ The group
vielbein is given by
\begin{eqnarray*}
e_\alpha{}^a (\theta) T_a = T_\alpha + {1 \over 2!} [T_\alpha, \theta \cdot T ] + {1 \over 3!} [( T_\alpha , \theta \cdot T]  , \theta \cdot T ] + \cdots\ ,
\end{eqnarray*}
whereas the adjoint matrix representation $M^{\rm adj}(\theta)$ is given by
\begin{eqnarray*}
{\rm e}^{- \theta \cdot T} T_a {\rm e}^{\theta \cdot T} = M^{\rm adj} (\theta)_a{}^b T_b = T_a +  [T_a , \theta \cdot T ] + \cdots\ .
\end{eqnarray*}
One has $M^{\rm adj} (\theta)_a{}^b = (\exp \theta^c f_{\cdot \; c}{}^\cdot )_a{}^b$.  There is a relation between the group vielbein and the adjoint matrix:  $\left( \theta^\beta {\partial \over \partial \theta^\beta} + 1 \right) e_\alpha{}^b(\theta) = (M^{\rm adj} (\theta) )_\alpha{}^b$}),
one obtains
\begin{eqnarray}
{\partial \over \partial \theta^\alpha} A_\mu (\theta) =  [A_\mu (\theta),
e_\alpha{}^a (\theta) T_a ]
\end{eqnarray}
For $\Lambda_{(\alpha)}$ we take now $ \Lambda_{(\alpha)}  (\theta ) =- {\rho^2 \over x^2 + \rho^2} e_\alpha{}^a (\theta) T_a$, and then we obtain for the
gauge zero modes at arbitrary $\theta$
\eqa
 Z_{\mu  (\alpha)} (\theta) &=& D_\mu (A (\theta)) \left( {x^2 \over x^2 + \rho^2} e_\alpha{}^a (\theta) T_a \right) \nonumber\\
& = &U^{-1} \left[ D_\mu (A (\theta =0)) \left( {x^2 \over x^2 + \rho^2} \partial_\alpha U U^{-1} \right) \right] U\ .
\eqae
We define\footnote{ The functions $e_\alpha{}^a(\theta)$ are sometimes called
the left-invariant one-forms, while $f_\alpha{}^a(\theta)$ are the
right-invariant one-forms.} $\partial_\alpha U U^{-1} = f_\alpha{}^a (\theta)
T_a$.  Note that $\tr\, \partial_\alpha U U^{-1} \partial_\beta U U^{-1} = \tr\, (U^{-1} \del_\a U \; U^{-1} \del_\beta U) = e_\alpha{}^a  e_\beta{}^b \tr\, T_a T_b = f_\alpha{}^a f_\beta{}^b \tr\, T_a T_b$.
Hence the left-invariant metric $e_\alpha{}^a e_\beta{}^b\delta_{ab}$ is equal
to the right-invariant metric. There is a geometrical interpretation of these
results \cite{West,vanN}.

There are only two differences with the $\theta =0$ case\\
(i) the factors $U(\theta)$ and $U^{-1} (\theta)$ in front and at the back; these drop out in the trace\\
(ii) the factors of $f_\alpha{}^a$ multiplying $T_a$.
Taking the trace one obtains the group metric
\begin{equation}
U^{\alpha \beta} (\theta) = \langle Z^{(\alpha)}_\mu \mid Z^{(\beta)}_\mu \rangle = e_\alpha{}^a  (\theta) e_\beta{}^b (\theta) U_{ab} (\theta =0) = e_\alpha{}^a (\theta) e_\beta{}^a (\theta) \; ( \ft12  \rho^2
S_{\rm cl})\  .
\end{equation}
Hence, in the square root of the determinant of $U$ one finds a factor $\det e_\alpha{}^a$ (because $\det (e_\alpha{}^a \delta_{ab} e_\beta{}^b)  = (\det e_\alpha{}^a)^2$), and this yields the Haar measure
\begin{eqnarray}
\mu (\theta) = \det e_\alpha{}^a (\theta) {\rm d}^3 \theta\ .
\end{eqnarray}
Using this measure one can calculate the group volume $V$ of $SU(2)$, $V = \int (\det e_\alpha{}^a) {\rm d}^3 \theta$, which is independent of the choice of coordinates $\theta$. (We chose the parametrization
$U(\theta)={\exp \theta^aT_a}$, but any other parametrization yields the same
result.)

We have now calculated all norms.  It is fairly easy to prove
that there is no mixing between the different modes, for example
$U^{\mu(\rho)} = U^{\mu}{}_a = U^{(\rho)}{}_a = 0$.   Thus the matrix
$U^{ij}$ for $SU(2)$ is eight by eight, with non-vanishing entries
along the block-diagonal
\begin{equation}
U^{ij}=\begin{pmatrix}\delta^{\mu\nu}S_{\rm cl} & & \cr
& 2 S_{\rm cl} & \cr
& & \ft12 g_{\alpha \beta} (\theta) \rho^2S_{\rm cl}
\end{pmatrix}_{8 \times 8}\ ,
\end{equation}
The square root of
the determinant is
\begin{equation} {\sqrt U} = \ft12 S_{\rm cl}^4 \rho^3 {\sqrt {\det
g_{\a\b} (\theta)}} = \frac{2^{11} \pi^8\rho^3}{\ggg^8}\, {\sqrt {\det
g_{\a\b} (\theta)}}\quad  ({\rm for} \; SU(2)) \ .
\end{equation}

Let us now consider the remaining generators of $SU(N)$ by first
analyzing the example of $SU (3)$. For simplicity, we restrict ourselves
again to lowest order in $\theta^a$.
There are seven gauge
orientation zero modes, three of which are given by  \eqn{su2zm}
by taking for $T_a$ the first three Gell-Mann matrices
$\lambda_1,\lambda_2,\lambda_3$ multiplied by $- {i \over 2}$. For
the other four zero modes, corresponding to
$\lambda_4,\dots,\lambda_7$, the formula \eqn{angle-zm} still
holds, but we have to change the gauge transformation in order to
keep the zero mode in background gauge,
\begin{equation}
\label{L-su3} \Lambda_k =  \left[ \sqrt{\frac{x^2}{x^2 + \rho^2}}
- 1 \right] T_k \ , \qquad k = 4, 5, 6, 7 \ ,
\end{equation} with
$T_k = (-i/2) \lambda_k$. The difference in $x$-dependence of the
gauge transformations \eqn{L-su2} and \eqn{L-su3} is due to the
change in commutation relations. Namely, $\sum_{a = 1}^{3}
[\lambda_a, [\lambda_a, \lambda_\beta]] = - (3/4) \lambda_\beta$
for $\beta = 4, 5, 6, 7$, whereas it is $-2 \lambda_\beta$ for
$\beta = 1, 2, 3$. (These are the values of the Casimir operator of $SU(2)$
on doublets and triplets, respectively). As argued before, there is no gauge
orientation
zero mode associated with $\lambda_8$, since it commutes with the
$SU(2)$ embedding. The zero modes are then
\begin{equation}
\label{su3zm} Z_{\mu\,(k)} =  D_\mu \left[ \sqrt{\frac{x^2}{x^2 +
\rho^2}} T_k \right] \ , \qquad  k = 4, 5, 6, 7 \ ,
\end{equation}
with norms\footnote{ These zero modes are given by $Z_{\mu\,(k)}=
\rho^2 x^\nu / (\sqrt{x^2} (x^2 + \rho^2)^{3/2}) (\delta_{\mu\nu} T_k + 2
\eta_{a \mu\nu} [T_a , T_k])$ in the singular gauge, see \eqref{expression}.
For the first three zero modes we found instead $Z_{\mu\,(a)}=
\rho^2 x^\nu / (x^2 + \rho^2)^{2} (2\delta_{\mu\nu} T_a + 2
\eta_{b \mu\nu} [T_b , T_a])$ with $[T_b , T_a] = \e_{bac} T_c$.  
The norm of \eqref{su3zm} is proportional
to $\tr\,T_k T_l + 4 \tr\, [T_a , T_k ] [ T_a , T_l ]= 4 \tr\, T_k T_l$,
where we used \eqref{eta-eta1}.
}
\begin{equation}
U_{kl} = \ft14 \delta_{kl} \rho^2 S_{\rm cl} \ ,
\end{equation}
and are orthogonal to \eqn{su2zm}, such that
$U_{ka} = 0$.  This construction easily generalizes to $SU (N)$.
One first chooses an $SU (2)$ embedding, and this singles out 3
generators. The other generators can then be split into $2 (N -
2)$ doublets under this $SU (2)$ and the rest are singlets. There
are no zero modes associated with the singlets, since they commute
with the $SU(2)$ chosen. For the doublets, each associated zero
mode has the form as in \eqn{su3zm}, with the same norm $\ft14
\rho^2 S_{\rm cl}$. This counting indeed leads to $4 N - 5$ gauge
orientation zero modes. Straightforward calculation for the
square-root of the complete determinant then yields an extra factor
$(\frac{1}{4}\rho^2S_{\rm cl})^{2(N-2)}$, and so
\begin{equation} {\sqrt U} = \frac{2^{2 N + 7}}{\rho^5} \left(
\frac{\pi \rho}{\ggg} \right)^{4N}   \quad ({\rm for} \; SU(N)) \
. \label{ppppp}
\end{equation}
This result is a factor $2^{4N-5}$ smaller than
\cite{Bernard,BelVanNie00a}, since we chose $U(\theta) = \exp
\theta^a T_a$ instead of $\exp (2\theta^a T_a)$.
This ends the discussion about the (bosonic)
zero mode normalization.

\subsection{Construction of the fermionic zero modes}

In this subsection we will explicitly construct the fermionic zero modes
(normalizable solutions of the Dirac equation) in the background of a single
anti-instanton.
For an $SU(2)$ adjoint fermion, there are 4 zero modes according to
\eqref{adjoint}, and these can be
written as follows \cite{Shi83}
\begin{equation} \label{xi-eta}
\lambda^\alpha = - \ft12 \sigma_{\rho\sigma\ \, \beta}^{\phan{ii}\alpha}
\left( \xi^\beta - \sigma_\nu^{\beta \gamma'}\bar\eta_{\gamma'}
(x - x_0)^\nu \right) F_{\rho\sigma} \ .
\end{equation}
The $SU(2)$ indices $u$ and $v$ are carried by $(\lambda^\alpha)^u{}_v$
and $(F_{\rho\sigma})^u{}_v$.

To prove that these spinors are solution of the Dirac equation,
use ${\bar \sigma}_\mu\sigma_{\rho\sigma}=\delta_{\mu\rho}{\bar \sigma}_\sigma
-\delta_{\mu\sigma}{\bar \sigma}_\rho-\epsilon_{\mu\rho\sigma\tau}
{\bar \sigma}_\tau$. Then $\not\!\!\!\bar{D}  \lambda$ vanishes since
$D_\mu F_{\rho\sigma}$ vanishes when contracted with $\eta_{\mu\rho},
\eta_{\mu\sigma}$ or $\epsilon_{\mu\rho\sigma\tau}$.
Actually,
this expression also solves the Dirac equation for higher order
$k$, but there are then additional solutions, $4|k|$ in total for
$SU(2)$, see \eqn{adjoint}.   The four fermionic collective
coordinates are denoted by $\xi^\alpha$ and $\bar\eta_{\gamma'}$,
where $\alpha, \gamma' = 1,2$ are spinor indices in Euclidean
space\footnote{ To check that the expression with $\bar\eta$ is a
solution, one may use that $\bar\sigma_\rho \sigma_{\mu\nu}
\sigma_\rho =0$.  Note that one may change the value of $x_0$ in
\eqn{xi-eta} while keeping $F_{\mu\nu}$ fixed, because the
difference is a solution with $\xi^\beta$.}. They are the fermionic
partners of the translational and dilatational collective
coordinates in the bosonic sector. These solutions take the same
form in any gauge, one just takes the corresponding gauge for the
field strength.   The canonical dimension of $\xi$ and $\bar\eta$
is $- 1/2$ and $1/2$, respectively.

For $SU(N)$ (and always $k = -1$) there are a further set of $2 \times (N - 2)$
 zero modes in the adjoint representation, and their explicit form depends on
the gauge chosen.  In regular gauge, with color indices $u,v = 1,\dots,N$
explicitly written, the gauge field is given by \eqn{overr} (setting $x_0=0$,
otherwise replace $x \rightarrow x - x_0$)
\begin{equation} {{A_\mu}}^u{}_v =
A_\mu^a \left( T_a \right)^u{}_v = - \frac{ \sigma_{\mu\nu\ \, v}^{\phan{ii}u}
x_\nu}{x^2 + \rho^2} \ , \qquad \sigma_{\mu\nu\ \, v}^{\phan{ii}u} =
\left( \begin{array}{cc} 0 & 0 \\ 0 & \sigma_{\mu\nu\ \, \beta}^{\phan{ii}
\alpha} \end{array} \right)\ .
\end{equation}
Then the corresponding fermionic
instanton in the adjoint representation reads \begin{equation} \lambda^{\alpha
\; u}{}_v = \frac{\rho}{\sqrt{ ( x^2 + \rho^2 )^3}} \left( \mu^u
{\delta^\alpha{}_v} + \epsilon^{\alpha u} \bar\mu_v \right) \ .
\label{nieuwenhuizen}
\end{equation}
Here we have introduced Grassmann collective coordinates
\begin{eqnarray} \mu^u = (\mu^1, \dots , \mu^{N - 2}, 0, 0) \ ;
\epsilon^{\alpha u} = \left(  \begin{array}{cc} 0, \dots , 0,  & \\   &
\epsilon^{\alpha\beta'}  \\ 0, \dots , 0, & \end{array} \right) \;
\mbox{with} \; N - 2 + \beta' = u \ ,
\end{eqnarray}
and similarly for $\bar\mu_v$ and $\delta^\alpha{}_v$.
Thus the $SU(N)$ structure for the fermionic instanton is as follows
\begin{equation}
\lambda \propto 
\begin{pmatrix}
0 & \mu \cr {\bar \mu} & \xi,{\bar \eta}
\end{pmatrix}\ .
\end{equation}
The canonical dimension
of $\mu$ and ${\bar \mu}$ is $- 1/2$.
To prove that $(\lambda^\alpha)^u{}_v$ in \eqn{nieuwenhuizen} satisfies the
Dirac equation $\bar\sigma^\mu  (\partial_\mu \lambda + [A_\mu , \lambda ])
=0$, note that the terms $(A_\mu)^u{}_w \mu^w$ and $\bar\mu_w (A_\mu)^w{}_v$
vanish due to the index structure of $A_\mu$ and $\mu , \bar\mu$.
Because $A_\mu$ has only nonzero entries in the lower right block, there
cannot be fermionic instantons in the upper left block.

In singular gauge, the gauge field is given by \eqn{expression}
\begin{equation} A_{\mu\, u}{}^v = - \frac{\rho^2}{x^2(x^2 + \rho^2)}
\bar\sigma_{\mu\nu\, u}{}^v x_\nu\ .
\label{PPvN}
\end{equation}
Notice that the position of the color indices is different from
that in regular gauge. This is due to the natural position of
indices on the sigma matrices\footnote{ To be very precise, we
could have used different Pauli matrices $(\tau^a)^u{}_v$ for the
internal $SU(2)$ generators.  Then we could have defined a matrix
$(\bar\sigma_{\mu\nu})^u{}_v$ by
$\bar\sigma_{\mu\nu} = i \eta_{a\mu\nu} \tau^a$, and the $SU(2)$
indices in \eqn{PPvN} and \eqn{ferm-angles} would have appeared in
the same position as in \eqn{nieuwenhuizen}. It is simpler to work
with only one kind of Pauli matrices.}. The fermionic
anti-instanton in singular gauge reads \cite{Cor79}
\begin{equation}
\label{ferm-angles} {\lambda^{\alpha}}_u{}^v =
\frac{\rho}{\sqrt{x^2 (x^2 + \rho^2)^3}} \left( \mu_u x^{\alpha v}
+ {x^\alpha}_u \bar\mu^v \right) \ ,
\end{equation}
where for fixed $\alpha$, the $N$-component vectors $\mu_u$ and $x^{\alpha
v}$ are given by
\begin{equation} \mu_u = \left( \mu_1 , \dots ,
\mu_{N - 2} , 0 , 0 \right) \ , \qquad x^{\alpha v} = \left( 0 ,
\dots , 0 , x^\mu \sigma_\mu^{\alpha\beta'} \right)
\quad\mbox{with}\quad N - 2 + \beta' = v \ .
\end{equation}
Further, ${x^\alpha}_u = x^{\alpha v} \epsilon_{vu}$ and
$\bar\mu^v$ also has $N - 2$ nonvanishing components. The
particular choice of zeros in the last two entries corresponds to
the choice of  embedding the $SU(2)$ instanton in the lower-right
block of $SU(N)$. Notice that the adjoint field $\lambda$ is
indeed traceless in its color indices. This follows from the
observation that $\mu$ and ${\bar \mu}$ only appear at the
off-diagonal blocks inside $SU(N)$. In general $\mu$ and ${\bar \mu}$ are
independent, but if there is a reality
condition on $\lambda$ in Euclidean space, the $\mu$ and $\bar\mu$
are related by complex conjugation. We will discuss this in a
concrete example when we discuss instantons in $\cN = 4$ super Yang-Mills
theory.  We should also mention that while the bosonic collective
coordinates are related to the rigid symmetries of the theory,
this is not obviously true for the fermionic collective
coordinates, although, as we will see later, the $\xi$ and
$\bar\eta$ collective coordinates can be obtained from  ordinary
supersymmetry and conformal supersymmetry in super Yang-Mills theories.

A similar construction holds for a fermion in the fundamental
representation. Now there is only one fermionic collective
coordinate, see \eqn{adjoint}, which we denote by $\cK$. The
explicit expression for $k=-1$ in singular gauge is\footnote{ The color
index should again be written as $(\lambda^\alpha)_{u'}$ because
$\lambda^{{\rm reg}, u}
= (U)^{uv'} \lambda^{\rm sing}_{v'}$ with $U^{uv'} = \sigma^{uv'}_\mu
x_\mu / {\sqrt {x^2}}$.  However, we drop these primes.
The proof that \eqn{van} satisfies the Dirac equation
uses $\bar\sigma_{\mu \alpha' \beta} \sigma^\beta_{\rho u}
x^\mu x^\rho = \epsilon_{\alpha' u} x^2$ and
$(\bar\sigma_{\mu\rho})_{\alpha' v} (\bar\sigma_{\mu\nu})_u{}^v
= - ( \bar\sigma_{\mu\rho})_{\alpha'}{}^v
(\bar\sigma_{\mu\nu})_{vu} = 3 \delta_{\rho \nu}
\epsilon_{\alpha' u}$.}
\begin{equation} ({\lambda^\alpha})_u =
\frac{\rho}{\sqrt{x^2 (x^2 + \rho^2)^3}}\, {x^\alpha}_u\, \cK \ .
\label{van}
\end{equation}
In regular gauge it is given by
\begin{equation}\label{fund-zmode}
(\lambda^\alpha)^u = {\epsilon^{\alpha u} \over (x^2 +
\rho^2)^{3/2}} \cK\ .
\end{equation}

The Dirac equation for $(\l^\a)^u$ is proportional to
\eqa
-3 x^\mu \bar\s_{\mu, \a' \b} \e^{\b u} - \bar\s_{\mu, \a' \b} \e^{\b v} (\s_{\mu\nu})^u{}_v x^\nu
\eqae
and to show that this vanishes one may use $(\s_{\mu\nu})^u{}_v \e^{\b v} = (\s_{\mu\nu})^{u \b}$ and the symmetry of the Lorentz generators $(\s_{\mu\nu})^{u \b} = (\s_{\mu\nu})^{\b u}$ and $\bar\s_\mu \s_{\mu\nu} = 3 \bar\s_\nu$.

\section{The measure for zero modes}

Having determined the bosonic and fermionic zero modes for $k=\pm 1$
instantons with $SU(N)$ gauge group, we now discuss the measure for the
zero mode sector of path integrals.
The one-loop corrections due to the nonzero modes, will be discussed in the
next section.

\subsection{The measure for the bosonic collective coordinates}

We now construct the measure on the moduli space of bosonic collective
coordinates, and show that the matrix $U$ plays the role of a Jacobian. We
first illustrate the idea for a generic system without gauge invariance, with
fields $\phi^A$, and action $S[\phi]$ (for example, the kink in one dimension).
  We expand around the instanton solution
  \begin{equation} \phi^A (x) =
\phi^A_{\rm cl} \left( x, \gamma \right) + \phi^A_{\rm qu} \left( x,\gamma
\right) \ .
\end{equation}
The collective coordinates are denoted by $\gamma$
and, for notational simplicity, we assume there is only one. At this point the
fields $\phi^A_{\rm qu}$ can still depend on the collective coordinate, as
they can include zero modes. The action, up to terms quadratic in the quantum
fields, is
\begin{equation} \label{ActionExpand} S = S_{\rm cl} + \ft12
\phi^A_{\rm qu} M_{AB} \left( \phi_{\rm cl} \right) \phi^B_{\rm qu} \ .
\end{equation} The operator $M$ has zero modes given by
\begin{equation}
Z^A = \frac{\partial \phi^A_{\rm cl}}{\partial \gamma}\ ,
\end{equation}
since, as we explained in \eqref{zmode},
$M_{AB} Z^B$ is just the derivative of the
field equation $\partial S_{\rm cl} / \partial \phi^A_{\rm cl}$ with respect
to the collective coordinate. More generally, if the operator $M$ is hermitian
(or rather self-adjoint\footnote{ More precisely, if there is an inner
product $(\phi_1 , \phi_2 ) = \int \phi^A_1 H_{AB} \phi^B_2 {\rm d}^4 x$ with
real $\phi_1,\phi_2$ and with metric $H_{AB}$, and $H_{AB} H^{BC} = \delta_A{}^C$, then one may define
$\phi^A H_{AB} = \phi_B$ so that $(\phi_1 , \phi_2 ) = \int \phi_{1\,A}
\phi^A_2 {\rm d}^4 x$.  If one further defines $H^{BC} M_{CD} = M^B{}_D$,
then $M^A{}_B$ is hermitian if $(\phi_1 , M \phi_2 ) = (M \phi_1 , \phi_2 )$.
 The need for a matrix to define an inner product is familiar from spinors,
but for bosons the metric is in general trivial $(H_{AB} = \delta_{AB})$.}),
it has a complete set of eigenfunctions $F_\alpha$ with eigenvalues
$\epsilon_\alpha$,
\begin{equation} M_{AB} F^B_\alpha = \epsilon_\alpha
F^A_\alpha\ . \end{equation}
One of the solutions is of course the zero
mode $Z=F_0$ with $\epsilon_0=0$. Any function can be expanded into a
basis of eigenfunctions, in particular the quantum fields,
\begin{equation}
\phi^A_{\rm qu} = \sum_\alpha \xi_\alpha F^A_\alpha \ ,
\label{inner}
\end{equation}
with coefficients $\xi_\alpha$. The
eigenfunctions have norms, determined by their inner product
\begin{equation}
\langle F_\alpha | F_\beta \rangle = \int\,
{\rm d}^4x\, F_\alpha^A  (x) F_\beta^A (x) \ .
\end{equation}
The eigenfunctions can always be chosen orthogonal, such that
$\langle F_\alpha | F_\beta \rangle = \delta_{\alpha\beta}
u_\alpha$. The action then becomes
\begin{equation} \label{S} S =
S_{\rm cl} + \ft12 \sum_\alpha \xi_\alpha \xi_\alpha
\epsilon_\alpha u_\alpha\ .
\end{equation}
If there is a coupling
constant in front of the action \eqn{ActionExpand}, we rescale the
inner product with the coupling, such that \eqn{S} still holds.
This was done in \eqn{norms}.
The path-integral measure is now {\it defined}
as
\begin{equation}
\left[ {\rm d}\phi \right] \equiv
\prod_{\alpha = 0}^\infty  \sqrt{\frac{u_\alpha}{2\pi}}\, {\rm d} \xi_\alpha \ .
\label{petvan}
\end{equation}
We perform the Gaussian
integration over the $\xi_\alpha$ and get
\begin{equation}\label{pi-measure}
\int\,\left[ {\rm d}\phi \right] \, {\rm e}^{ - S[\phi]} = \int\,
\sqrt{\frac{u_0}{2\pi}} {\rm d} \xi_0\, {\rm e}^{-S_{\rm cl}}
({\det}'M)^{-1/2} \ .
\end{equation}
One sees that if there were no zero modes, the measure in \eqn{petvan} produces the correct
result with the determinant of $M$. In the case of zero modes, the
determinant of $M$ is zero, and the path integral would be
ill-defined. Instead, we must leave out the zero mode in $M$, take
the amputated determinant (denoted by $\det'$), and integrate over
the mode $\xi_0$.  By slightly changing some parameters in the
action (for example by adding a small mass term) the zero mode
turns into a non-zero mode, and then one needs $\sqrt{{u_0 \over 2
\pi}} {\rm d} \xi_0$ as measure.  So, continuity fixes the measure for
the zero modes as in \eqn{petvan}.

The next step is to convert the $\xi_0$ integral to an integral over the
collective coordinate
$\gamma$ \cite{GerSakTom}. This can be done by inserting unity into the path
integral. Consider the identity \begin{equation}
1 = \int\, {\rm d} \gamma\, \delta \left( f (\gamma) \right)
\frac{\partial f}{\partial \gamma} \ , \end{equation} which holds
for any (invertible) function $f(\gamma)$. Taking $f(\gamma) = -
\langle \phi - \phi_{\rm cl}(\gamma)| Z \rangle$, and recalling
that the original field $\phi$ is independent of $\gamma$, we get
\begin{equation} 1 = \int\, {\rm d} \gamma\, \left( u_0 - \left\langle
\phi_{\rm qu} \left| \frac{\partial Z}{\partial \gamma}
\right\rangle\right. \right) \delta \Big( \langle \phi_{\rm qu} |
Z \rangle \Big) = \int\, {\rm d} \gamma\, \left( u_0 - \left\langle
\phi_{\rm qu} \left| \frac{\partial Z}{\partial \gamma}
\right\rangle\right. \right) \delta \Big( \xi_0 u_0 \Big) \ .
\label{pvn} \end{equation} This trick is similar to the
Faddeev-Popov trick for gauge fixing. In the semiclassical
approximation, the term $\langle
\phi_{\rm qu} \left| \frac{\partial Z}{\partial \gamma}
\right\rangle$ is subleading
and we will neglect it\footnote{ It will contribute however to a
two-loop contribution. To see this, one first writes this term in the
exponential, where it enters without $\hbar$, so it is at least a
one-loop effect. Then $\phi_{\rm qu}$ has a part proportional to
the zero mode, which drops out by means of the delta function
insertion. The other part of $\phi_{\rm qu}$ is genuinely quantum
and contains a power of $\hbar$ (which we have suppressed).
Therefore, it contributes at two loops \cite{TwoLoop2} (see also
\cite{TwoLoop1} for related matters).}.
The integration over $\xi_0$ is now trivial and one obtains
\begin{equation}
\int\, \left[ {\rm d}\phi \right] \, {\rm e}^{-S} = \int\,
{\rm d}\gamma\, \sqrt{\frac{u_0}{2\pi}} {\rm e}^{-S_{\rm cl}} \left( {\det}'M
\right)^{-1/2} \ .
\end{equation}
For a system with more zero modes $Z^i$ with norms-squared $U^{ij}$, the
result is\footnote{ One obtains
from \eqn{pvn} $\det \langle
\partial_{\gamma_i} A^{\rm cl}_\mu | Z^{(j)} \rangle$ times $(\det
U^{ab})^{-1/2}$.  The matrix elements $\langle \partial_{\gamma_i}
A^{\rm cl}_\mu | Z^{(j)} \rangle$ are equal to $\langle Z^{(i)} |
Z^{(j)} \rangle = U^{ij}$ minus $\langle D_\mu \Lambda^{(a)} |
Z^{(b)}_\mu \rangle$.  The latter term can be partially
integrated, and vanishes since there are no boundary
contributions, neither in the singular nor in the regular gauge. (For the
regular gauge one needs an explicit calculation to check this statement.)}
\begin{equation} \label{measure}
\int\, \left[ {\rm d}\phi \right] \,
{\rm e}^{-S} = \int \, \prod_{i = 1} \frac{{\rm d}\gamma_i}{\sqrt{2\pi}}
\left( \det\, U \right)^{1/2} {\rm e}^{-S_{\rm cl}} \left( {\det}' M
\right)^{-1/2} \ .
\end{equation}
Notice that this result is
invariant under rescalings of $Z$, which can be seen as rescalings
of the collective coordinates. More generally, the matrix $U_{ij}$
can be interpreted as a metric on the moduli space of collective
coordinates. The measure is then invariant under general
coordinate transformations on the moduli space.

One can repeat the analysis for gauge theories to show that \eqn{measure}
also holds for Yang-Mills instantons in singular gauge. For  regular gauges,
there are some complications due to the fact that neither of the two terms in
\eqn{zeromode} does fall off fast at infinity, but only their sum is
convergent.  In singular gauge, each term separately falls off fast at
infinity.  For this reason, it is more convenient to work in singular gauge.
The measure for the bosonic collective coordinates for $k = 1$ $SU(N)$ YM
theories, without the determinant from integrating out the quantum
fluctuations which will be analyzed in the next section, becomes
\begin{equation} \label{bos-meas}
\frac{2^{4N+2} \pi^{4 N - 2}}{(N - 1)!(N - 2)!} \frac{1}{\ggg^{4N}}
\int\, {\rm d}^4 x_0 \,
\frac{{\rm d}\rho}{\rho^5} \rho^{4N}\ .
\end{equation}
This formula contains the square-root of the determinant of $U$ in
\eqref{ppppp}, $4N$ factors
of $1/\sqrt{2\pi}$, and we have also integrated out the gauge orientation zero
modes. This may be done only if we are evaluating gauge invariant correlation
functions. The result of this integration follows from the volume of the coset
space
\begin{equation} \label{VolCoset} {\rm Vol} \left\{ \frac{SU(N)}{SU(N - 2)
\times U(1)} \right\} = \frac{2^{4N-5}\pi^{2 N - 2}}{(N - 1)!(N - 2)!}\ ,
\end{equation}
which is a factor $2^{4N-5}$ larger than in \cite{Bernard,BelVanNie00a},
because we have used the normalization $\tr(T_aT_b)=-\ft12\delta_{ab}$,
while in \cite{Bernard,BelVanNie00a} $\tr(T_aT_b)=-2\delta_{ab}$
was used. We found in \eqref{ppppp} another factor $2^{-(4N-5)}$,
and indeed the result for the total measure in \eqref{bos-meas} is the
same as in \cite{Bernard,BelVanNie00a}.
The derivation of this formula can be found in Appendix \ref{Volume}, which is
a detailed version of  \cite{Bernard}.

\subsection{The measure for the fermionic collective coordinates}

We must also construct the measure on the  moduli space of fermionic collective coordinates.   Consider \eqn{xi-eta}.  The fermionic zero modes are linear in the Grassmann parameters $\xi^\alpha$ and $\bar\eta_{\alpha'}$.  Thus these $\xi^\alpha$ and $\bar\eta_{\alpha'}$ correspond to the coefficients $\xi^\alpha$ in \eqn{inner}.  One obtains the zero modes by differentiating $\lambda^\alpha$ in \eqn{xi-eta} w.r.t. $\xi^\alpha$ and $\bar\eta_{\alpha'}$, and for this reason one often calls these $\xi^\alpha$ and $\bar\eta_{\alpha'}$ the fermionic collective coordinates.  This is not quite correct, because collective coordinates appear in the classical solution (the instanton) but we shall use this terminology nevertheless because it is common practice. We use again the measure in
\eqn{petvan}. There are in this case no factors ${1 \over \sqrt{2 \pi}}$ because of the Grassmann integration, and instead of $(\det M')^{-1/2}$ we now obtain $(\det M')^{1/2}$ in \eqref{pi-measure}. Because the parameters $\xi^\alpha , \bar\eta_{\alpha}$, etc.  appear linearly in the zero modes, we do not need the Faddeev-Popov trick to convert the integration over zero modes into an integration over collective coordinates.
So {\it for fermions the
Grassmannian coefficients of the zero modes are at the same time
the collective coordinates}.

We shall discuss these issues in more detail when we come to supersymmetric gauge theories, but now we turn to computing the norms of the fermionic zero modes.

For the zero modes with $\xi$ in \eqn{xi-eta}, one finds
\begin{equation} Z^\alpha_{(\beta)} = \frac{\partial
\lambda^\alpha}{\partial \xi^\beta} = - \ft12 \sigma_{\mu\nu\ \,
\beta}^{\phan{ii}\alpha} F_{\mu\nu} \ . \end{equation} The norms
of these two zero modes are given by \begin{equation}
{(U_\xi)_\beta}^\gamma = - \frac{2}{\ggg^2} \int \, {\rm d}^4 x \, \tr\,
\left\{ Z_{\alpha (\beta)} Z^{\alpha (\gamma)} \right\} = 4 S_{\rm
cl} {\delta_\beta}^\gamma \ , \end{equation} where we have used
the definition in \eqn{norms} and contracted the spinor indices
with the usual metric for spinors. This produces a term in the
measure\footnote{ Sometimes one finds in the literature that
$U_\xi=2 S_{\rm cl}$. This is true when one uses the conventions
for Grassmann integration $\int {\rm d}^2\xi\,\xi^\alpha\xi^\beta=\ft12
\epsilon^{\alpha\beta}$. In our conventions ${\rm d}^2\xi\equiv
{\rm d}\xi^1{\rm d}\xi^2$.} \begin{equation} \int \, {\rm d} \xi^1 {\rm d}
\xi^2\, \left(4 S_{\rm cl} \right)^{-1}\ . \label{xi-meas}
\end{equation}
The result \eqn{xi-meas} actually holds for any $k$.  We get the square root of the determinant in the denominator for fermions.  One really gets the square root of the super determinant of the matrix of inner product, but because there is no mixing between bosonic and fermionic moduli, the superdeterminant factorizes into the bosonic determinant divided by the fermionic determinant.

For the $\bar\eta$ zero modes, we obtain, using some algebra for
the  $\sigma$-matrices,
 \eqa
 Z^{\alpha \beta'} = \partial \lambda^\alpha / \partial \bar\eta_{\beta'} = \ft12 (\sigma_{\mu\nu} \sigma_\rho)^{\alpha \beta'} F_{\mu\nu} x_\rho\ ,
\eqae
and
\begin{equation}
(U_{\bar \eta})_{\alpha'}{}^{\beta'} = 8 S_{\rm cl} \delta_{\alpha'}{}^{\beta'} \rho^2 \ ,
\end{equation}
so that the corresponding measure is
\begin{equation}
\int\,{\rm d}{\bar \eta}_1{\rm d}{\bar \eta}_2\,(8 \rho^2 S_{\rm cl})^{-1}\ ,
\end{equation}
which only holds for $k = 1$.

Finally we compute the Jacobian for the fermionic ``gauge orientation'' zero modes.  For convenience, we take the solutions in regular gauge (the Jacobian is gauge invariant anyway), and find from \eqn{nieuwenhuizen}
\begin{equation} {\left( Z^\alpha_{({\mu^w})} \right)^u}_v = \frac{\rho}{\sqrt{(x^2 + \rho^2)^3}} \, {\delta^\alpha}_v \, {\Delta^u}_w\ , \qquad {\left( Z^\alpha_{({\bar \mu}_w)} \right)^u}_v = \frac{\rho}{\sqrt{(x^2 + \rho^2)^3}} \, \epsilon^{\alpha u} {\Delta^w}_v\  ,
\end{equation}
where the $N$ by $N$ matrix $\Delta$ is the unity matrix in the $(N-2)$ by $(N-2)$ upper diagonal block, and zero elsewhere.  So $\Delta$ restricts the values of $u, w$ and $v$ to up to $N-2$ while in $\delta^\alpha{}_v$ and $\epsilon^{\alpha v}$ the index runs over the next two values.  Consequently, the norms of $Z_{\mu}$ and $Z_{\bar\mu}$ are easily seen to be zero, but the nonvanishing inner product is
\begin{equation}
(U_{\mu\bar \mu})^u{}_v = - \frac{2}{\ggg^2} \int\, {\rm d}^4 x \, \tr\, Z^\alpha_{(\bar\mu_u)} Z_{\alpha\, (\mu^v)} = \frac{2\pi^2}{\ggg^2} \Delta^u{}_v\ ,
\end{equation}
where we have used the integral \eqn{Integral}. It also follows from the index structure that the $\xi$ and $\bar\eta$ zero modes are orthogonal to the $\mu$ zero modes, so there is no mixing in the Jacobian.

Putting everything together, the fermionic part of the measure for $\cN$ adjoint fermions coupled to $SU(N)$ YM theory, with $k=1$, is given by
\begin{equation} \label{ferm-meas} \int\, \left(\prod_{A = 1}^{\cN} {\rm d}^2 \, \xi^A\right) \left( \frac{\ggg^2}{32\pi^2} \right)^\cN  \left(\prod_{A = 1}^{\cN} {\rm d}^2\, \bar\eta^A \right)\left( \frac{\ggg^2}{64\pi^2\rho^2} \right)^\cN  \prod_{A = 1}^{\cN} \left(\prod_{u = 1}^{N - 2} \, {\rm d} \mu^{A,u} \, {\rm d} \bar\mu_u^A \right)\left( \frac{\ggg^2}{2\pi^2} \right)^{\cN (N - 2)}.
\end{equation}
Similarly, one can include fermions in the fundamental representation, for which the Jacobian factor is
\begin{equation}
U_{\cK} \equiv \int\,{\rm d}^4x\,Z^\alpha{}_u Z_\alpha{}^u = \pi^2 \ ,
\label{scalars}
\end{equation}
for each species.  Here $\cK$ is the Grassmann collective coordinate of \eqn{fund-zmode}.  Hence in this case the fermionic part of the measure is
\eqa
\int \left( \prod^{N_f}_{A=1} {\rm d} \cK^A \right)
\left( \sqrt{{1 \over \pi^2}} \right)^{N_f}
\label{part}
\eqae
for $N_f$ fundamental Weyl spinors coupled to $SU(N)$ YM theory with $k=1$.

Note that we did not put a factor ${1 \over g^2}$ in front of the integral in (\ref{scalars}), whereas we used such a factor for fermions in the adjoint representation.  The reason we do not use such a factor for fermions in the fundamental representation has to do with the action.  One finds a factor ${1 \over g^2}$ in front of the Yang-Mills action, and therefore also, by susy, in front of the Dirac action for gluinos.  However, in the matter action the $g$-dependence has been absorbed by the gluons, so there is no factor ${1 \over g^2}$ in front of the matter fermions.  The measure of the zero modes uses the metric of the collective coordinates. In soliton physics (and instantons can be considered as solitons in one higher dimension) one obtains this metric if one lets the collective coordinates become time dependent and integrates over ${\rm d}^4x$ in the action one ends up with a quantum mechanical action of the
\eqa
\cl = (U^{ij} \dot\g_i \dot\g_j + U_{\a\b} \dot\xi^\a \dot\xi^\b + U^{\a' \b'} \dot{\bar{y}}_{\a'} \dot{\bar{y}}_{\b'}  + U_{AB} \dot{K}^A \dot{K}^B
\eqae
Since $U^{ij} , U_{\a\b}$ and $U^{\a' \b'}$ are produced by the Yang-Mills action and its susy partner, while $U_{AB}$ is due to the matter action, there is no $g$-dependence in (\ref{part}).

\section{One loop determinants} 

Having determined the measure on the moduli space of collective coordinates,
we now compute the determinants that arise by Gaussian integration over the quantum fluctuations. Before doing so, we extend the model by adding real scalar fields and Majorana fermions in the adjoint representation.
The action is
\begin{equation}
S = - \frac{1}{\ggg^2} \int \, {\rm d}^4 x\, \tr\, \left\{\ft12 F_{\mu\nu}
F_{\mu\nu} +\left( D_\mu \phi \right) \left( D_\mu \phi \right) - i
\bar\lambda \not\!\!\bar D \lambda - i \lambda \not\!\!D \bar\lambda \right\}
 \ . \label{scal-ferm}
\end{equation}
Here, $\lambda$ is a two-component Weyl spinor which we take in the adjoint
representation\footnote{ As before $\lambda = \left( \begin{array}{cc}
\lambda^\alpha \\ \bar\lambda_{\dot\alpha} \end{array} \right)$, but the
4-component Majorana spinor $\bar\lambda$ is defined by $\lambda^T C$ both in Minkowski and in Euclidean space, where $C$ is the
charge conjugation matrix, $C =\left(  \begin{array}{cc}
\epsilon_{\alpha \beta} & 0 \\ 0 & \epsilon^{\dot\alpha \dot\beta}
\end{array} \right)$.
Then $\bar\lambda =  \left( \lambda_\alpha \ , - \bar\lambda^{\dot\alpha}
\right)$ and Lorentz (or rather $SO(4)$) invariance is  preserved in
Euclidean space because the relation $C \g^\mu =- \g^{\mu, T} C$ holds in both spaces.   In Euclidean space
we denote the indices of $\bar \lambda$ by $\alpha'$ instead of $\dot
\alpha$.}. In Minkowski space there is a reality condition between the two
complex 2-component spinors $\lambda$ and $\bar\lambda$, and as a result
${\bar \lambda}_{\dot \alpha}$ transforms in the complex conjugate of the
representation of $\lambda_\alpha$, but in Euclidean space this reality
condition is dropped.  So $\l^\a$ and $\bar\l_{\a'}$ are independent complex variabes.  For the Grassmann integration this makes no difference.
  Written with indices the Euclidean Dirac action in \eqref{scal-ferm}
reads $\{-i \lambda_\alpha (\sigma^\mu)^{\alpha \beta'} D_\mu
\bar\lambda_{\beta'} - i \bar\lambda^{\alpha'}
(\bar\sigma^\mu)_{\alpha' \beta} D_\mu \lambda^\beta\}$ where
$\l_\a = \l^\b \e_{\b\a}$ and $\bar\lambda^{\alpha'} = \epsilon^{\alpha'  \beta'} \bar\lambda_{\beta'}$.
Generalization to fundamental fermions is straightforward.
The anti-instanton solution around which we will expand is
 \begin{equation}
\label{cl-conf} A_\mu^{\rm cl}\ , \qquad \phi_{\rm cl} = 0\ , \qquad
\lambda_{\rm cl}=0\ , \qquad \bar\lambda_{\rm cl} = 0\ ,
\end{equation}
where $A_{\mu}^{\rm cl}$ is the anti-instanton. This background
represents an exact solution to the field equations.
The bosonic and fermionic zero modes are taken care of by the measure for the
collective coordinates, while in the orthogonal space of nonzero modes,
one can define propagators and vertices, and perform perturbation theory
around the (anti-) instanton.

After expanding $A_\mu = A_\mu^{\rm cl} + A_\mu^{\rm qu}$, and similarly for the other fields, we add gauge fixing and ghost terms
\begin{equation}
S_{\rm gf} = -\frac{1}{\ggg^2} \int\, {\rm d}^4x \,\tr\, \left\{ \left( D_\mu^{\rm cl} \, A_\mu^{\rm qu} \right)^2 - 2\, b \, D^2_{\rm cl} \, c \right\}\ , \end{equation}
such that the total gauge field action is given by \eqn{gf-action}. The integration over $A_\mu$ gives
\begin{equation} \label{det-gauge}
\left[ {\det}' \Delta_{\mu\nu} \right]^{-1/2} \ , \qquad \Delta_{\mu\nu} = - D^2 \delta_{\mu\nu} - 2 F_{\mu\nu}\ ,
\end{equation}
where the prime stands for the amputated determinant, with zero eigenvalues left out. We have suppressed the subscript `cl' and Lie algebra indices.  Integration over the scalar fields results in
\begin{equation} \left[ \det \Delta_\phi \right]^{-1/2}\ , \qquad \Delta_\phi = - D^2\ ,
\end{equation}
and the ghost system yields similarly
\begin{equation}
\left[ \det \Delta_{\rm gh} \right]\ , \qquad \Delta_{\rm gh} = - D^2\ .
\end{equation}

For the fermions $\lambda$ and $\bar\lambda$, we need a bit more explanation. Since neither ${\not\!\! D}$ nor ${\not\!\!\bar D}$ is hermitean(even worse, $\not\!\! D$ maps antichiral spinors into chiral spinors), we cannot evaluate the determinants in terms of their  eigenvalues. But both products
 \begin{equation} \Delta_- = - \not\!\! D \not\!\!\bar D = - D^2 - \ft12 \sigma_{\mu\nu} F_{\mu\nu}\ , \qquad \Delta_+ = - \not\!\!\bar D \not\!\!D = - D^2\ ,
\label{peter}
\end{equation}
with spinor indices still suppressed,
are hermitean.  Let us label the nonzero modes by a subscript $i$.  Then we can expand $\lambda$ in terms of commuting
eigenfunctions $F_i$ of $\Delta_-$ with anticommuting coefficients $\xi_i$,
and $\bar\lambda$ in terms of eigenfunctions $\bar F_i$ of $\Delta_+$
with coefficients $\bar\xi_i$. We have seen  before that both
operators have the same spectrum of non-zero eigenvalues
$\epsilon_i$, and the relation between the eigenfunctions is $\bar
F_i = {1 \over \sqrt{\epsilon_i}} \bar{\rlap{\,/}D} F_i$ and
$F_i = {-1 \over \sqrt{\epsilon_i}} \rlap{\,/}D \bar F_i$.
(The minus sign is needed in order that $\bar F_i = {1 \over
\sqrt{\epsilon_i}} \bar{\rlap{\,/}D}  F_i = {1 \over \epsilon_i}
(- \bar{\rlap{\,/}D}  \rlap{\,/}D ) \bar F_i = \bar F_i)$.
Defining the path integral over $\lambda$ and $\bar\lambda$ as the
integration over $\xi_i$ and $\bar\xi_i$, one gets the determinant
over the nonzero eigenvalues\footnote{ Namely, the action becomes
\begin{equation}
-\frac{1}{g^2}\tr\,\int {\rm d}^4 x \left[ -i \left( \bar\xi_i \bar{F}_i \not\!\!\bar D \xi_j {(- \not\!\! D \bar{F}_j) \over \sqrt{\epsilon_j}} \right) -i \xi_j F_j \not\!\! D \bar\xi_i {\not\!\!\bar D F_i \over \sqrt{\epsilon_i}} \right]
=- \frac{i}{2} \bar\xi_i \xi_j \sqrt{\epsilon_j} \left\langle  \bar{F}_i^a | \bar{F}_j^a \right\rangle + \frac{i}{2} \xi_j \bar\xi_i \left\langle F_j^a | F_i^a \right\rangle \sqrt{\epsilon_i}\ .
\end{equation}
Next we use that the norms of $\bar{F}_i$ and $F_i$ are equal:
\begin{eqnarray*}
\left\langle \bar{F}_i | \bar{F}_j \right\rangle &=& {1 \over \sqrt{\epsilon_i}} \left\langle \not\!\!\bar D F_i | {1 \over \sqrt{\epsilon_j}} \not\!\!\bar D F_j \right\rangle  = {1 \over \sqrt{\epsilon_i \epsilon_j}} \left\langle F_i | - \not\!\! D \not\!\!\bar D F_j \right\rangle \nonumber\\
&=& \sqrt{{\epsilon_j \over \epsilon_i}} \left\langle F_i | F_j
\right\rangle = {1 \over \sqrt{\epsilon_i \epsilon_j}} \left\langle - \not\!\! D \not\!\!\bar D F_i | F_j \right\rangle = \sqrt{{\epsilon_i \over \epsilon_j}} \left\langle F_i | F_j \right\rangle\ .
\end{eqnarray*}
Hence, as expected, the $F_i$ and $\bar{F}_j$ for different eigenvalues are orthogonal to each other, and the norms of $F_i$ and $\bar{F}_j$ are the same.  Denoting $\frac{1}{g^2}\int\,{\rm d}^4x\,(F^a_i)^*F^a_i=<F_i^a | F_i^a >$ by $u_i$, one finds for the path integral
\begin{eqnarray*}
\int\, {\rm d} \bar\xi_i d \xi_i \,{\rm e}^{i \xi_i \bar\xi_i u_i \sqrt{\epsilon_i}} = iu_i \sqrt{\epsilon_i} \ .
\end{eqnarray*}
Hence the measure is ${{\rm d} \xi_i \over \sqrt{u_i}} {{\rm d} \bar\xi_j \over \sqrt{u_j}}$, and the one-loop determinant is $\prod_i (\epsilon_i)^{1/2}$.}.
The result for the integration over the fermions can be written in symmetrized
form as
\begin{equation}
\left[ {\det}' \Delta_- \right]^{1/4} \left[ \det \Delta_+ \right]^{1/4} \ .
\end{equation}
As stated before, since all the eigenvalues of both $\Delta_-$ and $\Delta_+$ are the same, the determinants are formally equal. This result can also be obtained by writing the spinors in terms of Dirac fermions; the determinant we have to compute is then
\begin{equation}
\left[ {\det}' \, \Delta_D^2 \right]^{1/2}\ , \qquad \Delta_D = \left( \begin{array}{cc} 0 & \not\!\! D \\ \not\!\!\bar D & 0 \end{array} \right)\ .
\end{equation}

One would expect that in a supersymmetric model with vectors, spinors and
scalars, the sum of all zero point energies cancel. These zero point energies
correspond to the one-loop determinants in an external Yang-Mills field.
So this suggests that all one-loop determinants are related, and since the
one-loop determinants of fermions depend on $\Delta_+$ and $\Delta_-$, one
would expect that the determinants for the bosons can be expressed in terms of
the determinants of $\Delta_-$ and $\Delta_+$.  For the ghosts and adjoint
scalars this is obvious,
\begin{equation}
\det \, \Delta_{\phi} = \det \,
\Delta_{\rm gh} = \left[ \det \, \Delta_+ \right]^{1/2}\ .
\end{equation}
We get $\det \Delta_\phi = \det (-D^2) = \det \Delta^{1/2}_+$ and $\det \Delta_{\rm gh} = \det (-D^2 ) = \det \Delta^{1/2}_+$ because the spinor space is two-dimensional.

For the vector fields,  we rewrite the operator $\Delta_{\mu\nu}$ in 
(\ref{det-gauge}) in terms of the fermion operator $\Delta_-$.  Using $\tr({\bar\sigma}_\mu\sigma_\nu)=2\delta_{\mu\nu}$
and $\tr({\bar \sigma}_\mu\sigma_{\rho\sigma}\sigma_\nu)
 = 2 (\d_{\mu\rho} \d_{\s\nu} - \d_{\mu\s} \d_{\rho\nu} - \e_{\mu\rho\s\nu})$ we obtain the following
identity for $\Delta_{\mu\nu}=-\delta_{\mu\nu}D^2
-2F_{\mu\nu}$,
\begin{eqnarray} \label{smartequation}
\Delta_{\mu\nu} &=& \ft12 \tr \left\{ \bar\sigma_\mu \Delta_- \sigma_\nu \right\} = \ft12 \bar\sigma_{\mu\, \alpha'\beta}\, \Big( {\Delta_-}^\beta{}_\gamma \Big)\, \sigma_\nu^{\gamma\a'} \nn
&=& \ft12 (\bar\s_{\mu\a' \b}) ( {\Delta_-}^\beta{}_\gamma \d^{\a'}{}_{\d'} ) ( \bar\s_\nu{}^{\g\d'})
\end{eqnarray}
where $(\Delta_-)^\beta{} _\gamma \delta^{\alpha'}{}_{\delta'}$ is block-diagonal on the basis $\beta \alpha' = \gamma \delta' =
(11), (21), (12) , (22)$.
\eqa
{\Delta_-}^\beta{}_\gamma \d^{\a'}{}_{\d'} = \left( \begin{array}{llll} {\Delta_-}^1{}_1 & {\Delta_-}^1{}_2 & 0 & 0 \\ {\Delta_-}{}^2{}_1 & {\Delta_-}^2{}_2 & 0 & 0 \\ 0 & 0 & {\Delta_-}^1{}_1 & {\Delta_-}^1{}_2 \\ 0 & 0 & {\Delta_-}^2{}_1 & {\Delta_-}^2{}_2  \end{array} \right)
\eqae
This proves that\footnote{ Consider ${\bar \sigma}_{\mu,\alpha'\beta}$ and
$\sigma_\nu^{\gamma\delta'}$ as $4\times 4$ matrices. Then on the right-hand side of \eqref{smartequation} one has the product of three $4 \times 4$ matrices.  For fixed $\mu$ and $\nu$ one has $\bar\sigma^\mu_{\alpha' \beta} \sigma^{\beta \alpha'}_\nu  = 2\delta_{\mu}{}^{\nu}$, hence $\det [\bar\sigma_{\mu, \alpha' \beta}] =4$.}
\begin{equation}
{\det}' \Delta_{\mu\nu} = \left[ {\det}' \Delta_- \right]^2\ .
\end{equation}

Now we can put everything together. The one-loop determinant for a
Yang-Mills system,
including the ghosts, coupled to $n$ real
adjoint scalars and $\cN$ Weyl spinors
(or Majorana spinors) also in the adjoint representation is
\begin{equation}
\left[ {\det}' \Delta_- \right]^{- 1 + \cN/4} \left[ {\det} \Delta_+ \right]^{\ft14 ( 2 + \cN - n)}\ .
\end{equation}
This expression simplifies to the ratio of the determinants when
$\cN - \ft{n}2 = 1$. Particular cases are
\begin{eqnarray} \label{susydets}
&& \cN = 1 \quad n = 0 \quad \rightarrow \quad \left[ \frac{\det \Delta_+}
{{\det}' \Delta_-} \right]^{3/4}\ , \nonumber\\ &&\cN = 2 \quad n = 2 \quad \rightarrow \quad \left[ \frac{\det \Delta_+}{{\det}' \Delta_-} \right]^{1/2}\ , \nonumber\\ && \cN = 4 \quad n = 6 \quad \rightarrow \quad \left[ \frac{\det \Delta_+}{{\det}' \Delta_-} \right]^{0}\ .
\end{eqnarray}
These cases correspond to supersymmetric Yang-Mills theories with
$\cN$-extended supersymmetry. Notice that for $\cN = 4$, the determinants
of $\Delta_+$ and $\Delta_-$ separately
cancel, so there is no one-loop contribution.

For $\cN = 1, 2$ the determinants formally give unity since
the non-zero eigenvalues are the same. However, one must first regularize
the theory to define the determinants properly.  After regularization, the renormalization procedure must
be carried out and counterterms must be added. The counterterms are the same
as in the theory without instantons and their finite as well as infinite parts must be specified
by physical renormalization conditions. The ratios of products of non-zero
eigenvalues can be written as the exponent of the
difference of two infinite sums
\begin{equation}
\frac{\det \Delta_+}{\det^\prime \Delta_-} = \exp \left( \sum_n \omega_n^{(+)} - \sum_n \omega_n^{(-)} \right) \ ,
\end{equation}
with eigenvalues $\lambda_n = \exp \omega_n$. The frequencies $\omega_n^{(+)}$
and $\omega_n^{(-)}$ can be discretized by putting the system in a box of size
$R$ and imposing suitable boundary conditions on the quantum fields at
$R$ (for example, $\phi (R) = 0$, or $\frac{{\rm d}}{{\rm d}R} \phi (R) = 0$,
or a combination thereof \cite{Hooft1}). These boundary conditions may be
different for different fields. The sums over $\omega_n^{(+)}$ and
$\omega_n^{(-)}$ are divergent; their difference is still divergent (although
less divergent than each sum separately) but after adding counterterms
$\Delta S$ one obtains a finite answer. The problem is that one can combine
the terms in both series in different ways, giving different answers.
By combining $\omega_n^{(+)}$ with $\omega_n^{(-)}$ for each fixed $n$, one
would find that the ratio $\left(\det\Delta_+/\det'\Delta_-\right)$ equals
unity. However, other values could result by using different ways to regulate
these sums. We have discussed before that for susy instantons the
densities of nonzero modes are equal, hence for susy instantons the
contributions in \eqref{susydets} from the one-loop determinants cancel.
This makes these models simpler to deal with than non-susy models. For
ordinary (nonsusy) Yang-Mills theory, the results for the effective
action due to different regularization schemes differ at most by a local
finite counterterm. In the background field formalism we are using, this
counterterm must be background gauge invariant, and since we consider only
vacuum expectation values of the effective action, only one candidate is
possible: it is proportional to the gauge action $\int {\rm d}^4 x\,
\tr\, F^2$ and
multiplied by the one-loop beta-function for the various fields which can run in the loop,
\begin{equation}
\Delta S \propto \beta (\ggg) \int {\rm d}^4 x\, \tr\, F^2 \ \ln \frac{\mu^2}{\mu_0^2} \ .
\end{equation}
The factor $\ln \left( \mu^2/ \mu_0^2 \right)$ parametrizes the freedom in choosing different renormalization schemes.

A particular regularization scheme used in \cite{Hooft1} is Pauli-Villars regularization. In this case 't Hooft first used $x$-dependent regulator masses to compute the ratios of the one-loop determinants $\Delta$ in the instanton background and $\Delta^{(0)}$ in the trivial vacuum. Then he argued that the difference between using the $x$-dependent masses and using the more usual constant masses, was of the form $\Delta S$ given above. The final result for pure YM $SU(N)$ in the $| k = 1 |$ sector is \cite{Hooft1, Bernard}
\begin{equation}
\left[ \frac{{\det}' \Delta_-}{\det \Delta_-^{(0)}} \right]^{-1} \left[ \frac{\det \Delta_+}{\det \Delta_+^{(0)}} \right]^{1/2} = \exp \left\{  \ft23 N\, \ln (\mu \rho) - \alpha(1) - 2 ( N - 2 ) \alpha \left( \ft12 \right) \right\}\ .
\label{petervan} \end{equation} Here we have normalized the
determinants against the vacuum, indicated by the superscript
$(0)$.   From the unregularized zero mode sector one obtains a factor $\rho^{4N}$, see (\ref{bos-meas}), and Pauli-Villars regularization of the $4N$ zero modes yields a factor $M^{4N}_{PV}$.  All
together one obtains $\frac{8\pi^2}{g_0^2} + \frac{22}{3} \ln (M_{PV}  \rho)$ in the
exponent for $SU(2)$, where $g_0$ is the unrenormalized coupling constant. Subtracting
$\frac{22}{3}\ln(M_{PV} \rho_0)$ to renormalize at mass scale $1/\rho_0$, one is
left for the effective action with $\frac{8\pi^2}{g_0^2}-\frac{11}{3}
\ln (\rho/\rho_0)\equiv \frac{8\pi^2}{g^2(\rho)}$.
Replacing $\ln (\rho/\rho_0)$ by $\ln (\mu / \mu_0)$, this
is the correct one-loop
renormalization equation for the running of the coupling constant.
For supersymmetric theories, the nonzero mode corrections to the effective
action cancel, and performing the same renormalization procedure as for the
non-supersymmetric case, one now obtains only from the zero modes the
correct $\beta$ function. For ${\cal N}=4$ one finds a vanishing
$\beta$ function.

The fluctuations of the $SU(2)$ part of the gauge fields and the
Faddeev-Popov ghosts yield the term $\alpha (1)$ in \eqref{petervan},
while the fluctuations of the $2 (N-2)$ doublets (corresponding to
$\lambda_4 , \cdots \lambda_7$ for $SU(3)$) yields the term with
$\alpha \left( \ft12 \right)$.  The numerical values of the
function $\alpha(t)$ are related to the Riemann zeta function, and
take the values $\alpha \left( \ft12 \right) = 0.145 873$ and
$\alpha(1) = 0.443 307$. Notice that this expression for the
determinant depends on $\rho$, and therefore changes the
$\rho$-dependence of the integrand of the collective coordinate
measure. Combined with \eqn{bos-meas} one correctly reproduces the
$\beta$-function of $SU (N)$ YM theory.  The calculation of the contribution
of the nonzero modes can be simplified by using a so-called $O(5)$
formalism \cite{FOre} which uses the conformal symmetries of
instantons, in addition to the nonconformal symmetries.  One still
has to regulate the sums over zero-point energies, and both
Pauli-Villars regularization \cite{FOre} and zeta-function
regularization \cite{Zschadha} have been applied to the $O(5)$
formulation.

\subsection{The exact $\beta$ function for SYM theories}

In supersymmetric gauge theories, the contributions to the one-loop
partition function by the nonzero modes in the bosonic and fermionic loops
cancel each other \cite{Z}. Although this has only been shown to occur in a
gravitational background without winding, we assume here that still occurs in
an instanton background.
 Actually, all contributions from the nonzero mode sector cancel:  higher-loops as well as possible nonperturbative corrections.  The zero mode sector can be regularized by
Pauli-Villars fields, and since the partition function yields a physical observable, namely the cosmological constant (the sum over zero-point energies), the result for the partition function should not depend on the regularization parameter $M_{PV}$ (the Pauli-Villars mass).  From this observation one can derive a differential equation for the coupling constant $g(M_{PV})$, which yields the exact $\beta$ function:  it contains all perturbative contributions \cite{B}.

Before going on we should comment on the fact that from the 3-loop level on the result for the $\beta$ function depends on the regularization scheme chosen.  It is sometimes claimed that therefore higher-loop results for the $\beta$ function have no meaning.  This is incorrect: given a particular scheme, all orders in perturbation theory of $\beta$ have meaning.  In the derivation below of the $\beta$ function we shall find an all-order result, but it is not (yet?) known which regularization scheme for Feynman graphs would reproduce these results.  So the all-order expression for $\beta$ has in principle meaning, but in practice one cannot do much with it.  One can only say:  there must exist a regularization scheme which, if used for the calculation of higher-loop Feynman graphs, will produce the all-order result for $\beta$ obtained below.

We begin with pure supersymmetric gauge theory.  We recall that the measure of the zero modes of a single instanton or anti-instanton $(k= \pm 1)$ for $N=1$ susy with gauge group $SU(N)$ and one Majorana or Weyl fermion in the adjoint representation is given by
\begin{eqnarray}
&& {\rm d} {\cal M}_{k=\pm1} =
{\rm e}^{- {8 \pi^2 \over g^2}} \left[ {{\rm d}^4 x
{\rm d} \rho \over \rho^5}  2^{4N+2}
\left( { \rho \over g} \right)^{4N} \left( {M_{PV} \over \sqrt{2 \pi}} \right)^{4N} {\pi^{4N-2} \over (N-1)! (N-2)!} \right] \times \nonumber\\
&& \left[ {{\rm d} \xi_1 {\rm d} \xi_2 \over 4 S_{\rm cl} M_{PV}} {{\rm d} \bar\eta_1 {\rm d} \bar\eta_2 \over 8 \rho^2 S_{\rm cl} M_{PV}} {\prod^{N-2}_{u=1}  {\rm d} \mu^u {\rm d} \bar{\mu}_u \over \left( {1 \over 4} S_{\rm cl} M_{PV} \right)^{N-2}} \right]  \; {\rm Vol} \; \left\{ {SU(N) \over SU(N-2) \otimes U(1)} \right\}
\end{eqnarray}
where $S_{\rm cl} = 8 \pi^2/g^2$.  The volume of the gauge group was given in (\ref{VolCoset}) but because it does not depend on $g$ or $M_{PV}$ it will play no role below.  Note that this measure is dimensionless; ${\rm d}^4x{\rm d}\rho/\rho^5$ is
dimensionless, and the remaining $\rho$ and $M_{PV}$ occur only in the
combination $\rho M_{PV}$.  Also $d^2 \xi / M_{PV}$ and $d^2 \bar{\eta} / ( \rho^2 M_{PV})$ are dimensionless.  The prefactor
${\rm e}^{- 8 \pi^2 / g^2}$ is of course the classical action for the
one-instanton background, and we have left out the term with the term with the theta-angle.
In the first square brackets we find the product of the measure for the bosonic zero modes in (\ref{bos-meas}) and factors $\sqrt{{M_{PV}^2 \over 2 \pi}}$ for each bosonic zero mode from the corresponding Pauli-Villars modes.\footnote{  If the one-loop determinant for the bosonic fields is $(\det M_b)^{-1/2}$ and for the fermionic fields $\det M_f$, then the Pauli-Villars method yields further determinants $\det (M_b + M^2_{PV})^{+1/2}$ and $\det (M_f +M_{PV})^{-1}$.  The zero modes are eigenfunctions of $M_b$ and $M_f$ with eigenvalue zero, so their Pauli-Villars counterparts become nonzero modes with eigenvalues $M^2_{PV}$ and $M_{PV}$.}
The second expression in square brackets contains the contribution to the measure from the fermionic zero modes given by (\ref{ferm-meas}),  with factors ${1 \over \sqrt{M_{PV}}}$ for each fermionic zero mode.  Clearly, each bosonic zero mode contributes a factor $M_{PV}/g$ and each fermionic zero mode contributes a factor $g/ \sqrt{M}_{PV}$.

The dependence of ${\rm d} {\cal M}$ on $M_{PV}$ and $g$ is thus as follows
\begin{eqnarray}
{\rm d} {\cal M} \propto {\rm e}^{-{8 \pi^2 \over g^2}} (M_{PV})^{3N} \left( {1 \over g} \right)^{2N}\ ,
\label{bare}
\end{eqnarray}
where $g$ depends on $M_{PV}$, so $g = g (M_{PV})$. So $g$ is the bare coupling
constant in the regularized theory, and $g$ and $M_{PV}$ vary such that the renormalized coupling constant $g_R$ is kept fixed.  Usually one considers the renormalized coupling constant as a function of the renormalization mass $\mu$, and then the bare coupling constant $g$ satisfies $\mu {\del \over \del \mu} g=0$.  Using dimensional regularization and $g=Z_g (g_{\rm ren}) g_{\rm ren} \mu^{\e/2}$ with $\e =4 -n$ yields then the $\beta$ function.  If one uses Pauli-Villars regularization there are two masses which play a role: the cut-off (regulator) mass $M_{PV}$ and the physical renormalization mass $\mu$.  The bare coupling depends on one of them, the renormalized coupling on the other.
\eqa
\begin{array}{ll}  g= g (M_{PV} ) &   g_R = g_R (\mu)  \\ M_{PV} {\del \over \del M_{PV}} g (M_{PV}) = \beta (g) & \mu \del/\del \mu \;\;  g_R (\mu) = \beta (g_R) \\ \mu {\del \over \del \mu} g (M_{PV}) =0 & M_{PV} {\del \over \del M_{PV}} g_R (\mu) =0  \end{array}
\eqae
Physical quantities depend on $\mu$ but not on $M_{PV}$.  If one wants to apply the renormalization group to the measure, one must use the approach based on $(M_{PV} \del/\del M_{PV}) g_R (\mu) \newline =0$ because the regularized measure depends on $M_{PV}$, not on $\mu$.  The results for the $\beta$ function obtained from both schemes differ by a sign, because the logarithms in the theory depend on $\ln (M_{PV} / \mu)$.

Equating the derivative of the logarithm of the measure w.r.t. in (\ref{bare}) $M_{PV}$
to zero yields then
\begin{eqnarray}\label{beta-cond}
M_{PV} {\partial \over \partial M_{PV}} \left( -{ 8 \pi^2 \over g^2} + 3 N \ln M_{PV} -2 N \ln g \right) =0\ .
\end{eqnarray}
Hence
\begin{eqnarray}
M_{PV} {\partial \over \partial M_{PV}} g \equiv \beta = \left( {3N \over {2 N \over g} - {16 \pi^2 \over g^3}} \right)\ ,
\end{eqnarray}
or, written in terms of $\alpha = {g^2 \over 4 \pi}$
\begin{eqnarray}
{g \over 2 \pi} \beta = M_{PV} {\partial \over \partial M_{PV}} \alpha = {-3N \alpha^2 \over 2 \pi} {1 \over 1- {\alpha N \over 2 \pi}}\ .
\end{eqnarray}
This is the $\beta$-function for pure ${\cal N}=1$ supersymmetric Yang-Mills
theory.   It is straightforward to extend this result to pure ${\cal N}$-extended
supersymmetry with ${\cal N}$ Majorana or Weyl fermions in the adjoint representation.
One finds for $SU(N)$
\begin{equation}
M_{PV} {\partial \over \partial M_{PV}} \alpha = -\frac{\alpha^2}{2\pi}
\frac{4N-{\cal N}N}{1-\frac{\alpha}{2\pi}(2N-{\cal N}N)}\ .
\end{equation}

It is clear that for ${\cal N} = 2$ there is only a one-loop contribution to $\beta$, and for ${\cal N} = 4$ the $\beta$ function vanishes altogether.  These are well-known properties of pure extended susy gauge theories.  For ${\cal N} = 1$ one finds agreement for one- and two- loops.  Beyond two loops the result for the beta function becomes scheme dependent, so it becomes then pointless to investigate whether agreement holds.

Let us now add matter.  In susy QCD with $N_f$ flavours the matter part consists of $N_f$ pairs of chiral superfields $Q^i$ and $\tilde{Q}_i$ with $i=1, N_f$ in the $\underline{N}$ and $\underline{N}^\ast$ representations of $SU(N)$.  Each fermion in $Q$ and $\tilde{Q}$ has one zero mode, see \eqref{adjoint} and (\ref{van}), while the scalars do not have any zero modes.  So the zero mode measure for the matter part is according to (\ref{part})
\begin{eqnarray}
{\rm d}{\cal M} \; {\rm (matter)} \; = \left( {1 \over \pi^2} \right)^{2 N_f}  {1 \over (M_{PV})^{2N_f}} \prod^{N_f}_{u=1} {\rm d}K^u {\rm d} \tilde{K}_u\ .
\end{eqnarray}

Renormalization leads to a further term in the measure, and thus in the $\beta$ function.  In susy only the kinetic term $\bar\phi\,{\rm e}^V \phi$ of the matter fields gets a $Z$ factor
\begin{eqnarray}
\cL = Z \bar\phi_{\rm ren} e^{V {\rm ren}} \phi_{\rm ren} \ , \qquad
\phi = \sqrt{Z} \phi_{\rm ren}\ .
\end{eqnarray}
and rather than a factor ${1 \over \sqrt{M_{PV}}}$ for each fermion with one flavor, we now get in the measure a factor $(ZM_{PV})^{-1/2}$ for each zero mode.  (The Pauli-Villars field operator becomes $ZM_f + M_{PV}$, so the zero modes continue to produce a factor $M^{-1/2}_{PV}$ in the Pauli-Villars sector, {\bf not} $(ZM_{PV})^{-1/2}$.  In the nonzero mode sector one can neglect the dependence on $M_{PV}$, and here the $Z$ factors of bosons and fermions cancel due to susy).

For the gauge multiplet we factorized out a factor $1/g^2$ in front of the
action of all fields of the gauge multiplet, so that the fields $gA_\mu=
{\tilde A}_\mu$ do not renormalize. (We use here the background formalism
in which $Z_g=Z_A^{-1/2}$, where $Z_A$ is the wave function renormalization
constant for the background fields.) Thus the renormalization of the gauge
multiplet is taken care of by the renormalization of the factor $1/g^2$
in \eqref{beta-cond}.

{}From here on we proceed as before.   The measure for gauge group $SU(N)$ with $N_f$ flavors is now given by
\eqa
dM_{k= \pm 1} =e^{-8 \pi^2/g^2} (M_{PV})^{3N} \left( {1 \over g} \right)^{2N} \left( {1 \over ZM_{PV}} \right)^{N_f}
\eqae
We denote the anomalous dimension by $\gamma_i$ where\footnote{  Often one defines $\g = \mu {\del \over \del \mu} \ln \sqrt{Z}$; here we follow \cite{B}}
\begin{eqnarray}
 \gamma_i  &=& \mu  {\del \over \del \mu} \ln Z_i \nn
&=&  - M_{PV} {\partial \over \partial M_{PV}} \ln Z_i = \gamma \qquad
{\rm (the \; same \; for} \; i=1,..., N_f)\ ,
\end{eqnarray}
and obtain
\begin{eqnarray}
{g \over 2 \pi} \beta = M_{PV} {\partial \over \partial M_{PV}} \alpha =- {\alpha^2 \over 2 \pi} \left( {3N -N_f (1- \gamma) \over 1- {\alpha N \over 2 \pi}} \right)\ .
\label{wwww}
\end{eqnarray}
Expanding in terms of $\alpha$, the result agrees with the results in the
literature for the one-loop and two-loop $\beta$ functions for $N=1$ susy QCD with $N_f$ pairs of chiral fields $Q^i$ and $\tilde{Q}_i$.  Namely, the one- and two-loop $\beta$ function for an $N=1$ vector multiplet coupled to a chiral multiplet in a representation $R$, including the effects of the Yukawa couplings whose coupling constant is also $g$ (in fact, the renormalized coupling constant
$g_R$), is given by \cite{J}
\eqa
&& {g \over 2 \pi} \mu {\del \over \del \mu} g = {\a^2 \over 2 \pi} ( -3 C_2 (G) + T (R)) \nn
&& + {\a^3 \over 8 \pi^2} ( -6 C^2_2 (G) + 2 C_2 (G) T (R) + 4 C_2 (R) T (R))
\eqae
For $N_f$ pairs of chiral matter fields $\sum T (R) = N_f$, and $C_2 (G) =N$ for $SU(N)$.  Using also that the anomalous dimension $\g = \mu {\del \over \del \mu} \ln Z$ for a complex fermion in the fundamental representation {\bf N} of $SU(N)$ is equal to $- \a C_2 (R) /  \pi$, we indeed find agreement.\footnote{With the usual normalization $\g = \mu {\del \over \del \mu} \ln Z^{1/2}$ is equal to $\g = {- \a \over 2 \pi} C_2 (R)$ \cite{J}.}

The $\beta$ function in \eqref{wwww} can be rewritten such that only the numbers of zero modes appear.
\begin{equation}
\beta(\alpha)=-\frac{\alpha^2}{2\pi}\left( n_g -\ft12 n_f - \ft12
\sum_g \gamma_g + \ft12 \sum_f \gamma_f\right)\ .
\end{equation}
Here $n_g$ is the number of bosonic zero modes $(4N), n_f$ the total (gluino and matter) number of fermionic zero modes $(2N+2N_f)$,  and the sums $\sum_g$ and $\sum_f$ run over the gluon and fermion zero modes.  For gluons and gluinos $\l$, the anomalous dimension is the same (due to susy) and proportional to the $\beta$ function
\eqa
\g_g = \g_\l = \beta/ \a \; .
\eqae
Substitution of this result yields back (\ref{wwww}).
  This result does not yet agree with the results in the literature for the
$\beta$-function of gauge fields minimally coupled to scalars and fermions,
because in supersymmetry the Yukawa couplings between scalars and fermions have not an independent coupling constant $\lambda$ but rather $\lambda=g^2$.
At the two-loop level one therefore gets extra contributions which one must
add to the results from the literature, and then one gets complete agreement.

\section{$\cN = 4$ supersymmetric Yang-Mills theory}

An interesting field theory with instantons is the $\cN = 4$ super
Yang-Mills theory \cite{Sch77}. The action is of course well known
in Minkowski space, but instantons require the formulation in $\cN =
4$ Euclidean space. Due to absence of a real representation of Dirac
matrices in four-dimensional Euclidean space, one cannot
straightforwardly define Majorana spinors in Euclidean space. This
complicates the construction of Euclidean Lagrangians for
supersymmetric models \cite{Zum,Nic,Nie96}. For $\cN = 2, 4$
theories, one can replace the Majorana condition by the so-called
symplectic Majorana condition and then one can define (symplectic)
Majorana spinors in Euclidean space. Equivalently, one can work with
complex (Dirac) spinors \cite{Bla97,BVV}.  In the following
 subsection we write down the action in Minkowski space-time and discuss the
reality conditions on the fields. Next we construct the hermitean $\cN = 4$
Euclidean model via the dimensional reduction of ten-dimensional $\cN = 1$
super Yang-Mills theory along the time direction. One can also define a continuous Wick
rotation for the spinors directly in four dimensions \cite{Nie96}.

\subsection{Minkowskian $\cN = 4$ SYM}

The $\cN = 4$ action in Minkowski space-time with the signature
$\eta^{\mu\nu} = {\rm diag}(-,+,+,+)$ is given by
\begin{eqnarray} \label{action}
S\!\!\!&=&\!\!\! \frac{1}{\ggg^2} \int\, {\rm d}^4x \, \tr \, \left\{ {1 \over 2} F_{\mu\nu} F^{\mu\nu} - i \bar\lambda_A^{\dot\alpha} \not\!\!{\bar D}_{\dot\alpha\beta} \lambda^{\beta, A} - i \lambda_{\alpha}^A \not\!\!D^{\alpha\dot\beta} \bar\lambda_{A\dot\beta} + {1 \over 2} \left( D_\mu {\bar \phi}_{AB} \right) \left( D^\mu \phi^{AB} \right) \right. \nonumber\\
&-& \left. \sqrt{2} {\bar \phi}_{AB} \left\{ \lambda^{\alpha,A}, \lambda^B_{\alpha}\right\} - \sqrt{2} \phi^{AB} \left\{ \bar\lambda_A^{\dot\alpha} , \bar\lambda_{\dot\alpha, B} \right\} + {1 \over 8} \left[ \phi^{AB}, \phi^{CD} \right] \left[ \bar\phi_{AB} , \bar\phi_{CD} \right] \right\}.
\end{eqnarray}
The on-shell $\cN = 4$ supermultiplet consists of a real gauge field
$A_\mu$, four complex Weyl spinors $\lambda^{\alpha,A}$ (equivalently,
four Majorana spinors) and an antisymmetric complex scalar $\phi^{AB}$ with
labels $A,B = 1, \dots, 4$ of the internal $R$ symmetry group $SU (4)$.  The
reality conditions on the components of this  multiplet are\footnote{ Unless
specified otherwise, equations which involve complex conjugation of fields
will be understood as not Lie algebra valued, i.e.\ they hold for the
components $\lambda^{a, \alpha, A}$, etc.} the Majorana conditions
$\left( \lambda^{\alpha,A} \right)^\ast = - \bar\lambda^{\dot\alpha}_A$
and $(\lambda^A_\alpha)^\ast = \bar\lambda_{\dot\alpha,A}$ and
\begin{equation} \label{MinkRealScal}
{\bar \phi}_{AB} \equiv \left( \phi^{AB} \right)^\ast = \ft12 \epsilon_{ABCD} \phi^{CD} \ .
\end{equation}
These conditions are invariant under $SU(4)$ transformations. The
sigma matrices are defined by $\sigma^{\mu\, \alpha\dot \beta} = (1,
\tau^i)$, ${\bar \sigma}^\mu_{\dot\alpha\beta} = (-1, \tau^i)$ for
$\mu=0,1,2,3$ and complex conjugation gives $\left(
\sigma_{\mu}^{\alpha \dot\beta} \right)^\ast =
\sigma_\mu^{\beta\dot\alpha} = {\bar \sigma}^{\dot \alpha
\beta}_\mu= \epsilon^{\dot\alpha\dot\gamma} \epsilon^{\beta\delta}
\bar\sigma_{\mu\, \dot\gamma\delta}$, with
$\epsilon^{\dot\alpha\dot\beta} = \epsilon_{\dot\alpha\dot\beta} = -
\epsilon^{\alpha\beta} = - \epsilon_{\alpha\beta}$. Since
$\phi^{AB}$ is antisymmetric, one can express it on a basis spanned
by the real eta-matrices  (see Appendix \ref{HooftSpinor})
\begin{equation}
\phi^{AB} = \frac{1}{\sqrt 2} \left\{ S^i \eta^{iAB} + i P^i \bar\eta^{iAB} \right\}, \qquad \bar\phi_{AB} = \frac{1}{\sqrt 2} \left\{ S^i \eta^{i}_{AB} - i P^i \bar\eta^{i}_{AB} \right\} \ ,
\end{equation}
in terms of real scalars $S^i$ and real pseudoscalars $P^i$, $i = 1,
2, 3$. Because $\eta^{iAB}$ is selfdual and $\bar\eta^{iAB}$
anti-selfdual, $\eta^{iAB} = \eta^i{}_{AB}$ and $\bar\eta^{iAB} =-
\bar\eta^i{}_{AB}$.  Then the reality conditions are fulfilled and
the kinetic terms for the $(S,P)$ fields take the standard form. The
action in (\ref{action}) is invariant under the following
supersymmetry transformation laws with parameters $\zeta^A_\a$ and
$\bar\zeta_{\dot\a A}$
\begin{eqnarray} \label{susy-Mink}
\delta A_\mu \!\!\!&=&\!\!\! - i {\bar\zeta}^{\dot\alpha}_A \bar\sigma_{\mu\, \dot\alpha\beta} \lambda^{\beta, A} + i \bar\lambda_{\dot\beta, A} \sigma_\mu^{\alpha\dot\beta} \zeta^A_\alpha\ , \nonumber\\ \delta \phi^{AB} \!\!\!&=&\!\!\! \sqrt{2} \Big( \zeta^{\alpha,A} \lambda^B_\alpha - \zeta^{\alpha,B} \lambda^A_\alpha + \epsilon^{ABCD} \bar\zeta^{\dot\alpha}_C \bar\lambda_{\dot\alpha,D} \Big)\ , \nonumber\\ \delta \lambda^{\alpha,A} \!\!\!&=&\!\!\! - \ft12 \sigma^{\mu\nu\, \alpha}_{\phan{mn}\beta} F_{\mu\nu} \zeta^{\beta,A} - i \sqrt{2} \bar\zeta_{\dot\alpha,B} \not\!\! D^{\alpha\dot\alpha} \phi^{AB} + \left[ \phi^{AB}, \bar\phi_{BC} \right] \zeta^{\alpha,C} \ ,
\end{eqnarray}
which are consistent with the reality conditions. Let us turn now to the discussion of the Euclidean version of this model and discuss the differences with the Minkowski theory.

\subsection{Euclidean $\cN = 4$ SYM} \label{EuclidN4}

To find out the $\cN = 4$ supersymmetric YM model in Euclidean $d =
(4, 0)$ space, we follow the same procedure as in \cite{Sch77}. We
start with the $\cN = 1$ SYM model in $d = (9, 1)$ Minkowski
space-time, but contrary to the original papers we reduce it on a
six-torus with one time and five space coordinates \cite{Bla97,BVV}.
As opposed to the action in  (\ref{action}) with the $SU(4)=SO(6)$
$R$-symmetry group, this reduction leads to a model with an internal
non-compact $SO(5, 1)$ $R$-symmetry group in Euclidean space. As we
will see, the reality conditions on bosons and fermions will both
use an internal metric for this non-compact internal symmetry group.

The $\cN = 1$ Lagrangian in $d = (9,1)$ dimensions reads
\begin{equation} \cL_{10} = \frac{1}{\ggg^2_{10}} {\rm tr} \left\{ \ft12 F_{MN} F^{MN} + \bar \mP \mG^M D_M \mP \right\} \ ,
 \end{equation}
with the field strength $F_{MN} = \partial_M A_N - \partial_N A_M + \left[ A_M, A_N \right]$ and the Majorana-Weyl spinor $\mP$ defined by the conditions
\begin{equation} \label{MajoranaWeyl}
 \mG^{11} \mP = \mP \ , \qquad \mP^T C^-_{10} = \mP^\dagger i \mG^0 \equiv \bar \mP \ .
\end{equation}
Here the hermitean matrix $\mG^{11} \equiv \ast\!\mG$ is a product of all Dirac matrices, $\mG^{11} = \mG^0 \dots \mG^9 $, normalized to $\left( \ast\!\mG \right)^2 = + 1$.  Furthermore, $C^-_{10}$ is the charge conjugation matrix, satisfying $C^-_{10} \mG_M   =- \mG^T_M C^-_{10}$.  The $\mG$-matrices obey the Clifford algebra $\left\{ \mG^M, \mG^N \right\} = 2 \eta^{MN}$ with metric $\eta^{MN} = {\rm diag} (-, +, \dots, +)$. The Lagrangian transforms into a total derivative
under the standard transformation rules\footnote{  After partial integrations, the Yang-Mills action transforms into $[ \zeta \mG_N  \psi^a) D_M F^{a,MN}$ and the Dirac action varies into $- \bar\psi^a \mG^M ( -\frac12 \mG^{PQ} D_M F_{PQ} \zeta )$.  The sum of these two variations cancels if one uses the Bianchi identity $D_{[M} F_{PQ]} =0$.  The variation of $A_M$ in the covariant derivative in the Dirac action cancels separately due to the 3-spinor identity $(\bar\psi^a \mG^M \psi^b) (\bar\zeta \mG_M \psi^c) f_{abc} =0$ which holds in 3,4,6 and 10 dimensions.}
\begin{equation}
\delta A_M = \bar \zeta \mG_M \mP \ , \qquad \delta {\mit\Psi} = - \ft12 F_{MN} \mG^{MN} \zeta \ ,
\end{equation}
with $\mG^{MN} = \ft12 [\mG^M \mG^N - \mG^M \mG^N]$.   The susy parameter is a Majorana-Weyl spinor, $\bar\zeta = \zeta^T C^-_{10} = \zeta^\dagger i \mG^0$ and $\star \mG \zeta = \zeta$.  To proceed with the dimensional reduction we choose a particular representation of the gamma matrices in $d = (9, 1)$, namely
\begin{equation} \label{10Gammas}
\mG^M = \left\{ \hat\gamma^a \otimes \gamma^5, \1_{[8] \times [8]} \otimes \gamma^\mu \right\}, \qquad \mG^{11} = \mG^0 \dots \mG^9 = \hat\gamma^7 \otimes \gamma^5 ,
\end{equation}
where the $8 \times 8$ Dirac matrices $\hat\gamma^a$ and $\hat\gamma^7$ of $d = (5,1)$ with $a = 1, \dots, 6$ can be conveniently defined by means of 't Hooft symbols as follows
\begin{equation} \hat\gamma^a = \left( \begin{array}{cc} 0 & \mS^{a, AB} \\ \bar\mS^a_{AB} & 0 \end{array} \right) , \qquad \hat\gamma^7 = \hat\gamma^1 \dots \hat\gamma^6 = \left( \begin{array}{cc} 1 & 0 \\ 0 & - 1 \end{array} \right) \ ,
\end{equation}
In Euclidean $d=(6,0)$ one defines
\eqa
\mS^{a,AB} &=& (\eta^{kAB} , i \bar\eta^{k, AB}) \nn
\bar\mS^a_{AB} &=& (- \eta^k{}_{AB} , i \bar\eta^k{}_{AB})
\eqae
but in Minkowski space one puts a factor $-i$ in front of the first one.  So explicitly
$\mS^{a, AB} = \left\{ - i \eta^{1,AB}, \eta^{2,AB},
\eta^{3,AB}, i \bar\eta^{k,AB} \right\}$, $\bar\mS^a_{AB} =
\left\{ i \eta^1_{AB}, - \eta^2_{AB},  - \eta^3_{AB}, i \bar\eta_{AB}^k \right\}$ so that $\ft12 \epsilon_{ABCD} \mS^{a\, CD} = - \bar\mS^a_{AB}$.   The first three matrices $\hat\g^1, \hat\g^2, \hat\g^3$ are symmetric while the latter three matrices $\hat\g^4, \hat\g^5, \hat\g^6$ are antisymmetric.  Meanwhile $\gamma^\mu$ and $\gamma^5$ are the usual Dirac matrices of $d = (4, 0)$ introduced in \eqn{gamma-sigma}. Note that in this construction we implicitly associated one of the Dirac matrices, namely $\hat\gamma^1$, in $6$ dimensions with the time direction and thus it is anti-hermitean and has square $- 1$; all others  (as well as all Dirac matrices in $d = (4,0)$) are again hermitean with square $+ 1$.  Let us briefly discuss the charge conjugation matrices in $d = (9, 1)$, $d = (5, 1)$ and $d = (4, 0)$. One can prove by means of finite group theory \cite{Nie81} that all their properties are representation independent. In general there are two charge conjugation matrices $C^+$ and $C^-$ in even dimensions, satisfying $C^\pm \mG^\mu = \pm \left( \mG^\mu \right)^T C^\pm$, and $C^+ = C^- \ast\!\mG$.  These charge conjugation matrices do not depend on the signature of space-time and obey the relation $C^- \ast\!\mG = \pm \left( \ast \mG \right)^T C^-$ with $-$ sign in $d = 10,\, 6$ and $+$ sign in $d=4$. The transposition depends on the dimension and leads  to $\left( C^\pm \right)^T = \pm C^\pm$ in $d = 10$, $\left( C^\pm \right)^T = \mp C^\pm$ in $d = 6$, and finally $\left( C^\pm \right)^T = - C^\pm$ for $d = 4$. Explicitly, the charge conjugation matrix $C^-_{10}$ is given by $C_{6}^- \otimes C_{4}^-$ where
\begin{equation}
C_4^- = \gamma^4 \gamma^2 = \left( \begin{array}{cc} \epsilon_{\alpha\beta} & 0\\ 0 & \epsilon^{\alpha'\beta'} \end{array} \right), \qquad C_6^- = i \hat\gamma^4 \hat\gamma^5 \hat\gamma^6 = \left( \begin{array}{cc} 0 & \delta_{AB}  \\ \delta^{AB} & 0 \end{array} \right) \ .
\end{equation}

Upon compactification to Euclidean $d = (4, 0)$ space, the 10-dimensional Lorentz group $SO(9,1)$ reduces to $SO(4) \times SO(5,1)$ with compact space-time group $SO(4)$ and $R$-symmetry group $SO(5,1)$. In these conventions a Weyl spinor $\mP$ in ten dimensions with 16 (complex) nonvanishing components decomposes as follows into $8$ and $4$ component chiral-chiral and antichiral-antichiral spinors \begin{equation} \mP = \left( \begin{array}{c} 1 \\ 0 \end{array} \right) \otimes \left( \begin{array}{c} \lambda^{\alpha, A} \\ 0 \end{array} \right) + \left( \begin{array}{c} 0 \\ 1 \end{array} \right) \otimes \left( \begin{array}{c} 0 \\ \bar\lambda_{\alpha', A} \end{array} \right) \ ,
\end{equation}
or more explicitly
\begin{eqnarray}
\psi^T &=& [ ( \lambda^{\alpha 1} , 0) ; (\lambda^{\alpha 2} , 0) ; (\lambda^{\alpha 3} , 0) ; ( \lambda^{\alpha 4} , 0) ; (0,0); (0,0); (0,0), (0,0)] \nonumber\\
&+& [ (0,0); (0,0); (0,0); (0,0); (0 \bar\lambda_{\alpha' 1}); (0, \bar\lambda_{\alpha' 2}); (0, \bar\lambda_{\alpha' 3}) ; (0, \bar\lambda_{\alpha' 4})]\ .
\end{eqnarray}
Here $\lambda^{\alpha, A}$ ($\alpha = 1,2$) transforms only under the first $SU(2)$ in $SO(4) = SU(2) \times SU(2)$, while $\bar\lambda_{\alpha', A}$ changes only under the second $SU (2)$. Furthermore, $\bar\lambda_{\alpha',A}$ transforms in the complex conjugate of the $SO(5,1)$ representation of $\lambda^{\alpha,A}$.  To understand this latter statement, note that the $SO(5,1)$ generators are $\hat{M}^{ab} = \ft12 ( \hat\gamma^a \hat\gamma^b - \hat\gamma^b \hat\gamma^a)$, and $\hat\gamma^1$ is antihermitian.  Furthermore, $\hat\gamma^1, \hat\gamma^4, \hat\gamma^5$ and $\hat\gamma^6$ are purely imaginary.  Thus
\begin{eqnarray}
(\hat\gamma^a)^\ast =- \hat{S} \hat\gamma^a \hat{S}^{-1} ; \qquad \hat{S} = \hat\gamma^2 \hat\gamma^3\ .
\end{eqnarray}
This matrix $\hat{S}$ is not the charge conjugation matrix.  The Lorentz generators $\hat{M}^{ab}$ and $\hat{S}$ are block diagonal
\begin{eqnarray}
\hat{M}^{ab} = \left( \begin{array}{ll} \mS^{ab} & 0 \\ 0 & \bar\mS^{ab} \end{array} \right) ;\qquad \hat{S} = \left( \begin{array}{ll} S & 0 \\ 0 & S \end{array} \right)\ ,
\end{eqnarray}
where $\mS^{ab} = \ft12 ( \mS^a \bar\mS^b - \mS^b \bar\mS^a )$ and  $\bar\mS^{ab} = \ft12 ( \bar\mS^a \mS^b - \bar\mS^b \mS^a )$ while $S =- \eta^2 \eta^3 = \eta^1$.  It follows that
\begin{eqnarray}
&& S \mS^a S^{-1} =- ( \mS^a)^\ast \Rightarrow S \mS^{ab} S^{-1} = (\mS^{ab})^\ast \ ,\nonumber\\
&& S \bar\mS^b S^{-1} =- ( \mS^b)^\ast  \Rightarrow S \bar\mS^{ab} S^{-1} = (\bar\mS^{ab})^\ast\ .
\end{eqnarray}
Thus the two spinor representations of $SO(5,1)$ are each pseudoreal (they are not real since $S$ is antisymmetric), but they are not equivalent to each other.  For $SO(6) \simeq SU(4)$, the two spinor representations are of course complex and inequivalent to each other.  For $SO(3,1)$ the opposite is the case: there the two spinor representations
are complex, and equivalent to each other under complex conjugation, $(\sigma_{\mu\nu})^\ast = \sigma_2 \bar\sigma^{\mu\nu} \sigma_2$ because $\hat{S} = \gamma_2$ is off-diagonal.

Substituting these results, the Lagrangian reduces to
\begin{eqnarray} \label{N4susy}
\cL_E^{\cN = 4}\!\!\! &=&\!\!\! \frac{1}{\ggg^2} {\rm tr}\ \Bigg\{ \ft12 F_{\mu\nu} F_{\mu\nu} - i \bar\lambda^{\alpha'}_{A} \not\!\!{\bar D}_{\alpha'\beta} \lambda^{\beta, A} - i \lambda_{\alpha}^{A} \not\!\!D^{\alpha\beta'} \bar\lambda_{\beta', A} + \ft12 \left( D_\mu \bar\phi_{AB} \right) \left( D_\mu \phi^{AB} \right) \nonumber\\ &-&\!\!\! \sqrt{2} \bar\phi_{AB} \left\{ \lambda^{\alpha, A}, \lambda_{\alpha}^{B} \right\} - \sqrt{2} \phi^{AB} \left\{ \bar\lambda^{\alpha'}_{A}, \bar\lambda_{\alpha', B} \right\} + \ft18 \left[ \phi^{AB}, \phi^{CD} \right] \left[ \bar\phi_{AB}, \bar\phi_{CD} \right]  \Bigg\}\ , \nonumber\\ &&
\end{eqnarray}
where we still use the definition $\bar\phi_{AB} \equiv \ft12 \epsilon_{ABCD}
\phi^{CD}$. These scalars come from the ten-dimensional gauge field, and can
be grouped into $\phi^{AB} = \ft1{\sqrt{2}} \mS^{a\, AB} A_a $, where $A_a$
are the first six real components of the ten dimensional gauge field $A_M$.
Using $\eta^{ab} = \{{1 \over \sqrt{2}} \mS^{aAB}, {1 \over \sqrt{2}}
\bar\mS^b_{AB}\}$
with $\eta^{ab} = (-1, +1, +1, +1, +1, +1)$ the vector indices
are turned into $SU(4)$ indices.  Writing the action in terms of the 6
scalars $A_a$, one of these fields, say $A_0$, has a different sign in the
kinetic term, which reflects the $SO(5,1)$ symmetry of the theory. In the basis with the $\phi^{AB}$ fields, we obtain formally the same action for the Minkowski case by reducing on a torus with $6$ space coordinates, but the difference hides in the reality conditions which we will discuss in the next subsection.
The action is invariant under the dimensionally reduced supersymmetry transformation rules
\begin{eqnarray} \delta A_\mu \!\!\!&=&\!\!\! - i {\bar\zeta}^{\alpha'}_A \bar\sigma_{\mu\, \alpha'\beta} \lambda^{\beta, A} + i \bar\lambda_{\beta', A} \sigma_\mu^{\alpha\beta'} \zeta^A_\alpha\ , \nonumber\\ \delta \phi^{AB} \!\!\!&=&\!\!\! \sqrt{2} \Big( \zeta^{\alpha,A} \lambda^B_\alpha - \zeta^{\alpha,B} \lambda^A_\alpha + \epsilon^{ABCD} \bar\zeta^{\alpha'}_C \bar\lambda_{\alpha',D} \Big)\ , \nonumber\\ \delta \lambda^{\alpha,A} \!\!\!&=&\!\!\! - \ft12 \sigma^{\mu\nu\, \alpha}_{\phan{mn}\beta} F_{\mu\nu} \zeta^{\beta,A} - i \sqrt{2} \bar\zeta_{\alpha',B} \not\!\! D^{\alpha\alpha'} \phi^{AB} + \left[ \phi^{AB}, \bar\phi_{BC} \right] \zeta^{\alpha,C}\ , \nonumber\\ \delta \bar\lambda_{\alpha',A} \!\!\!&=&\!\!\! - \ft12 \bar\sigma^{\mu\nu\phan{ii}\beta'}_{\phan{m}\alpha'} F_{\mu\nu} \bar\zeta_{\beta',A} + i \sqrt{2} \zeta^{\alpha,B} \not\!\!{\bar D}_{\alpha'\alpha} \bar\phi_{AB} + \left[ \bar\phi_{AB}, \phi^{BC} \right] \bar\zeta_{\alpha',C}\ . \end{eqnarray} Again, these rules are formally the same as in \eqn{susy-Mink}.  Note that the indices $A,B$ are lowered by complex conjugation, but the spinor indices $\alpha$ and $\alpha'$ are lowered by $\epsilon$- symbols.

\subsection{Involution in Euclidean space}

The Majorana-Weyl condition (\ref{MajoranaWeyl}) on $\mP$ leads in four-dimensional Euclidean space to reality conditions on $\lambda^\alpha$ which are independent of those on $\bar\lambda_{\alpha'}$, namely, \begin{equation} \label{RealityConds1} \left( \lambda^{\alpha, A} \right)^\ast = - \lambda^{\beta, B} \epsilon_{\beta\alpha} \eta^1_{BA} \ , \qquad \left( \bar\lambda_{\alpha', A} \right)^\ast = - \bar\lambda_{\beta', B} \epsilon^{\beta'\alpha'} \eta^{1, BA} \ . \end{equation} These reality conditions are consistent and define a symplectic Majorana spinor in Euclidean space. The $SU (2) \times SU (2)$ covariance of (\ref{RealityConds1}) is obvious from the pseudoreality of the ${\bf 2}$ of $SU (2)$, but covariance under $SO(5,1)$ can also be checked (use $[\eta^a, \bar\eta^b] = 0$). Since the first $\mS$ matrix has an extra factor $i$ in order that $\left( \mG^0 \right)^2 = - 1$, see (\ref{10Gammas}), the reality condition on $\phi^{AB}$ involves $\eta^1_{AB}$ \begin{equation} \label{RealityConds2} \left( \phi^{AB} \right)^\ast = \eta^1_{AC} \phi^{CD} \eta^1_{DB}\ . \end{equation} The Euclidean action in (\ref{N4susy}) is hermitean under the reality conditions in (\ref{RealityConds1}) and (\ref{RealityConds2}). For the $\sigma$-matrices, we have under complex conjugation \begin{equation} \left( \sigma_{\mu}^{\alpha\beta'} \right)^\ast = \sigma_{\mu\, \alpha\beta'} \ , \qquad \left( \bar\sigma_{\mu\, \alpha'\beta} \right)^\ast = \bar\sigma_{\mu}^{\alpha'\beta}\ .
\end{equation}
Due to the nature of the Lorentz group the involution cannot change one type of indices into another, as opposed to the Minkowskian case.

\section{Large instantons and the Higgs effect} 

We have seen in previous sections that the instanton measure on the
moduli space for
 pure $SU(2)$ gauge theory with one anti-instanton $(k=-1)$ is given by (dropping overall multiplicative factors of two and $\pi$)
\begin{equation}
{\rm d} {\cal M} \propto {\rm d}^4 x_0 {{\rm d} \rho \rho^3 \over g^8}
M^8_{PV} {\rm e}^{-\frac{8\pi^2}{g^2}} = \left( {{\rm d}^4 x_0 {\rm d} \rho
\over \rho^5 g^8} \right) {\rm e}^{-\frac{8\pi^2}{g^2}+4 \ln (\rho M_{PV})^2}
\ .
\end{equation}
The one-loop corrections coming from the determinants further modify the factor 4 into $4-\ft13 = \frac{11}{3}$, see \eqn{petervan}, and in addition yield some constants in the exponent.  The integral over $\rho$, the instanton size, is clearly nonsingular for small $\rho$ as long as asymptotic freedom holds\footnote{ One integrates $\rho$ up to the renormalization scale $\mu$, and instantons with scale $\rho$ yield the prefactor $\exp(-8\pi^2/g^2)$. The $g^2$ in this prefactor depends on $\rho$, not on $\mu$. One finds then that $g^2(\rho)=8\pi^2/(-\beta_1\ln(\rho\Lambda))$ where $\beta_1=-\frac{11}{3}C_2(G)+\cdots$ is negative if asymptotic freedom holds. So, if $-\beta_1+3\geq 0$, there is no singularity at $\rho=0$.} , but for large $\rho$ it diverges severely.  However, in a Higgs model, the mass term for the gauge bosons ($\cL =- \ft12 A^2_\mu g^2 v^2$ if there are no instantons) yields further terms of the form
\begin{eqnarray}\label{L-Higgs}
-\frac{1}{\hbar}\cL_{\rm cl} \; {\rm (Higgs)} \; =-\frac{1}{\hbar}2\pi^2
v^2 \rho^2 + \ldots \ .
\end{eqnarray}
Thus for spontaneously broken gauge theories the $\rho$ integral acquires a Gaussian cut-off, and yields a finite result.  This solves the large-$\rho$ problem for the electroweak interactions.  For QCD the situation is more complicated; in fact, the large-$\rho$ problem is presumably intimately related to confinement.  We now give some details.

The Higgs action for an $SU(2)$ Higgs doublet is given by
\begin{eqnarray}
&& \cL_H = D_\mu \varphi^\ast D_\mu \varphi + \lambda (\varphi^\ast \varphi - v^2)^2 \nonumber\\
&& D_\mu \varphi = \partial_\mu \varphi + A_\mu \varphi ;\qquad
 \varphi = \left( \begin{array}{cc} \varphi^+ \\ \varphi^0 \end{array} \right)
 ;\qquad A_\mu = A_\mu^a {\tau_a \over 2i}\ .
\end{eqnarray}
With $< \varphi^0 > = v$, the ordinary Higgs effect in Minkowski space gives a mass term $\cl =- \frac14 A^2_\mu v^2$ for the vector bosons\footnote{ One usually decomposes $\varphi^0$ into $\varphi^0=\frac{1}{\sqrt 2}(\sigma-i\chi_3)$, see below, with $<\sigma>=v_\sigma$. Then $v^2=\ft12 v_\sigma^2$, and the mass of the vector boson is $m_A = \frac12 gv$.}.  We could also discuss other representations for the Higgs field but the analysis is very similar, and a doublet is of course the most interesting case.
In Euclidean space we take for $A_\mu$ the regular selfdual instanton solution with $k=1$
\begin{equation}
A_\mu =- {\bar\sigma_{\mu\nu} x^\nu \over x^2 + \rho^2}\ .
\end{equation}
We next solve the $\varphi$ field equation in this instanton background.
For general $\lambda$, an exact solution to the coupled equations seems out of reach. We therefore
drop the potential term and only require that $| \varphi | \rightarrow | v |$ at large $| x |$~\footnote{ This is 't Hooft's approach \cite{Hooft1}. Note that the field equation for $A_\mu$ is not restricted due to the backreaction of the Higgs field. Affleck \cite{Affleck} considered instead the case $v=0$, $\lambda$ arbitrary, in which case the usual instanton solution together with $\varphi=0$ solves the coupled equations. Both approached yield equivalent results.}.
 So this is not an exact solution, but the first term in an approximate solution.   We shall discuss the higher-order terms later.  As we shall show, the solution to the equation $D^\mu D_\mu \varphi =0, | \varphi (|x| \rightarrow \infty )| = v$ is of the form $\varphi = f (r^2) \left( \begin{array} {cc} -i x_4 + x_3 \\ x^1 + i x^2 \end{array} \right)$.  This clearly looks awkward, and a more covariant way to construct the solution is to write $\varphi$ as
\begin{eqnarray}\label{eqn-phi}
\varphi = \left( \begin{array}{cr}  \varphi^+  & -   (\varphi^0)^\ast \\ \varphi^0 & (\varphi^+)^\ast \end{array} \right) \left( \begin{array}{cc} 1 \\ 0 \end{array} \right)\ ,
\end{eqnarray}
and to make the ansatz
\begin{eqnarray}
\varphi = v f(x^2) \left( \bar\sigma_\mu x_\mu / \sqrt{x^2} \right)\left( \begin{array}{cc} 1 \\ 0 \end{array} \right)\ ,
\end{eqnarray}
with $f(x^2) \rightarrow 1$ as $x^2 \rightarrow \infty$.  (Recall that one can always write ${\varphi^+ \choose \varphi^0}$ as ${1 \over \sqrt{2}} (\sigma + i \vec\chi \cdot \vec\sigma) {1 \choose 0}$ and this yields the form of $\varphi$ given in \eqref{eqn-phi} up to an inessential factor $i$).

The function $f(x^2)$ satisfies a second-order differential equation, but we do not analyze this equation, but present the result and check that it solves $D^\mu D_\mu \varphi =0$:
\begin{eqnarray}\label{higgs-inst}
\varphi = v \sqrt{{x^2 \over x^2 + \rho^2}} {\bar\sigma_\mu  x_\mu \over \sqrt{x^2}} \left( \begin{array}{cc} 1 \\ 0 \end{array} \right) = {v \over \sqrt{x^2 + \rho^2}} \bar\sigma_\mu x_\mu  \left( \begin{array}{cc} 1 \\ 0 \end{array} \right)\ .
\end{eqnarray}

The boundary condition is clearly satisfied because ${\bar\sigma_\mu x_\mu \over \sqrt{x^2}} {1 \choose 0}$ has unit norm.  It is straightforward to check that this expression for $\varphi$ satisfies the field equation.  Namely, omitting
the overall factor $v$ and the spinor ${1 \choose 0}$ one finds
\begin{eqnarray}
D_\mu \varphi &=& \partial_\mu \varphi + A_\mu \varphi = {\bar\sigma_\mu (x^2 + \rho^2) - x_\mu x_\nu \bar\sigma_\nu \over (x+ \rho^2)^{3/2}} - {(\bar\sigma_{\mu \nu} x_\nu) (\bar\sigma_\rho x^\rho) \over (x^2 + \rho^2)^{3/2}} \nonumber\\
&=&  \bar\sigma_\mu \rho^2 / (x^2 + \rho^2)^{3/2} \nonumber\\
D_\mu D_\mu \varphi &=& \partial_\mu D_\mu \varphi + A_\mu D_\mu \varphi =  {- 3 \bar\sigma_\mu x_\mu \rho^2 - (\bar\sigma_{\mu \nu} x_\nu ) \bar\sigma_\mu \rho^2 \over (x^2 + \rho^2)^{5/2}} =0\ .
\end{eqnarray}
Having found the solution of the field equation of the Higgs scalar in the background of an instanton, we now substitute it into the action to find the corrections to the classical action.  The kinetic term only yields a surface integral due to partial integration
\begin{eqnarray}
&&  \int D_\mu \varphi^\dagger D_\mu \varphi \; {\rm d}^4 x = \int {\rm d} \Omega_\mu (\varphi^\dagger D_\mu \varphi) \nonumber\\
&& = \lim_{x^2 \rightarrow \infty}  2 \pi^2 (x^2)^{3/2} v^2 \left( \begin{array}{cc} 1 \\ 0 \end{array} \right)^T   {\sigma_\nu x_\nu \over \sqrt{x^2 + \rho^2}} {1 \over \sqrt{x^2}} {x_\tau \bar\sigma_\tau \rho^2 \over (x^2 + \rho^2)^{3/2}} \left(  \begin{array}{cc} 1 \\ 0 \end{array} \right) = 2 \pi^2 v^2 \rho^2 \ . \nn &&
\end{eqnarray}
This is the extra term mentioned in \eqref{L-Higgs}.

However, the contribution of the term with the potential is divergent
\begin{eqnarray}
\lambda \int (\varphi^\ast \varphi - v^2)^2 {\rm d}^4 x =  \lambda \int \left( {v^2 \rho^2 \over x^2 + \rho^2} \right)^2 {\rm d}^4 x = \infty\ .
\end{eqnarray}
The reason for this divergence is clear:  we did not solve the full field equation, but rather took the instanton solution  of pure Yang-Mills theory, and solved the field equation for the scalar in this background, omitting the potential term.

We enter here the difficult area of ``constrained instantons"
\cite{Affleck, niels}.  There does not exist an exact and stable solution of
the coupled field equations, as can be shown as follows. Suppose there was a
solution with $\varphi \neq 0$, and a finite but nonvanishing action for the scalars.  If one
replaces $A_\mu (x)$ by $a A_\mu (ax)$ and $\varphi (x)$ by $\varphi (ax)$
(which preserves the boundary condition $| \varphi | \rightarrow v$) then
the action becomes upon also setting $ax =y$
\begin{eqnarray}
&& S_{\rm cl}  (a) = \int {\rm d}^4 y \left[ - {1 \over 2 g^2} \tr F^2_{\mu\nu} (y) + {1 \over a^2} \mid D_\mu \varphi (y) \mid^2 \right. \nonumber\\
&& \hspace{.65in} \left. + {1 \over a^4} \lambda (\varphi^\ast (y) \varphi (y) - v^2)^2  \right]\ .
\end{eqnarray}
Note that all three terms in the action are positive.
Replacing $A_\mu (x)$ by $a A_\mu (ax)$  for $a$ near unity amounts to a
particular small variation of $A_\mu$, and similarly for $\varphi$.
So one can make the value of the action slightly smaller by making $a$
slightly larger then unity. This proves that no solution exists. In fact, if
$a$ tends to infinity, we approach
the bound $S=8\pi^2/g^2$, but this bound can never be reached.
The expression for $a A_\mu (ax)$ is equal to the instanton solution with
$\rho$ replaced by $\rho /a$, and for $a\rightarrow \infty$ we get a zero-size
instanton.
That leaves open the possibility that a local minimum might stil
exist, but detailed analysis shows that this is not the case.
This scaling argument is called Derrick's theorem \cite{Der64}, and often
yields valuable information without having to perform integrals.

One can still use an approximate solution to find a large part of the
contributions to the path integral, and this approximate solution is obtained
by first inserting a constraint into the path integral which yields an exact
solution, and then to integrate over this constraint. The idea is as follows. There are one or at most a finite number of directions in field space along which the action decreases (``destabilizing directions'', in our $SU(2)$ model the directions parametrized by $a$). Deformations in all
other directions increase the action. The constraint prevents deformations in the destabilizing directions, and on first minimizes the action with the constraint present. The solution is called the constrained instanton. It looks like the instanton for pure Yang-Mills theory at short distances but decays exponentially at large distances. It has a particular value of $\rho$. Finally, one integrates with the measure for the zero modes over all values of $\rho$. The expectation is that this should capture most of the path integral, even though one is not expanding around a solution of the theory without constraint. For the $SU(2)$ instanton one may add a term $\sigma_1\int {\rm d}^4x [\tr F^3 -c_1\rho^{-2}]$ to the action to constrain deformations in the direction of the gauge zero mode $(\partial/\partial \rho)A_\mu^{{\rm cl}}$, and a term $\sigma_2[\int {\rm d}^4x (\varphi^*\varphi -v^2)^3-c_2 \rho^{-2}]$ to freeze deformations in the directions
of the matter zero mode $(\partial/\partial \rho)\varphi^{{\rm cl}}$, with
$\varphi^{{\rm cl}}$ given by \eqref{higgs-inst}. One might fix the values of $c_1$ and $c_2$ such that the constraint is satisfied for the instanton solution
and $\varphi$ in \eqref{higgs-inst}. The Lagrange multipliers $\sigma_1$ and $\sigma_2$ are then fixed order by order in perturbation theory, by requiring suitable boundary conditions for the deformations.

The result is that one can make an expansion of the full approximate solution in terms of $\rho v$ and finds
then the following results in the singular gauge \cite{Affleck, niels,ndorey}: \\
(i) inside a core of radius $\rho = {1 \over m_W}$ where $m_W=gv$, the approximate solution
given in \eqref{higgs-inst} is still valid \\
(ii) far away the solution decays exponentially, $A_\mu \sim \exp (-m_W |x|)$
and $| \varphi -v | \sim \exp (- m_H |x|)$ with $m_H=2{\sqrt \lambda}v$.\\
(iii) the integral over $| D_\mu \varphi |^2$ has the same leading term
$2 \pi^2 \rho^2 v^2 + \cO \Big(\lambda (v \rho)^4 \ln (v \rho \sqrt{\lambda})
\Big)$, but the potential term is now convergent and  yields a result
$\cO \Big(\lambda (v\rho )^4 \ln (v \rho \sqrt{\lambda})\Big)$.

Hence, the Higgs effect indeed solves the large $\rho$ problem, and
asymptotic freedom solves the small $\rho$ problem.  Constrained instantons
are also relevant for ${\cal N} = 1, 2$ SYM theories.  They can also be
studied in the context of topological YM theories, as was discussed e.g. in
\cite{BFT}.

\section{Instantons as most probable tunnelling paths}

Instantons of nonabelian gauge theories can be interpreted as amplitudes for
tunnelling between vacua in Minkowski space with different winding numbers
$Q$. We shall determine a path in Minkowski spacetime which yields the ``most
probable barrier tunnelling amplitude".  We follow closely \cite{bitar}, but
related work is found in \cite{bender,gervais}.

We begin with one particular path $A_{I, \mu} = \{ \vec{A}_I (\vec{x}, t) , A_{I,0} ( \vec{x}, t) \}$ from which we construct a class of paths which all differ by how fast one goes from one
configuration at $t_1$ to the next at $t_2$.  Namely, we make a coordinate transformation from $t$ to $\l (t)$ {\bf in Minkowski spacetime}
and consider the following collection of paths
\eqa \label{paths}
\vec{A}^{(\l)}_I ( \vec{x}, t) = \vec{A_I} ( \vec{x}, \l (t))\ ;\qquad  A^{(\l)}_{I,0} ( \vec{x}, t) = A_{I,0} (\vec{x} , \l (t)) \dot\l (t)
\eqae
(Often one works in the temporal gauge $A^{(\l)}_0 = 0$ because this makes the physical interpretation clearer.  All our results are, however,
gauge invariant).  The case $\l (t) = t$ yields the original path, but different $\l (t)$ yield paths which all run through the same sequence of
3-geometries $\vec{A}_I (\vec{x}, t_1), \vec{A}_I (x, t_2),  \vec{A}_I (\vec{x}, t_3) \ldots$ but at different speeds.  The variable $\l (t)$ can be
considered as a kind of collective coordinate which measures a kind of continuous winding number because we will start with one winding number and end up with another winding number.  For $t$ between $t_1$ and $t_2$ this continuous winding number is due to an integral $\int d^3 x \int^t_{t_{1}} dt' \del_\mu j^\mu$ over a surface where $A_\mu$ is not everywhere pure gauge.  Only for $t = t_1$ and $t = t_2$ does $A_\mu$ everywhere on the surface become pure gauge and only at these times the winding number is an integer. These initial and final configurations describe vacua of the theory in Minkowski spacetime. We can also consider another particular path $A_{II , \mu} = \{ \vec{A}_{II} ( \vec{x},
t) , A_{II,0} ( \vec{x}, t) \}$, and then we can in the same way create a second class of paths, parametrized again by the function $\l (t)$.  In this way we generate an infinite collection of classes of paths.

For a given class $A_\mu^{(\l)} ( \vec{x}, t)$, we can substitute $\vec{A}^{(\l)}$ and $A^{(\l)}_0$ into the action, and then we obtain, as we shall show,
the Lagrangian for a point particle (one dynamical degree of freedom)
\eqa
 L = {1 \over 2} m (\l) \dot{\l}^2 - V (\l)
\eqae
where $m (\l)$ and $V (\l)$ depend on the choice for $A_\mu$.  We shall then determine for which $m (\l)$ and $V (\l)$ the tunnelling rate is
maximal.  The solution of this problem in Minkowski space involves instantons in Euclidean space.  A crucial role is played by the notion of a
winding number in Minkowski space, so we first discuss this subject.

One can define a winding number $Q$ in Minkowski space in the same way as in Euclidean space because $Q$ does not depend on the metric (in technical terms it is an affine quantity)
\eqa
Q &=& {-1 \over 64 \pi^2} \int\limits^{\s_2}_{\s_1} F^a_{\mu\nu} F^a_{\rho\s} \e^{\mu\nu\rho\s}\, {\rm d}^4 x \nn
   &=& {1 \over 32 \pi^2} \int (\tr\, F_{\mu\nu} \e^{\mu\nu\rho\s} F_{\rho\s} ) \,{\rm d}^4 x \nn
   &=& {- 1 \over 4 \pi^2} \int \tr\, \vec{E} \cdot \vec{B}\, {\rm d}^4 x
   \label{shingles}
\eqae
where ${\rm d}^4 x = {\rm d}^3 x {\rm d}t$ and $\e^{0123} =+1$, and we used that $A_\mu = A^a_\mu T_a$ with $T_a = - {i \over 2} \s_a$ so that $\tr(T_a T_b)
=- {1 \over 2} \delta_{ab}$
and the structure constants are given by $[T_a, T_b] = \e_{ab}{}^c T_c$, so $f_{ab}{}^c = \e_{ab}{}^c$.  Furthermore, by definition $E_j = - F_{0j}$ and $B_j = {1
\over 2} \e_{jkl} F_{kl}$.  Because we are (and stay all the time) in Minkowski space, $\e^{0123} = - \e_{0123} = +1$ and $-F_{\mu\nu}
F^{\mu\nu} = 2 F^2_{0i} -F^2_{ij}$.  The integral is taken between two 3-dimensional hypersurfaces $\s_1$ and $\s_2$ at $t_1$ and $t_2$.

If at $t_1$ the configuration $A_\mu (\vec{x}, t)$ describes a
vacuum, it has by definition vanishing energy.  Since the energy
density\footnote{ The gravitational stress tensor is $T_{\mu\nu} =
F^a_{\mu\rho} F^{a \rho}_\nu - \frac14 \eta_{\mu\nu} F^a_{\rho\s}
F^{a,\rho\s}$ and $T_{00} = \frac12 (E^a)^2 + \frac12 (B^a)^2$.  One
can also obtain $T_{\mu\nu}$ from canonical methods as follows.
Evaluating $H= p \dot{q} -L$ with $q = A_j$ and $p = - E_j$ one
finds upon using that $\dot{A}_j = F_{0j} + D_j A_0$ and partially
integrating that $H = \int \left[ {1 \over 2} \left\{ (E_j^a)^2 +
(B^a_j)^2 \right\} + A_0^a (D^j E_j^a) \right] {\rm d}^3 x$ plus a
boundary term.  For solutions of the field equations such as the
vacuum, $D^j E_j =0$.  For configurations with finite energy $(E =
\cO {1 \over r^2})$ the boundary term vanishes when $A_0$ falls off
like $\cO ( {1 \over r})$.  Moreover in the temporal gauge the last
term vanishes.  Actually, according to the Dirac formalism, the
Gausz operator $D^j E_j$ is a first-class constraint, and should be
omitted from the Hamiltonian.  Thus, $H= \int \left[ {1 \over 2}
(E^a_j)^2 + {1 \over 2} (B^a_j)^2 \right] {\rm d}^3x$ also according
to canonical methods.} is given by $\ch = {1 \over 2} (\vec{E}^a)^2
+ {1 \over 2} ( \vec{B}^a)^2$, vanishing energy means $F^a_{\mu\nu}
=0$, hence $A_\mu$ is pure gauge at $t =t_1$ \eqa A_\mu (\vec{x},
t_1) = {\rm e}^{- \a (\vec{x}, t_1)} \del_\mu {\rm e}^{\a ( \vec{x},
t_1)}\ . \eqae Similarly, at $t_2$ we have $A_\mu ( \vec{x}, t_2) =
{\rm e}^{- \beta ( \vec{x}, t_2)} \del_\mu {\rm e}^{\beta (\vec{x},
t_2)}$.  We now choose the temporal gauge \eqa A_0 ( \vec{x}, t) =0\
. \eqae Having fixed $A_0 =0$, there are still residual
space-dependent gauge transformations possible because they preserve
the gauge $A_0 =0$.  To check this statement is easy: \eqa A'_0 (
\vec{x}, t) = {\rm e}^{-\ggg (\vec{x})} \del_0 {\rm e}^{\ggg
(\vec{x})} =0\ . \eqae We use these residual gauge transformations
to set $\a ( \vec{x}, t_1) =0$.\footnote{ With $A_j (\vec{x}, t_1) =
{\rm e}^{-\a (\vec{r}, t_1)} \del_j {\rm e}^{\a ( \vec{x}, t_1)}$ we
get $A'_j = {\rm e}^{- \ggg (\vec{x})} {\rm e}^{- \a ( \vec{x},
t_1)} \del_j ({\rm e}^{\a (\vec{x}, t_1)} {\rm e}^{\ggg ( \vec{x})}
)$ and clearly $A'_j = 0$ if we take ${\rm e}^{\ggg ( \vec{x})}$ to
be the inverse of ${\rm e}^{\a ( \vec{x}, t_1)}$.}  Then $A_\mu (
\vec{x}, t_1) =0$ for all $\mu$ and all $\vec{r}$.

Note that even if there is winding in the vacuum at $t=t_1$ (such winding at one fixed time is discussed below (\ref{mapss})), one can still gauge it away by a time-independent gauge transformation, but then the winding at $t=t_2$ increases by just the same amount.  This is as it should be, because the total winding is gauge-invariant.

We shall consider paths from $\s_1$ to $\s_2$ which at every time
$t$ have finite energy (finite integral $\int (E^2 + B^2) {\rm d}^3
x$). This means that the energy density for fixed $t$ must tend to
zero for $| \vec{x} | \rightarrow \infty$ (to make the integral
$\int (E^2 + B^2) {\rm d}^3 x$ convergent), hence at large $|
\vec{x} |$ the gauge fields become pure gauge \eqa A_\mu ( \vec{x},
t) {\under{| \vec{x} | \rightarrow \infty}{\hbox to
33pt{\rightarrowfill}}} \,\,{\rm e}^{- \a (\vec{x}, t)} \del_\mu
{\rm e}^{\a (\vec{x}, t)} \ . \eqae But since $A_0 ( \vec{x}, t )
=0$, we see that $\a ( \vec{x}, t)$ is independent of $t$.  Because
$\a ( \vec{x}, t_1) =0$ we obtain $\a ( \vec{x}, t) = 0$ for all $t$
and $| \vec{x} | \rightarrow \infty$.  This means in particular that
at $t_2$ for large $| \vec{x} |$ the gauge fields tend to zero \eqa
A_j ( \vec{x}, t_2) {\under{| \vec{x} | \rightarrow \infty}{\hbox to
33pt{\rightarrowfill}}}\,\,  0 \eqae

The fact that for large $| \vec{x} |$ all $A_j$ vanish allows us to compactify the 3-dimensional spacelike hypersurfaces at fixed $t$ into spheres
$S_3$.  The north pole of each sphere corresponds to all points with $| \vec{x} | = \infty$, and at this point on $S_3$ all $A_j$ vanish.  Thus, {\bf all
3-spaces at fixed $t$ compactify to a sphere $S_3$}.  We summarize the results in a figure
\eqa
\includegraphics[height=70mm,width=120mm]{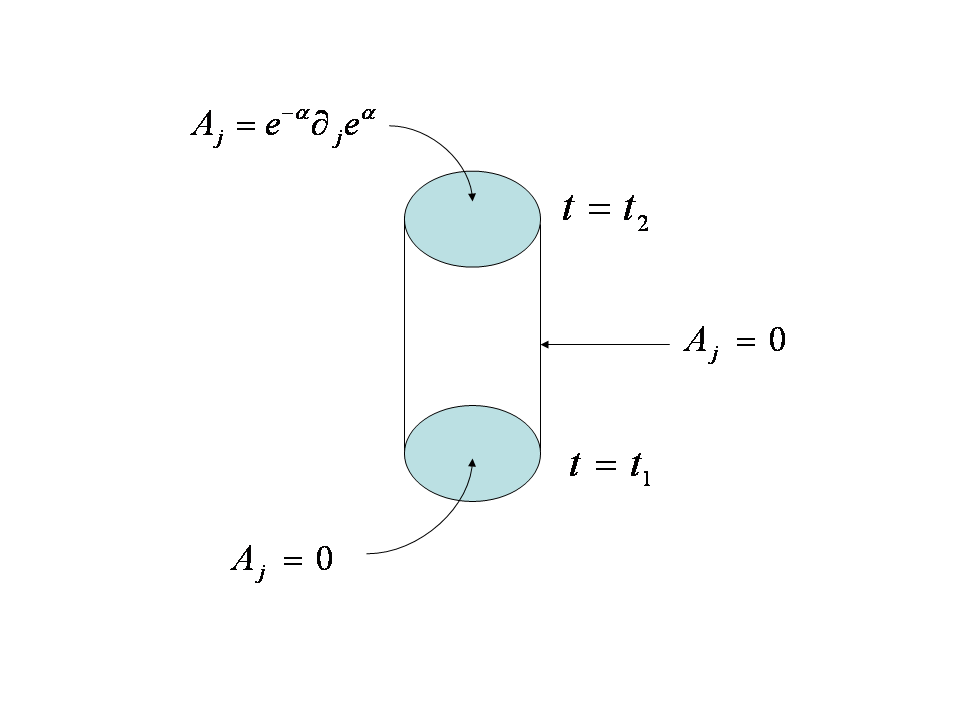}
\eqae
Everywhere on the boundary of this cylinder the gauge fields vanish, except at the disk at $t = t_2$, but there $A_0 =0$ and $A_j$ are only pure gauge.

We now return to $Q$.  First of all, $Q$ can be written as a total derivative, using the same algebra as in Euclidean space
\eqa
Q= {1 \over 8 \pi^2} \e^{\mu\nu\rho\s} \int \del_\mu \tr\, [ A_\nu \del_\rho A_\s + {2 \over 3} A_\nu A_\rho A_\s ] {\rm d}^4 x\ .
\label{two8}
\eqae
(we recall that $\tr\, [A_\mu A_\nu A_\rho A_\s \e^{\mu\nu\rho\s}] =0$).
Furthermore, since on the boundary $F_{\mu\nu} =0$, we can replace
$\del_\rho A_\s$ by $-A_\rho A_\s$ in (\ref{two8}).  We then find
\eqa
Q= {-1 \over 24 \pi^2} \e^{\mu\nu\rho\s} \int {\rm d} \s_\mu
\tr\, [ A_\nu A_\rho A_\s] \qquad A_\nu = {\rm e}^{-\a}
\del_\nu {\rm e}^\a\ .
\eqae
Since $A_0 =0$, there is no contribution from the sides of the cylinder, and
since $A_j =0$ at the bottom, there is also no contribution from the
bottom.  Hence in the gauge we have chosen, all contributions to the winding
come from the top of the cylinder:
\eqa
Q = {-1 \over 24 \pi^2} \e^{0ijk} \int Tr ({\rm e}^{-\a} \del_i {\rm e}^\a)
({\rm e}^{-\a}  \del_j {\rm e}^\a) ({\rm e}^{-\a} \del_k {\rm e}^\a)
{\rm d}^3 x\ .
\eqae

At the top of the cylinder the 3-space $t=t_2$ compactifies to a sphere $S_3$ (space).
The map from this 3-sphere into the group $SU(2)$ is a map
from one $S_3$ to another $S_3$\footnote{ The matrix elements of
any $2 \times 2$ complex matrix can be written as $g = a_\mu
\s^\mu$ with $\s^\mu = \{ \vec{\s} , I \}$ and $\mu = 1,2,3,0$.
Unitarity requires that $g^\dagger = a^\ast_\mu \s^\mu$ equals
$g^{-1}$, hence $g^\dagger g = \sum | a_\mu |^2 + (a^\ast_j a_k i
\e_{jkl} + a^\ast_0 a_{l} + a^\ast_l a_0) \s^l =1$.  Hence $| a_0
|^2  + | a_k |^2 =1$ and the coefficients of $\s_l$ must vanish.
The determinant yields $\det g = a^2_0 - a^2_k$, and since also $|
a_0 |^2 + | a_k |^2 =1$, requiring $\det g=1$ leads to $a_k = \pm i
| a_k |$ and $a_0 = \pm | a_0 |$.  Then we are left with $g =
a_0 I + i a_k \s_k$ with real $a_0$ and $a_k$ satisfying
$a^2_0 + a^2_k =1$ which defines $S_3$.}
because (i) we can always compactify the ${\bf R}^3$ with coordinates
$\vec{x}$ to an $S_3$ and (ii) the gauge fields at $| \vec{x} | =
\infty$ are equal (and vanish)
\eqa
\includegraphics[height=70mm,width=120mm]{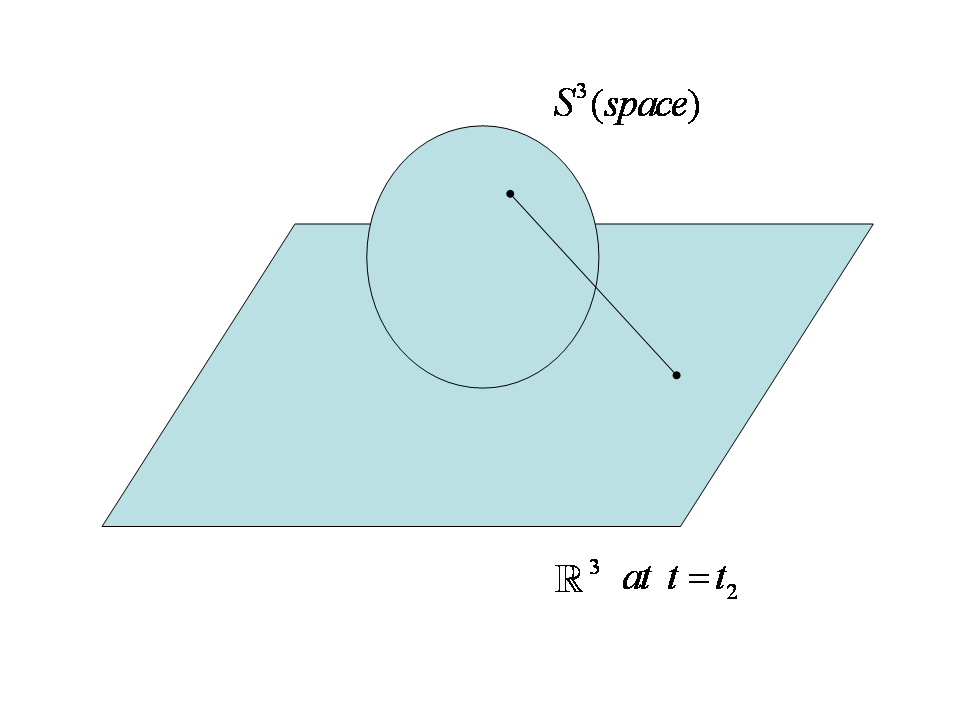}
\label{mapss}
\eqae

The maps $S_3$ (space) $\rightarrow S_3$ (group) in Minkowski space
fall into equivalence classes with a winding number $k \in {\bf Z}$,
just as the maps of instantons in Euclidean space give maps from
$S_3$ (space) $\rightarrow S_3$ (group).  In the latter case $S_3$
(space) is the boundary of all of ${\bf R}^4$ while here it is the
compactification of the whole ${\bf R}^3$ at $t=t_2$. It follows
that \eqa Q = \pm k \qquad k \in {\bf Z}\ . \eqae

We now can draw a picture of the energy $H = \int \ch {\rm d}^3 x$ at times $t$ as we move from $t=t_1$ to $t=t_2$.  Initially and at the end
one has $H=0$, but in between we must have $H > 0$ (note that $\ch \geq 0$) for the following reason.
\eqa
\includegraphics[height=60mm,width=120mm]{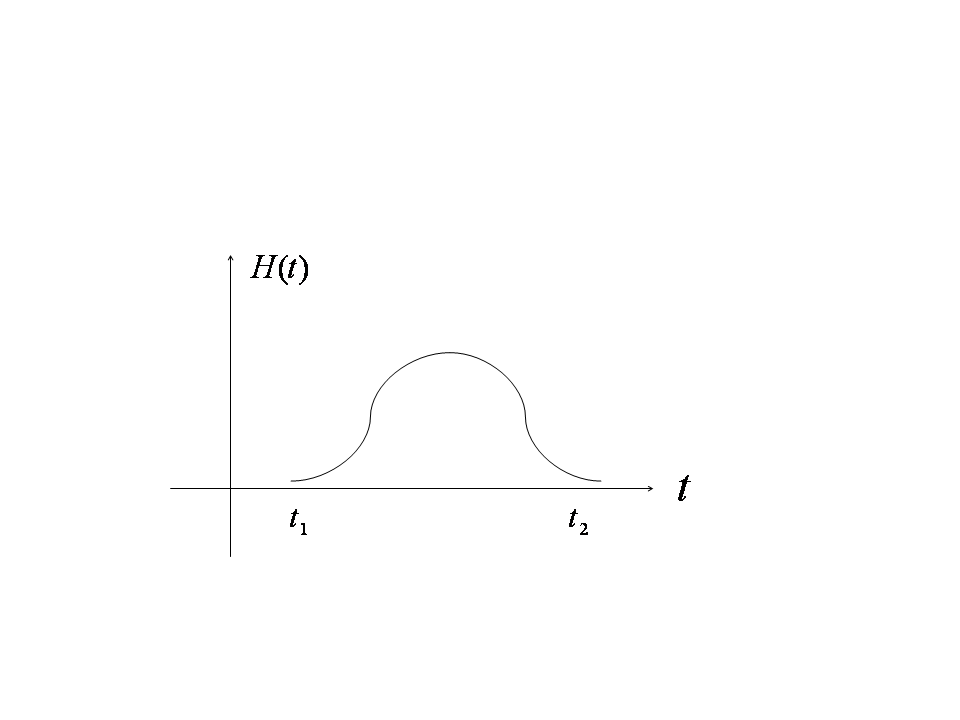}
\label{two13}
\eqae
There are no paths possible which connect the
vacuum at $t_1$ to the vacuum at $t_2$ which are solutions of the
field equations because if $F_{\mu\nu} = 0$ on $\s_1$ (or $\s_2$)
and the field equations are satisfied, one has $F_{\mu\nu} =0$
everywhere\footnote{ For the proof, note that if at $t_1$ one has
$F_{\mu\nu} =0$ and at all $t$ one has $D^\mu F_{\mu \nu} =0$,
then $\del^\mu F_{\mu j} = \del_0 F_{0j} =0$ at $t_1$.
Furthermore, the Bianchi identity $D_0 F_{ij} + D_i F_{j0} + D_j
F_{0i} =0$ yields $\del_0 F_{ij} =0$.  Hence $\del_0 F_{\mu\nu}
=0$ at $t_1$.  Also $\del_j F_{\mu\nu} =0$ because $F_{\mu\nu} =0$
at $t =t_1$ for all $x$.  Hence $\del_\rho F_{\mu\nu} =0$ at
$t=t_1$ for all $\rho, \mu, \nu$. We can rewrite this as $(D_\rho
F_{\mu\nu}) =0$ at $t = t_1$.  Next we repeat this analysis by
noting that also $D_\mu (D_\rho F^{\mu\nu}) =0$ at $t =t_1$,
because $D_\mu (D_\rho F^{\mu\nu} ) = [D_\mu, D_\rho ] F^{\mu\nu}
+ D_\rho (D_\mu F^{\mu\nu})$ and $D_\mu F^{\mu\nu} =0$ everywhere.
This shows that $\del_0 (D_\rho F^{0j})=0$ at $t = t_1$.  To also
show that $\del_0 (D_\rho F_{ij}) =0$ at $t = t_1$ we rewrite
$\del_0 (D_\rho F_{ij} ) = - \del_0 D_i F_{j \rho} - \del_0 D_j
F_{\rho i}$ and then use $D_0 (D_i F_{\mu\nu}) = [D_0 , D_i ]
F_{\mu\nu} + D_i (D_0 F_{\mu\nu}) =0$.  In this way we get
$\del^n_0 F_{\mu\nu} =0$ for any $n$.  Hence $F_{\mu\nu} =0$ at
all $t$.}.  But, if $F_{\mu\nu}$ would vanish everywhere, $Q \sim
\int E \cdot B {\rm d}^4 x$ would vanish, hence one could not change the
winding number.  The conclusion is that paths which go from one
vacuum with winding number zero to another vacuum with
nonvanishing winding number necessarily have positive energy at
some intermediate times.

We are now ready to define a subset of paths which depend on one collective
coordinate, and to which (we claim) we can
restrict our attention.  Consider first one given path corresponding to a fixed field configuration $A_j ( \vec{x}, t)$.  Instead
of this single path, we consider the set of paths $A^{(\l)}_j ( \vec{x}, t)$,
as defined in (\ref{paths}).
Each path is labeled by a different function $\l (t)$, and is defined by
\eqa
A^{(\l)}_j ( \vec{x}, t) = A_j ( \vec{x}, \l (t)) \ .
\eqae
As we already discussed, for $\l (t) =t$ we recover the original path, but for different $\l (t)$ we obtain paths which run through the same
3-dimensional configurations $\vec{A} ( \vec{x}, t_1), \vec{A} ( \vec{x}, t_2) , \vec{A} (\vec{x}, t_3) \ldots$ at different speeds.
For example if $\l (t)$ is constant for some time interval, the corresponding $\vec{A} ( \vec{x}, t)$ do not change, but if $\l (t)$ changes rapidly, the sequence of $A (\vec{x}, t)$ is traversed rapidly.

Each path $A_j ( \vec{x}, \l (t))$ should begin at $A_j ( \vec{x}, t_1)$ and end at $A_j ( \vec{x}, t_2)$, so we require $\l (t_1) =t$, and $\l (t_2) =
t_2$, but between $t_1$ and $t_2$ the function $\l (t)$ is arbitrary.  We shall later take $t_1 =- \infty$ and $t_2 = + \infty$, and then
also require that $\l (t_1) =- \infty$ and $\l (t_2) = + \infty$.  Given a path $A^{(\l)}_j ( \vec{x}, t)$ we can compute the electric and magnetic fields
\eqa
&& - E_j = F_{0j} = \del_0 A^{(\l)}_j ( \vec{x}, t) = {\del A_j \over \del \l} ( \vec{x}, \l (t)) \dot\l \; \mbox{because} \; A_0 ( \vec{x} , t) =0 \nn
&& B_i = {1 \over 2} \e_{ijk} F_{jk} = {1 \over 2} \e_{ijk} ( \del_j A_k ( \vec{x}, \l (t)) + A_j ( \vec{x}, \l (t)) A_k ( \vec{x}, \l (t)) - j \leftrightarrow k) \ . \nn &&
\eqae

The Lagrangian $L = \int \cl {\rm d}^3 x$ with $\cl = {1 \over 2g^2}
\tr\,  F^2_{\mu\nu} = {-1 \over g^2} \tr\, ( \vec{E}^2 - \vec{B}^2)$
can then be written as
\eqa
L &=& {1 \over 2} m (\l) \dot{\l}^2 - V (\l) \ ,\nn
m (\l) &=& {-2 \over g^2} \int \tr\,  \left( {\del \vec{A} \over \del \l} \right)^2 {\rm d}^3 x  \geq 0 \ ,\nn
V (\l) &=& - {1 \over g^2} \int \tr\, \vec{B}^2  \;  {\rm d}^3 x  \geq 0
\label{two16}
\eqae
The momentum conjugate to $\l (t)$ is $p (\l) =  {\del \over \mathstrut \del \dot{\l}} L = m (\l) \dot{\l}$.  Hence
\eqa
H = {(p (\l))^2 \over 2m (\l)} + V (\l)\ .
\eqae
For a given path $A^{(\l)}_j ( \vec{x}, t)$ one can plot $H$ as a function of $t$, and one finds then the profile in figure (\ref{two13}).

We have thus isolated a class of paths $A^{(\l)} (\vec{x}, t)$ which depends on one collective coordinate $\l (t)$.  For one given $A (\vec{x},
t)$, this still yields an infinite set of paths, but all these paths run through the same set of 3-configurations $A_j ( \vec{x}, t_1), A_j (
\vec{x}, t_2), \ldots$.  These are, of course, infinitely many other collective coordinates which describe a general path $A_j (x,t)$, but the
idea is that $\l (t)$ is the relevant coordinate to describe tunnelling, while the other collective coordinates describe variations away from
the paths $A^{(\l)}_j ( \vec{x}, t)$ which give only small corrections to the results obtained from $\l (t)$.  It is, of course, difficult to prove this assertion; one could begin with two collective coordinates as a start, but even this would lead to a complicated analysis.

The action for $\l (t)$ in (\ref{two16}) can be viewed as the action for one point particle.  This particle feels the potential barrier $V(\l)$, and to
go from the vacuum at $t=t_1$ with $V (\l) =m (\l) =0$ to the vacuum at $t_2$ with also $V(\l) = m (\l) =0$, we need tunnelling.  The tunnelling rate $R$ in quantum mechanics is proportional to ${\rm e}^{-2R}$ where
\eqa
 R = \int\limits^{\l_2}_{\l_1} {\rm d} \l \sqrt{2 m (\l) (V (\l) -E)}\ ,
\label{two18}
\eqae
with $\l (t_1) \equiv \l_1 = t_1, \l (t_2) \equiv \l_2 = t_2$ and $V (\l (t_1)) = V (\l (t_2)) =0$ and
$m (\l (t_1)) = m ( \l (t_2)) =0$.
We also set $E=0$
because we consider tunnelling from one vacuum (with $E=0$) to another.

There is, of course, an important difference with ordinary quantum mechanics.  The point particle $\l (t)$ feels a potential $V(\l)$, but both are derived from the same object, the fields $A_j (x, \l (t))$.  In addition the mass is here ``position"-dependent, $m =m (\l)$.  One can show that in quantum mechanics the formula for $R$ also holds if the mass $m (\l)$ depends on the point particle $\l (t)$). The crucial step is now to pose the question: {\bf for which set of paths
$\vec{A} ( \vec{x}, \l (t))$ is the tunnelling rate maximal?}  The tunnelling rate for the quantum mechanical particle $\l (t)$ can
be described by Minkowski path integrals, so we ask: for which $\vec{A} (\vec{x}, t)$ is there least destructive interference of the associated paths $A( \vec{x}, \l (t))$ in the path integral?  Clearly, $V(\l)$ should be as small as possible, but it cannot be too small because it must produce winding.

The tunnelling rate is ${\rm e}^{-2R}$ where according to (\ref{two18})
\eqa
&& R = \int\limits^{\l_2}_{\l_1} {\rm d} \l \; 2 \left[ \left( {1 \over g^2} \int \tr\, \left( {\del \vec{A} \over \del \l} \right)^2 {\rm d}^3 x \right) \left( {1
\over g^2} \int \tr\, \vec{B}^2 {\rm d}^3 x \right) \right]^{1/2}  \nn
&& = {2 \over g^2} \int\limits^{t_2}_{t_1} {\rm d}t [ ( \tr\, \int \vec{E}^2
{\rm d}^3 x) (\tr\, \int \vec{B}^2 {\rm d}^3 x) ]^{1/2}\ .
\eqae
We replaced $d \l$ by $d t \dot{\l}$ and brought $\dot{\l}$ inside the square root.  The fields $\vec{E}$ and $\vec{B}$ still depend on $\l (t)$.
Since $\tr\, \int \vec{a} (\vec{x}) \vec{b} (\vec{x}) d^3 x$ is an inner product, while $\int \tr\, \vec{E} \cdot \vec{B}$ is proportional to the winding number according to (\ref{shingles}), we have the triangle inequality
\eqa
R \geq {2 \over g^2} \Big| \int\limits^{t_2}_{t_1} (\tr\, \vec{E} \cdot \vec{B} ) {\rm d}^4 x \Big| = {8 \pi^2 \over g^2} \mid Q \mid \ .
\eqae
Hence the tunnelling amplitude is bounded from above by
\eqa
{\rm e}^{-R} \leq -{\rm e}^{-{8 \pi^2 \over g^2} \mid  Q \mid}\ .
\eqae

The inequality is saturated when $\vec{E}$ is parallel to $\vec{B}$
at each vector $\vec x$ and at each time $t: \vec{E} ( \vec{x}, t) =
\a (t) \vec{B} ( \vec{x}, t)$. The claim is that among all paths
with the same $Q$, the paths with the smallest $R$ are the paths
with $\vec{E}$ parallel to $\vec{B}$.

Let us discuss the meaning of this result.  Paths which interpolate between vacua with different winding number must produce
 electric and magnetic fields $\vec{E}$ and $\vec{B}$ in between at finite $\vec{x}$ and $t$ which cannot be too small, namely $| \int (
E^a_j B^a_j) {\rm d}^4 x | $ should be equal to $8 \pi^2 | Q |$.  On the other hand, the tunnelling rate is proportional to the length of $E^a$ times the
length
of $B^a$, so to make the tunnelling rate as large as possible, the product of these lengths should be as small as possible.  One could set up
a variational problem for $R$ under the constraint that $\int \tr\, \vec{E} \cdot \vec{B}\, {\rm d}^4 x$ be equal to $4 \pi^2 Q$, but we shall not work this
out.

The bound is reached, namely the tunnelling rate is maximal, when
the set of paths $A_j ( \vec{x}, \l (t))$ produces parallel electric
and magnetic fields \eqa \vec{E} (\vec{x}, \l (t)) = \a (t) \vec{B}
( \vec{x}, \l (t) )\ . \eqae Of course, $\a (t)$ can also be viewed
as a function of $\l (t)$ because $\l (t)$ is just another
parametrization of the time interval.  Note that this condition does
not change if one changes the parametrization from $\l (t)$ to
another function $\l' (t)$, because under such reparametrizations
$\vec{E}$ scales by a constant factor $\partial \l' / \partial \l$,
which cancels the Jacobian in \eqn{two18} for this change of
integration variables.  We use this scaling property to select a
particular $\l_0 (t)$ such that $\vec{E} ( \vec{x} , \l_0 (t) ) =
\pm \vec{B} (\vec{x} , \l_0 (t))$.  The property of $\vec{E}$ and
$\vec{B}$ being parallel is also a gauge-invariant property, and
$\cl$ and $R$ are of course gauge-invariant.    So, our
characterization of paths with maximal tunnelling rate is
gauge-invariant, as it should be.  Thus the use of temporal gauge
did not restrict the generality of the results.

We now can establish the connection between tunnelling and
instantons. The fields for which $\vec{E}$ and $\vec{B}$ in
Minkowski space are parallel are closely connected to instantons in
Euclidean space.  Namely, among the class of paths $\vec{A} (
\vec{x}, \l (t))$ parametrized by $\l (t)$, there is the path
$\vec{E} ( \vec{x}, \l_0 (t)) = \vec{B} ( \vec{x}, \l_0 (t))$ (and
another path with another $\l'_0 (t)$ such that $\vec{E} (\vec{x},
\l_0 (t)) =- \vec{B} (\vec{x}, $ $\l_0 (t))$).  If we then define
Euclidean gauge fields $A^E_\mu (x,t)$ by $A^E_j (\vec{x}, t) = A_j
(\vec{x}, \l_0 (t))$ and $A^E_4 ( \vec{x} , t) = A_0 (\vec{x} ,
\lambda (t)) {d \lambda \over dt}$ then this $A^E_\mu (\vec{x}, t)$
is self dual.  The parameter $t$ is Minkowski time, but in the
expressions for $A^E_j ( \vec{x}, t)$ we should interpret $t$ as the
Euclidean time.

Summarizing:  the most probable tunnelling paths are given by the set of paths $A_j (\vec{x}, \l (t))$ with parallel $\vec{E}$ and
$\vec{B}$ fields.  A given class of paths with $\vec{E}$ parallel to $\vec{B}$ contains one path which, when viewed as a configuration in Euclidean space,  is an
instanton.  Conversely, given an instanton $A^E_\mu (\vec{x}, t)$ in Euclidean space, one can construct a corresponding set of paths
$A^M_\mu (x, \l (t))$ in Minkowski space by setting
\eqa
A^{M, (\l)}_j (x, t) &=& A^E_j ( \vec{x}, \l (t)) \nn
A^{(M, (\l)}_0 (x, t) &=& A^E_4 ( \vec{x}, \l (t)) \dot\l\ .
\eqae

As an example we take the $Q = - 1$ anti-instanton solution in regular gauge, $A_\mu =- \s_{\mu\nu} x^\nu / (x^2 + \rho^2)$, see \eqn{overr}, which yields the following set of paths in Minkowski space
\eqa \label{A-lambda}
\left. \begin{array}{ll} A^{(\l)}_0 ( \vec{x}, t) = {-i \vec{x} \cdot \vec{\s} \over \vec{x}^2 + \l (t)^2 + \rho^2} \dot{\l} (t) & \\
\\ \vec{A}^{(\l)} (\vec{x}, t) = {i \l (t) \vec{\s} - i \vec{x} \times \vec{\s} \over \vec{x}^2 + \l (t)^2 + \rho^2} & \end{array} \right\} \begin{array}{ll} \l (t \rightarrow - \infty) =- \infty \\ \l ( t \rightarrow + \infty) = + \infty \; . \end{array}
\eqae
We are clearly not in the temporal gauge, but since our results are
gauge-invariant, it does not matter which gauge we use.    We still have
$A_\mu \rightarrow 0$ at large $| \vec{x} |$, so that we still have the
notion of winding as a map from $S_3$ (space) into $S_3$ (group) at each time.

Straightforward calculation yields for the curvatures in Minkowski space
\eqa
F_{01} &=& \del_0 A_1 - \del_1 A_0 + [A_0, A_1 ] = {2 i \rho^2 \s_1 \over (\vec{x}^2 + \l^2 + \rho^2)^2} \dot{\l}\ , \nn
F_{23} &=& \del_2 A_3 - \del_3 A_2 + [A_2 , A_3] = {2 i \rho^2 \s_1 \over (\vec{x}^2 + \l^2 + \rho^2)^2}\ .
\eqae
Hence
\eqa
\vec{E} = {- 2i \rho^2 \vec{\s} \over ( \vec{x}^2 + \l^2 + \rho^2)^2} \dot{\l}
\ ;\qquad \vec{B} = {2 i \rho^2 \vec{\s} \over (\vec{x}^2 + \l^2 + \rho^2)^2}
\ ,
\eqae
which depend on $x^2 = \vec{x}^2 + \l (t)^2$ ({\bf not} on $\vec{x}^2 - t^2$).  Hence, $\vec{E}$ is indeed parallel to $\vec{B}$ (in fact, anti-parallel).

The winding number $Q$ can be written in two ways
\eqa
Q &=& {- 1
\over 4 \pi^2} \int\limits^\infty_{-\infty} [ \tr\, \vec{E} \cdot
\vec{B} {\rm d}^3 x ] {\rm d}t \nn &=& {-1 \over 24 \pi^2} \e^{\mu\nu\rho\s}
\int \del_\mu \tr\,  [A_\nu A_\rho A_\s ] {\rm d}^4 x\ .
\eqae
In the latter expression $Q$ receives only a contribution from the
boundary,\footnote{ For example, the contribution to $Q$ from the
surface at $t = t_1$ is proportional to $\int {t ( \vec{x}^2 +
t^2) {\rm d}^3 x \over (t^2 + \vec{x}^2 + \rho^2)^3}$ which is
nonvanishing.  On the other hand, the contribution to $Q$ from the
sides of the cylinder converges for large $| t |$.} but in the
former expression we compute $Q$ by integrating over all space and
time.  It is then natural to define a $t$-dependent function by
integrating only up to a time $t$
\eqa
q (t) &=& {- 1 \over 4
\pi^2} \int^t_{-\infty} \left[ \int \tr\, \vec{E} \cdot \vec{B}
{\rm d}^3 x \right] \nn &=& -{1 \over 4 \pi^2} \int^\l_{-\infty} {\rm d} \l
\int {\rm d}^3 x {24 \rho^4 \over [ \vec{x}^2 + \l^2 + \rho^2 ]^4} \nn
&=& - {3 \over 4} \int^\l_{-\infty} {\rho^4 {\rm d} \l \over (\lambda^2
+ \rho^2)^{5/2}} \nn &=& - {3 \over 4} \int^{\l /\rho}_{-\infty}
{{\rm d}y \over (y^2 + 1)^{5/2}} \nn &=& - {3 \over 4} \left(t - {1
\over 3} t^3 \right) \bigg|^x_{-1} \qquad {\rm with} \; x= {\l
\over \sqrt{\l^2 + \rho^2}}\ . \eqae
Clearly, $q(t)$ is
gauge-invariant and has the following form \eqa
\includegraphics[height=70mm,width=120mm]{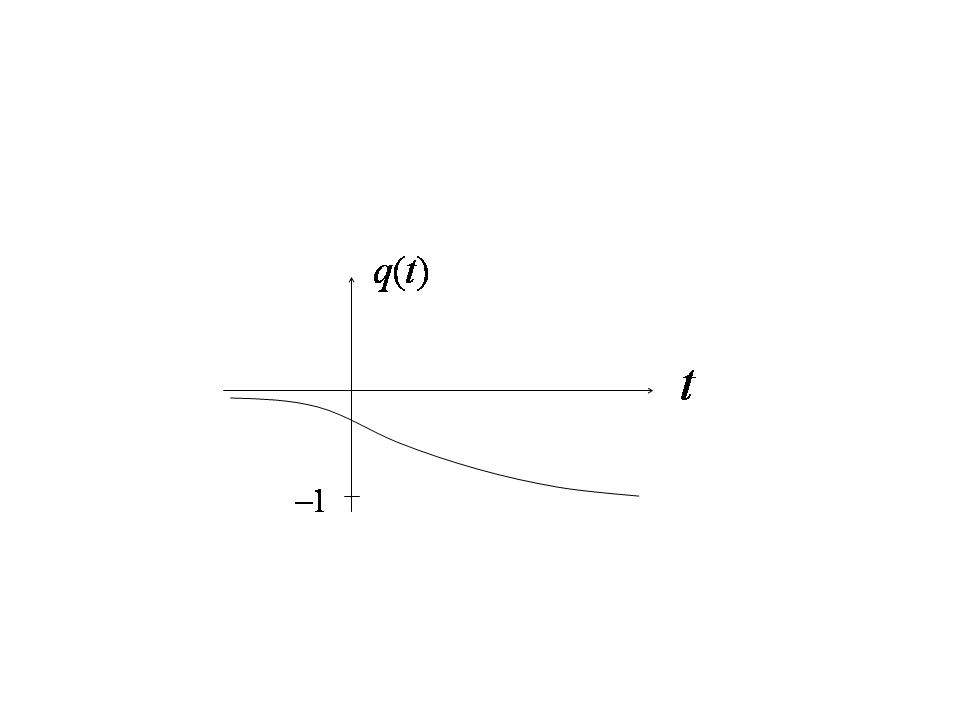}
\eqae
It only receives contributions from regions where $\vec{E}$ and $\vec{B}$ are nonvanishing, hence where $A^a_\mu$ is not pure gauge.

To obtain the action for $\l (t)$ in this example we evaluate
\eqa
\cl =- {1 \over g^2} \tr\, (\vec{E}^2 - \vec{B}^2) = {24 \over g^2} \rho^4 \left[ {\dot{\l}^2 \over (x^2 + \rho^2)^4} - {1 \over (x^2 + \rho^2)^4}
\right]\ .
\eqae
Doing the space integral we obtain
\eqa
L = {1 \over 2} m (\l) \dot{\l}^2 - V (\l) = {3 \pi^2 \rho^4 \over g^2 (\l^2 + \rho^2)^{5/2}} ( \dot{\l}^2 -1)\ ,
\eqae
where we used
\eqa
\int {{\rm d}^3 x \over ( \vec{x}^2 + \l^2 + \rho^2)^4} = {4 \pi \over (\l^2 + \rho^2)^{5/2}} {1 \over 2} \int\limits^\infty_{- \infty} {y^2 {\rm d}y \over (y^2 + 1)^4}  = {4 \pi^2 \over 32} {1 \over (\lambda^2 + \rho^2)^{5/2}}\ .
\eqae

In this example, we were dealing with a gauge with $A_0 \not= 0$.  We can map to a gauge in which  $A_0 =0$ by a suitable large gauge transformation
\eqa
A'_{\mu} &=& U^{-1} (\del_\mu + A_\mu )U \nn
U &=& \exp \left[ {i \vec{x} \cdot \vec{\s} \over \sqrt{\vec{x}^2 + \rho^2}} \; {\rm arctg} \; {\l (t) \over \sqrt{\vec{x}^2 + \rho^2}} \right]\ .
\eqae
Indeed, using the expression for $A_0$ in (\ref{A-lambda})
\eqa
A_0 = {-i \vec{x} \cdot \vec{\s} \over \vec{x}^2 + \l (t)^2 + \rho^2} \dot{\l} (t)\ ,
\eqae
one finds that $A'_0 = U^{-1} (\partial_t + A_0 ) U = U^{-1} \del_t U + A_0$ vanishes
\eqa
A'_0 = {i \vec{x} \cdot \vec{\s} \over \sqrt{\vec{x}^2 + \rho^2}} {1 \over 1 + {\l (t)^2 \over \vec{x}^2 + \rho^2}}  {\dot{\l} (t) \over
\sqrt{\vec{x}^2 + \rho^2}} + A_0 =0\ ,
\eqae
where we used that $A_0$ commutes with $U$.  Of course, $Q$ is gauge invariant because it can be written as a trace over $\vec{E} \cdot \vec{B}$ but it is instructive to see what happens if one writes $Q$ as a surface integral and makes a gauge transformation with $U$.   On the boundary of Minkowski space the $A_\mu =
V^{-1} \del_\mu V$ transform into $(VU)^{-1} \del_\mu VU$ and the winding number of $VU$ is the sum of the winding numbers of
$V$ and $U$.  However, $U$ is connected to the identity element: $U \equiv \exp \a [ {i \vec{x} \cdot \vec{\s} \over \sqrt{\vec{x}^2 +
\rho^2}}$ arctg ${\l (t) \over \sqrt{\vec{x}^2 + \rho^2}} ]$ traces an orbit as $\a$ runs from 0 to 1 which begins at the identity element
and ends at $U$.  Thus $U$ does not produce any winding, and thus the answer for $Q$ from the total derivative is the same, whether one uses a gauge in which $A_0$ vanishes or a gauge in which $A_0$ is nonvanishing.  Note, however, that
when $A_0 \not= 0$ one gets contributions from the timelike part of the boundary of the spacetime cylinder.

\section{False vacua and phase transitions}

In spontaneously broken gauge theories, the potential has a local maximum and an absolute minimum.  These extrema form a metastable
and a stable vacuum, respectively.  If a system is in the metastable vacuum at all points in spacetime, it could at some point and at some
time, say $\vec{x} = 0$ and $t = 0$, make a quantum fluctuation to the stable vacuum.  This transition costs energy, but if the region around $x$
(``the bubble") is large enough, the energy needed for creation of a bubble (this energy is located in the boundaries of the bubble) is less
than the energy gained by tunnelling to the lower vacuum (this energy is liberated in the volume of the bubble), and then the bubble will rapidly
expand.  In fact, since the rate of energy production increases the larger the bubble, the bubble will spread through space, with
accelerating speed, converting the false vacuum to a true vacuum.  As an application of this process one may consider the universe just
after the Big Bang; at high temperature the universe is in the symmetric vacuum, but as cooling due to expansion sets in the potential
develops a lower (true) vacuum, and if for some reason the universe remains stuck in the false vacuum, one can study the decay of the
universe towards the true (asymmetric) vacuum.  We shall consider another example: the perturbed double-well potential, with two classically stable minima, but one minimum (the true vacuum) below the other minimum (the false vacuum).  We shall study the decay of the false vacuum in this model into the true vacuum \cite{coleman, callen}.  We follow \cite{katzz}.

As a preliminary to the calculation of the phase transition in field theory, we first revert to quantum mechanics and study the double-well.   Let us pretend that we do not know that there are big differences between the
double-well potential and the following potential.
\eqa
\includegraphics[height=70mm,width=120mm]{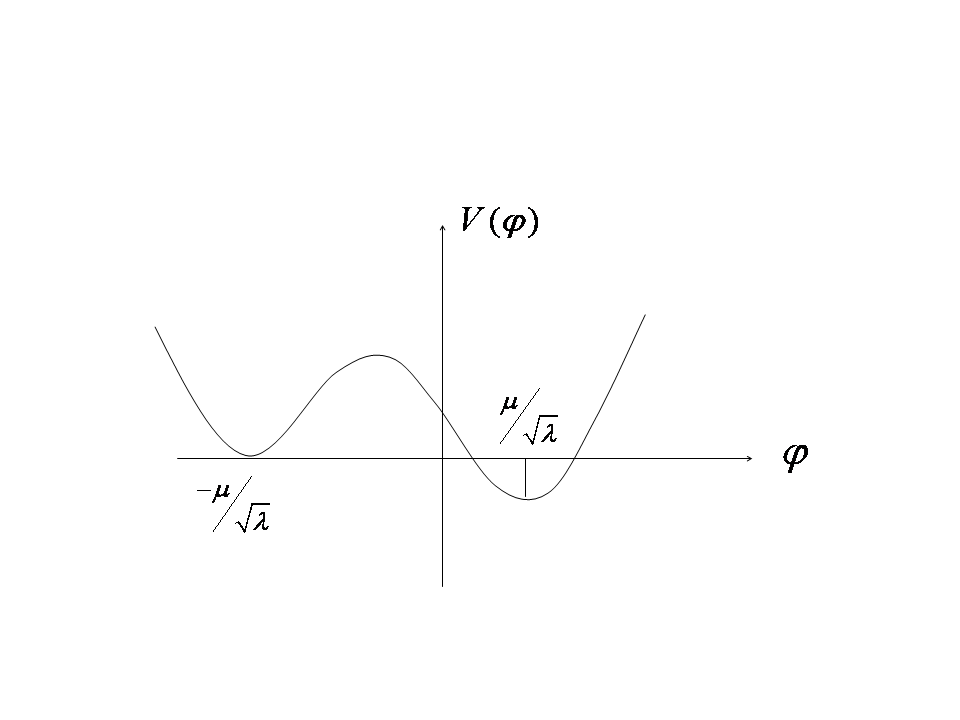}
\eqae
We can then repeat the calculation of the nonperturbative corrections to the energy of the ground state.  Already at this point it is clear
that we should not blindly repeat all steps because previously we were dealing with two perturbatively degenerate vacua, and the
kink-instantons provided the energy shift between both vacua.  In the present case, the degeneracy is already broken at the classical level.  Proceeding nevertheless we find a classical solution of the Euclidean
equation $- {\del^2 x \over \del t^2} + {\del V \over \del x} =0$ describing a point particle $x(t)$ in the inverted potential and use path integral methods.
\eqa
\includegraphics[height=70mm,width=120mm]{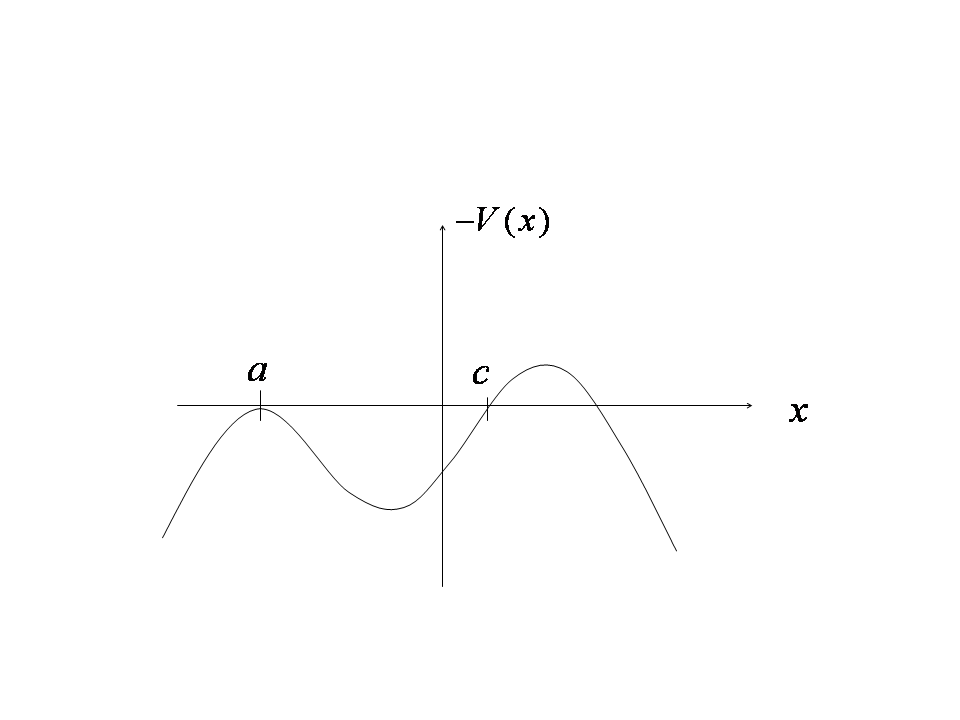}
\eqae
The particle starts at $t =- \infty$ in the point $x=a$, rolls to the point $x=c$, ``bounces" at time $t=X$, and ends up at $t= + \infty$ at the
same point $x=a$.  Clearly, $X$ is the collective coordinate for this classical solution $x_{cl} (t)$.  We then get for the ``one-bounce solution"
\eqa
T_{00} &\equiv& < x =a \mid {\rm e}^{-{1 \over \hbar} H \tau_0} \mid x=a > =
{\rm e}^{{1 \over \hbar} S_{cl}} \tau_0 \sqrt{-S_{cl}}\, I_0\ , \nn
I_0 &=& \cn \int\limits_{n.z.} {\rm d}q (\tau) {\rm e}^{{1 \over \hbar}
S^{(2)}_E} \quad {\rm with} \; q ( \pm \tau_{0/2}) =0\ ,
\eqae
where we used the Faddeev-Popov trick, and ``n.z." indicates that the path integral is over the solutions of the field equation for the fluctuations about
$x_{cl} (t)$ in the space orthogonal to the almost-zero mode.  Assuming again that $I_0$ can be written as a factor $K$ times the path integral of the harmonic oscillator we get
\eqa
I_0 = K \sqrt{{\o \over \pi \hbar}} {\rm e}^{-{1 \over 2} \o \tau_0}\ ;\qquad
 K= \sqrt{{\det ( - \del^2_t + \o^2 ) \over \det' (- \del^2_t + V'' (x_{cl}))}}
\eqae
Continuing without further thought we would sum over multi-bounces and obtain
\eqa
T_{00} &=& \sqrt{{\o \over \pi \hbar}} {\rm e}^{-{1 \over 2} \o \tau_0} \sum^\infty_{n=0} {( \sqrt{-S_{cl}} \tau_0 K {\rm e}^{{1 \over \hbar} S_{cl}})^n
\over n!} \nn
&=& {\o \over \pi \hbar} {\rm e}^{-{1 \over 2} \o \tau_0} \exp (K \tau_0 {\rm e}^{{1 \over \hbar} S_{cl}})\ .
\eqae
Using the same arguments as used before for the unperturbed double-well potential, we would conclude that the ground state energy is given by
\eqa
E_0 = {1 \over 2} \hbar \o - \hbar K {\rm e}^{{1 \over \hbar} S_{cl}}\ .
\eqae

However, at this point we note that there are problems with this result\\
(i) first a small problem: the nonperturbative correction is exponentially suppressed, hence it should be neglected compared to the perturbative correction.\\
(ii) a more serious problem (actually a virture, as we shall see) is that $K$ has a negative eigenvalue.  This is easy to prove: ${d \over dX} x_{cl} (t-X)$ is the zero mode fluctuation.  It has a mode because $x_{cl}$ bounces: unlike the kink,  $x_{cl} (\tau)$ moves first forward and then backwards, yielding a kind of kink-antikink solution.
\eqa
\includegraphics[height=70mm,width=120mm]{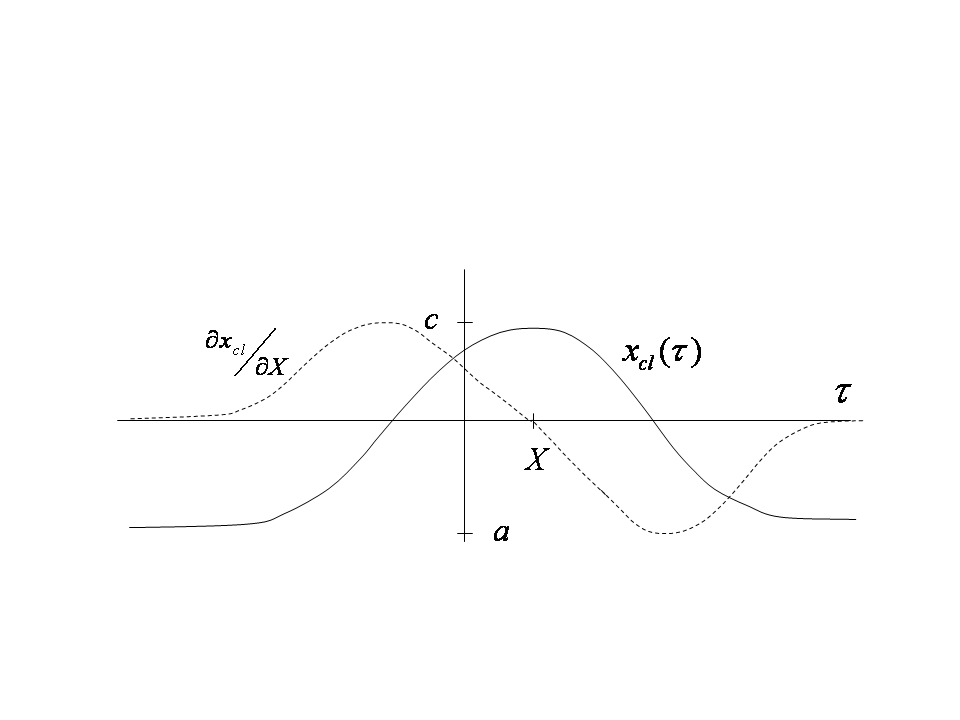}
\eqae
Hence there exists one mode for the fluctuations with lower eigenvalue and without a node, and since ${\del \over \del X} x_{cl} (x-X)$
has zero eigenvalue, there exists an eigenfunction for the fluctuation with negative eigenvalue.  Thus the nonperturbative correction is
imaginary, reflecting the fact that the perturbative ground state near $x=0$ is nonperturbatively unstable
\eqa
Im E_0 = \hbar \mid K \mid {\rm e}^{{1 \over \hbar} S_{cl}} \equiv \G/2\ .
\eqae
So, instantons (or rather bounces, still solutions of the classical field equations with finite action) yield in this case the width $\G$ of the
unstable state.

Having seen that in quantum mechanics the path integral approach to nonperturbative corrections to the vacuum energy leads to the correct result that the ground
state is unstable, we now return to the problem of phase transitions.

As a toy model for studying such decays we need a system with at least one space coordinate because bubbles have a finite extension in
space.  The simplest choice is a $1+1$ dimensional field theory.  We choose the double-well potential with an extra term to destroy the
degenacy between both minima.  Since the double-well potential is symmetric under $\varphi \rightarrow - \varphi$, the extra term should
be antisymmetric, and if it is to be a small perturbation compared to the leading $\l \varphi^4$ term, we need either a term linear in
$\varphi$ or cubic in $\varphi$, or both.  It simplifies the mathematics if we keep the local minima of the perturbed potential at the same
place as the minima of the unperturbated potential, namely at $\varphi = \pm \mu / \sqrt{\l}$.  We are then led to  the following model
\eqa
\cl = {1 \over 2} \dot\varphi^2 - {1 \over 2} (\varphi')^2 - {\l \over 4} \left( \varphi^2 - {\mu^2 \over \l} \right)^2 - B \left( {1 \over 3}
\varphi^3 - {\mu^2 \over \l} \varphi \right) + {2 \over 3} B \left( {\mu \over \sqrt{\l}} \right)^3\ ,
\eqae
where we take $B$ small and positive.  For constant $\varphi$, the solutions of the classical field equations occur at
\eqa
{\del V \over \del \varphi} = \l \varphi \left( \varphi^2 -{\mu^2 \over \l} \right) + B \left( \varphi^2 - {\mu^2 \over \l} \right) =0\ ,
\eqae
and from this result it is clear that the values $\varphi = \pm \mu / \sqrt{\l}$ are indeed extrema.  The potential has the following form
\eqa
\includegraphics[height=70mm,width=120mm]{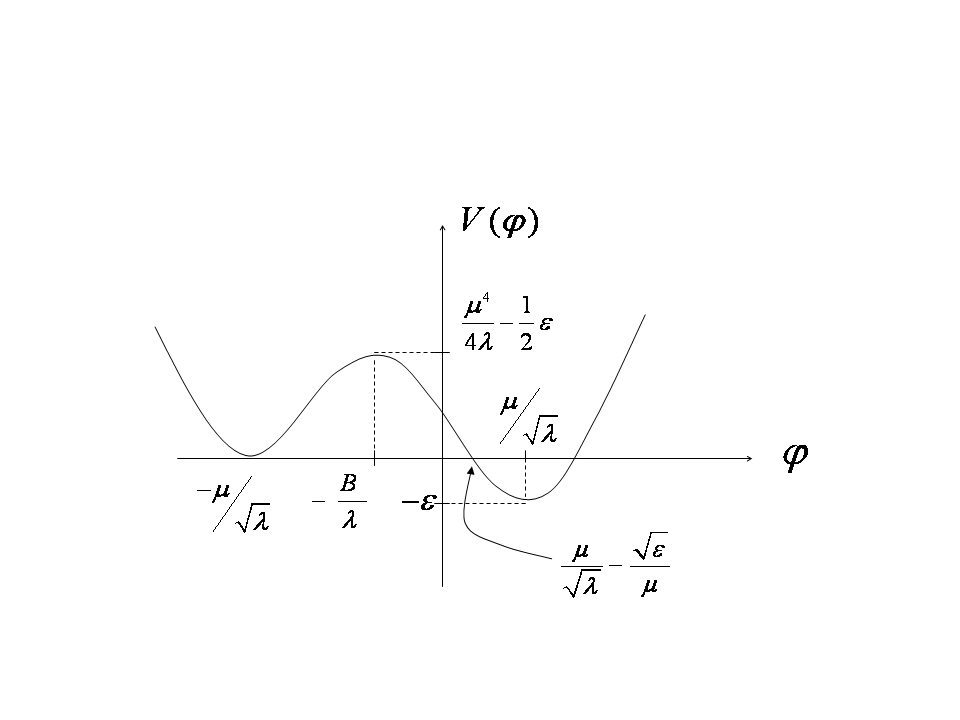}
\eqae
It vanishes at $\varphi =- \mu / \sqrt{\l}$ because we added the constant ${2 \over 3} B ( \mu / \sqrt{\l})^3$, but at $\varphi = \mu /
\sqrt{\l}$ it is negative.  Thus $\varphi =- \mu / \sqrt{\l}$ is the unstable vacuum and $\varphi = \mu / \sqrt{\l}$ is the stable vacuum.  The value of
the potential at the stable minimum is
\eqa
V \left( \varphi = \mu / \sqrt{\l} \right) = - \e = - {4 \over 3} B \left( \mu / \sqrt{\l} \right)^3\ .
\eqae
There is a relative maximum a bit below the maximum of the symmetric potential $V(B =0, \varphi)$ at $\varphi =0$; for small $B$ it occurs
at $\varphi \simeq - B/\l$ and its value is ${1 \over 4} \mu^4 / \l - {1 \over 2} \e + \co (B^2)$.  These results are intuitively clear: if one pulls $\varphi$ down at $\mu/ \sqrt{\l}$ by an amount $\e$, then the maximum at $\varphi =0$ is pulled down half as much, and moves of course a bit to the left.

In addition to the three solutions of the classical field equations with constant $\varphi (\varphi =- \mu / \sqrt{\l} , \varphi = \mu / \sqrt{\l}$, and $\varphi \sim - B /  \l)$, there is an exact kink-antikink solution.  This is clear by inspection of the inverted potential
\eqa
\includegraphics[height=70mm,width=120mm]{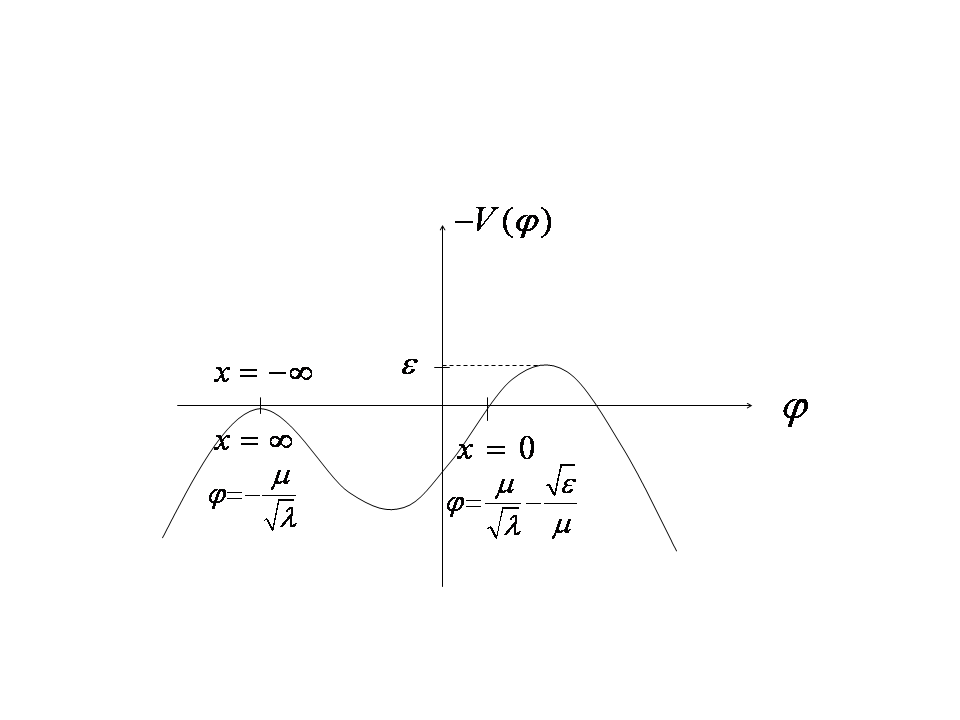}
\label{one11thirteenn}
\eqae
A ball at rest at $\varphi = - \mu / \sqrt{\l}$ at $x =- \infty$ starts rolling down to the hill and up the other hill; it reaches the point where
$V (\varphi) =0$ at $x =0$ and then returns and comes to rest at $\varphi =- \mu / \sqrt{\l}$ at $x = + \infty$.  The classical solution
$\varphi_{cl} (x)$ is thus a soliton of the following form
\eqa
\includegraphics[height=70mm,width=120mm]{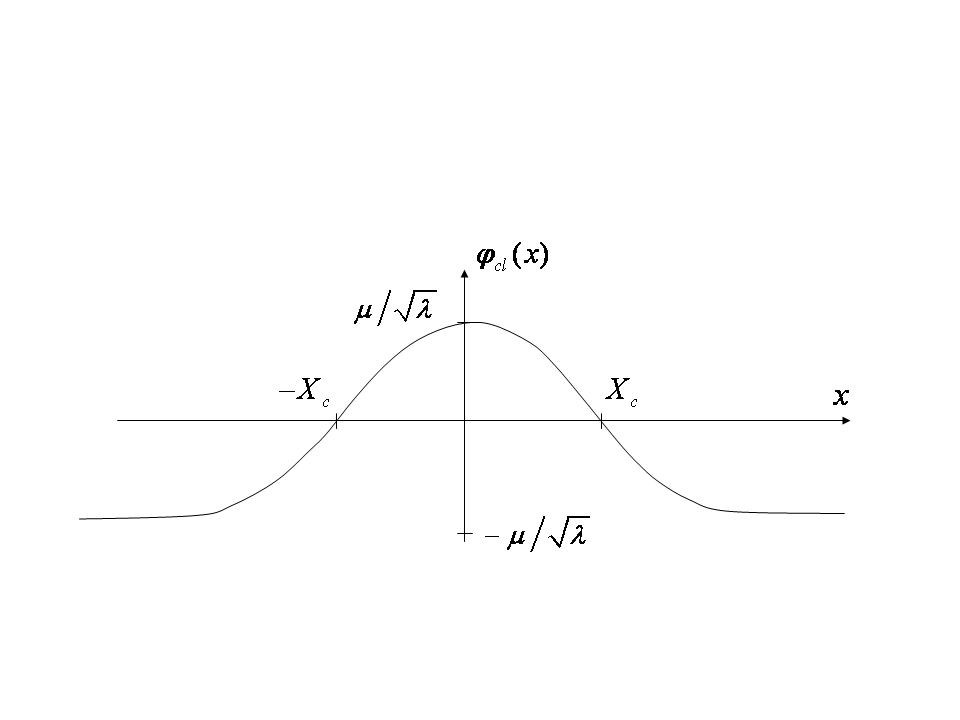}
\eqae
We approximate $\varphi_{cl} (x)$ by the following expression
\eqa
\varphi_{cl} (x) = {\mu \over \sqrt{\l}} \left[ \tanh \left( {m \over 2} (x + X_c)  \right) - \tanh \left( {m \over 2} (x-X_c)  \right) -1 \right]\ .
\label{one11fifteeen}
\eqae
This is a static soliton in $1+1$ dimensions, which can also be viewed as an instanton in $x$-space.  (In the quantum mechanical models we
considered previously, we dealth with instantons in Euclidean time).  Near $x=- X_c$ the antikink is exponentially suppressed and the mass of
the kink is $M$.  Between the kink and antikink $\varphi$ is equal to $\mu / \sqrt{\l}$ (up to exponentially suppressed corrections), and near
$x= X_c$ we have an antikink with mass $M$.  For large $x$ we find the correct asymptotic value $\varphi_{cl} (x \rightarrow \pm \infty) =
- \mu/ \sqrt{\l}$.  We fix the value of $X_c$ such that the total energy of $\varphi_{cl} (x)$ (which is the energy of the ball rolling up and down the hills in (\ref{one11thirteenn})) vanishes
\eqa
E = 2 M - 2 \e X_c = 0\ ,
\eqae
where $M = {m^3 \over 3 \l}$ is the classical mass of a single kink.  Hence, the separation between the kink and antikink is given by $2X$ with $X = M/ \e$.

The exact solution begins at $V=0$, climbs the hill, and comes down on the other side where it reaches the value $V=0$, and then it returns,
climbing the hill once more, and ending at $V=0$.  The approximate solution comes down to $V=- \e$ after climbing the hill, but it has more
energy in the kink (and antikink) region, such that in both cases the total energy is zero.

We now compute the transition amplitude from the unstable vacuum $\varphi =- \mu / \sqrt{\l}$ to the kink-antikink solution (the bubble).
Once a bubble has formed, it will rapidly grow (the kink and antikink move increasingly fast away from each other, i.e., $X$ exponentially
increase).

This is a tunnelling process because classically it is forbidden but quantum mechanically allowed.  If the field $\varphi$ at $x=0$ starts
making a transition from the metastable vacuum to the stable vacuum, it must first climb the potential barrier, but when it comes down in
the true vacuum energy density $- \e$ is gained.  However, as we already mentioned, it takes energy to distort the field in order to go from
one vacuum to another; this is just the energy (mass) of a kink and of an antikink.  These energies are located at the boundary of the bubble
(around the centers of the kink and the antikink).  Once in a while there occurs a quantum mechanical transition to a bubble which is large
enough that $\e 2 X$ is larger than $2M$; in that case the bubble does not collapse but grows increasingly rapidly.

Note that we do not tunnel from the state $\varphi (x) =- \mu / \sqrt{\l}$ to the state $\varphi (x) = \mu / \sqrt{\l}$ because the energy
difference of these states is infinite (namely $\e$ times the volume of $x$-space, so $2L \e$ with $L \rightarrow \infty$).  When we
discussed the unperturbed kink, the vacua $\varphi = \pm \mu / \sqrt{x}$ were exactly degenerate, and in such cases the true vacuum is a
linear combination of these vacua which can be determined by tunnelling from one vacuum to another.

The intermediate configuration with the kink and antikink moving away from each other can be described by Lorentz boosting the kink to a velocity $-v$ and the antikink to a velocity $+v$
\eqa
\varphi_{cl} (x,t) = {\mu \over \sqrt{\l}} \left[ \tanh {m \over 2} \left( {x + X_c + vt \over \sqrt{1- v^2}} \right) - \tanh {m \over 2} \left( {x- X- vt
\over \sqrt{1^2 - v^2}} \right) - 1 \right]\ .
\eqae
For constant $\dot{X}$ the boost of the kink is again a solution because the field equation use relativistically invariant.  However, since
$\dot{X}$ itself is expected to change with time, we denote $X + \dot{X} t$ by $\l (t)$ and obtain then
\eqa
\varphi_{cl} (x,t) = {\mu \over \sqrt{\l}} \left[ \tanh {m \over 2} \left( {x + \l (t) \over \sqrt{1- \dot{\l}^2}} \right) - \tanh {m \over 2} \left( {x- \l
(t) \over \sqrt{1 - \dot{\l}^2}} \right) -1 \right]\ .
\eqae

The distance between the kink and antikink is now $2 \l (t)$.
The Lagrangian for this approximate solution is obtained by substituting $\varphi_{cl}$ into the action.  The calculation of the first two
terms is straightforward.  Taking twice the result for a single kink yields
\eqa
&& \int\limits^\infty_{-\infty} \left[ {1 \over 2} \dot\varphi^2 - {1 \over 2} \left( {\del \varphi \over \del x} \right)^2 \right] {\rm d}x =2
\int\limits^\infty_{-\infty} {\rm d}x {1 \over 2} {1 \over \cosh^4 \left( {m \over 2} {x + \l \over \sqrt{1- \dot\l^2}} \right)} \nn
&& \left[ {\mu^2 \over \l} {m^2 \over 4} \left( {\dot\l \over \sqrt{1 - \dot\l^2}} + { \dot\l \ddot\l (x + \l ) \over (1- \dot\l^2 )^{3/2}}
\right)^2 - {\mu^2 \over \l} {m^2 \over 4} {1 \over 1- \dot\l^2} \right]\ .
\eqae
The calculation of the contribution from the nonderivative terms splits into two parts: from the region between the kink and antikink we
obtain a term $\e 2 \l$, while from each of the two walls we find a term ${1 \over 2} M \sqrt{1 - \dot\l^2}$ as we now explain.  Around $x =-
\l$ and $x= + \l$, the integral $\int V (\varphi) {\rm d} x$ with $\varphi = {\mu \over \sqrt{\l}} \tanh {m \over 2} {x+ \l \over \sqrt{1- \dot\l^2}}$ can
be evaluated as follows.  The integral $\int^\infty_{-\infty} {1 \over 2} U^2 (\varphi) {\rm d}x$ with $\varphi = {\mu \over \sqrt{\l}} \tanh {m
\over 2} ( x^\ast + \l^\ast)$ with $x^\ast = {x \over \sqrt{1- \dot\l^2}}$ and $\l^\ast = {\l \over \sqrt{1 - \dot\l^2}}$ is equal to $( \int {1
\over 2} U^2 (\varphi (x^\ast)) {\rm d}x^\ast) \sqrt{1- \dot\l^2}$.  From equipartion of energy for a static kink we know that the integral $\int {1
\over 2} U^2 (\varphi (y)) {\rm d}y$ equals ${1 \over 2} M$.  Thus $\int\limits_{\rm around \; kink} ( {1 \over 2} U^2) (\varphi_{cl} ) {\rm d}x = {1 \over 2}
M \sqrt{1 -\dot\l^2}$.

Hence, neglecting term with $\ddot{\l}$, we find
\eqa
&& L =- {m^4 \over 8 \l} \left( {2 \over m} \sqrt{1 - \dot\l^2} \int\limits^\infty_{-\infty} {{\rm d}y \over \cosh^4 y} \right) + \e 2 \l - M
\sqrt{1- \dot\l^2} \nn
&& = - 2 M \sqrt{1 - \dot\l^2} + \e 2 \l \ .
\eqae
The Hamiltonian follows from $p = {\del L \over \del \dot\l} = {2 M \dot\l \over \sqrt{1 - \dot\l^2}}$ and reads
\eqa
H = {2 M \over \sqrt{1 - \dot\l^2}} - \e 2 \l = \sqrt{p^2 + 4 M^2} - \e 2 \l\ .
\eqae
We can split $H$ into a kinetic term $K$ and a potential term $V$
\eqa
&& K= \sqrt{p^2 + 4 M^2} - 2 M = {1 \over 2} p^2 / M + \co (p^4)\ , \nn
&& V ({\l}) = 2 M - \e 2 \l\ .
\eqae
This formula for $V (\l)$ is valid when the bubble is reasonably large: when $\l$ is larger than the kink size (when the bubble is larger than
the thickness if its walls).  For smaller $x$ we expect that $V (\l)$ rises from $0$ till a maximum value when the bubble is formed, and then decreases as the bubble gets larger
\eqa
\includegraphics[height=70mm,width=120mm]{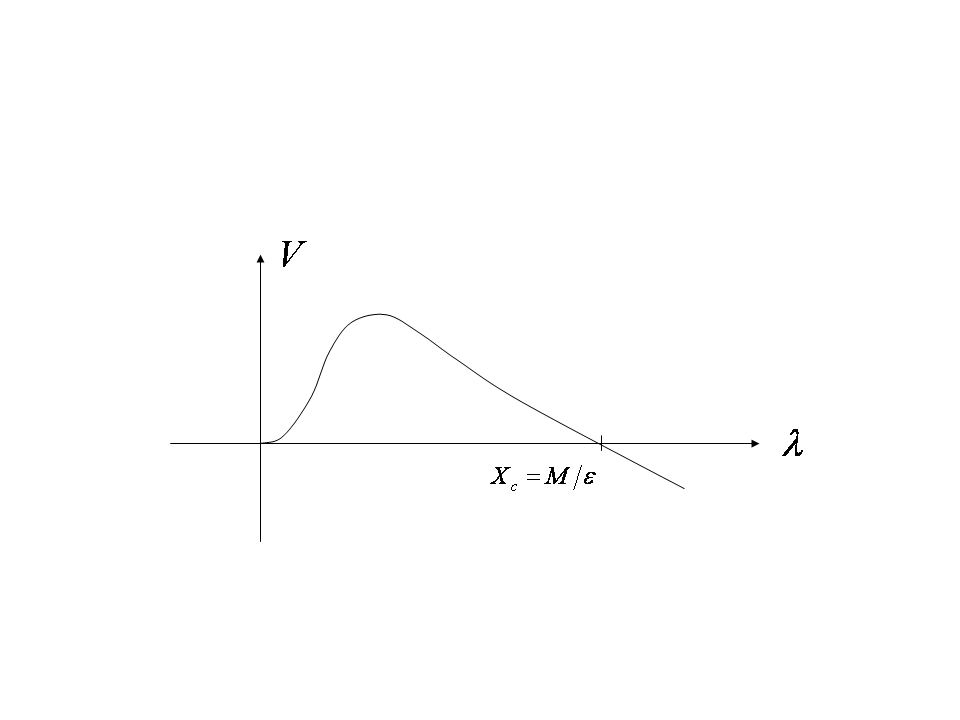}
\eqae
The value $X_c$ corresponds to the classical solution, with energy $E=0$ and constant $X$, corresponding to the ball rolling in the inverted
potential.  For this case, $p=0$.  Quantum fluctuations with $X < X_c$ produce only bubbles which collapse since their potential energy is
positive, but bubbles with $X = X_c$ are metastable (they have constant $X=X_c$ so $p=0$), while for $X>X_c$ the bubble expands.

We now treat $H$ as the Hamiltonian of a point particle which sees the potential $V (\l)$ and has energy zero.  We find with the WKB approximation for the tunneling amplitude
\eqa
A &=& \exp\Big[- \int\limits^{X_c}_0 \mid p \mid {\rm d} \l\Big] \nn
&=& \exp\Big[- \int\limits^{X_c}_0 \sqrt{4 M^2 - (\e 2 \l)^2} {\rm d} \l\Big]\ ,
\eqae
where we used that $H =0= \sqrt{p^2 + 4 M^2} - \e 2 \l$.  Since $X_c = {M \over \e}$, we have
\eqa
A &=& \exp\Big[- {2 M^2 \over \e} \int\limits^1_0 \sqrt{1 - {\e^2 \over M^2} \l^2}
{\rm d} \left( {\e \over M} \l \right)\Big] \nn
&=& \exp\Big[- {2 M^2 \over \e} \int\limits^1_0 \sqrt{1-y^2} {\rm d}y\Big] =
\exp(- {\pi M^2 \over 2 \e})\ .
\eqae
Hence, the rate of the transition to the true vacuum is $\exp - {\pi M^2 \over \e}$ per second and per unit volume.  (To evaluate the integral we set $y = \cos \varphi$).

We end this section with a few comments.
\noindent 1.  The decay of the false vacuum per unit time and per unit volume is of the form $\G /V = A {\rm e}^{-B / \hbar} (1+ \co (\hbar ) )$.  We computed $B$.  For $A$ see \cite{callen, bitar}.

\noindent 2.  We used energy conservation to determine how fast a bubble expands.  However, we neglected radiation of mesons.  In general, when the false vacuum collapses to the true vacuum, mesons will be created, and thus the bubble will expand less rapidly.

\noindent 3. Above we considered the critical bubble: a static solution of the classical field equations which describes a bubble which has just the correct form and
size that it is metastable.  For larger sizes there is no static solution, but one can consider the creation at $t=0$ of a large bubble which then expands.
This is an initial value problem: $\varphi (x)$ is given and also ${\del \varphi \over \del t} =0$ at $t=0$.  One can define the size of a bubble for example
as $Q = \int^\infty_{-\infty} ( \varphi + {\mu \over \sqrt{\l}} )^2 {\rm d}x$.  Far away, $\varphi = - {\mu \over \sqrt{\l}}$, so $Q$ is finite for bubbles.
A problem we now want to solve is: given the size $Q$ of a bubble, for which shape is its action minimal.  (Minimal action in Euclidean space means
maximal tunnelling rate).  This will yield a one-parameter parametrization of bubbles; the parameter is a collective coordinate $\l (t)$, and having found the solution, we can then compare our ansatz in (\ref{one11fifteeen}) and see how good the ansatz was.
Mathematically, we can formulate this problem as a variational problem with a constraint.  Introducing a constant Lagrange multiplier $\a$ we consider
the action for the variational problem
\eqa
&& \cl =- {1 \over 2} ( \del_x \varphi )^2 - {\l \over 4} \left( \varphi^2 - {\mu^2 \over \l} \right)^2 - B \left( {1 \over 3} \varphi^3 - {\mu^2 \over \l}
\varphi \right) \nn
&& \hspace{.75in} + {2 \over 3} B \left( {\mu \over \sqrt{\l}} \right)^3 + {1 \over 2} \a \left( \varphi + {\mu \over \sqrt{\l}} \right)^2\ .
\eqae
The equation of motion
\eqa
{\del \over \del x} { \del \cl \over \del \varphi_x} - {\del \cl \over \del \varphi} =0\ ,
\eqae
has a first integral due to equipartition of energy
\eqa
{1 \over 2} \left( {d \varphi \over dx} \right)^2 = {\l \over 4} \left( \varphi^2 - {\mu^2 \over \l} \right)^2 + B \left( {1 \over 3} \varphi^3 - {\mu^2 \over
\l} \varphi \right) - {2 \over 3} B \left( {\mu \over \sqrt{\l}} \right)^3 - {1 \over 2} \a \left( \varphi + {\mu \over \sqrt{\l}} \right)^2 \ .\nn &&
\eqae
The integration constant vanishes for bubbles.  Introducing a field $\tilde{\varphi}$ which vanishes for large $x$
\eqa
\tilde{\varphi} = \varphi + {\mu \over \sqrt{\l}} \ ,\qquad \varphi =
\tilde{\varphi} - {\mu \over \sqrt{\l}}\ ,
\eqae
we obtain
\eqa
{d \tilde{\varphi} \over d x} = \sqrt{{\l \over 2}} \tilde\varphi \sqrt{\left( \tilde\varphi - {2 \mu \over \sqrt{\l}} \right)^2 + {4 B \over \l} \left( {1 \over
3} \tilde\varphi - {\mu \over \sqrt{\l}} \right) - {2 \a \over \l}}\ .
\eqae
For $\a = B =0$ the solution is the kink, but for $\a \not= 0$ we get bubbles.  One can actually solve this equation exactly by using (see Gradhstein and
Resznik, page 84, 2.266)
\eqa
 \int {{\rm d}y \over y \sqrt{a + by + cy^2}} = {1 \over \sqrt{a}} {\rm arc} \cosh {2a + by \over y \sqrt{-4ac}}\ ,
\label{arccosh}
\eqae
which holds if $a>0$ and $b^2 -4ac >0$.  This corresponds to $0< \a < 2 \mu^2$.  By writing the differential equation as
\eqa
\int {\rm d} \left( \sqrt{{\l \over 2}} x \right) = \int {{\rm d} \tilde{\varphi} \over \tilde{\varphi} \sqrt{\left( {4 \mu^2 \over \l} - {2 \a \over \l} - {4B \mu \over \l
\sqrt{\l}} \right) + \left( - {4 \mu \over \sqrt{\l}} + {4 B \over 3 \l} \right) \tilde{\varphi} + \tilde{\varphi}^2}}\ ,
\label{holdds}
\eqae
we obtain for the bubble with fixed size and minimum action
\eqa
&& \cosh \left( \sqrt{a} \sqrt{{\l \over 2}} x \right) = {2 a + b \tilde{\varphi} \over \tilde{\varphi} \sqrt{4 a c -b^2}} \nn
&& \tilde{\varphi} = {2a \over \sqrt{4ac - b^2} \cosh \left( \sqrt{a} \sqrt{{\l \over 2}} x - x_0 \right) -b} \simeq {{2 \over \sqrt{\l}} ( 2 \mu^2 - \a) \over
\sqrt{\a} \cosh ( \sqrt{2} \mu (x-x_0) ) + 2 \mu} \ .\nn &&
\eqae
The constant $\a$ lies in the domain $0 < \a < 2 \mu^2$.  For $\a =0$ we find $\tilde{\varphi} = {2 \mu \over \sqrt{\l}} \left(  {\rm or} \; \varphi = {\mu
\over
\sqrt{\l}} \right)$ while for $\a = 2 \mu^2$ we find $\tilde{\varphi} =0 \left( {\rm or} \; \varphi =- {\mu \over \sqrt{\l}} \right)$.  In between, we
have bubbles of finite extent; for small $\a$ the function $\tilde{\varphi}$ remains constant for a long time (the bubble) and then it falls rapidly off to
zero (due to the $\cosh$).  This is the same behaviour as displayed by our ansatz in (\ref{one11fifteeen}).

\section{The strong CP problem} 

The vacua $|n>$ of Yang-Mills theory in Minkowski space with winding number $n$ all have the same energy, namely zero (because they are vacua).  We recall that at fixed time space was compactified to an $S^3$ which was mapped to the $S^3$ of the group manifold of $SU(2)$.  Since there is tunnelling as we have discussed, the physical vacuum is a linear combination of all of them.  Since they all appear on equal footing, we expect that the generator $T$ for large gauge transformations which change the winding number, defined by  $T | n> = | n+1>$, commutes with the Hamiltonian.  Hence $T$ maps the physical vacuum into itself.  It follows that $T | {\rm vac} > = {\rm e}^{i \varphi} | {\rm vac} >$ with $\varphi$ some phase.  The solution of this equation is
\begin{eqnarray}\label{theta}
\mid {\rm vac}   > \equiv \mid  \theta > = \sum_n {\rm e}^{i n \theta} | n >\ .
\end{eqnarray}
Indeed $T | \theta > = \sum {\rm e}^{in \theta} \mid n+1 > = {\rm e}^{-i \theta} \mid \theta >$.

Instead of using the infinite set of states in (\ref{theta}), we can work
with the ordinary vacuum $\mid 0 >$ if at the same time we add a term
\begin{eqnarray}\label{theta-term}
\cL_\theta = - \theta_{\rm QCD} {g^2 \over 16 \pi^2} \tr\, F_{\mu\nu}\ {^*\!F_{\mu\nu}}
\end{eqnarray}
to the action.  This term yields a factor ${\rm e}^{i n \theta}$ in the action
${\rm e}^{{i \over \hbar}S}$ if one is in the vacuum with winding number $n$.
We shall set $\hbar=1$. Note that we are in Minkowski space and that
$\cL_\theta$ is hermitian.

Strictly speaking, we should first make a Wick rotation to Euclidean space because we can only discuss instantons in Euclidean space, but $\cl_\theta$ has the same form in Euclidean space: one gets a factor $i$ from ${\rm d}^4 x$ and another factor $i$ from $F_{0i}$.  Together with the factor ${i \over \hbar}$ in ${\rm e}^{{i \over \hbar} S}$ one gets the same factor ${\rm e}^{in \theta}$ in Euclidean space.   The $\theta$-term is a total derivative, and usually one discards total derivatives in Lagrangians because fields vanish at infinity, but for instanton backgrounds one finds of course a nonvanishing contribution due to winding.

The $\theta$-term clearly violates parity P.  It conserves charge conjugation symmetry $C$, hence it violates CP.  The strong interactions described by QCD are not supposed to violate P or PC, hence $\theta_{\rm QCD}$ should be very small.  However, the observed $\theta$ parameter contains more than only $\theta_{\rm QCD}$.  There is a second origin for a $\theta$-angle coming from the electroweak sector:  the manipulations leading to the CKM matrix.
  Recall that the mass terms of the quarks in the Standard Model come from Yukawa couplings
\begin{eqnarray}
&& \cL = - \sum_{m,n} \left[ g (qu)_{mn} \left( \begin{array}{ll} \bar{q}_{L,m} \\ \bar{q}^\prime_{L,m} \end{array} \right)^T \left( \begin{array}{ll} (h^0)^\ast  \\ - (h^+)^\ast \end{array} \right) q_{R,n} \right. \nonumber\\
&& \qquad \left. + g^\prime (qu)_{mn} \left( \begin{array}{ll} \bar{q}_{L,m} \\ {\bar q}^\prime_{L,m} \end{array} \right)^T \left( \begin{array}{ll} h^+ \\ h^0 \end{array} \right) q^\prime_{R,n} \right] + h.c.\ ,
\end{eqnarray}
where $g (qu)$ are the Yukawa couplings to quarks, and $h^+, h^0$ are the two components of the complex $SU(2)$ Higgs doublet. Furthermore $m=1,2,3$ labels the families, so $q_1$ denotes the up quark while $q^\prime_1$ denotes the down quark.   When $h^0$ gets a vacuum expectation value $< h^0 > = {1 \over \sqrt{2}} v$,  one obtains mass matrices $M$ for the $(u, c, t)$ quarks and $M'$ for the $(d,s, b)$ quarks, where
\begin{eqnarray}
M_{mn} =  {v \over \sqrt{2}} g_{mn} \ ,\qquad  M^\prime_{mn} = {v \over \sqrt{2}} g^\prime_{mn}\ .
\end{eqnarray}
These matrices are in general arbitrary complex $3 \times 3$
matrices. One diagonalizes them with $3 \times 3$ unitary matrices
which are different for left- and right- handed quarks\footnote{ A
complex matrix $M$ can always be written as $VH$ where $V$ is
unitary and $H$ hermitian.  This is the generalization to matrices
of the decomposition $z = {\rm e}^{i \varphi} \rho$ of complex numbers.
Then $H$ can be diagonalized by a unitary matrix, $H = U_R D
U^{-1}_R$, and $U_L$ is given by $V U_R$.}
\begin{eqnarray}
&& U_L M U^{-1}_R = \ {\rm diag} (m_u , m_c , m_t  ) \equiv D\ , \nonumber\\
&& U^\prime_L M^\prime U^{\prime -1}_R = {\rm diag} (m_d , m_s , m_b )
\equiv D^\prime\ .
\end{eqnarray}
The mass matrix for the quarks becomes then diagonal with real masses
\begin{eqnarray}
\bar{q}_{L,m} M_{mn} q_{Rn} = \overline{(U_L q_L)} D(U_R q_R)
\end{eqnarray}
and similarly for $\bar{q}^\prime_{L,m} M^\prime_{mn} q^\prime_{Rn}$.  So, the physical quarks are $Q_L = U_L q_L$ and $Q_R = U_R q_R$, and similarly for $Q^\prime_L$ and $Q^\prime_R$.

If one rescales $q_L$ to $Q_L$, and $q_R$ to $Q_R$,  three things happen\\
(i) the quark mass terms are diagonalized as we have discussed, yielding real physical quark masses \\
(ii) a phase $\delta$ appears in the CKM matrix which describes electroweak CP violation\\
(iii) a new term is produced in the action by the Jacobian for these chiral rescalings.  This new term is again proportional to $\int F_{\mu\nu} {^*\!F}_{\mu\nu} {\rm d}^4 x$ with a coefficient which we call $- \theta_{\rm EW}$.  Hence, now the action contains the sum $\theta = \theta_{\rm QCD} + \theta_{\rm EW}$
\begin{eqnarray}
\cl_\theta = - (\theta_{\rm QCD} + \theta_{\rm EW}) {g^2 \over 32 \pi^2} \int
(F_{\mu\nu}^a  {^*\!F}^a_{\mu\nu} ) {\rm d}^4x\ .
\end{eqnarray}
There is no reason that $\theta_{\rm strong} = \theta_{\rm QCD} + \theta_{\rm EW}$ vanishes, yet, as we now discuss, this seems to be the case.

We can make a final chiral rescaling of the 3 light quarks (the $u,d$ and $s$ quarks) such that the $\theta$-term is entirely removed.
 Rescaling the left-handed quarks by $U(1)$ factors ${\rm e}^{i \varphi_u},
{\rm e}^{i \varphi_d} $ and ${\rm e}^{i \varphi_s} $, the Jacobians for these
rescalings yield a term
\begin{eqnarray}
- (\varphi_u + \varphi_d + \varphi_s) {g^2_s \over 16 \pi^2} \tr\, F_{\mu\nu} {^*\!F}_{\mu\nu}\ ,
\end{eqnarray}
which cancels the $\theta$-term if $\varphi_u + \varphi_d + \varphi_s = \theta_{\rm strong}$.
Because the action is invariant except for the mass terms, only the transformation of the mass terms yields a new term in the action. In the diagonal mass term
\begin{eqnarray}
m_u \bar{u} u + m_d \bar{d} d + m_s \bar{s} s\ ,
\end{eqnarray}
the rescalings yield, to first order in $\varphi_u , \varphi_d , \varphi_s$, a new term in the action
\begin{eqnarray}
\cL_{\rm CP \; violation} = i \varphi_u m_u \bar{u} \gamma_5 u + i \varphi_d m_d \bar{d} \gamma_5 d  + i \varphi_s m_s \bar{s} \gamma_5 s\ .
\end{eqnarray}
The $\varphi$'s are only constrained by $\varphi_u + \varphi_d + \varphi_s =  \theta_{\rm strong}$, so we can still choose them such that this new term is $SU(3)_V$ invariant.  Namely if $\varphi_u = {\theta m_d m_s \over m_u m_d + m_u m_s + m_d m_s}$, and cyclically for $\varphi_d$ and $\varphi_s$, then
\begin{eqnarray}\label{massterm}
\cL_{\rm CP \; violation} =  {i \theta_{\rm strong}  m_u m_d m_s \over m_u m_d + m_u m_s + m_d m_s} (\bar{u} \gamma_5 u + \bar{d} \gamma_5 d+ \bar{s} \gamma_5 s)\ .
\end{eqnarray}
This term is hermitian and $SU(3)_V$ invariant, but it violates P, and since it conserves C, it also violates CP. The original $\theta$-term in the action
in (\ref{theta-term}) has been transformed into the masslike terms in
(\ref{massterm}). No longer does one have to deal with total derivatives, but
an ordinary extra masslike term has appeared in the QCD action.
There is no reason that $\theta_{\rm strong}$ should be small, but one can compute the electric dipole moment of the neutron which is nonzero if $\theta_{\rm strong}$ is nonzero, and since experimentally the electric dipole moment has a very small upper bound, one finds that $\theta_{\rm strong}$ is incredibly small
\begin{eqnarray}
\theta_{\rm strong} < 10^{-9}\ .
\end{eqnarray}
The problem why $\theta_{\rm strong}$ is so small is called the strong CP problem.  Note that it has nothing to do with the CP violation due to the phase $\delta$ in the CKM matrix, which is an electroweak effect. Also the electroweak
CP violation is very small;  it can be parametrized by the area of the unitarity triangles (each of the 6 unitarity triangles has the same area $2J$ in the Standard Model)
\begin{eqnarray}
J = (3.0 \pm 0.3) 10^{-5} \, .
\end{eqnarray}

\section{The $U(1)$ problem}

In this section we discuss an application of instantons in QCD.

In the 1960's, in the absence of a renormalizable theory of the strong
interactions, current algebra was developed as a method to derive information
about matrix elements of currents, mostly the vector and axial-vector
Noether currents which correspond to the (approximate) rigid flavor
symmetry of the up, down and strange quarks. In terms of modern QCD,
the action for the strong interactions reads
\begin{equation}
{\cal L}=-\frac{1}{4} (F_{\mu\nu}^a)^2 - \sum {\bar \psi}_i  \not\!\!{\bar D} \psi^i\ ,
\end{equation}
where $i=1,...,N_f$ labels the flavors. One can consider either two very
light quarks ($u$ and $d$), or three rather light quarks ($u$, $d$ and $s$).
Decomposing the massless quarks into left-handed and right-handed parts,
their action becomes
\begin{equation}
{\cal L}({\rm quarks})={\bar \psi}_{i,L}  \not\!\!{\bar D} \psi^i_L-
{\bar \psi}_{i,R}  \not\!\!{\bar D} \psi^i_R\ .
\end{equation}
It has clearly a rigid $U_L(N_f)\times U_R(N_f)$ symmetry group, where
$U_L$ acts only on $\psi^i_L$ and $U_R$ only on $\psi^i_R$. Instead of
$U_L$ and $U_R$ we consider the vector part $U_V(N_f)$ and the axial
vector part $U_A(N_f)$. The vector part transforms $\psi^i_L$ and
$\psi^i_R$ the same way, while they transform oppositely under $U_A$. The total
number of symmetries and group parameters has not changed, but physically
$U_V$ and $U_A$ are very different. The $SU(2)_V$ part of the symmetry is realized in Nature, and yields the $SU(2)$ classification scheme
for quark hadroscopy.  The $U(1)_V$ corresponds to baryon-number conservation which is also (very well) satisfied.  The $SU(2)_A$ symmetry is spontaneously broken, and the corresponding Goldstone bosons form an $SU(2)$ multiplet of pseudoscalars (the pions and the $\eta$ meson).  One might be inclined to apply the same reasoning to the $U(1)_A$ symmetry, and argue that it, too, must be spontaneously broken because there is no doubling of multiplets with opposite parity observed in nature.  However, the $U(1)_A$ symmetry is explicitly violated by the presence of instantons in QCD, leading to the instanton-induced six-fermion interaction in the effective action.  This solves ``the $U(1)$ problem" that no isoscalar Goldstone boson exists in Nature \cite{tHooft86}. There is a pseudoscalar meson, the $\eta$ with a mass of $478 MeV$.  It cannot be the Goldstone boson because from current algebra S. Weinberg has shown that the mass of such a Goldstone boson has to be smaller than $\sqrt{3} m_\pi$, far below the mass of the $\eta$ meson \cite{WeinbergSphys}.  (The $\eta$ meson can still be made of a quark and an antiquark, so the usual $SU(2)$ scheme is still applicable - only this $\eta$ meson is not a Goldstone boson)\footnote{ One can extend this discussion
to $U_L(3)\times U_R(3)$ with pions, kaons and $\eta$ now 8 Goldstone bosons,
and the $\eta'$ with mass $958 MeV$ taking the place of $\eta$. This $\eta'$
is an $SU(3)$ singlet.}.

The axial-vector isoscalar current associated with the $U_A(1)$ symmetry is
$j_\mu^5= \sum_i^2 \bar\psi_i \gamma_5 \gamma_\mu \psi^i$. It has an Adler-Bell-Jackiw chiral anomaly
\begin{eqnarray}
\partial^\mu j^{(5)}_\mu =  N_f {g^2_3 \over 32 \pi^2} \epsilon^{\mu\nu\rho\sigma}F^a_{\mu\nu}
{F_{\rho\sigma}^a}\ ,
\label{lightest}
\end{eqnarray}
where $F^a_{\mu\nu}$ denotes the field strengths of the gluons, and $g_3$ is
the QCD coupling constant. In a QCD instanton background, integration
over spacetime yields
\begin{eqnarray}
\int {\partial \over \partial t} Q^5  {\rm d}t = N_f k\ ,
\end{eqnarray}
where $k$ is the winding number. To make sense of this equation one should
first integrate in Euclidean space to obtain a non-vanishing expression
for the right -hand side in terms of the winding number $k$ of the QCD
instanton, and then Wick-rotate so that the left-hand side can be written
as $\int {\rm d}^4x \partial_\mu j^\mu_5 \propto \int {\partial \over \partial t} Q^5  {\rm d}t$. The conclusion is that $Q^5$ is not conserved because
$k$ can be different from zero. For further discussion of the $U(1)$ problem
we refer to \cite{tHooft86}, \cite{Weinb-book,CL}, or to the lecture notes
by Coleman in \cite{Rev}.

\section{Baryon decay}

In this section we present an application of instantons to the gauge fields
of the electroweak sector of the Standard Model.

In an instanton background with winding number $k$, massless (or approximately massless) fermions in the  fundamental representation of $SU(N)$ have $|k|$ zero modes, see \eqn{adjoint}.  In the electroweak $SU(2)_w \times U(1)$ theory  (the subscript $w$ stands for weak), quarks and leptons are in the fundamental representation (doublets) of $SU(2)_w$.  In Euclidean space the integration over zero modes of these quarks and leptons has dynamical consequences which we shall derive, but of course real quarks and leptons live in Minkowski space and not in Euclidean space.  We assume that the Green functions in Minkowski spacetime can be obtained from those in Euclidean space by analytic continuation.  Ideally we should prove that the Euclidean results give the main contribution to processes in Minkowski space in the same way as this was shown for tunnelling, but as far as we know this has not been done.  Since processes involving electroweak instantons are suppressed by a factor $\exp \left( - {1 \over \hbar} {8 \pi^2 \over g^2_2} |k| \right)$ with $g_2$ the electroweak $SU(2)$ coupling constant, we only consider instantons with $|k| =1$ made from $W^+ , W^-$ and $W^0$ bosons.  Then the left-handed quark doublets ${u \choose d'}$ and ${c \choose s'}$ each have 3 zero modes because there are 3 colors, while the lepton doublets ${\nu_e \choose e^-}_L$ and ${\nu_\mu \choose \mu^-}_L$ each have one zero mode.  The primes on $d'$ and $s'$ denote Cabibbo-rotated quarks
\eqa
&& d' = d \cos \theta_c + s \sin \theta_c \nn
&& s' = s \cos \theta_c - d \sin \theta_c
\eqae
with $\theta_c = 13^0$ the Cabibbo angle.
As we shall explain, this Cabibbo rotation makes it possible for a neutron and a proton (six quarks together) to decay into two antileptons \cite{Hooft2}
\begin{equation}
p + n \rightarrow e^+ + {\bar \nu}_\mu \quad ({\rm or}\,\,\,\,\, \mu^+ + {\bar \nu}_e)
\ .
\end{equation}
In these instanton- induced processes, the electron number $E$, muon number $M$, up plus down number, and charm plus strangeness number change as follows
\begin{eqnarray}
\Delta E = \Delta M=1\ ,\quad \Delta u + \Delta d' = 3\ ,\quad \Delta c + \Delta s' =3\ .
\end{eqnarray}
The decay of a proton with $(u, u, d)$ and neutron with $(u, d,d)$ quarks into
$e^+$ and $\bar\nu_\mu$, or into $\mu^+$ and $\bar\nu_e$, can be described by a local vertex operator with 3 up-quark fields with different colors from ${u \choose d'}$ doublets, and 3 down-quark fields also with different colors from ${c \choose s'}$ doublets, and further one field from each of the two lepton doublets.  This operator is of course nonrenormalizable, but it can be used in effective field theories for phenomenological purposes. Although this efective operator
is derived from field theory in the sector with an instanton, once it is obtained one can add it to the effective action and then forget about the existence of instantons. We now derive these results.

The $U(1)_A$ symmetry has at the perturbative level an anomaly.  There are triangle graphs with an anomaly: one vertex of the triangle graph is given by $j^{(5)}_\mu = \sum_s \bar\psi^s \gamma_5 \gamma_\mu \psi^s$ (where $s=1, \ldots, N_f$ and $N_f$ is the number of flavors, 3 in our case if we restrict our attention to the lightest quarks $u, d$ and $s$).  The one-loop perturbative chiral anomaly is then given by
\begin{eqnarray}
\partial^\mu j^{(5)}_\mu = i N_f {g^2_2 \over 16 \pi^2} G^a_{\mu\nu}
{^*\!G_{\mu\nu}^a}\ ,
\label{lightest2}
\end{eqnarray}
where $G^a_{\mu\nu}$ is the $W$-boson field strength and $g_2$ the coupling constant of the $SU(2)$ weak interactions.  (This is thus the abelian flavor $U(1)_A$ anomaly.  The nonabelian anomaly for the rigid flavor group vanishes because it is proportional to the trace of $T_a$ of the flavor group, which vanishes).

If one integrates over space and time, the anomaly equation becomes
\begin{eqnarray}
\int {{\rm d} \over {\rm d}t} Q^{(5)} \equiv \int dt {{\rm d} \over {\rm d}t} \int  {\rm d}^3 x (ij^5_0) = 2 N_f k\ .
\end{eqnarray}
The instanton number $k$ counts the number of
left-handed fermions minus the number of right-handed fermions, and in
ordinary perturbation theory (with $k=0$) for massless quarks, this difference
is thus conserved. However, in an instanton background ($k\neq 0$),
the chiral charge of the vacuum at $t=- \infty$ changes to a different chiral charge of the vacuum at $t = + \infty$:  $\Delta Q^{(5)} = 2 N_f k$.
The conclusion is that the perturbative anomaly, and the violation of the axial charge which occurs when one tunnels from one vacuum to another, are related!   Both are different aspects of the same chiral anomaly.  The perturbative anomaly occurs when fields are small, so the winding vanishes and one is in the $k=0$ sector.  The nonperturbative anomaly is due to the same axial-vector current but now in the background of instantons which cannot be viewed as small and tending to zero at infinity, since they must produce winding.

One may at this point wonder whether the Higgs effect which gives the
$W$-bosons a mass, in such a way that they vanish exponentially at large
distances, does at the same time destroy the concept of winding. There is
no contradiction. When we discussed the large instanton problem, we chose
the regular gauge for the instanton to simplify the calculations. However,
exponential fall-off only occurs in the singular gauge. In that case, the
winding takes place at the origin, as we discussed in the introduction. In this
section we use the regular gauge and then there is winding at infinity even in
Higgs models.

To saturate the integrations over the Grassmann collective
coordinates, one needs 6 chiral quark fields in a correlator (one for
each zero mode).  Each field has a mode expansion into a zero mode
and all nonzero modes, but the integration measure ${\rm d}{\cal K}$ over
the Grassmann variable ${\cal K}$ in the mode expansion picks out only the zero
mode. Then the integration over collective coordinates
gives as result the product of the 6 zero mode functions.  If we
put one $SU(2)_w$ doublet with one up quark and one down quark at a point $x_1$, a second
pair a $x_2$, and a third pair at $x_3$,\footnote{  A massless complex Dirac spinor contains two Weyl spinors which are decoupled from each other $\bar\psi_D \rlap{\,/}D \psi_D = \bar\psi_L \rlap{\,/}D \psi_L + \bar\psi_K \rlap{\,/}D \psi_R$.  Each has a zero mode.  However, since only left-handed quarks couple to the $W$ gauge fields, only left-handed quarks feel the presence of instantons, and so we neglect the right-handed quarks in this discussion.}  and the instanton is at
$x_0$, we find from \eqn{van} for large separations $(x^2 >>
\rho^2)$ the factor\footnote{ The down quark is contained in the
$s'$ of the doublet $(c, s')$.  This is an $SU(2)_w$ doublet, and
the instanton is an $SU(2)_w$ instanton.   Although the $c$ quark
is heavier than the $s$ quark, one can still view them as massless
compared to the scale $250 GeV$ of electroweak interactions.  Massive
spinors in an instanton background have no zero modes, as one may
show by adding a mass term to \eqn{dictionary} and \eqn{DbarD}.
We assume that for such a broken $SU(2)_w$ doublet there still
exists approximately a zero mode.}
\begin{eqnarray}
\prod^3_{i=1} {1 \over (x_i - x_0)^6}\ .
\end{eqnarray}
So if one computes some correlator in a theory with instantons, six quark fields from the correlator are needed to saturate the Grassmann integrals, and the remaining fields are then treated as in ordinary field theory (with propagators and vertices).  Thus instantons induce a term proportional to $\prod^3_{i=1} {1 \over (x_i - x_0)^6}$ in the effective action which describes the annihilation of 6 quarks.  Further there are $\sigma$ matrices and other constants which are also due to the zero mode function.

One can now construct an effective local 6-quark vertex $V$ at a point $x_0$ which yields the same results in a theory without instantons as one obtains in a theory with instantons if one integrates over the fermionic collective coordinates of the quarks.  This vertex must contain 6 quark fields which contain the 6 different collective coordinates, hence it has the form $V=u_L^{\alpha, 1} u_L^{\beta ,2} u_L^{\gamma , 3} d_L^{\delta , 1} d_L^{\epsilon , 2} d_L^{\zeta , 3} T_{\alpha \beta \gamma \delta \epsilon \zeta}$ where $T$ is a numerical tensor.   Contraction of 6 ``probe-quarks" at positions $x_1 , x_2 , x_3$ with $V$ at $x_0$ using ordinary flat space propagators ${1 \over (x_i - x_0)^3}$ for massless quarks in a trivial vacuum precisely reproduces the result for the correlation function of the 6 probe-quarks in an instanton background centered around $x_0$, provided the form of $T$ is correctly chosen.

These new vertices lead to anomalies in the baryon currents and lepton currents.   In particular, the rigid $U(1)_A$ symmetry is {\bf explicitly} broken by the presence of the interaction $V$ in the action, and as we discussed in the previous section, this solves the $U(1)_A$ broken.  As we already mentioned, a proton and a neutron (two baryons equal six quarks) may annihilate to form two antileptons (an $e^+$ or a $\mu^+$, and an anti-neutrino).  However, due to the incredibly small prefactor $\exp \left( - {8 \pi^2 \over g^2_2} |k| \right)$, where $g_2$ is the $SU(2)$ weak coupling constant,
these processes are not observable.

\section{Discussion} 

In this chapter we have reviewed the general properties of single Yang-Mills 
instantons, and have given tools to compute non-perturbative effects in (non-) supersymmetric gauge theories.  However, we have not discussed several other important or interesting topics:

$\bullet$ Perturbation theory around the instanton: the methods
described here enable us to compute non-perturbative effects in the
semi-classical approximation where the coupling constant is small.
It is in many cases important to go beyond this limit, and to study
subleading corrections that arise from higher order perturbation
theory around the instanton \cite{TwoLoop1,TwoLoop2}. Apart from a
brief discussion about the one-loop determinants in section 7, we
have not really addressed these issues.

$\bullet$ Multi-instantons: we have completely omitted a discussion
of multi-instantons. These can be constructed using the ADHM
formalism \cite{ADHM}. The main difficulty lies in the explicit
construction of the collective coordinates in an instanton solution
and of the measure of collective coordinates beyond instanton number
$k=2$. However, it was demonstrated that certain simplifications
occur in the large $N$ limit of ${\cal N}=4$ SYM theories
\cite{DorHolKhoMatVan99}, where one can actually sum over all
multi-instantons to get exact results for certain correlation
functions. For reviews on the ADHM construction in super Yang-Mills theories, 
see e.g. \cite{ndorey,DorHolKhoMatVan99,KMS}.  
The same techniques were later applied for ${\cal N}=2,1$ SYM
\cite{N=2largeN,N=1largeN}, and it would be interesting to study the
consequences of multi-instantons for large $N$ non-supersymmetric
theories. For a review on instantons in QCD, see for instance \cite{ShuSch97}.
\\

\appendix

\section{Winding number} \label{Winding} 

For a gauge field configuration with finite classical gauge action the field strength must
tend to zero faster than $x^{-2}$ at large $x$. For vanishing $F_{\mu\nu}$,
the potential $A_\mu$ becomes then pure gauge, $A_\mu \stackrel{x \to \infty}
{ \longrightarrow} U^{-1} \partial_\mu U$. All configurations of $A_\mu$ which
become pure gauge at infinity fall into equivalence classes, where each class
has a definite winding number. As we now show, this winding number is given
by
\begin{equation}
k = - \frac{1}{16 \pi^2} \int {\rm d}^4 x\, {\rm tr}\, {^\ast F}_{\mu\nu} F_{\mu\nu} \ ,
\label{cycliccity}
\end{equation}
where ${^\ast F}_{\mu\nu}=\ft12 \epsilon_{\mu\nu\rho\sigma}F_{\rho\sigma}$ and
$T_a$ are the generators in the fundamental representation of $SU(N)$, antihermitean $N \times N$ matrices satisfying  $\tr\, T_a T_b = - \ft12 \delta_{ab}$. This is the normalization we adopt for the fundamental representation. The key observation is that ${^\ast F}_{\mu\nu} F_{\mu\nu}$ is a total derivative of a gauge variant current\footnote{ Note that ${^\ast F}_{\mu\nu} F_{\mu\nu}$ is equal to $2 \epsilon_{\mu\nu\rho\sigma} \left\{ \partial_\mu A_\nu \partial_\rho A_\sigma + 2 \partial_\mu A_\nu A_\rho A_\sigma + A_\mu A_\nu A_\rho A_\sigma \right\}$ but the last term vanishes in the trace due to the cyclicity of the trace.}
\begin{equation}
{\rm tr}\, {^\ast F}_{\mu\nu} F_{\mu\nu} = 2 \partial_\mu {\rm tr}\, \epsilon_{\mu\nu\rho\sigma} \left\{ A_\nu \partial_\rho A_\sigma + \ft23 A_\nu A_\rho A_\sigma \right\} \ .
\end{equation}
According to Stokes' theorem, the four-dimensional space integral becomes an integral over the three-di\-men\-si\-onal boundary at infinity if one uses the regular gauge in which there are no singularities at the orgin. Since  $F_{\mu\nu}$ vanishes at large $x$, one may replace $\partial_\rho A_\sigma$ by $-A_\rho A_\sigma$, and since $A_\mu$ becomes a pure gauge at large $x$, one obtains
\begin{equation} \label{k-charge} k = \frac{1}{24 \pi^2} \oint_{S^3({\rm space})} {\rm d} {\mit\Omega}_\mu \epsilon_{\mu\nu\rho\sigma} {\rm tr}\, \left\{ \left( U^{-1} \partial_\nu U \right) \left( U^{-1} \partial_\rho U \right) \left( U^{-1} \partial_\sigma U \right) \right\} \ ,
\end{equation}
where the integration is over a large three-sphere, $S^3({\rm space})$, in
four-dimensional Euclidean space. To each point $x^\mu$ on this large
three-sphere in space
corresponds a group element $U$ in the gauge group $G$. If $G = SU (2)$, the
group manifold is also a three-sphere\footnote{ The elements of $SU (2)$ can
be written in the fundamental representation as $U = a_0 \1 + i \sum_k a_k
\tau_k$ where $\tau_k$ are the Pauli matrices and $a_0$ and $a_k$ are real coefficients satisfying the condition
$a_0^2 + \sum_k a_k^2 = 1$.  This defines a sphere $S^3({\rm group})$.  (If
the $a$'s are not real but carry a common phase, one obtains the elements of
$U(2)$).} $S^3({\rm group})$.  Then $U (x)$ maps $S^3({\rm space})$ into
$S^3({\rm group})$,\footnote{ There is actually a complication.  Far away $A_\mu = U^{-1} \del_\mu U$ but in order that $U$ be only a function on $S^3({\rm space})$ it should only depend on the 3 polar angles but not on the radius.  Hence $A_r = U^{-1} \del_r U$ should vanish.  We can make a gauge transformation with a group element $V$ such that $A_r^\prime = V^{-1} (\del_r + A_r) V$ vanishes.  The $V$ which achieves this is the path ordered integral along the radius from the origin, $V=P \exp - \int^r A_r dr$.  Note that $U$ is only defined for large $r$, but $V$ must be defined everywhere, and $V \not= U$.  In fact, $V$ does not have winding since it can be continuously deformed to the unit group element.  The winding number is computed in the text for $UV$, but since $k$ in (\ref{cycliccity}) is gauge invariant, $k$ is also the winding number of the original gauge field $A_\mu$.} and as we now show, $k$ is an integer which counts how
many times $S^3({\rm space})$ is wrapped around $S^3({\rm group})$.
Choose a parametrization of the group elements of $SU (2)$ in terms of group
parameters\footnote{ For example, Euler angles, or Lie parameters $U = a_0 \1
+ i \sum_k a_k \tau_k$ with $a_0 = \sqrt{1 - \sum_k a_k^2}$.} $\xi^i (x)$
($i = 1,2,3$). Then the functions $\xi^i (x)$ map $x$ into $SU (2)$. Consider
a small surface element of $S^3({\rm space})$. According to the chain rule
\begin{eqnarray} && {\rm tr}\, \left\{ \left( U^{-1} \partial_\nu U \right) \left( U^{-1} \partial_\rho U \right) \left( U^{-1} \partial_\sigma U \right) \right\} \nonumber\\
&&  = \frac{\partial \xi^i}{\partial x_\nu} \frac{\partial
\xi^j}{\partial x_\rho} \frac{\partial \xi^k}{\partial x_\sigma}
{\rm tr}\, \left\{ \left( U^{-1} \partial_i U \right) \left(
U^{-1} \partial_j U \right) \left( U^{-1} \partial_k U \right)
\right\} \ ,
\end{eqnarray}
and using\footnote{ For example, if the
surface element points in the $x$-direction we have
${\mit\Delta\Omega} = {\mit\Delta} y {\mit\Delta}
z {\mit\Delta}\tau$ if $\epsilon_{1234} =1$.} \begin{equation}
{\mit\Delta\Omega}_\mu = \ft16 \epsilon_{\mu\alpha\beta\gamma}
{\mit\Delta} x_\alpha {\mit\Delta} x_\beta {\mit\Delta} x_\gamma\ ,
\end{equation} with $\ft16 \epsilon_{\mu\nu\rho\sigma}
\epsilon_{\mu\alpha\beta\gamma} = \delta^{\nu\rho\sigma}_{
[\alpha\beta\gamma]}$ and $\Delta \xi^{[i} \Delta \xi^j \Delta \xi^{k]} =
\epsilon^{ijk} \Delta^3 \xi$, we obtain for the contribution
${\mit\Delta} k$ of the small surface element to $k$
\begin{equation}
{\mit\Delta} k = \frac{1}{24\pi^2} \epsilon^{ijk}
{\rm tr}\, \left\{ \left( U^{-1} \partial_i U \right) \left(
U^{-1} \partial_j U \right) \left( U^{-1} \partial_k U \right)
\right\} \Delta^3 \xi \ ,
\end{equation}
where $k = \oint_{S^3({\rm space})} {\mit\Delta} k$.
The elements $\left( U^{-1} (\xi)
\partial_i U  (\xi) \right)$ lie in the Lie algebra, and define
the group vielbein $e_i^a (\xi)$ by \begin{equation} \left( U^{-1}
\partial_i U \right) = e_i^a (\xi) T_a \ . \end{equation} With
$\epsilon^{ijk} e_i^a e_j^b e_k^c = \left( \det e \right) \,
\epsilon^{abc}$, we obtain for the contribution to $k$ from a
surface element ${\mit\Delta} {\mit\Omega}_\mu$ \begin{equation}
{\mit\Delta} k = \frac{1}{24 \pi^2} \left( \det e \right) \tr
\left( \epsilon^{abc} T_a T_b T_c \right) \Delta^3 \xi = - \frac{1}{16
\pi^2} \left( \det e \right) \Delta^3 \xi \ .
\end{equation}
We used that for $SU(2)$ we have $[T_a,T_b]=\epsilon_{abc} T_c$.
As we have demonstrated, the original integral over the physical space is
reduced to one over the group with measure $\left( \det e \right)
{\rm d}^3 \xi$. The volume of a surface element of $S^3({\rm group})$
with coordinates ${\rm d} \xi^i$ is proportional to $\left( \det e
\right) {\rm d}^3 \xi$ (called the Haar measure).  Since this expression
is a scalar in general relativity,\footnote{ Under a change of
coordinates $\xi = \xi (\xi')$ at the point $\xi$, the vielbein transforms
as $e^a_i(\xi) = {\partial \xi^{'j} \over \partial \xi^i} e^{\prime a}_j (\xi
(\xi'))$, hence $\det e (\xi) = \left( \det {\partial \xi' \over
\partial \xi} \right) \det (e' (\xi'))$, while ${\rm d}^3 \xi$ is equal
to $| \det {\partial \xi \over \partial \xi'} | {\rm d}^3 \xi'$.  For
small coordinate transformations $\det \partial \xi /
\partial \xi'$ is positive, hence $\det e \; {\rm d}^3 \xi$ is
invariant.} we know that the value of the volume does not depend
on which coordinates one uses except for an overall normalization.
We fix this overall normalization of the group volume such that
near $\xi = 0$ the volume is $\Delta^3 \xi$. Since $e_i^a = \delta_i^a$
near $\xi = 0$, we have there the usual Euclidean measure ${\rm d}^3 \xi$.
Each small patch on $S^3({\rm space})$ corresponds to a small
patch on $S^3({\rm group})$, $\Delta k \sim {\rm Vol} (\Delta^3\xi)$.
Since the $U$'s fall into homotopy
classes, integrating once over $S^3({\rm space})$ we cover
$S^3({\rm group})$ an integer number of times. To check the
proportionality factor in ${\mit\Delta} k \sim {\rm Vol}\, \left(
\Delta^3 \xi \right)$, we consider the fundamental map
\begin{equation}
\label{Rotation} U (x) = i x_\mu \s_\mu / \sqrt{x^2} \ , \qquad
U^{-1} (x) = -i x_\mu \bar{\s}_\mu / \sqrt{x^2} \ .
\end{equation}
where $\s_\mu$ denotes the $2 \times 2$ matrices $(\vec\sigma,
i)$ with $\vec\sigma$ the Pauli matrices, and $\bar\s_\mu = (\vec\s , -i)$.
  This is clearly a one-to-one map from
$S^3({\rm space})$ to $S^3({\rm group})$ and should therefore
yield $|k| = 1$. Direct calculation gives
\begin{equation} U^{-1}
\partial_\mu U =  {-x_\mu \over x^2} + {x_\nu \bar\sigma_\nu
\over x^2} \s_\mu = - \sigma_{\mu\nu} x_\nu /x^2 \ ,
 \end{equation}
where $\sigma_{\mu\nu}$ is defined in \eqref{lor-gen}.
Substitution into \eqn{k-charge} leads to $k = -
\frac{1}{2\pi^2} \oint {\rm d} {\mit\Omega}_\mu$ $x_\mu/x^4 = -1$ making
use of  \eqn{AntiComm}.\footnote{ Only the commutator of the first
two matrices in $\tr\ (\sigma_{\nu \alpha} \sigma_{\rho \beta}
\sigma_{\sigma \gamma}) x^\alpha x^\beta x^\gamma$ contributes
because the anticommutator is proportional to the unit matrix.  In
the result only the anticommutator gives a nonvanishing result, because the commutator yields term proportional to $\s_{\a\b}$ whose trace vanishes.}
To obtain $k = 1$ one has to make the change $\s  \leftrightarrow
\bar{\s}$ or $x \leftrightarrow - x$ in Eq.\ (\ref{Rotation}).

Let us comment on the origin of the winding number of the instanton in
the singular gauge. In this case $A^{\rm sing}_\mu$ vanishes fast
at infinity, but becomes pure gauge near $x = 0$.  In the region
between a small sphere in the vicinity of $x = 0$ and a large
sphere at $x = \infty$ we have an expression for $k$ in terms of
a total derivative, but now for $A^{\rm sing}_\mu$ the only
contribution to the topological charge comes from the boundary
near $x = 0$:
\begin{equation}
k = - \frac{1}{24 \pi^2}
\oint_{S_{x \to 0}^3({\rm space})} {\rm d} {\mit\Omega}_\mu
\epsilon_{\mu\nu\rho\sigma} {\rm tr}\, \left\{ \left( U^{-1}
\partial_\nu U \right) \left( U^{-1} \partial_\rho U \right)
\left( U^{-1} \partial_\sigma U \right) \right\} \ .
\end{equation}
The extra minus sign is due to the fact that the
normal to the $S^3({\rm space})$ at $x = 0$ points inward.
 Furthermore, $A^{\rm sing}_\mu \sim U^{-1} \partial_\mu U = -
\bar\sigma_{\mu\nu} x_\nu / x^2$ near $x = 0$, while $A^{\rm
reg}_\mu \sim U \partial_\mu U^{-1} = - \sigma_{\mu\nu} x_\nu /
x^2$ for $x \sim \infty$. There is a second extra minus sign in the
evaluation of $k$ from the trace of Lorentz generators.
As a result $k_{\rm sing} = k_{\rm reg}$, as it should
be since $k$ is a gauge invariant object. The gauge transformation
which maps $A^{\rm reg}_\mu$ to $A^{\rm sing}_\mu$ transfers the
winding from a large to a small $S^3({\rm space})$.

 \section{'t Hooft symbols and Euclidean spinors} \label{HooftSpinor}

In this appendix we give a list of conventions and formulae useful for
instanton calculus.  Let us first discuss the structure of Lorentz algebra
$so (3,1)$ in Minkowski space-time. The generators can be represented by
$L_{\mu\nu} =\ft12 (x_\mu \partial_\nu - x_\nu \partial_\mu)$ and form the
algebra
$[L_{\mu\nu}, L_{\rho\sigma}] =  -  \eta_{\mu\rho} L_{\nu\sigma}  -
 \eta_{\nu\sigma} L_{\mu\rho} +  \eta_{\mu\sigma} L_{\nu\rho} +
\eta_{\nu\rho} L_{\mu\sigma}$, with the signature $\eta_{\mu\nu} = {\rm diag}
(-,+,+,+)$. The spatial rotations $J_i \equiv \ft12 \epsilon_{ijk} L_{jk}$
and boosts $K_i \equiv L_{0i}$ satisfy the algebra $[J_i,J_j]=-\epsilon_{ijk}
J_k, [J_i,K_j]=[K_i,J_j]=-\epsilon_{ijk}K_k$ and $[K_i,K_j]=\epsilon_{ijk}
J_k$.

There exist two 2-component spinor representations, which
we denote by $\lambda^\alpha$ and ${\bar \chi}_{\dot\alpha}$ ($\alpha =1,2$
and $\dot\alpha =1,2$). The generators for these spinor representations
are $\sigma^{\mu\nu}$ and ${\bar \sigma}^{\mu\nu}$, where
$\sigma_{\mu\nu} \equiv \ft12 (\sigma_\mu \bar\sigma_\nu - \sigma_\nu
\bar\sigma_\mu), \bar\sigma_{\mu\nu} =
\ft12 (\bar\sigma_\mu \sigma_\nu - \bar\sigma_\nu \sigma_\mu)$, with
$\sigma_\mu^{\alpha\dot\beta} = (\vec \tau, I),
\bar\sigma_{\mu\, \dot\alpha \beta} = (\vec \tau, -I), \mu=1,2,3,0$, and $I$
denotes the identity matrix. The matrices $\tau^i$ with $i=1,2,3$ are the usual Pauli matrices. They consist
of $\sigma^{ij}=i\epsilon^{ijk}\tau^k$ and $\sigma^{0i}=\tau^i$ for
$\lambda^\alpha$, and ${\bar \sigma}^{ij}=i\epsilon^{ijk}\tau^k$ and
${\bar \sigma}^{0i}=-\tau^i$
for ${\bar \chi}_{\dot\alpha}$. The rotation generators $\s^{ij}$ are clearly antihermitian, but the boost generators are hermitian.

Under a rotation or boost, both spinors
simultaneously transform. Most importantly, the two spinor representations
are complex. In fact, they are each other's complex conjugate up to a
similarity transformation: $(\sigma^{\mu\nu})^*=\sigma_2
{\bar \sigma}^{\mu\nu}\sigma_2$.  The matrices $i \tau^k$ and $\tau^k$ form the $2 \times 2$ defining representation of the group $Sl(2,C)$, which is
the covering group of $SO(3,1)$.

The situation differs for Euclidean space ($\delta_{\mu\nu} =
{\rm diag} (+,+,+,+)$) with $SO (4)$  instead of the
Lorentz group $SO(3,1)$.  Now $[L_{\mu\nu},L_{\rho\sigma}]=\delta_{\nu\rho}
L_{\mu\sigma}+{\mbox {3 terms}}$, and
$[J_i , J_j ] =- \e_{ijk} J_k , [J_i , K_j ] =- \e_{ijk} K_k$ but $[K_i , K_j ] = - \e_{ijk}  J_k$ where obviously $J_i  \equiv \ft12 \epsilon_{ijk} L_{jk}$ and boosts $K_i \equiv L_{i4}$. The linear combinations of $(ij)$ and  $(4,i)$-plane rotations
\begin{equation} M_i \equiv \frac{1}{2}( J_i + K_i )\ ,
\qquad N_i \equiv \frac{1}{2}( J_i - K_i )\ ,
 \end{equation}
give the algebras of commuting $SU (2)$ subgroups of $SO (4) = SU (2)
\times SU (2)$ in view of the anti-hermiticity $M_i^\dagger = -M_i$,
$N_i^\dagger = -N_i$. We now denote the two spinor representations
by $\lambda^\alpha$ and ${\bar \chi}_{\alpha'}$. Because $M$ and $N$
are represented by generators $i{\vec \sigma}_M$ and
$i{\vec \sigma}_N$ which act in different spaces, one can transform
$\lambda^\alpha$ while ${\bar \chi}_{\alpha'}$ stays fixed, or vice versa.
The two spinor representations in Euclidean space are each pseudoreal:  as we shall discuss $(\s^{\mu\nu})^\ast = \s_2 \s^{\mu\nu} \s_2$ and $(\bar\s_{\mu\nu})^\ast = \s_2 \bar\s_{\mu\nu} \s_2$.

It is an easy exercise to check that we can represent the operators
$M$ and $N$ by
\begin{equation}
M_i =  \bar\eta_{i\mu\nu}  \ , \qquad\mbox{and}\qquad N_i =
\eta_{i\mu\nu}  \ ,
\end{equation}
where we introduced 't Hooft symbols \cite{Hooft1}
\begin{eqnarray} && \eta_{a\mu\nu}
\equiv \epsilon_{a\mu\nu} + \delta_{a\mu} \delta_{\nu4} - \delta_{a\nu}
\delta_{4\mu} , \  {\rm or} \  \eta_{aij} = \e_{aij} , \eta_{aj4} = \d_{aj} \nonumber\\
&& \bar\eta_{a\mu\nu} \equiv \epsilon_{a\mu\nu}
- \delta_{a\mu} \delta_{\nu4} + \delta_{a\nu} \delta_{4\mu} ,  \  {\rm or} \  \bar\eta_{aij} = \e_{aij} , \bar\eta_{aj4} = - \d_{aj}
\end{eqnarray}
and $\bar\eta_{a\mu\nu} = (-1)^{\delta_{4\mu} + \delta_{4\nu}} \eta_{a\mu\nu}$.
 They form a basis of anti-symmetric 4 by 4 matrices and are (anti-)selfdual
in vector indices ($\epsilon_{1234} = 1$) \begin{equation} \eta_{a\mu\nu} =
\ft12 \epsilon_{\mu\nu\rho\sigma} \eta_{a\rho\sigma} \ , \qquad
\bar\eta_{a\mu\nu} = - \ft12 \epsilon_{\mu\nu\rho\sigma}
\bar\eta_{a\rho\sigma} \ .
\end{equation}
The $\eta$-symbols obey the following relations
\begin{eqnarray} \label{eta-eta1} &&\epsilon_{abc} \eta_{b\mu\nu}
\eta_{c\rho\sigma} = \delta_{\mu\rho} \eta_{a\nu\sigma} + \delta_{\nu\sigma}
\eta_{a\mu\rho} - \delta_{\mu\sigma} \eta_{a\nu\rho} - \delta_{\nu\rho}
\eta_{a\mu\sigma} \ , \nonumber\\ &&\eta_{a\mu\nu} \eta_{a\rho\sigma} =
\delta_{\mu\rho} \delta_{\nu\sigma} - \delta_{\mu\sigma} \delta_{\nu\rho} +
\epsilon_{\mu\nu\rho\sigma} \ , \nonumber\\ \label{eta-eta3} &&\eta_{a\mu\rho}
\eta_{b\mu\sigma} = \delta_{ab} \delta_{\rho\sigma} + \epsilon_{abc}
\eta_{c\rho\sigma} \ , \nonumber\\ &&\epsilon_{\mu\nu\rho\tau}
\eta_{a\sigma\tau} = \delta_{\sigma\mu} \eta_{a\nu\rho} +
\delta_{\sigma\rho} \eta_{a\mu\nu} - \delta_{\sigma\nu} \eta_{a\mu\rho} \ ,
\nonumber\\ &&\eta_{a\mu\nu} \eta_{a\mu\nu} = 12 \ ,\quad \eta_{a\mu\nu}
\eta_{b\mu\nu} = 4 \delta_{ab} \ ,\quad \eta_{a\mu\rho} \eta_{a\mu\sigma}
= 3 \delta_{\rho\sigma} \ .
\end{eqnarray}
The same holds for $\bar\eta$ except for the terms with
$\epsilon_{\mu\nu\rho\sigma}$,
\begin{eqnarray}
\bar\eta_{a\mu\nu} \bar\eta_{a\rho\sigma} &=&
\delta_{\mu\rho} \delta_{\nu\sigma} - \delta_{\mu\sigma} \delta_{\nu\rho} -
\epsilon_{\mu\nu\rho\sigma} \ ,\nonumber\\
\epsilon_{\mu\nu\rho\sigma}
{\bar \eta}_{a\sigma\tau} &=& -\delta_{\sigma\mu} {\bar \eta}_{a\nu\rho} -
\delta_{\sigma\rho} {\bar \eta}_{a\mu\nu} + \delta_{\sigma\nu}
{\bar \eta}_{a\mu\rho} \ .
\end{eqnarray}
Obviously $\eta_{a\mu\nu}
\bar\eta_{b\mu\nu} = 0$ due to different duality properties.  In matrix
notation, we have
\begin{eqnarray}
[\eta_a,\eta_b]=-2\epsilon_{abc}\eta_c \ ,&\qquad&
[{\bar \eta}_a,{\bar \eta}_b]=-2\epsilon_{abc}{\bar \eta}_c\ ,\nonumber\\
\{\eta_a,\eta_b\}= -2 \delta_{ab} \ ,&\qquad& \{{\bar \eta}_a,{\bar \eta}_b\}
= -2 \delta_{ab}\ ,
\end{eqnarray}
and the two sets of matrices commute, i.e. $[\eta_a,{\bar \eta}_b]=0$ (this
is equivalent to the statement that the generators $M$ and $N$ commute).

The two inequivalent spinor representations of the Euclidean Lorentz
algebra are given by
\begin{equation} \label{lor-gen}
\sigma_{\mu\nu} \equiv \ft12 [\sigma_\mu \bar\sigma_\nu - \sigma_\nu
\bar\sigma_\mu] \ , \qquad \bar\sigma_{\mu\nu} =
\ft12 [\bar\sigma_\mu \sigma_\nu - \bar\sigma_\nu \sigma_\mu] \ ,
\end{equation}
in terms of Euclidean matrices
\begin{equation} \label{Eucl-sigma}
\sigma_\mu^{\alpha\beta'} = (\tau^a, i) \ , \qquad
\bar\sigma_{\mu\, \alpha'\beta} = (\tau^a, -i) \ ,
\qquad \mu=1,2,3,4\ ,
\end{equation}
obeying the Clifford algebra $\sigma_\mu \bar\sigma_\nu + \sigma_\nu \bar\sigma_\mu = 2 \delta_{\mu\nu}$. Since $\sigma_{\mu\nu}$ contains $\sigma_{ij}=\epsilon_{ijk} i
\tau^k$ and $\sigma_{i4}=-i\tau_i$, while ${\bar \sigma}_{\mu\nu}$ contains
${\bar \sigma}_{ij}=\epsilon_{ijk} i \tau^k$ and ${\bar \sigma}_{i4}=i\tau_i$, they are not  each
others complex conjugate, contrary to the Minkowski case.
 Rather, they are pseudo-real, meaning that their complex-conjugates are related to themselves by a similarily transformation
 \eqa
 \s_{\mu\nu}^\ast = \s_2 \s_{\mu\nu} \s_2 ; (\bar\s_{\mu\nu})^\ast = \s_2 \bar\s_{\mu\nu} \s_2 \ .
 \eqae

To prove these, and other, spinor relations, one needs some formulas which we now present.  As in Minkowski space, also in Euclidean space $\s_\mu$ and $\bar\s_\mu$ are related by transposition
 \eqa
 \s_\mu{}^{\a\a'} = \bar\s_\mu{}^{\a'\a}
 \eqae
 where $\bar\s_\mu{}^{\a'\a}$ is obtained from $\bar\s^\mu{}_{\b'\b}$ by raising indices
 \eqa
 \bar\s_\mu{}^{\a'\a} \equiv \e^{\a'\b'} \e^{\a\b} \bar\s^\mu_{\b'\b}
 \eqae
 We use everywhere the north-west convention for raising and lowering the spinor indices
 \begin{equation}
\epsilon^{\alpha\beta} \xi_\beta = \xi^\alpha \ , \qquad \bar\xi^{\beta'} \epsilon_{\beta'\alpha'} = \bar\xi_{\alpha'} \ ,
 \end{equation}
 with $\epsilon_{\alpha\beta} = - \epsilon_{\alpha'\beta'},
\epsilon_{\alpha\beta} = \epsilon^{\alpha\beta}$, and $\epsilon_{\alpha'\beta'} = \epsilon^{\alpha'\beta'}$.  However, the relation between $\s_\mu$ and $\bar\s_\mu$ under complex conjugation is different (as expected because $\s^0 = I$ but $\s^4 =i I$). In Minkowski space we have  $(\s_\mu^{\a \dot\b})^\ast = \bar\s^{\dot\b \a}_\mu$, while in
 Euclidean space $(\s_\mu^{\a\b'})^\ast = {\bar \sigma}_{\mu,\beta'\alpha}=
\s_{\mu , \a\b'}$ and $ (\bar\s_{\mu , \a'\b})^\ast = \sigma_\mu^{\beta\alpha'}
=\bar\s_\mu^{\a'\b}$.

 Let us now apply these formulas to give another proof that $\s_{\mu\nu}$ and $\bar\s_{\mu\nu}$ are pseudoreal in Euclidean space
 \eqa
 && ((\s_{\mu\nu})^\a{}_\b)^\ast = {1 \over 2} (\s_\mu{}^{\a\b'})^\ast (\bar\s_{\nu,\b'\b})^\ast - \mu \leftrightarrow \nu \nn
 && = {1 \over 2} \s_{\mu, \a\b'} \bar\s_\nu^{\b'\b} - \mu \leftrightarrow \nu =- {1 \over 2} \s_{\mu , \a}{}^{\b'} \bar\s_{\nu , \b'}{}^\b - \mu \leftrightarrow \nu \nn
 && = -\e_{\g \a} (\s_{\mu\nu})^\g{}_\d  \e^{\b \d} = (-i \s_2) (- \s_{\mu\nu}) (- i \s_2) = \s_2 \s_{\mu\nu} \s_2
 \eqae
 and idem for $\bar\s_{\mu\nu}$.

The two spinor and vector representations of the $su (2)$ algebra are all given
in terms of anti-hermitian 2x2 matrices $\sigma_{\mu\nu},
{\bar \sigma}_{\mu\nu}$ and $i\tau^a$ and they are
related by the 't Hooft symbols,
\begin{equation}
\bar\sigma_{\mu\nu} = i \eta_{a\mu\nu} \tau^a \ , \qquad \sigma_{\mu\nu} = i \bar\eta_{a\mu\nu} \tau^a \ .
\label{obeyingg}
\end{equation}
Furthermore, $\bar\s_{\mu\nu}$ is selfdual whereas $\s_{\mu\nu}$ is anti-selfdual.
Some frequently used identities are
\begin{eqnarray}
\bar\sigma_\mu \sigma_{\nu\rho} = \delta_{\mu\nu} \bar\sigma_\rho - \delta_{\mu\rho} \bar\sigma_\nu - \epsilon_{\mu\nu\rho\sigma} \bar\sigma_\sigma \ , \quad \sigma_\mu \bar\sigma_{\nu\rho} = \delta_{\mu\nu} \sigma_\rho - \delta_{\mu\rho} \sigma_\nu + \epsilon_{\mu\nu\rho\sigma} \sigma_\sigma \ , \nonumber\\ \sigma_{\mu\nu} \sigma_\rho = \delta_{\nu\rho} \sigma_\mu - \delta_{\mu\rho} \sigma_\nu + \epsilon_{\mu\nu\rho\sigma} \sigma_\sigma \ , \quad \bar\sigma_{\mu\nu} \bar\sigma_\rho = \delta_{\nu\rho} \bar\sigma_\mu - \delta_{\mu\rho} \bar\sigma_\nu - \epsilon_{\mu\nu\rho\sigma} \bar\sigma_\sigma \ .
\end{eqnarray}
The Lorentz generators are antisymmetric in vector and symmetric in spinor
indices
\begin{equation} \label{symm-sigma}
\sigma_{\mu\nu\, \alpha\beta} = - \sigma_{\nu\mu\,
\alpha\beta} \ , \qquad \sigma_{\mu\nu\, \alpha\beta} = \sigma_{\mu\nu\,
\beta\alpha} \ ,
\end{equation}
and obey the algebra \begin{eqnarray} \label{Sigma-Algebra} [\sigma_{\mu\nu}, \sigma_{\rho\sigma}] \!\!\!&=&\!\!\! - 2 \left\{ \delta_{\mu\rho} \sigma_{\nu\sigma} + \delta_{\nu\sigma} \sigma_{\mu\rho} - \delta_{\mu\sigma} \sigma_{\nu\rho} - \delta_{\nu\rho} \sigma_{\mu\sigma} \right\} \ , \nonumber\\ \label{AntiComm} \{\sigma_{\mu\nu}, \sigma_{\rho\sigma}\} \!\!\!&=&\!\!\! - 2 \left\{ \delta_{\mu\rho} \delta_{\nu\sigma} - \delta_{\mu\sigma} \delta_{\nu\rho} - \epsilon_{\mu\nu\rho\sigma} \right\} \ .
\end{eqnarray}
The same relations hold for $\bar\sigma_{\mu\nu}$ but with $+ \epsilon_{\mu\nu\rho\sigma}$. In spinor algebra the following contractions are frequently used \begin{equation} \sigma_\mu^{\alpha\alpha'} \bar\sigma_{\mu\, \beta'\beta} = 2 \delta_\alpha^{\phan{i}\beta} \delta_{\alpha'}^{\phan{i}\beta'} \ , \qquad \sigma_{\rho\sigma\ \, \beta}^{\phan{ii}\alpha} \sigma_{\rho\sigma\ \, \delta}^{\phan{ii}\gamma} = 4 \left\{ \delta_\beta^{\phan{i}\alpha} \delta_\delta^{\phan{i}\gamma} - 2 \delta_\delta^{\phan{i}\alpha} \delta_\beta^{\phan{i}\gamma} \right\} \ .
\end{equation}
so that $\xi_{(1)}^\alpha \xi_{(2)\alpha} = \xi_{(2)}^\alpha \xi_{(1)\alpha}$.
For hermitean conjugation we define
$\left( \xi_{(1)}^\alpha \xi_{(2)\alpha} \right)^\dagger =
(\xi_{(2)\alpha})^\dagger (\xi_{(1)}^\alpha)^\dagger$
\begin{equation}
\left( \sigma_{\mu}^{\alpha\beta'} \right)^\ast = \sigma_{\mu\, \alpha\beta'} , \qquad \left( \bar\sigma_{\mu\, \alpha'\beta} \right)^\ast = \bar\sigma_{\mu}^{\alpha'\beta}\ .
\end{equation}

Throughout the paper we frequently use the following integral formula \begin{equation} \label{Integral} \int d^4 x \frac{\left( x^2 \right)^n}{\left( x^2 + \rho^2 \right)^m} = \pi^2 \left( \rho^2 \right)^{n - m + 2} \frac{{\mit\Gamma} (n + 2){\mit\Gamma} (m - n - 2)}{{\mit\Gamma}(m)} \ , \end{equation} which converges for $m - n > 2$.

\section{The volume of the gauge orientation moduli space} \label{Volume} 

The purpose of this appendix\footnote{ We thank R. Roiban for help in writing this appendix.} is to prove equation \eqn{VolCoset}. Let us consider an instanton in $SU (N)$ gauge theory. Deformations of this configuration which are still self-dual and not a gauge transformation are parametrized by collective coordinates. Constant gauge transformations $A_\mu \to U^{-1} A_\mu U$ preserve self-duality and transversality but not all constant $SU (N)$ matrices $U$ change $A_\mu$. Those $U$ which keep $A_\mu$ fixed form the stability subgroup $H$ of the instanton, hence we want to determine the volume of the coset space $SU (N)/H$.  If the instanton is embedded in the lower-right $2 \times 2$ submatrix of the $N \times N$ $SU (N)$ matrix, then $H$ contains the $SU (N - 2)$ subgroup in the left-upper part, and a $U (1)$ subgroup with elements $\exp \left( \theta A \right)$ where $A$ is the diagonal matrix \begin{equation} A = \frac{i}{2} \sqrt{\frac{N - 2}{N}} {\rm diag} \left( \frac{2}{2 - N}, \dots, \frac{2}{2 - N}, 1, 1 \right) \ . \end{equation} All generators of $SU (N)$ (and also all generators of $SO (N)$ discussed below) are normalized according to $\tr \, T_a T_b = - \ft12 \delta_{ab}$, as in the main text.

At first sight one might expect the range of $\theta$ to be such
that the exponents of all entries cover the range $2\pi$ an integer
number of times. However, this is incorrect: only for the last two
entries of $\exp \left( \theta A \right)$ we must require
periodicity, because whatever happens in the other $N - 2$ diagonal
entries is already contained in the $SU (N - 2)$ part of the
stability subgroup. Thus all elements $h$ in $H$ are of the form
\cite{Bernard}
\begin{equation} h = {\rm e}^{\theta A} g, \qquad\mbox{with}\qquad g \in SU (N - 2) \qquad\mbox{and}\qquad 0 \leq \theta \leq \theta_{\rm max} = 4 \pi \sqrt{\frac{N}{N - 2}} \ .
\end{equation}
For $N = 3$ the range of $\theta$ is larger than required by periodicity of the first $N-2$ entries, for
$N=4$ it corresponds to periodicity of all entries, but for $N \geq 5$ the range of $\theta$ is less than
required for periodicity of the first $N-2$ entries.\footnote{ For example, consider $SU(5)$ with
$\exp [\frac{i \theta}{2} \sqrt{{3 \over 5}} \; {\rm diag} \; ( -\ft23 , - \ft23 , - \ft23 , 1, 1)]$.  When
$\theta$ runs from $0$ to $\sqrt{{5 \over 3}} 4\pi$, last two entries repeat, but the first three entries only
reach $\exp ( - 4 \pi i/3)$.  The first three entries form then an element of $SU(N-2) = SU(3)$, namely they yield
an element $z$ of the center $Z_3$.  So when $\theta$ ranges beyond $\sqrt{{5 \over 3}} 4 \pi$, these $SU(5)$
elements can be written as a product of $z$ and $\exp i \theta A$ with $\theta$ smaller than $\sqrt{{5 \over 3}}
4 \pi$.  So, the range of $\theta$ is bounded by $\sqrt{{5 \over 3}} 4 \pi$.}  Thus $H \neq SU(N) \times U (1)$
for $N \geq 5$. The first $N - 2$ entries of $\exp \left( k \theta_{\rm max} A \right)$ with integer $k$ are given
by $\exp \left( - i k \frac{4\pi}{N - 2} \right)$ and lie therefore in the center $Z_N$ of $SU (N - 2)$. So, the
$SU (N)$ group elements $h = \exp\left( \theta A \right) g$ with $0 \leq \theta \leq \theta_{\rm max}$ and $g$ in
$SU(N-2)$ form a subgroup $H$. We shall denote $H$ by $SU (N - 2) \times ``U (1)"$ where $``U (1)"$ denotes the
part of the $U (1)$ generated by $A$ which lies in $H$.  We now use three theorems to evaluate the volume of
$SU (N)/H$: \begin{eqnarray} ({\rm I})   && {\rm Vol}\, \frac{SU (N)}{SU (N - 2) \times ``U (1)"} =
\frac{{\rm Vol} \, \left( SU (N)/SU (N - 2) \right) }{{\rm Vol} \, ``U (1)"} \ , \nonumber\\ ({\rm II})  &&
{\rm Vol} \, \frac{SU (N)}{SU (N - 2)} = {\rm Vol} \, \frac{SU (N)}{SU (N - 1)} {\rm Vol} \,
\frac{SU (N - 1)}{SU (N - 2)} \ , \\ ({\rm III}) && {\rm Vol}\, \frac{SU (N)}{SU (N - 1)} = \frac{{\rm Vol} \,
SU (N)}{{\rm Vol} \, SU (N - 1)} \ . \nonumber \end{eqnarray} It is, in fact, easiest to first compute ${\rm Vol}
\left( SU (N)/ SU (N - 1) \right)$ and then to use this result for the evaluation of  Vol $SU(N) /H$ (with Vol
$SU(N)$ as a bonus).

In general the volume of a coset manifold $G/H$ is given by $V = \int \prod_{\mu} dx^\mu\,
\det\ e^m_\mu (x)$ where $x^\mu$ are the coordinates on the coset manifold and $e^m_\mu (x)$ are the coset vielbeins.
 One begins with ``coset representatives'' $L (x)$ which are group elements $g \in G$ such that every group element
 can be decomposed as $g = L(x) h$ with $h \in H$.  We denote the coset generators by $K_m$ and the subgroup
generators by $H_i$. Then $L^{-1} (x) \partial_\mu L (x) = e^m_\mu (x) K_m + \omega_\mu^i (x) H_i$.  We shall take the generators
$K_m$ and $H_i$ in the fundamental representation of $SU(N)$: antihermitian $N \times N$ matrices.  Under a
general coordinate transformation from $x^\mu$ to $y^\mu (x)$, the vielbein transforms as a covariant vector with
index $\mu$ but also as a contravariant vector with index $m$ at $x = 0$. Hence {\it $V$ does (only) depend on
the choice of the coordinates at the origin}.
At the origin, $L^{-1}\partial_\mu L = e^m_\mu(0)K_m$, and we fix the normalization
of $K_m$ by ${\rm tr}\, K_m^2 = - \ft12 $ for $K_m$
in the $N\times N$ matrix representation of $SU (N)$.

To find the volume of $SU (N)/SU (N - 1)$  we note that the group elements of $SU (N)$ have a natural action on the space ${\rm\bf C}^N$ and map a point $\left( z^1, \dots, z^N \right) \in {\rm\bf C}^N$ on the complex hypersphere $\sum_{i = 1}^{N}\left| z^i \right|^2 = 1$ into another point on the complex hypersphere. The ``south-pole'' $(0, \dots, 0, 1)$ is kept invariant by the subgroup $SU (N - 1)$, and points on the complex hypersphere are in one-to-one correspondence with the coset representatives $L (z)$ of $SU (N)/SU (N - 1)$. We use as generators for $SU (N)$ the generators for $SU (N - 1)$ in the upper-left block, and further the following coset generators: $N - 1$ pairs $T_{2 k}$ and $T_{2k + 1}$ each of them containing only two non-zero elements
\begin{equation} \left( \begin{array}{cccc} 0      & \ldots &        & 0 \\
         &            &       & \cdot \\ [-15pt]
         &            &        & \cdot \\ \vdots &        &        & i/2      \\        &        & \ddots & \vdots  \\ 0      &  i/2   & \ldots &   0
\\ \end{array} \right) \ , \qquad\qquad
\left( \begin{array}{cccc} 0      & \ldots &        & 0 \\
        &            &       & \cdot \\[-15pt]
         &            &        & \cdot \\ \vdots &        &        & 1/2      \\        &        & \ddots & \vdots  \\ 0      &    -1/2   & \ldots &   0
\\ \end{array} \right) \ ,
\end{equation}
and further one diagonal generator
\begin{equation} T_{N^2 - 1} = \frac{i}{2} \sqrt{\frac{2}{N (N - 1)}} {\rm diag} \left( - 1, \dots, - 1, N - 1 \right) .
\end{equation}
(For instance, for $SU (3)$ there are two pairs, proportional to the
usual $\lambda_4$ and $\lambda_5$ and $\lambda_6$ and $\lambda_7$,
and the diagonal hypercharge generator $\lambda_8$.)  The idea now
is to establish a natural one-to-one correspondence between points
in ${\bf C}^N$ and points in ${\bf R}^{2N}$, namely we write all
points $(x^1, \dots, x^{2N})$ in ${\rm\bf R}^{2N}$ as points in
${\rm\bf C}^N$ as follows: $(i x^1 + x^2, \dots, i x^{2N - 1} +
x^{2N})$. In particular the south pole $(0,0,...,0,1)$ in ${\rm\bf
R}^{2N}$ corresponds to the south pole $(0,0,...,0,1)$ in ${\rm\bf
C}^{N}$ and the sphere $\sum_{i = 1}^{2N} (x^i)^2 = 1$ in ${\bf
R}^{2N}$ corresponds to the hypersphere $\sum_{i = 1}^N |z^i|^2 = 1$
in ${\rm\bf C}^{N}$ .   Points on the sphere $S^{2N - 1}$ in ${\bf
R}^{2N}$ correspond one-to-one to coset elements of $SO (2N)/ SO (2N
- 1)$. The coset generators of $SO(2N)/SO(2N - 1)$ are antisymmetric
$2N \times 2N$ matrices $A_I$ $(I = 1, \dots, 2N - 1)$ with the
entry $+1/2$ in the last column and $-1/2$ in the last row. The
coset element $1 + \delta g = 1 + {\rm d}t^I A_I$ maps the south
pole $s = (0, \dots, 0, 1)$ in ${\rm\bf R}^{2N}$ to a point $s +
\delta s$ in ${\rm\bf R}^{2N}$ where $\delta s =1/2 ({\rm d}t^1,
\dots , {\rm d}t^{2N - 1}, 0)$. We know how points in ${\rm \bf
C}^N$ correspond to points in ${\rm \bf R}^{2N}$, so we can ask
which coset element in $SU(N)/SU(N-1)$ maps the south-pole in  ${\rm
\bf C}^N$ to the point in ${\rm \bf C}^N$ which corresponds to $s +
{\delta}s$. In ${\rm\bf C}^N$ the corresponding point is $s + \delta
s$ with $\delta s = 1/2(i {\rm d}t^1 + {\rm d}t^2, \dots, i {\rm
d}t^{2N - 1})$. The coset generators of $SU (N)/SU (N - 1)$ act in
${\rm\bf C}^N$ as follows: $g = 1 + {\rm d}x^\mu K_\mu$ maps the
south-pole $s$ to $s + \delta s$ where now $\delta s = 1/2( i {\rm
d}x^1 + {\rm d}x^2, \dots, i \sqrt{\ft{2(N - 1)}{N}}{\rm d}x^{2N -
1} )$.  We can cover $SO(2N) / SO(2N-1) = S^{2N - 1}$ with small
patches.  Similarly we cover $SU(N) / SU(N-1)$ with small patches.
Each patch of $S^{2N-1}$ can be brought by the action of a suitable
coset element to the south-pole, and then we can use the inverse of
this group element to map this patch back into the manifold $SU
(N)/SU (N - 1)$. In this way both $S^{2N - 1}$ and $SU (N)/SU (N -
1)$ are covered by patches which are in a one-to-one correspondence.
Each pair of patches has the same ratio of volumes since both
patches can be brought to the south pole by the same group element
and at the south pole the ratio of their volumes is the same. To
find the ratio of the volumes of $S^{2N - 1}$ and $SU (N)/SU (N -
1)$, it is then sufficient to consider a small patch near the south
pole.  Near the south pole the vielbeins become unit matrices for
coset manifolds, hence the volume of the patches near the south-pole
is simply the product of the coordinates of these patches.  Consider
then a small patch at the south pole of $S^{2N - 1}$ with
coordinates $\left( {\rm d}t^1, \dots, \right. $ $\left. {\rm
d}t^{2N - 1} \right)$ and volume ${\rm d}t^1 \dots {\rm d}t^{2N -
1}$. The same patch at the south pole in ${\bf C}^N$ has coordinates
${\rm d} x^\mu$ where $\left( i {\rm d}t^1 + {\rm d}t^2, \dots, i
{\rm d}t^{2N - 1}\right)$ $ = \left( i {\rm d}x^1 + {\rm d}x^2,
\dots, i \sqrt{\frac{2 (N - 1)}{N}} {\rm d}x^{2N - 1} \right)$. The
volume of a patch in $SU (N)/SU (N - 1)$ with coordinates ${\rm
d}x^1, \dots, {\rm d}x^{2N - 1}$ is ${\rm d}x^1 \dots {\rm d}x^{2N -
1}$. It follows that the volume of $SU (N)/SU (N - 1)$ equals the
volume of $S^{2N - 1}$ times $\sqrt{\frac{N}{2 (N -
1)}}$~\footnote{This result yields the same answer for
\eqref{VolCoset} as \cite{Bernard}, but it yields $\pi^N/(N\,N!)$
for the volume of the complex projective space
$CP(N)=SU(N+1)/(SU(N)\times U(1))$ which differs from the result
${\rm Vol} [U(N+1)/(U(N)\times U(1))]= {\rm Vol} S^{2N}$ given in
\cite{Gilmore}.},
\begin{equation} \label{SUNtoN1}
{\rm Vol} \ \frac{SU (N)}{SU (N - 1)} = \sqrt{\frac{N}{2(N - 1)}}\ {\rm Vol}\ S^{2N - 1} \ .
\end{equation}

From here the evaluation of ${\rm Vol}\ SU(N)/H$  is straightforward. Using
\begin{equation} {\rm Vol}\ S^{2N - 1} = \frac{2 \pi^N}{(N - 1)!}\,l \ ,
\end{equation}
where $l=1$ if one uses the normalization ${\rm tr}\, K_m^2=-2$, but
$l=2^{2N-1}$ with our normalization of ${\rm tr}\, K_m^2=-\ft12$,
we obtain
\begin{equation} \label{VolSUN}
{\rm Vol}\ SU(N) = \sqrt{N} \prod_{k = 2}^{N} \frac{\sqrt{2}\pi^k}{(k - 1)!}
2^{2k-1}\ .
\end{equation}
We assumed that ${\rm Vol}\,SU(1)=1$ which seems a natural value but must be, and will be, justified below.
Then
\begin{eqnarray} && {\rm Vol}\ H = {\rm Vol}\ SU (N - 2) {\rm Vol}\ ``U (1)" \ , \qquad {\rm Vol}\ ``U(1)" =
 4 \pi \sqrt{\frac{N}{N - 2}} \ , \nonumber\\ && {\rm Vol}\ SU (N)/H = \ft12
\frac{\pi^{2N - 2}}{(N - 1)!(N - 2)!}2^{2N-1}2^{2N-3} \ . \label{volcoset-app}
\end{eqnarray}
This then produces formula \eqn{VolCoset}.

As an application and check of this analysis let us derive a few
relations between the volumes of different groups. From now on till
the end of this appendix we adopt the normalization ${\rm
tr}(T_aT_b)=-2\delta_{ab}$ for the generators of all groups
involved. Let us check that ${\rm Vol}\ SU (2) = 2 {\rm Vol}\ SO
(3)$, ${\rm Vol}\ SU (4) = 2 {\rm Vol}\ SO (6)$ and ${\rm Vol}\ SO
(4) = \ft12 \left( {\rm Vol}\ SU (2) \right)^2$ (the latter will
follow from $SO (4) = SU (2) \times SU (2)/Z_2$).  We begin with the
usual formula for the surface of a sphere with unit radius (given
already above for odd $N$)
\begin{equation}
{\rm Vol}\ S^N = \frac{2 \pi^{(N + 1)/2}}{{\mit\Gamma} \left( \frac{N + 1}{2} \right)} \ .
\end{equation}
In particular ${\rm Vol}\ S^1=2\pi$ and
\begin{eqnarray}
&&{\rm Vol}\ S^2 = 4 \pi \ , \quad {\rm Vol}\ S^3 = 2 \pi^2 \ , \quad {\rm Vol}\ S^4 = \ft83 \pi^2 \ , \nonumber\\ &&{\rm Vol}\ S^5 = \pi^3, \quad {\rm Vol}\ S^6 = \ft{16}{15} \pi^3 \ , \quad {\rm Vol}\ S^7 = \ft13 \pi^4 \ .
\end{eqnarray}
Furthermore ${\rm Vol}\ SO (2) = 2 \pi$ since the $SO (2)$ generator with $\tr\, T^2=-2$ is $T =
\left( {0\, \ 1\atop - 1\, 0} \right)$ and $\exp (\theta T)$ is an ordinary rotation $\left( {\cos \theta\, \ \sin \theta \atop - \sin \theta \, \cos \theta} \right)$ for which $0 \leq
\theta \leq 2 \pi$. The vielbein is unity for an abelian group, and thus the Haar measure
is\footnote{ One clearly must specify the normalization
of the generators $T_a$; for example by choosing $T_a =
\left( {0\, \ \ \ \ft12 \atop - \ft12\, \ \ 0} \right)$, the range of $\theta$ becomes $0\leq \theta \leq 4\pi$, but
the Haar measure is still ${\rm d}\theta$.} simply ${\rm d}\theta$.

With ${\rm Vol}\, SO (N) = {\rm Vol} \, S^{N - 1} {\rm Vol} \, SO(N - 1)$ we obtain ${\rm Vol}\,SO(1)=1$ and
\begin{eqnarray} &&{\rm Vol}\ SO (2) = 2 \pi \ , \quad {\rm Vol}\ SO (3) = 8 \pi^2 \ , \quad {\rm Vol}\ SO (4) = 16 \pi^4 \ , \nonumber\\ &&\qquad\qquad{\rm Vol}\ SO (5) = \ft{128}{3} \pi^6 \ , \quad {\rm Vol}\ SO (6) = \ft{128}{3} \pi^9 \ .
\end{eqnarray}

 Now consider $SU (2)$. In the normalization $T_1 = - i \tau_1$, $T_2 = - i \tau_2$ and $T_3 = - i \tau_3$ (so that ${\rm tr}\ T_a T_b = - 2 \delta_{ab}$) we find by direct evaluation\footnote{ Parametrize $g={\rm e}^{\alpha T_3}{\rm e}^{\beta T_1}{\rm e}^{\gamma T_3}$, determine the range of $\alpha, \beta, \gamma$ and compute the group vielbeins.} using Euler angles ${\rm Vol}\ SU (2) = 2 \pi^2$.
This also agrees with (\ref{SUNtoN1}) and (\ref{VolSUN}) for $N = 2$, justifying our assumption that
${\rm Vol}\ SU (1) = 1$. For higher $N$ we get
\begin{equation} {\rm Vol}\ SU (2) = 2 \pi^2 \ ,
\qquad {\rm Vol}\ SU (3) = \sqrt{3} \pi^5 \ , \qquad {\rm Vol}\ SU (4) = \ft{\sqrt{2}}{3} \pi^9 \  .
\end{equation}

The group elements of $SU (2)$ can also be written as $g = x^4  + i \vec\tau \cdot \vec x$
with $\left( x^4 \right)^2 + \left( \vec x \right)^2 = 1$ which defines a sphere $S^3$. Since near the unit
element $g \approx 1 +i \vec\tau \cdot \delta\vec x$, the normalization of the generators is as before,
and hence for this parametrization ${\rm Vol}\ SU (2) = 2 \pi^2$. This is indeed equal to ${\rm Vol}\ S^3$.
In the mathematical literature one finds the statement that ${\rm Vol }\,SU(2)$ is twice ${\rm Vol}\,SO(3)$
because $SU(2)$ is the double covering group of $SO(3)$. However, we have just found that ${\rm Vol}\ SU(2)=\frac{1}{4}{\rm Vol}\ SO(3)$.
The reason is that in order to compare properties of different groups we should normalize the generators
such that the structure constants are the same (the Lie algebras are the same, although the group volumes
are not). In other words, we should use the normalization that the
{\bf adjoint} representations have the same
${\rm tr} \ T_a T_b$.  For $SU (2)$ the generators which lead to the same commutators as the usual $SO (3)$
rotation generators (with entries $+1$ and $-1$) are $T_a = \left\{ -\ft{i}2 \tau_1, -\ft{i}2 \tau_2,
-\ft{i}2 \tau_3 \right\}$. Then ${\rm tr}\ T_a T_b = - \ft12 \delta_{ab}$. In this normalization, the
range of each group coordinate is multiplied by 2, leading to ${\rm Vol}\ SU (2) = 2^3 \cdot 2 \pi^2 = 16
\pi^2$. Now indeed ${\rm Vol}\ SU (2) = 2{\rm Vol}\ SO (3)$.

For $SU (4)$ the generators with the same Lie
algebra as $SO (6)$ are the 15 antihermitean $4\times 4$ matrices
 $\ft14 ( \gamma_m \gamma_n$ $- \gamma_n \gamma_m )$, $i \gamma_m/2$, $\gamma_m\gamma_5/2$
and $i \gamma_5/2$, where $\gamma_m$ and $\gamma_5$ are the five $4 \times 4$ matrices $\gamma_M$
obeying the Clifford
algebra $\{ \gamma_M, \gamma_N \} = 2 \delta_{MN}$~\footnote{As Dirac
matrices in six dimensions we take $\gamma_m\otimes \tau_2, \gamma_5\otimes
\tau_2$ and $I\times \tau_3$.}. Now, ${\rm tr}\ T_a T_b = - \delta_{ab}$ (for example,
$\tr \left\{ \left( \ft12 \gamma_1 \gamma_2 \right) \left( \ft12 \gamma_1 \gamma_2 \right) \right\} = - 1$).
 Recall that originally we had chosen the normalization ${\rm tr}\ T_a T_b = - 2 \delta_{ab}$. We must thus
multiply the range of each coordinate by a factor $\sqrt{2}$, and hence we must multiply our original
result for ${\rm Vol}\ SU (4)$ by a factor $\left( \sqrt{2} \right)^{15}$. We find then indeed that
the relation ${\rm Vol}\ SU (4) = 2\, {\rm Vol}\ SO (6)$ is fulfilled.

Finally, we consider the relation $SO (4) = SU (2) \times SU (2)/ Z_2$. (The vector representation of $SO (4)$ corresponds to the representation $\left( \ft12,
\ft12 \right)$ of $SU (2) \times SU (2)$, but representations like $\left( \ft12, 0 \right)$ and
$\left( 0, \ft12 \right)$ are not representations of $SO (4)$ and hence we must divide by $Z_2$. The
reasoning is the same as for $SU (2)$ and $SO (3)$, or $SU (4)$ and $SO (6)$.) We choose the generators of
$SO (4)$ as follows \begin{equation} T_1^{(+)} = \frac{1}{\sqrt{2}} \left( L_{14} + L_{23} \right) \ ,
\quad T_2^{(+)} = \frac{1}{\sqrt{2}} \left( L_{31} + L_{24} \right) \ , \quad T_3^{(+)} = \frac{1}
{\sqrt{2}} \left( L_{12} + L_{34} \right) \ , \end{equation} and the same but with minus sign denoted by
$T_i^{(-)}$. Here $L_{mn}$ equals $+1$ in the $m^{\rm th}$ column and $n^{\rm th}$ row, and is
antisymmetric. Clearly $\tr\ T_a T_b = - 2 \delta_{ab}$. The structure constants follow from
\begin{equation}
\left[ \frac{1}{\sqrt{2}} \left( L_{12} + L_{34} \right), \frac{1}{\sqrt{2}}
\left( L_{14} + L_{23} \right), \right] = - \left( L_{31} + L_{24} \right) \ ,
\end{equation}
thus
\begin{equation}
\label{CRSO4} \left[ T_i^{(+)}, T_j^{(+)} \right] = - \sqrt{2} \epsilon_{ijk} T_k^{(+)}
\ , \quad \left[ T_i^{(-)}, T_j^{(-)} \right] = - \sqrt{2} \epsilon_{ijk} T_k^{(-)} \ , \quad
\left[ T_i^{(+)}, T_j^{(-)} \right] = 0 \ .
\end{equation}
We choose for the generators of $SU (2)
\times SU (2)$ the representation \begin{equation} T^{(+)}_i = \frac{i \tau_i}{\sqrt{2}} \otimes \1 \ ,
\qquad T^{(-)}_i = \1 \otimes \frac{i \tau_i}{\sqrt{2}} \ . \end{equation} Then we get the same commutation
relations as for $SO (4)$ generators (\ref{CRSO4}); however, the generators are normalized differently,
namely ${\rm tr}\ T_a T_b = - 2 \delta_{ab}$ for $SO (4)$ but ${\rm tr}\ T_a T_b = - \delta_{ab}$ for
$SU (2)$. With the normalization $\tr\ T_a T_b = - 2 \delta_{ab}$ we found ${\rm Vol}\ SU (2) = 2 \pi^2$.
In the \ present normalization we find ${\rm Vol}\ SU (2) = 2 \pi^2 \left( \sqrt{2} \right)^3$. The
relation ${\rm Vol}\ SO (4) = \ft12 \left( {\rm Vol} \ SU (2) \right)^2$ is now indeed satisfied
\begin{equation} {\rm Vol}\ SO (4) = 16 \pi^4 = \ft12 \left( {\rm Vol} \ SU (2) \right)^2 =
\frac{1}{2}\left( 2 \pi^2 \left( \sqrt{2} \right)^3 \right)^2 \ . \end{equation}

\section{Zero modes and conformal symmetries} \label{nofurther} 

The bosonic collective coordinates obtained for gauge group $SU(2)$ and the one-instanton solution could all be identified with rigid symmetries of the action:  $a_\mu$ with translations, $\rho$ with scale transformations and $\theta^a$ with rigid gauge symmetries.   Similarly, the fermionic collective coordinates for $SU(2) ( \xi^\alpha$ and $\bar\eta_{\dot\alpha}$ with $\alpha , \dot\alpha = 1,2)$ could be identified with ordinary supersymmetry and conformal supersymmetry.   However, the full conformal algebra in $4$ Euclidean dimensions is $SO(5,1)$, and its generators are $P_\mu , K_\mu , D, M_{\mu\nu}$, so one might expect that the conformal boost transformations $K_\mu$ and the Lorentz rotations $M_{\mu\nu}$ produce further collective coordinates.  As we now show, the transformations due to these symmetries can be undone by suitably chosen gauge transformations with constant gauge parameters \cite{JackReb}. So there are no further bosonic collective coordinates, as we already know from the index theorem discussed in the main text.

Consider first rigid Lorentz transformations.  Here one should not forget that in addition to a spin part which acts on the indices of a field they also contain an orbital part that acts on the coordinates:  $M_{\mu\nu} = \Sigma_{\mu\nu} + L_{\mu\nu}$.   For example, for a spinor one has $\delta (\lambda_{mn} ) \psi = \ft14 \lambda_{mn} \gamma_{mn} \psi + (\lambda_{mn} x_m \partial_n ) \psi$.    One may check that only with this orbital part present the Dirac action is Lorentz invariant.  In fact, starting with only the spin part or the orbital part, one can find the other part by requiring invariance of the action.  We begin by considering the field strength $F_{\mu\nu} = 2 \bar\sigma_{\mu\nu} \rho^2 / (x^2 + \rho^2)^2$ for an instanton with $k=1$ in the regular gauge.  Under a Lorentz transformation with parameter $\lambda_{\mu\nu} =- \lambda_{\nu\mu}$ one has $\delta_M A_\mu = \lambda_{\mu\nu} A_\nu + \lambda_{mn} x_m \partial_n A_\nu$.  (Note that coordinates transform opposite to fields: $\delta x^m =- \lambda_{mn}
x^n$. One may check this transformation rule by showing that the Maxwell
action is Lorentz invariant (use the Bianchi identities\footnote{ One has $\d_M \frac14 F^2_{\mu\nu} = F_{\mu\nu} \del_\mu (\l_{\nu\rho} A_\rho ) + F_{\mu\nu} \del_\mu (\l_{mn} x_m \del_n A_\nu) = F_{\mu\nu} \l_{mn} x_m \del_\mu \del_n A_\nu$.  Replacing $\del_\mu \del_n A_\nu$ by $- \del_n \del_\nu A_\mu - \del_\nu \del_\mu A_n$ yields $\del_\mu (\xi^\mu \cl)$.}), or just by writing
down the transformation law for a covariant vector in general relativity.
The Lagrangian transforms into $\partial_\mu(\xi^\mu {\cal L})$, where
$\xi^\mu=\lambda^{\rho\mu}x_\rho$.).
The field strength of the instanton transforms as follows
\begin{eqnarray}
\delta_M F_{\mu\nu} = \lambda_{\mu\rho} F_{\rho\nu} + \lambda_{\nu\rho} F_{\mu\rho}
\end{eqnarray}
There is no contribution from the orbital part because $x^2$ is
Lorentz invariant.  On the other hand, under a gauge
transformation with parameter $\Lambda_{\rho\sigma}$ we
obtain\footnote{ The usual form of an $SU(2)$ gauge transformation is
$\delta F_{\mu\nu}  = [ F_{\mu\nu}, \Lambda^a (x) {\tau^a \over
2i} ]$, but using $\eta_{a \mu\nu} \eta_{b \mu\nu} = 4
\delta_{ab}$ and $\bar\sigma_{\rho \sigma} = i \eta_{a \rho\sigma}
\tau^a$, this can be rewritten as $\delta F_{\mu\nu} = [
F_{\mu\nu} , \ft14  \Lambda^{\rho \sigma} \bar\sigma_{\rho\sigma}
]$ where $\Lambda_{\rho\sigma} = - \ft12 \eta_{a \rho\sigma}
\Lambda^a$.}
\begin{eqnarray}
\delta_{\rm gauge} F_{\mu\nu} = [ \bar\sigma_{\mu\nu} , \ft14 \Lambda_{\rho\sigma} \bar\sigma_{\rho\sigma} ]  (2 \rho^2 / (x^2 + \rho^2 )^2 )= \Lambda_{\nu\sigma} F_{\mu\sigma} - \Lambda_{\mu\sigma} F_{\nu\sigma}
\label{parameter}
\end{eqnarray}
Thus $F_{\mu\nu}$ is invariant under combined Lorentz and gauge transformations with opposite parameters, $\Lambda_{\rho\sigma} = - \lambda_{\rho\sigma}$.  Using $\bar\sigma_{\rho\sigma} = i \eta_{a \rho\sigma} \tau_a$, it is clear that the $SU(2)$ gauge parameter $\Lambda_a$ is proportional to $\eta_{a\rho\sigma} \lambda_{\rho\sigma}$.  Only the selfdual part of $\lambda_{\rho\sigma}$ contributes.  For an anti-instanton we would have needed the anti-selfdual part of $\lambda_{\rho\sigma}$.  So we have only proven that $F_{\mu\nu}$ is invariant under combined Lorentz and gauge transformation if the Lorentz parameter is self dual.   However, the anti-self dual part of $\lambda_{\rho\sigma}$ leaves $F_{\mu\nu}$ separately invariant, without the need to add compensating gauge transformations.  One can prove this directly, using that $\lambda_{\mu\rho} F_{\rho\nu} =- (\ast \lambda_{\mu\rho}) (\ast F_{\rho\nu})$ and then working out the product of two $\epsilon$-tensors and finally antisymmetrizing in $\mu\nu$, but it is already clear from the index structure:  $F_{\mu\nu}$ is proportional to $(\bar\sigma_{\mu\nu})_{\alpha'}{}^{\beta'}$ while an anti-selfdual $\lambda_{\rho\sigma}$ has in spinor notation only undotted indices.

Let us now repeat this exercise for the gauge field $A_\mu$.  One finds for the combined Lorentz and gauge transformation
\begin{eqnarray}
\delta A_\mu = \lambda_{\mu\nu} A_\nu + \lambda_{\rho\sigma} x_\rho \partial_\sigma A_\mu + [ A_\mu , \ft14 \Lambda_{\rho\sigma} \bar\sigma_{\rho\sigma} ]
\end{eqnarray}
The instanton field $A_\mu$ for $k=1$ in the regular gauge is given by $A_\mu = (- \bar\sigma_{\mu\nu} x^\nu) / (x^2 + \rho^2)$.  The orbital part with $\lambda_{\rho\sigma}$ now contributes, but there is no term $\partial_\mu \Lambda^a$ in the gauge transformation of $A_\mu$ since $\Lambda^a$ is constant.  One obtains
\begin{eqnarray}
\delta A_\mu = \lambda_{\mu\nu} A_\nu + {\bar\sigma_{\mu\nu} (\lambda_{\nu\rho} x_\rho ) \over x^2 + \rho^2} - {(\Lambda_{\nu\sigma} \bar\sigma_{\mu\sigma} - \Lambda_{\mu\sigma} \bar\sigma_{\nu\sigma} ) x^\nu \over x^2 + \rho^2}
\end{eqnarray}
For $\Lambda_{\mu\nu} =- \lambda_{\mu\nu}$ all terms again cancel.  Hence, Lorentz symmetry does not yield further zero modes.

In spinor notation these results are almost obvious.  In general the selfdual
part of a curvature reads in spinor notation
\begin{eqnarray}
(F_{\mu\nu})^u{}_v (\bar\sigma_{\mu\nu})_{\alpha'}{}^{\beta'}
\end{eqnarray}
where $u,v$ are the indices of  $(\tau^a)^u{}_v$, and $\alpha' , \beta'$ are the spinor indices.  If we raise/lower indices by $\epsilon$ tensors, we get for the instanton solution
\begin{eqnarray}
(F_{\mu\nu})^{uv} (\bar\sigma_{\mu\nu})_{\alpha' \beta'} \equiv F^{uv}_{\alpha'\beta'} \sim \delta^u_{\alpha'} \delta^v_{\beta'} + \delta^v_{\alpha'} \delta^u_{\beta'}
\end{eqnarray}
It is then clear that $F_{\mu\nu}$ is invariant under diagonal transformations of $SU(2)_R$ and $SU(2)_{\rm gauge}$, and separately invariant under $SU(2)_L$.  For an anti-instanton, the roles of $SU(2)_L$ and $SU(2)_R$ are interchanged.

We come now to the more complicated problem of conformal
transformations.  A conformal transformation of a field $\varphi$
with constant parameter $a^m$ is given by\footnote{ This formula
follows from $\delta (a^m K_m) \varphi (x) = [ \varphi (x), a^m
K_m]$, and $\varphi (x) = e^{-P \cdot x} \varphi (0) e^{P \cdot
x}$ with $[ \varphi (0) , P_\mu ] = \partial_\mu \varphi (0)$.
One may then use $e^{P \cdot x} K_m = (e^{P \cdot x} K_m e^{-P
\cdot x} ) e^{P \cdot x}$ and $[K_m , P_n ] = - 2 \delta_{mn} D-2
M_{mn} ; [P_m , D] = P_m ; [P_m , M_{rs} ] = \delta_{mr} P_s -
\delta_{ms} P_r$ and this yields \eqn{wayy}.  In the same way one
may derive the Lorentz transformation rule for a spinor $\psi
(x)$, with both spin and orbital parts, by using that the spin
part is given by $[ \psi (0) , \ft12 \lambda_{mn} M_{mn}] = \ft14
\lambda_{mn} \gamma_{mn} \psi (0)$.  One finds then the correct
result:  $\delta ( \ft12 \lambda_{mn} M_{mn}) \psi (x) = \ft14
\lambda_{mn} \gamma_{mn} \psi (x) + \lambda_{mn} x_m \partial_n
\psi (x)$.  Given the spin part of the transformation rule of the
field at the origin, one derives in this way the orbital part.
In this way one finds that the generators of the conformal algebra
act as follows on the coordinates: $\delta (P_m) x^n =
\delta_m{}^n , \delta (D) x^n = x^n , \delta (M_{st}) x^m = x_s
\delta_t{}^m - x_t \delta_s{}^m$ and $\delta (K_m) x^n = 2 x_m x^n
- x^2 \delta_m{}^n$.  Note that coordinates transform
contragradiently to fields.  For example, whereas $[\delta (K_m),
\delta (P_n)] \varphi =- \delta ([ K_m , P_n ]) \varphi$ (by
definition), one finds $[ \delta (K_m) , \delta (P_n) ] x^s =
\delta ([ K_m , P_n ]) x^s$.}
\begin{eqnarray}
\delta (a^m K_m) \varphi &=& (2a \cdot x \; x^m - a^m x^2 ) \partial_m \varphi + \delta (2a \cdot x D^{\rm (spin)})  \varphi \nonumber\\
&+& \delta (2 a_m x_n M^{\rm (spin)}_{mn} ) \varphi
\label{wayy}
\end{eqnarray}
where $D^{\rm spin}$ and $M^{\rm spin}_{mn}$ act only on $\varphi (0)$ and $\delta (a^m K_m) \varphi (x)$ is by definition $[ \varphi (x), a^m K_m]$.  As the notation indicates, only the spin parts of the dilatational generator $D$ and the Lorentz generators contribute.  For example
\begin{eqnarray}
\delta (D^{\rm spin}) A_\mu = [A_\mu ,  D^{\rm (spin)} ]  = A_\mu \ , \qquad  \delta ( \ft12 \lambda_{mn} M^{\rm (spin)}_{mn} ) A_\mu = \lambda_{\mu\nu} A_\nu\ .
\end{eqnarray}

Consider first $F_{\mu\nu}$.  We obtain
\begin{eqnarray}
&& \delta (a^m K_m) F_{\mu\nu} = ( 2a \cdot x x^m - a^m x^2) \partial_m F_{\mu\nu} + 4 a \cdot x F_{\mu\nu} \nonumber\\
&& \qquad + 4 \delta \left( \ft12 a_m x_n M^{\rm (spin)}_{mn} \right) F_{\mu\nu} \; \mbox{with} \; F_{\mu\nu} = {2 \bar\sigma_{\mu\nu} \rho^2 \over (x^2 + \rho^2)^2}\ .
\label{dilatational}
\end{eqnarray}
We already know that the last term can be canceled by a suitable gauge transformation (there are no contributions from $M^{\rm (orb)}_{mn}$ because $x^2$ is Lorentz invariant).  The first term gives $-4 {a \cdot x x^2 \over x^2 + \rho^2} F_{\mu\nu}$.  The first and second term together produce then ${4 a \cdot x \rho^2 \over x^2 + \rho^2} F_{\mu\nu}$.  But this is the opposite of a translation with parameter $a^m \rho^2$, namely
\begin{eqnarray}
\delta (a^m \rho^2 P_m) F_{\mu\nu} = {- 4 a \cdot x \rho^2 \over x^2 + \rho^2} F_{\mu\nu} ; \delta (P_m ) \varphi =  \partial_m \varphi\ .
\end{eqnarray}
Thus the following combination of symmetry transformations leaves $F_{\mu\nu}$ {\bf in}variant
\begin{eqnarray}
a^m K_m + \rho^2 a^m P_m + \delta_{\rm gauge} (\Lambda_{mn} = - 2a_m x_n + 2x_m a_n)
\end{eqnarray}

Let us now check that also $A_\mu$ itself is invariant under this combination of symmetries.  We find by direct evaluation, using $A_\mu = (- \bar\sigma_{\mu\nu} x^\nu) / (x^2 + \rho^2)$ and \eqn{dilatational} and \eqn{parameter}
\begin{eqnarray}
\delta A_\mu &=& \left( {-2 a \cdot x x^2 \over x^2 + \rho^2} A_\mu - \bar\sigma_{\mu\nu} {( 2 a \cdot x x_\nu - a_\nu x^2) \over x^2 + \rho^2} \right) + 2 a \cdot x A_\mu \nonumber\\
&+& ( 2 a_\mu x_\nu A_\nu - 2 x_\mu a_\nu A_\nu ) + \left( - {\rho^2 2 a \cdot x \over x^2 + \rho^2} A_\mu - {\bar\sigma_{\mu\nu} a_\nu \rho^2 \over x^2 + \rho^2} \right) \nonumber\\
&+& \partial_\mu (- a_\rho x_\sigma \bar\sigma_{\rho\sigma} ) + [ A_\mu , -a_\rho x_\sigma \bar\sigma_{\rho\sigma} ]\ .
\end{eqnarray}
As in the case of $F_{\mu\nu}$, the sum of the first, third and sixth term cancels.  This takes care of the dilatation term and the denominator of $A_\mu$.  We are left with terms from the numerator, and Lorentz and gauge terms
\begin{eqnarray}
&& (2 a \cdot x A_\mu + ( \bar\sigma_{\mu\rho} a_\rho) \left( {x^2 \over x^2 + \rho^2} \right) + (0 -2 x_\mu a_\nu A_\nu) +\nonumber\\
&& \left( { - \bar\sigma_{\mu\nu} a_\nu \rho^2 \over x^2 + \rho^2} \right) + ( \bar\sigma_{\mu\rho} a_\rho ) + {2 x^\nu \over x^2 + \rho^2} \left( \begin{array}{ll}  a_\nu x_\sigma \bar\sigma_{\mu\sigma} - a_\rho x_\nu \bar\sigma_{\mu\rho} \\ + 0 + a_\rho x_\mu \bar\sigma_{\nu\rho} \end{array} \right)
\end{eqnarray}
The terms denoted by ``0" vanish due to $x_\nu A_\nu =0$.  All other terms cancel in the following combinations\\
(i) the second, fourth, fifth, and seventh nonvanishing contributions sum up to zero.  These are the terms with $\bar\sigma_{\mu\rho} a_\rho$.  \\
(ii) the first and third-but-last nonvanishing term cancel each other.  Here conformal boosts cancel a gauge term.\\
(iii) the remaining Lorentz term $-2 x_\mu a_\nu A_\nu$ cancels the remaining gauge term $2x^\nu (x^2 + \rho^2)^{-1} (a_\rho x_\mu \bar\sigma_{\nu\rho})$.

Hence, conformal boosts do not lead to further zero modes either.

\end{document}